\documentclass[fleqn,10pt]{wlscirep}
\usepackage[utf8]{inputenc}
\usepackage[T1]{fontenc}
\usepackage{graphicx}
\usepackage{dcolumn}
\usepackage{subfig}
\usepackage{hyperref}
\usepackage{siunitx}
\usepackage[final]{changes}
\usepackage{blindtext}
\usepackage[capitalise]{cleveref}

\newcommand{\ub}{{\bf u}}

\newcommand{\Xb}{{\bf X}}

\newcommand{\xb}{{\bf x}}

\newcommand{\fb}{{\bf f}}
\newcommand{\Fb}{{\bf F}}
\newcommand{\partialb}{{\boldsymbol \partial}}

\newcommand{\Rey}{\mathit{Re}}
\newcommand{\Ca}{\mathit{Ca}}

\graphicspath{{figs/}{./}}

\title{Collective dynamics of dense hairy surfaces in turbulent flow}

\author[1,+]{Alessandro Monti}
\author[1,2,+]{Stefano Olivieri}
\author[1,*]{Marco E. Rosti}

\affil[1]{Complex Fluids and Flows Unit, Okinawa Institute of Science and Technology Graduate University, 1919-1 Tancha, Onna-son, Okinawa 904-0495, Japan}
\affil[2]{Department of Aerospace Engineering, Universidad Carlos III de Madrid, Avda. de la Universidad, 30, 28911 Leganés, Spain
}
\affil[*]{corresponding author: marco.rosti@oist.jp}
\affil[+]{these authors contributed equally to this work}

\begin{abstract}
\added{Flexible filamentous beds interacting with a turbulent flow represent a fundamental setting for many environmental phenomena, e.g., aquatic canopies in marine current.} Exploiting direct numerical simulations at high Reynolds number where the canopy stems are modelled individually, we provide evidence on the essential features of the \textit{honami/monami} collective motion experienced by hairy surfaces over a range of different flexibilities, i.e., Cauchy number. Our findings clearly confirm that the collective motion is essentially driven by fluid flow turbulence, with the canopy having in this respect a fully-passive behavior. Instead, some features pertaining to the structural response turn\deleted{s} out to manifest in the motion of the individual canopy elements when focusing, in particular, on the spanwise oscillation and/or on sufficiently small Cauchy numbers.
\end{abstract}
\begin{document}

\flushbottom
\maketitle
%
%
\thispagestyle{empty}


\section*{Introduction} 

The coherent waving motion of seagrass meadows in marine currents or plant crops under the action of wind is a fascinating example of the complex interaction between fluid flows and a multitude of deformable structures, namely hairy surfaces or filamentous beds. 
\added{More generally, the emergence of large-scale collective motion (e.g., the so-called \textit{honami} or \textit{monami} phenomena in terrestrial and aquatic canopy flows~\cite{ackerman1993reduced}, respectively) represents an important reason why this kind of flow-structure interaction (FSI) keeps attracting interest from both the fundamental and applied viewpoint~\cite{abdolahpour2018impact,wong2020shear,tschisgale2021large,houseago2022turbulence,wang_he_dey_fang_2022}.}
Such intriguing features indeed manifest in a vast number of problems pertaining to various fields, ranging from aforementioned environmental processes (e.g., terrestrial or aquatic canopy flows)~\cite{finnigan2000turbulence,nepf2012flow,abdolahpour2018impact,brunet2020turbulent,houseago2022turbulence} to engineering applications (e.g., energy harvesting and drag reduction)~\cite{hobeck2012artificial,sundin2019interaction} and, in a broader sense, biological systems (e.g., ciliated organisms)~\cite{alvarado2017nonlinear,loiseau2020active,deblois2023canopy}.  
 
 \begin{figure}
    \centering
    \includegraphics[width=.48\textwidth]{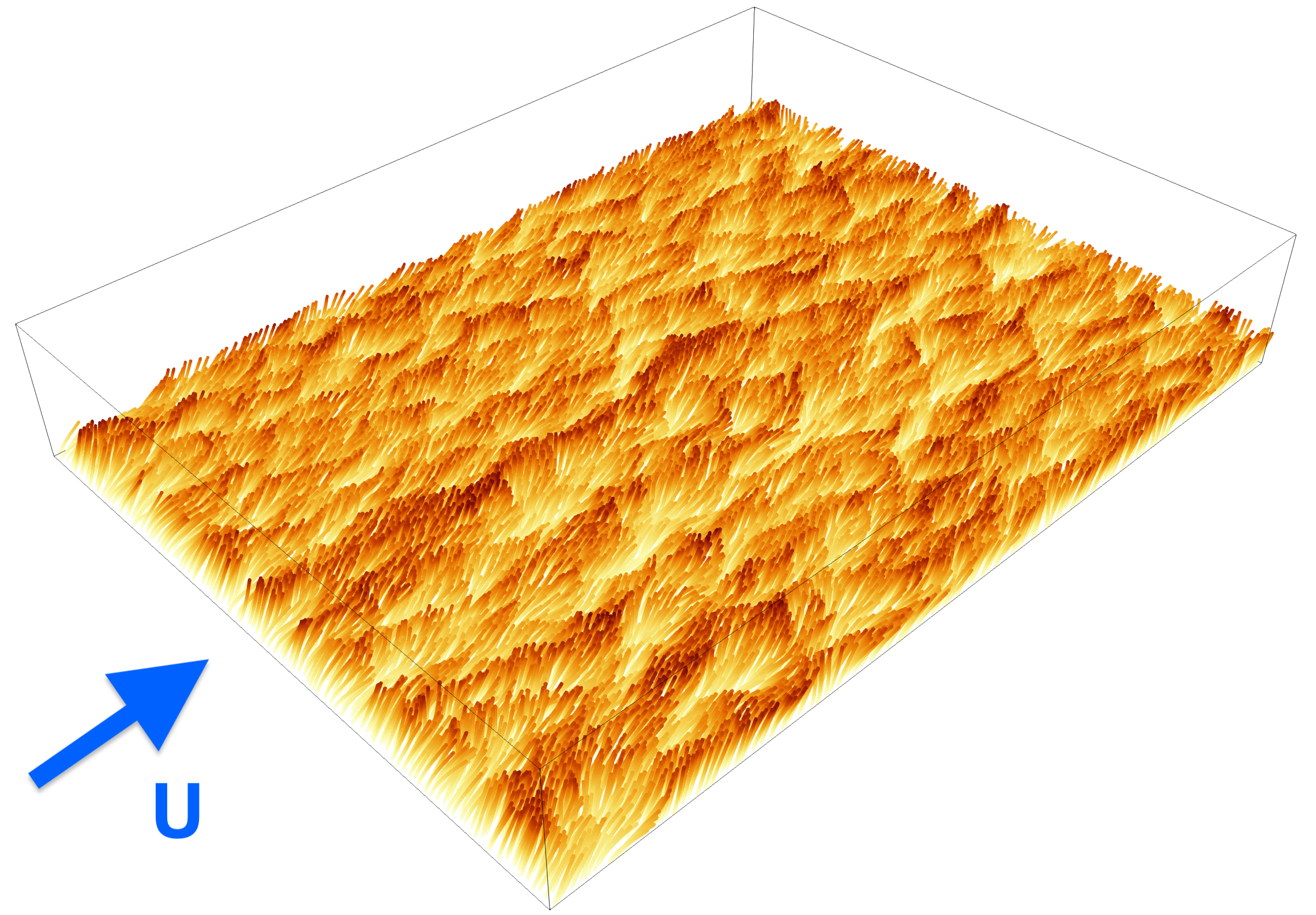}
    \includegraphics[width=.48\textwidth]{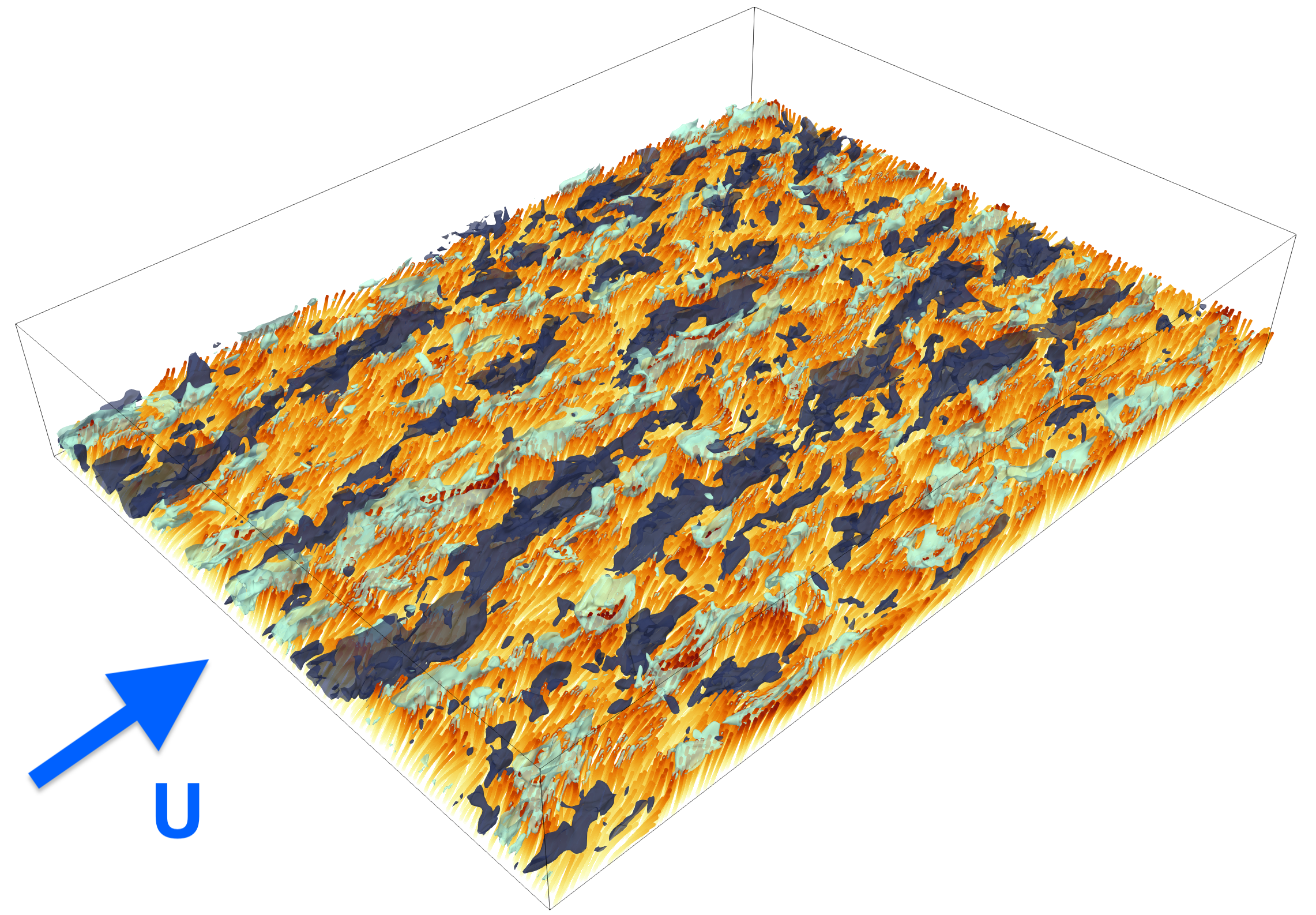}
    \caption{
    Snapshots of flexible-canopy filaments (left), colored by their vertical position, increasing from light to dark, in order to highlight the coherent waving motion, and with superimposed large-scale flow structures (right), depicted by isosurfaces of positive and negative streamwise velocity fluctuations at $u' = \pm 0.4 U$ (in dark and light blue, respectively) in the vicinity of the canopy edge. Results are obtained from the case $\Ca=50$.
    }
    \label{fig1}
\end{figure}
 
Environmental phenomena arguably represent the most classical example where collective motion can be observed and numerous studies have been devoted to understand the key features of the mutual coupling between the fluid and structural motion~\cite{py2006frequency,nepf2012flow}. \added{Along with the decreased drag due to the adaption of compliant stems to the action of the flow, the canopy flexibility is typically responsible for an attenuation of vertical mixing between the outer flow and the canopy}~\cite{ghisalberti2006structure,abdolahpour2018impact}.
Yet, several questions of paramount importance remain not fully understood.
Firstly, the mathematical description of canopy flows is still the subject of theoretical studies aiming at providing comprehensive models able to predict the onset of the honami/monami-like instability~\cite{raupach1996coherent,py2006frequency,gosselin2009destabilising,singh2016linear,zampogna2016fluid,zampogna2016instability,luminari2016drag,wong2020shear}.
On the other hand, despite their importance for understanding the incipient cause of such phenomena, these models typically rely on linear approximations, and it remains elusive to predict and measure the characteristic properties of fully-developed honami/monami in the nonlinear stage.
Furthermore, from the current literature it can be noticed conflicting evidence concerning the identification of the main features of their collective motion, such as its characteristic frequency and lengthscale~\cite{py2006frequency,tschisgale2021large,houseago2022turbulence}.
It is also important to highlight the technical challenges for experimental investigations, in particular, for measuring the spatial conformation of the fluid flow in the intra-canopy region~\cite{houseago2022turbulence}, as well as isolating the effect of the various governing parameters (e.g., Cauchy and Reynolds number, solid-to-fluid density ratio, reduced velocity).
Both issues, on the other hand, can be overcome by the support of computational studies based on direct or large-eddy numerical simulations~\cite{dupont2010modelling,dijkstra2010modeling,pan2014strong,marjoribanks2017does}, with more recent advancements based on fully-resolved approaches where the stems/elements of the canopy are resolved individually and not modelled, e.g., as a continuous porous medium~\cite{monti2019large,sundin2019interaction,monti2020genesis,tschisgale2021large,monti2022solidity,wang_he_dey_fang_2022}.

In this work, we aim at deciphering the highlighted issues by means of a high-fidelity computational approach to accurately and comprehensively access both the flow physics and canopy collective dynamics, as well as the motion of the individual stems.
Overall, we aim at establishing whether the fluid and structural dynamics act competitively or not, and which of the two has a dominant role. Moreover, we explore the connection between the individual motion of the elements (e.g., filaments or blades) and the collective coherent motion. 

\section*{Results}

\begin{figure}
    \centering
    \includegraphics[width=.2\textwidth]{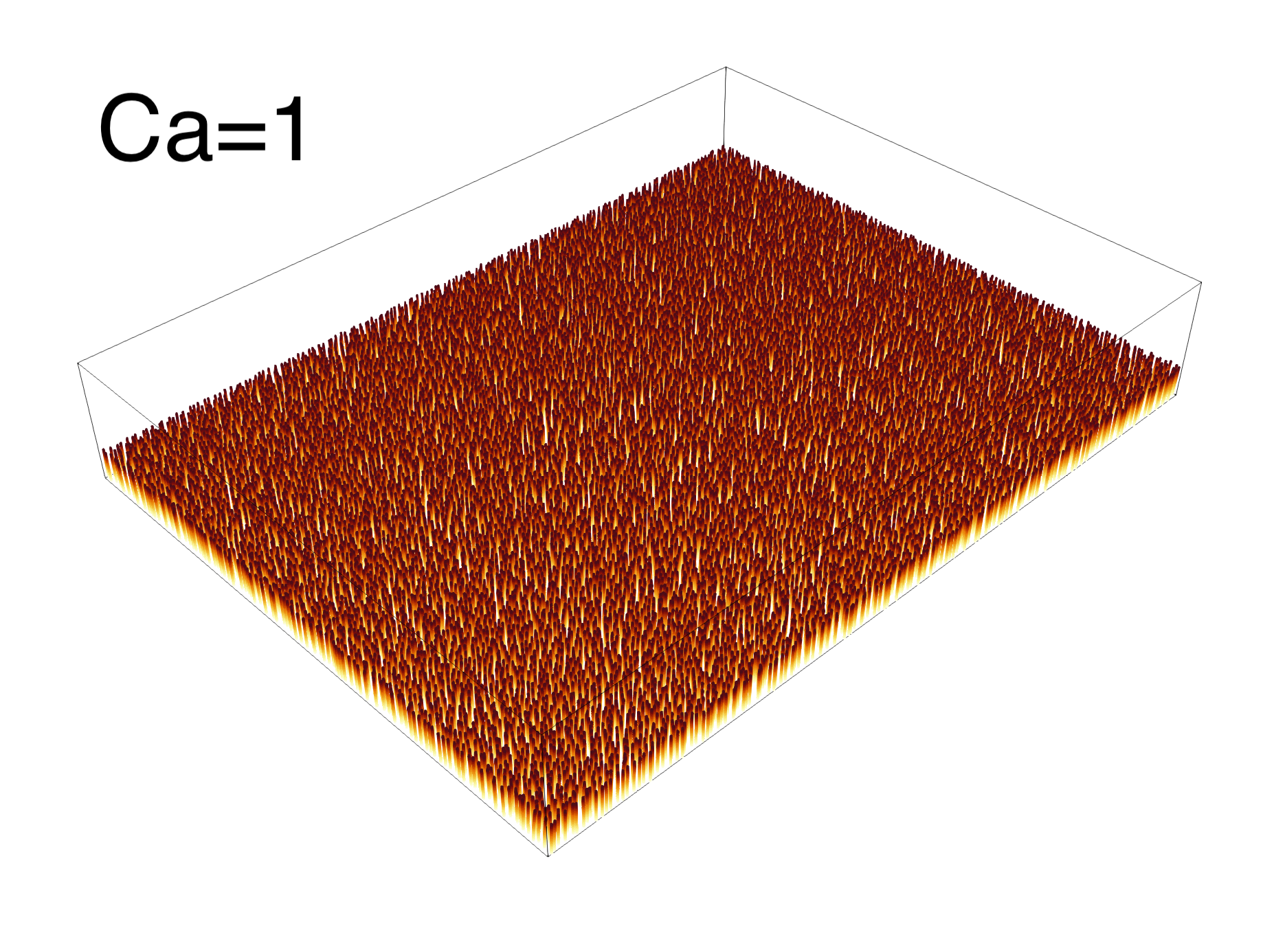}%
    \includegraphics[width=.2\textwidth]{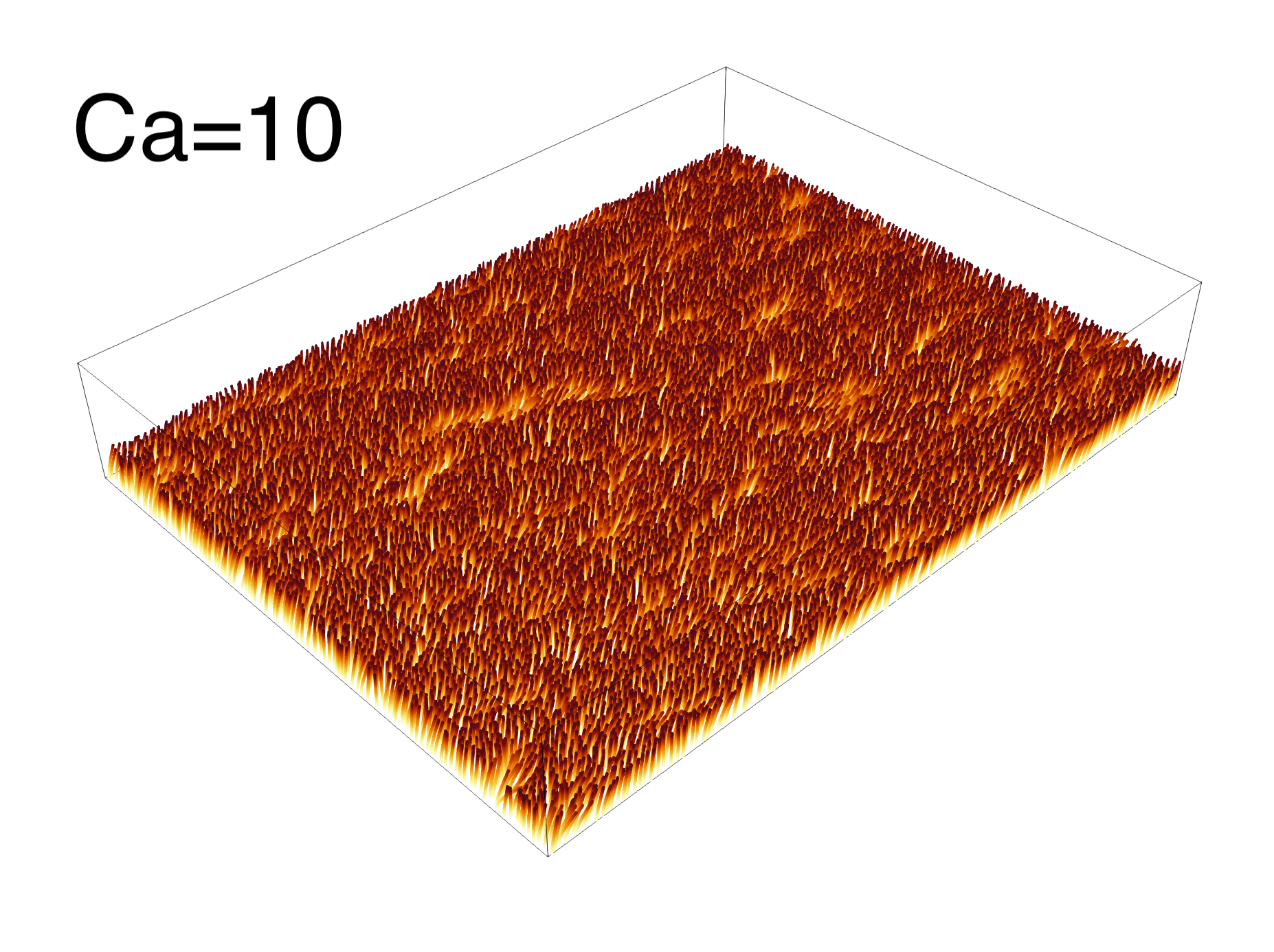}%
    \includegraphics[width=.2\textwidth]{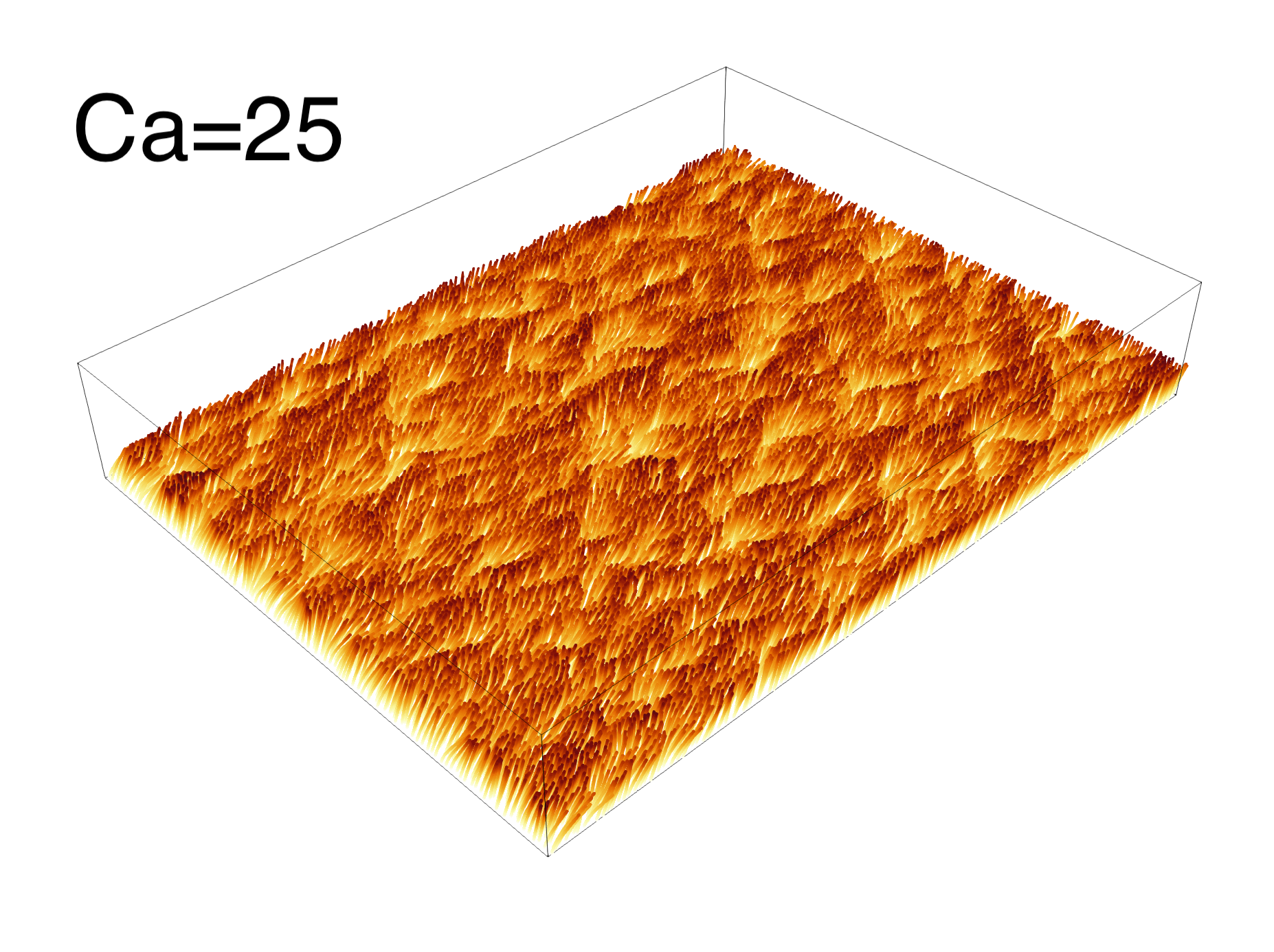}%
    \includegraphics[width=.2\textwidth]{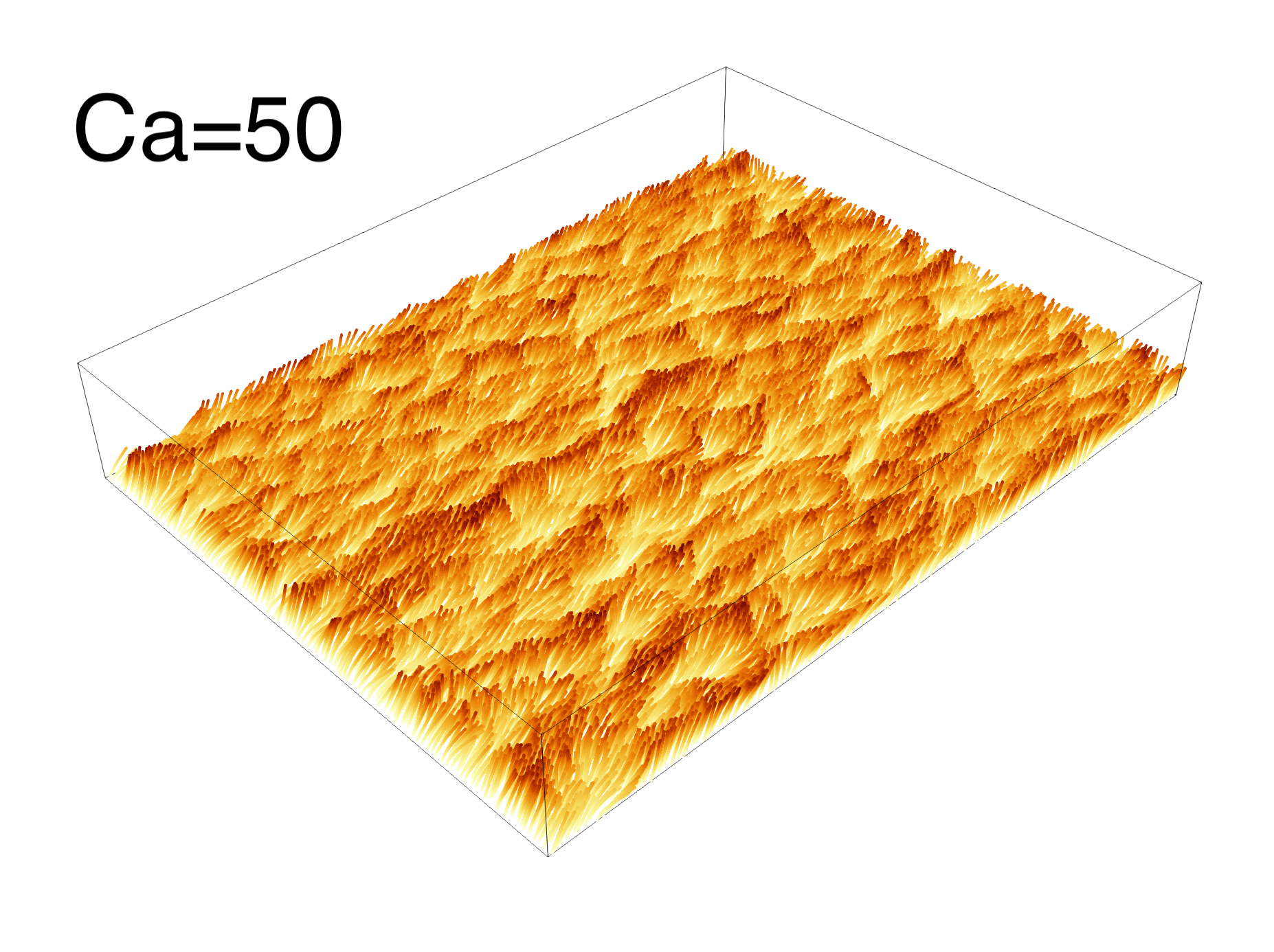}%
    \includegraphics[width=.2\textwidth]{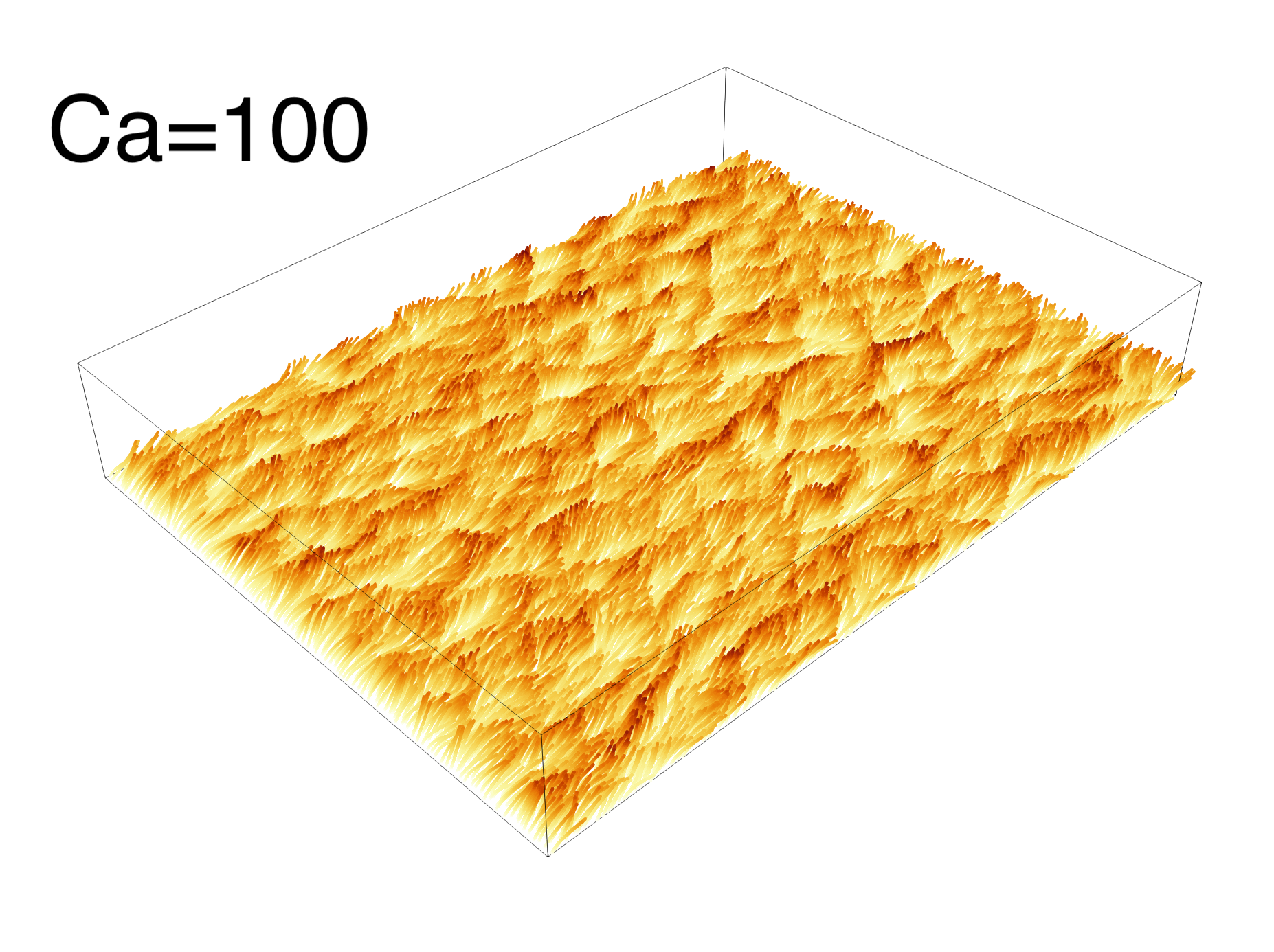}\\[12pt]
    \includegraphics[width=.2\textwidth]{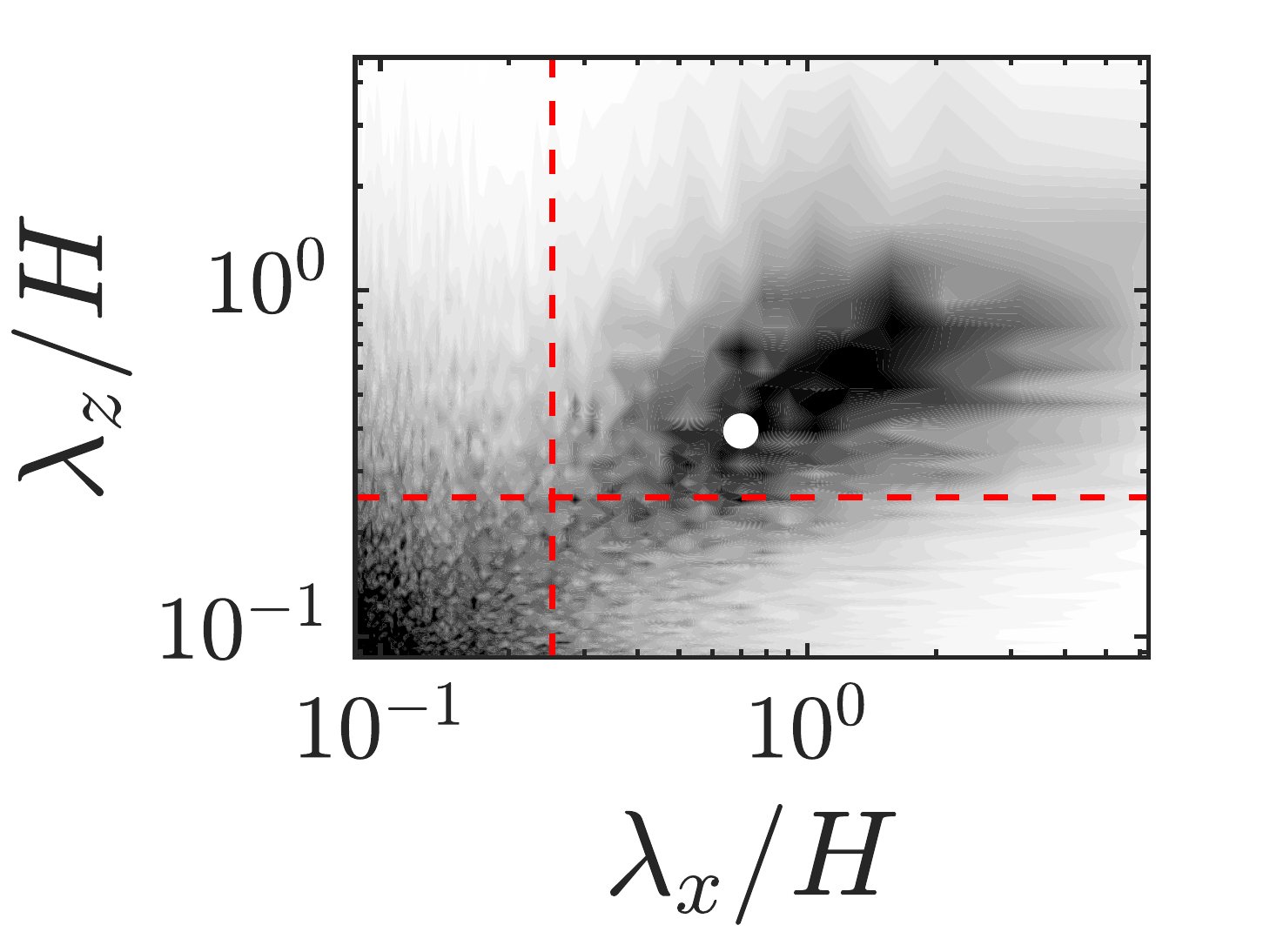}%
    \includegraphics[width=.2\textwidth]{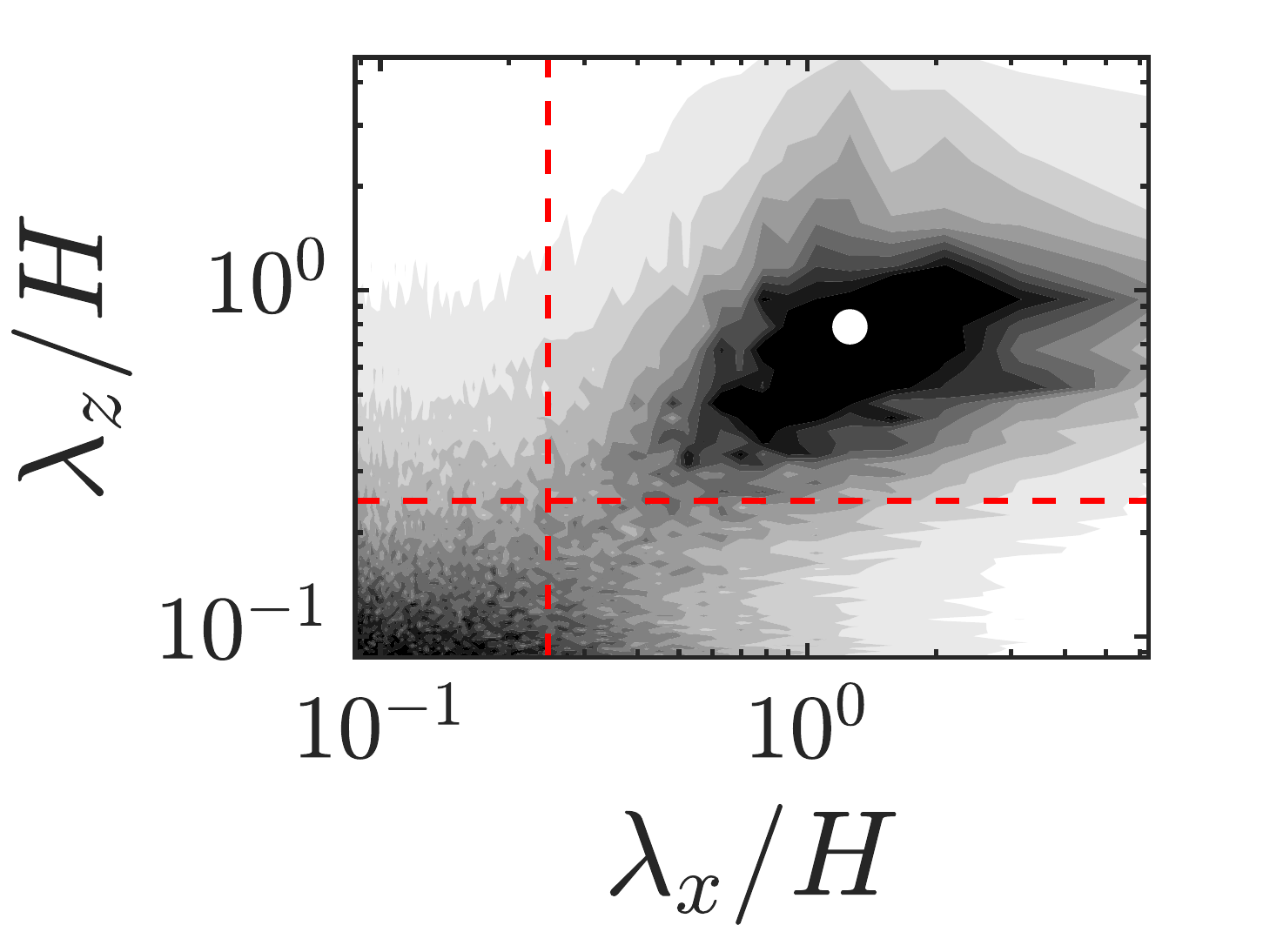}%
    \includegraphics[width=.2\textwidth]{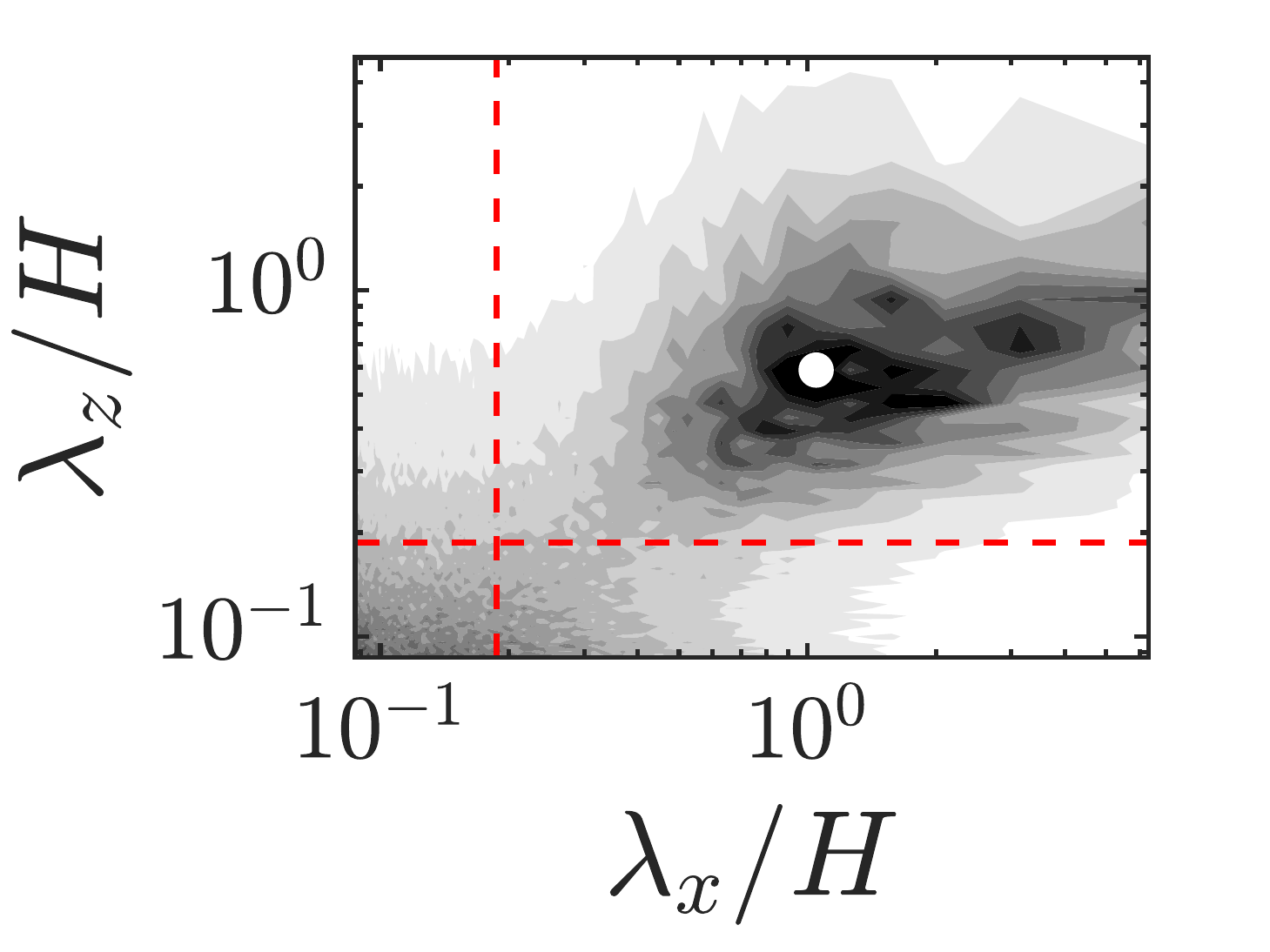}%
    \includegraphics[width=.2\textwidth]{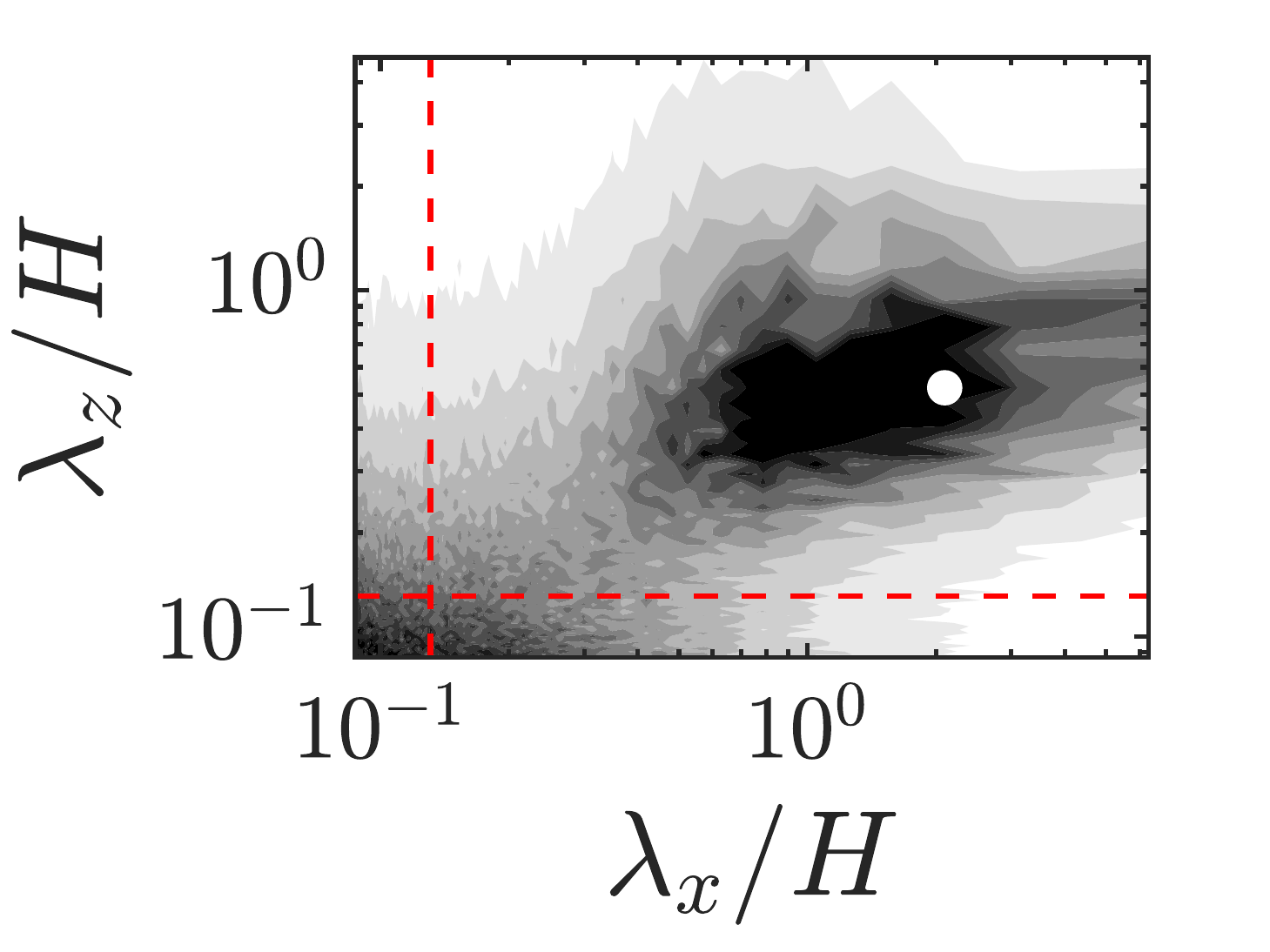}%
    \includegraphics[width=.2\textwidth]{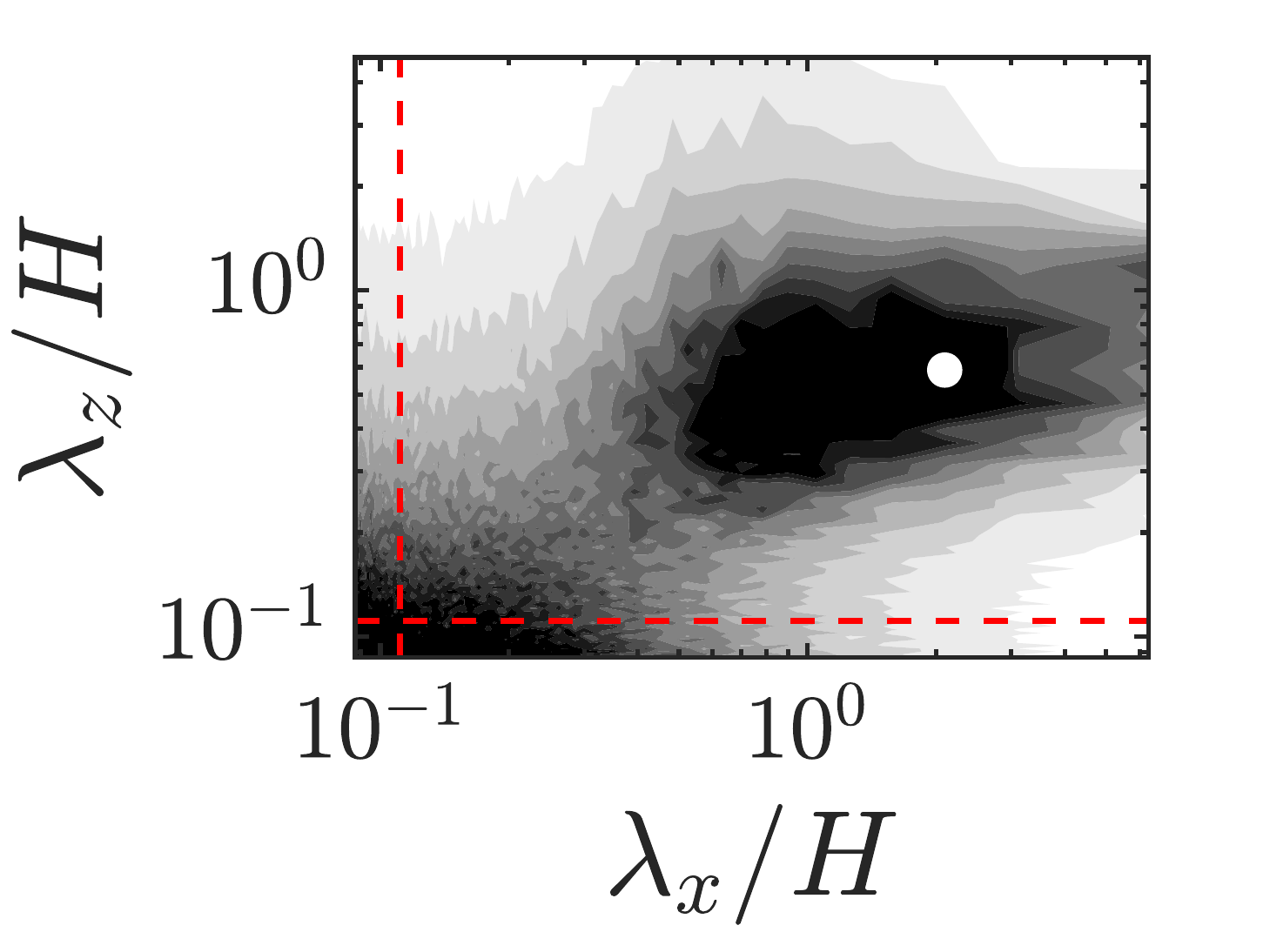}%
    \caption{
	    Top panels: snapshots of flexible-canopy filaments (colored by their vertical position, increasing from light to dark) from the performed DNS for different Cauchy number, increasing from left to right, ($\Ca=1, 10, 25, 50, 100$). Bottom panels: time average of the (premultiplied by the wavenumbers $4\pi^2/[\lambda_x\lambda_z]$) spectra of the vertical displacement of the filaments tip $Y$,
$4\pi^2\Phi_{YY}/(H^2\lambda_x\lambda_z)$,
for each corresponding case (with red lines indicating the wavelengths relative to the average canopy height \added{$\bar{Y}$} and the white marker denoting the peak location in the spectrum). 
The grey levels range in: $[0, 10^{-9}]$ with a $2.5\times 10^{-11}$ increment for $Ca=1$; $[0, 5\times 10^{-6}]$ with a $5\times 10^{-7}$ increment for $Ca=10$; $[0, 5\times 10^{-5}]$ with a $5\times 10^{-6}$ increment for the remaining $Ca$.
    }
    \label{figSpa}
\end{figure}


\added{To investigate the essential features of the FSI between a turbulent flow and a flexible canopy, we have performed high-fidelity, direct numerical simulations (DNS) where the canopy elements are modelled individually and where the two-way coupling between the fluid and the solid dynamics is realized by means of an immersed boundary method (IBM).  The considered setup is that of an open channel at constant flow rate with the flexible-canopy elements randomly distributed~\cite{monti2020genesis} and individually clamped at the bottom boundary, as shown in \cref{fig1} for a representative case.} \added{See Methods section for a complete description of the problem setup and numerical method.} 

We fix the hydrodynamic parameters to a reference configuration with bulk Reynolds number $\Rey = U H / \nu =5000$ (where $U$ is the imposed bulk flow velocity, $H$ the height of the open channel and $\nu$ the kinematic viscosity), with a submergence level $h/H$ equal to $0.25$ (where $h$ is the height of the canopy in the rigid vertical configuration) and with canopy solidity $\lambda = h \, d / \Delta S^2 \approx 1.43$ (with $d$ being the stem diameter and $\Delta S$ the average spacing between adjacent stems). \added{The latter choice corresponds (in the rigid case) to a configuration well within the dense regime, thus avoiding the transitional regime whose definition remains debated~\cite{nepf2012flow,monti2020genesis,nicholas2022numerical}.}
The selected solid-to-fluid density ratio is $\rho_s/\rho_f = \mathcal{O}(1)$ in order to consider the case of almost neutrally-buoyant filaments.
Since we aim at isolating the essential flow-structure interaction mechanisms, we neglect the presence of gravity (i.e., the buoyancy and Galileo number are set to zero).
\deleted{See Methods section for further information on the modeling and computational technique.} 

We have performed a set of simulations varying the Cauchy number $\Ca$, such parameter representing the ratio between the forcing exerted by the flow and the elastic restoring force exerted by the filaments, defined as
\begin{equation}
 \Ca = \frac{1}{2}  \frac{\rho_f \, d \, h^3 \, U^2}{\gamma},
\end{equation}
where $\gamma$ is the bending stiffness of the filaments. 
Specifically, the following values of the Cauchy number have been chosen: $\Ca = \{ 0, 1, 10, 25, 50, 100\}$, with $\Ca=0$ corresponding to the rigid canopy case where the stems cannot deform. 
Such range is expected to include the variety of possible dynamical regimes for flexible canopies, both in terms of the individual element motion (i.e., weak vs strong reconfiguration)~\cite{pan2014strong} as well as collective motion (i.e., uncorrelated gentle swaying vs honami/monami)~\cite{okamoto2009turbulence}.
Note that the analysis can also be expressed in terms of a variation of the reduced velocity $U_r = U / (f_{n,1}\, h) \sim \sqrt{\Ca}$, where $f_{n,1}$ is the natural frequency of the filaments. 

\added{For the chosen configurations, the (average) canopy height expressed in wall units is overall $\mathcal{O}(10^2)$, monotonically decreasing with the Cauchy number from $\sim280$ in the rigid-canopy case (i.e., $\mathit{Ca}=0$) to $\sim95$ in the most flexible case ($\mathit{Ca}=100$) (note that the friction velocity used here is that evaluated at the virtual wall origin, i.e. the virtual wall seen by the outer boundary-layer~\cite{monti2019large}).}
To give a qualitative indication, snapshots from the representative case at $\Ca=50$ are shown in~\cref{fig1}, from which it can be observed the large-scale coherent motion undergone by the filaments (left panel), and superimposed the large-scale flow structures in the proximity of the canopy top (right panel). As it will be shown, the two are strictly related and represent the key features of the underlying physical mechanisms.

%


 

\subsection*{A revised view of honami/monami}
\label{sec:rev}

\cref{figSpa} (top panels) shows some instantaneous configurations of the canopy for several values of $\Ca$ considered in our study. From the elevation of the filaments it can be observed the typical pattern of the honami/monami motion, which appears more pronounced as the Cauchy number is increased.
To extract the dominant lengthscales of the canopy coherent motion, we look at the (spatial) spectra of the canopy top surface, reported in~\cref{figSpa} (bottom panels), which is obtained by the surface enveloping the vertical position of the filament tips, $Y(x,z,t)$. The first insight is that a peak can be observed for all cases, and specifically also at the smallest Cauchy number,  i.e., $\Ca = 1$. Note that in this condition, often referred to as the “gently swaying” regime, the motion of the filaments is typically claimed to be uncorrelated~\cite{okamoto2009turbulence}. Secondly, the location of the peak is typically at ${O}(H)$, both in the streamwise and spanwise wavelength component, and does not appear to appreciably vary within the investigated range. Increasing $\Ca$, for the spanwise component $\lambda_z/H$ remains around a constant value of 0.6 (only for $\Ca=1$, it drops around 0.4), whereas a more pronounced variation is observed in the streamwise component with $\lambda_x/H$ approximately ranging between 0.7 and 2.
Nevertheless, the absence of a systematic modification in the peak location with $\Ca$ suggests that the dominant wavelength is essentially imposed by the flow rather than the structure or that, in other terms, the coherent patterns of the canopy collective deformation can be seen as a signature of the large-scale turbulent structures~\cite{raupach1996coherent,sundin2019interaction}.

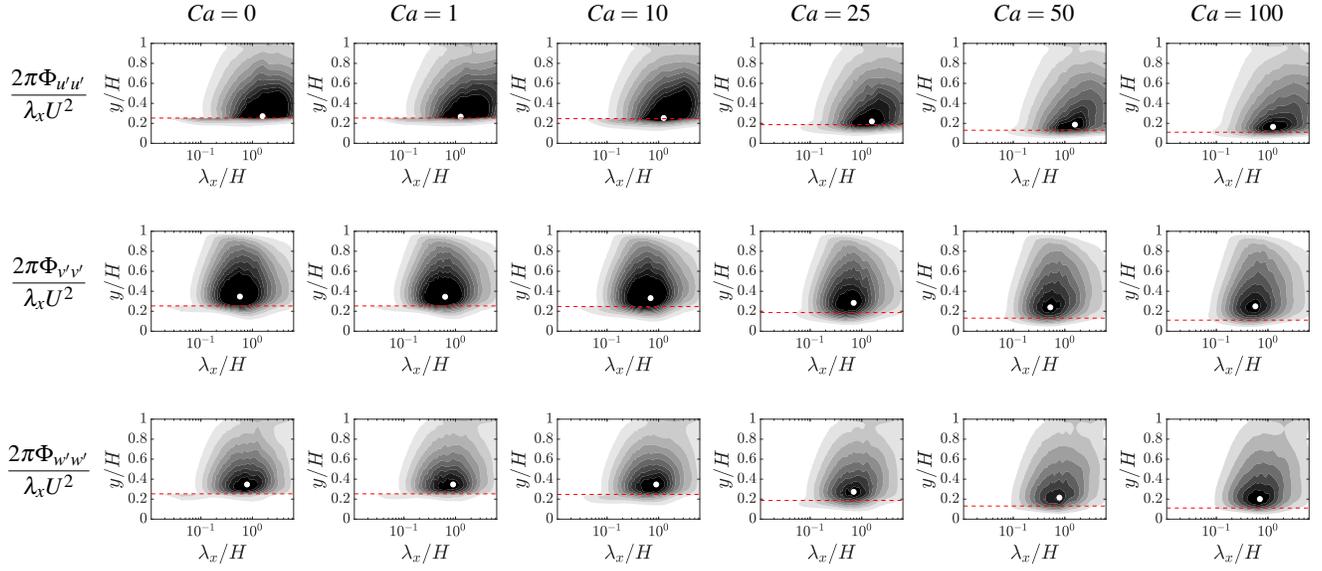
\begin{figure}
    \centering
    \begin{tikzpicture}
    \node[] at ( 1.6,2.4) {\small$Ca=0$};
    \node[] at ( 4.3,2.4) {\small$Ca=1$};
    \node[] at ( 7.0,2.4) {\small$Ca=10$};
    \node[] at ( 9.7,2.4) {\small$Ca=25$};
    \node[] at (12.4,2.4) {\small$Ca=50$};
    \node[] at (15.1,2.4) {\small$Ca=100$};

    \node[] at (-0.7,1.3) {\small$\dfrac{2\pi\Phi_{u'u'}}{\lambda_x U^2}$};
    \node[anchor=south west,inner sep=0] (image) at ( 0.0,0) {
    \includegraphics[width=.16\textwidth]{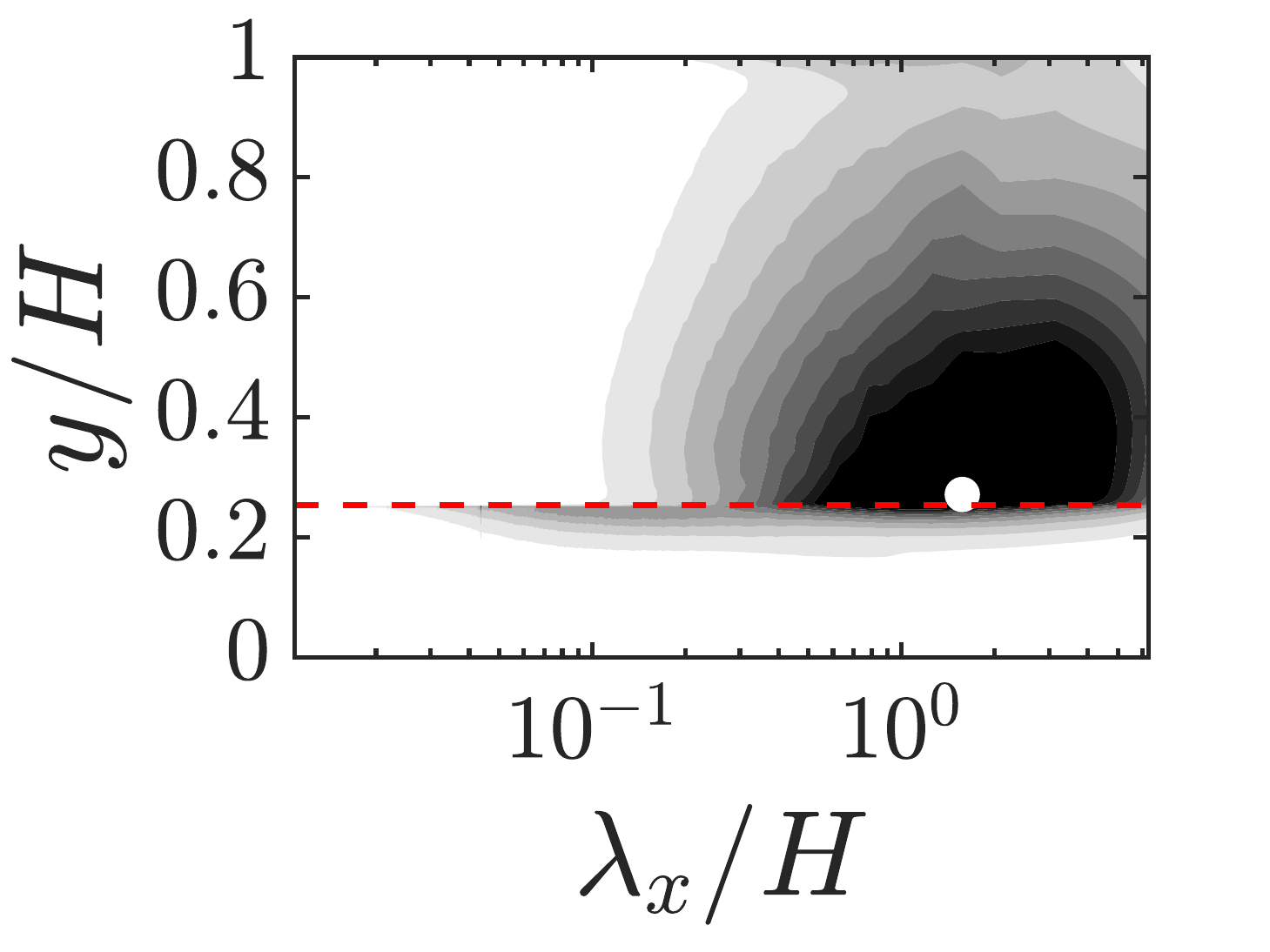}};
    \node[anchor=south west,inner sep=0] (image) at ( 2.7,0) {
    \includegraphics[width=.16\textwidth]{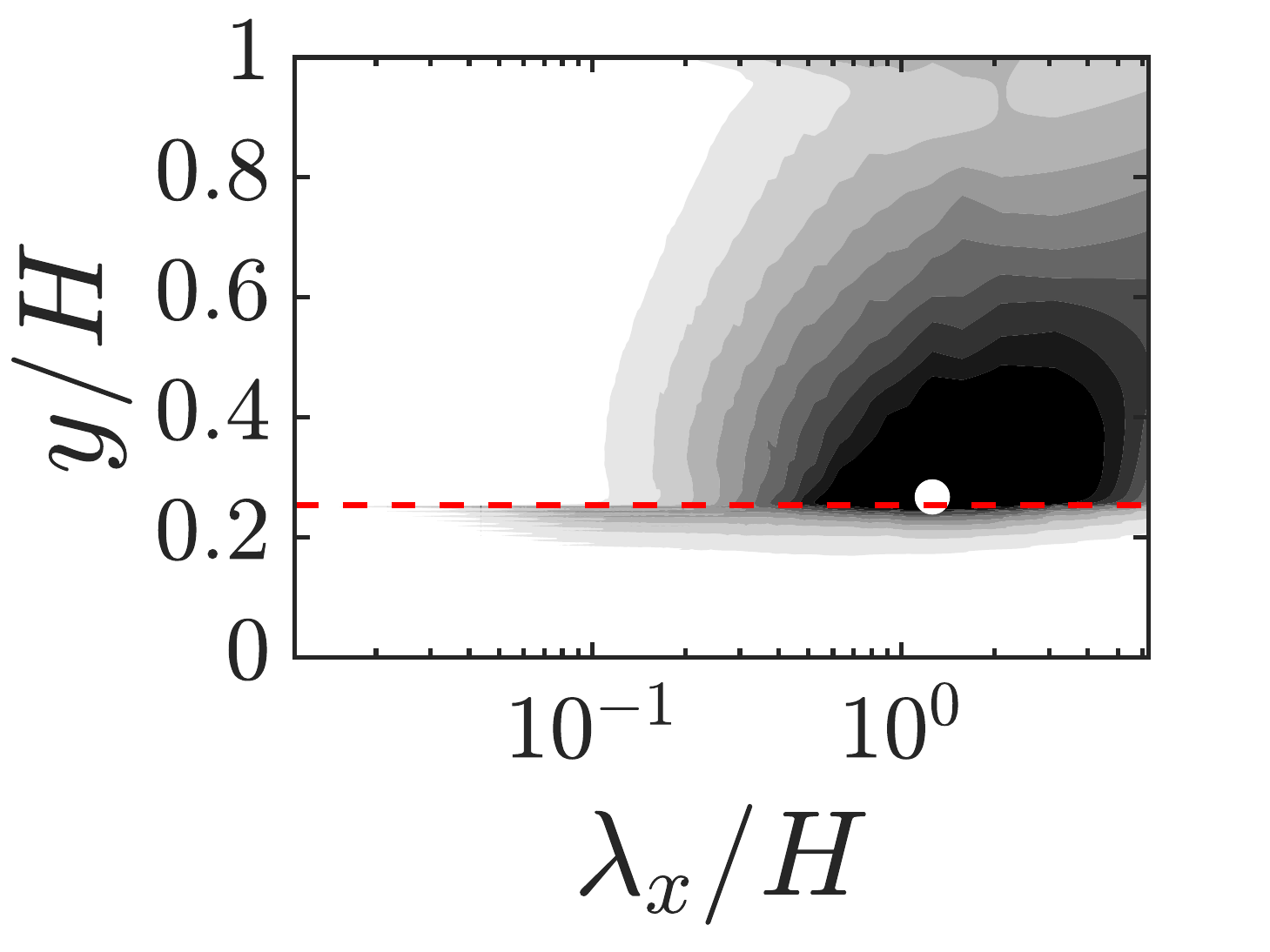}};
    \node[anchor=south west,inner sep=0] (image) at ( 5.4,0) {
    \includegraphics[width=.16\textwidth]{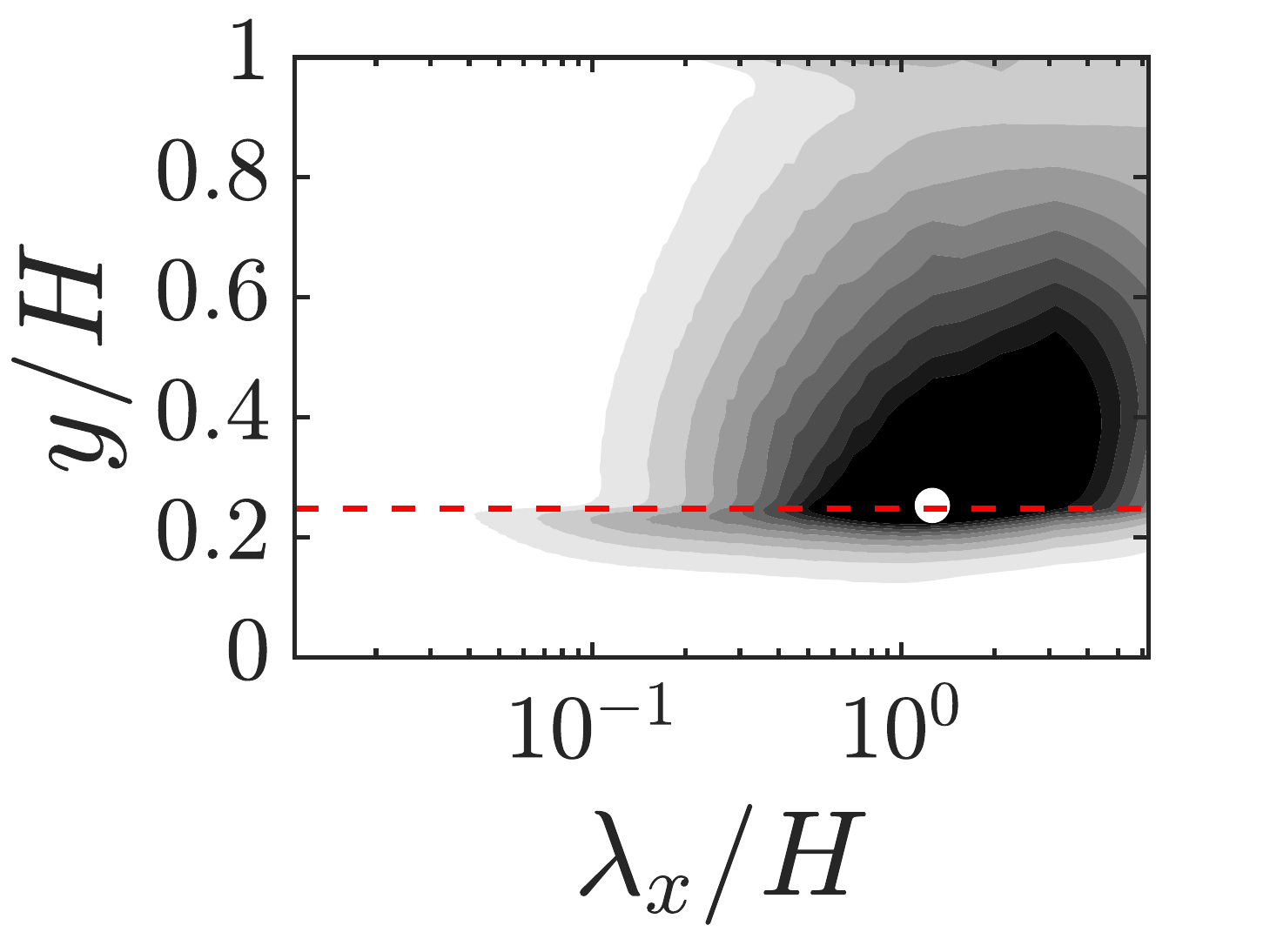}};
    \node[anchor=south west,inner sep=0] (image) at ( 8.1,0) {
    \includegraphics[width=.16\textwidth]{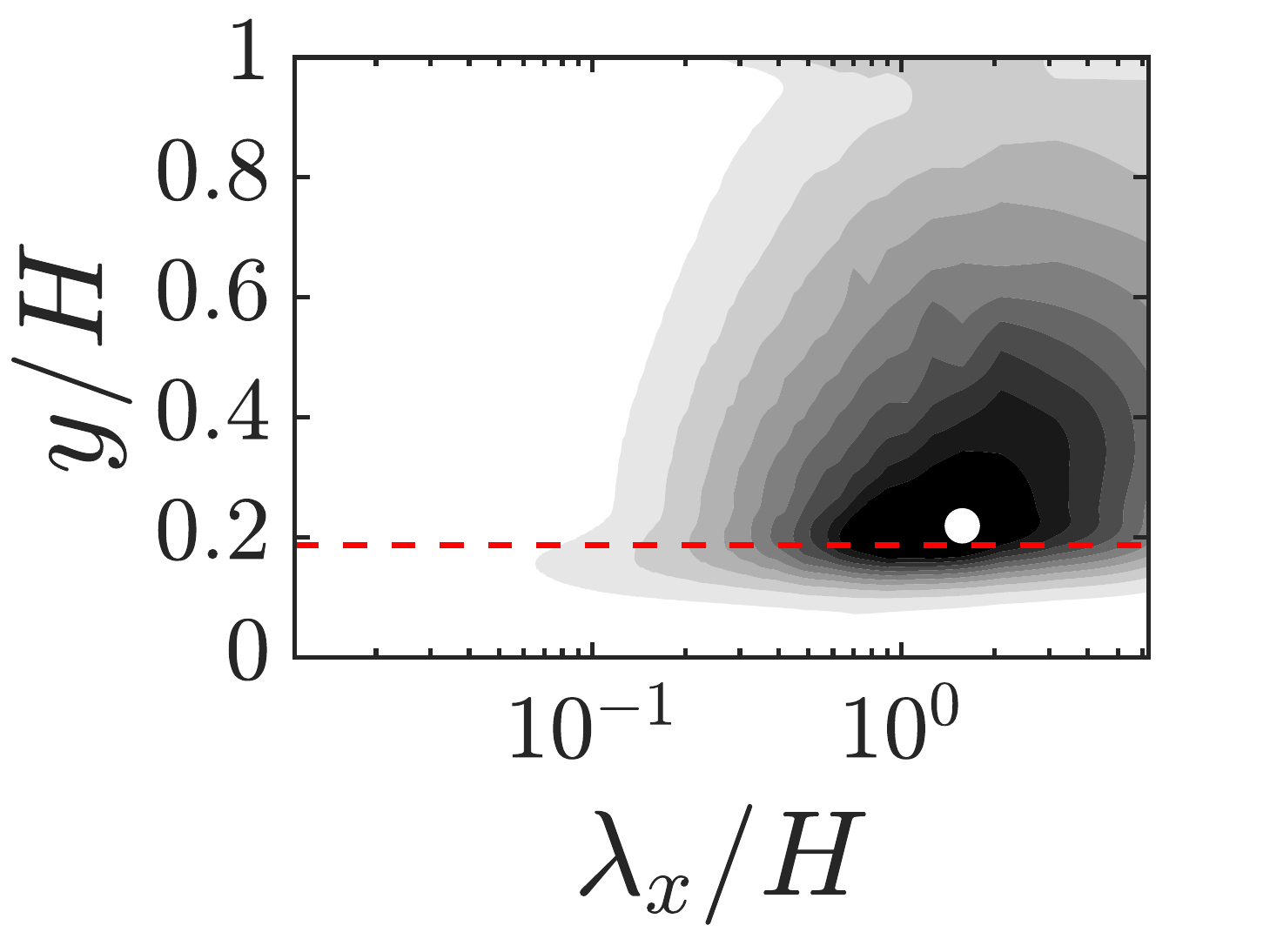}};
    \node[anchor=south west,inner sep=0] (image) at (10.8,0) {
    \includegraphics[width=.16\textwidth]{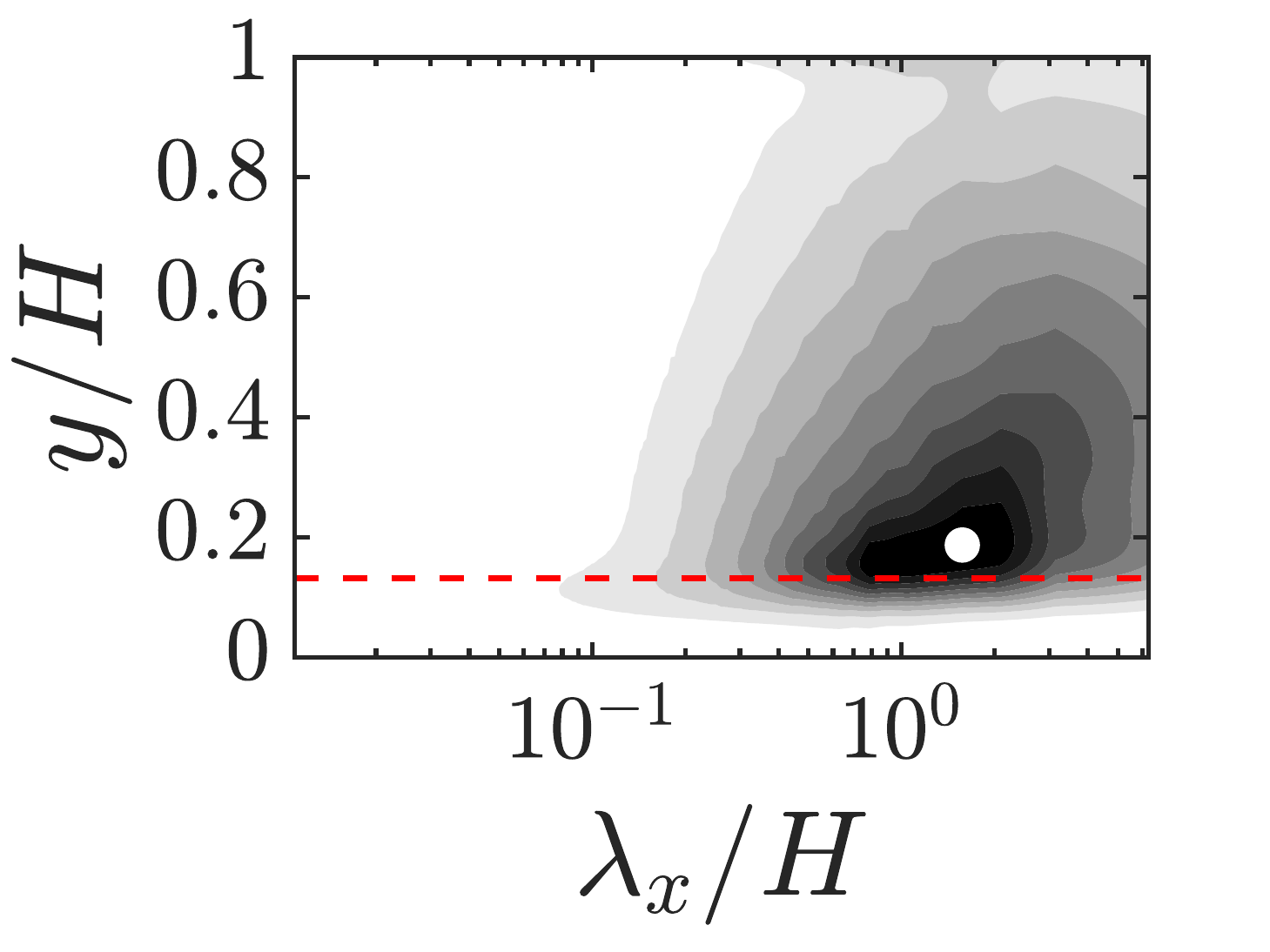}};
    \node[anchor=south west,inner sep=0] (image) at (13.5,0) {
    \includegraphics[width=.16\textwidth]{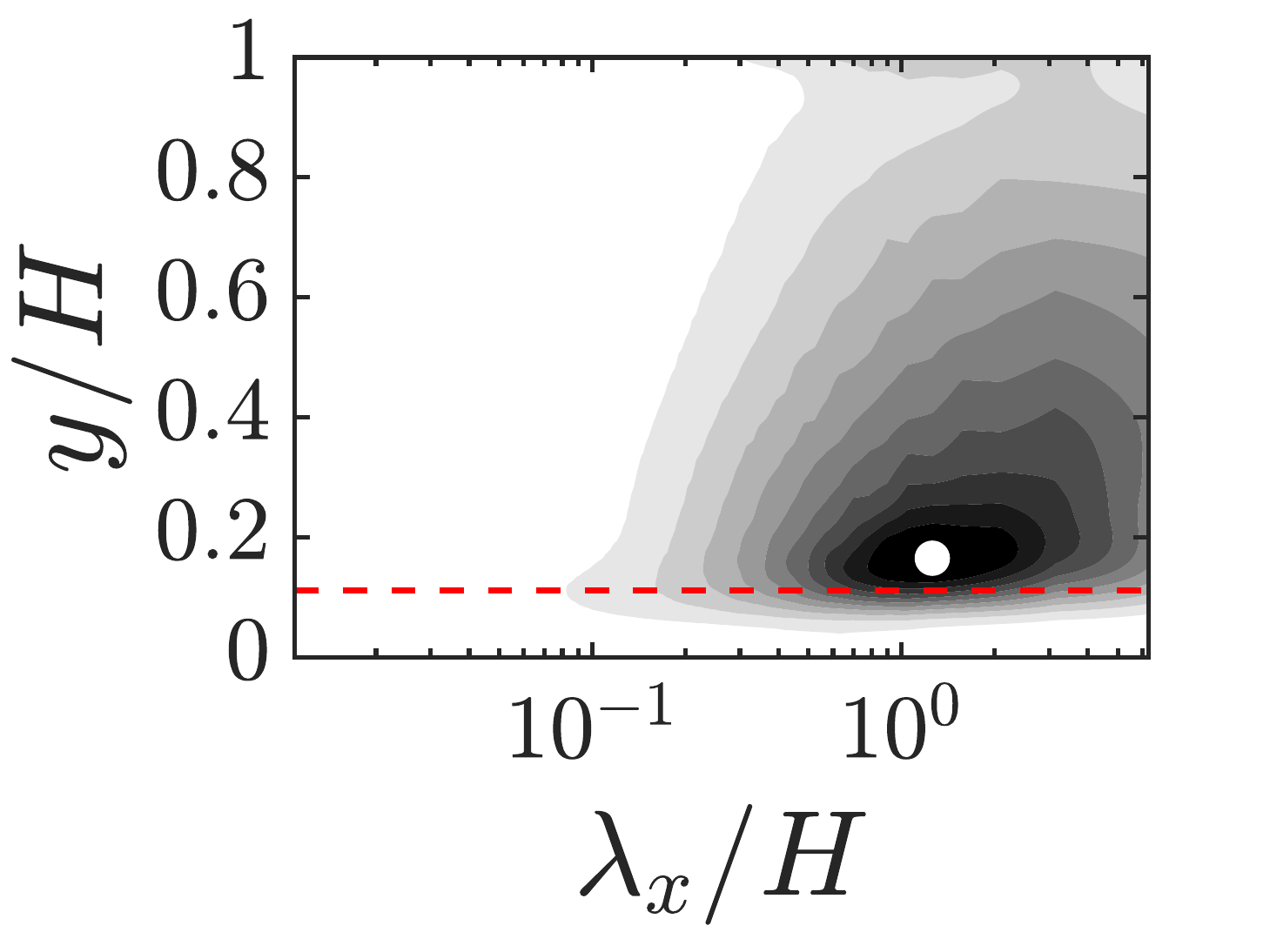}};

    \node[] at (-0.7,-1.2) {\small$\dfrac{2\pi\Phi_{v'v'}}{\lambda_x U^2}$};
    \node[anchor=south west,inner sep=0] (image) at ( 0.0,-2.50) {
    \includegraphics[width=.16\textwidth]{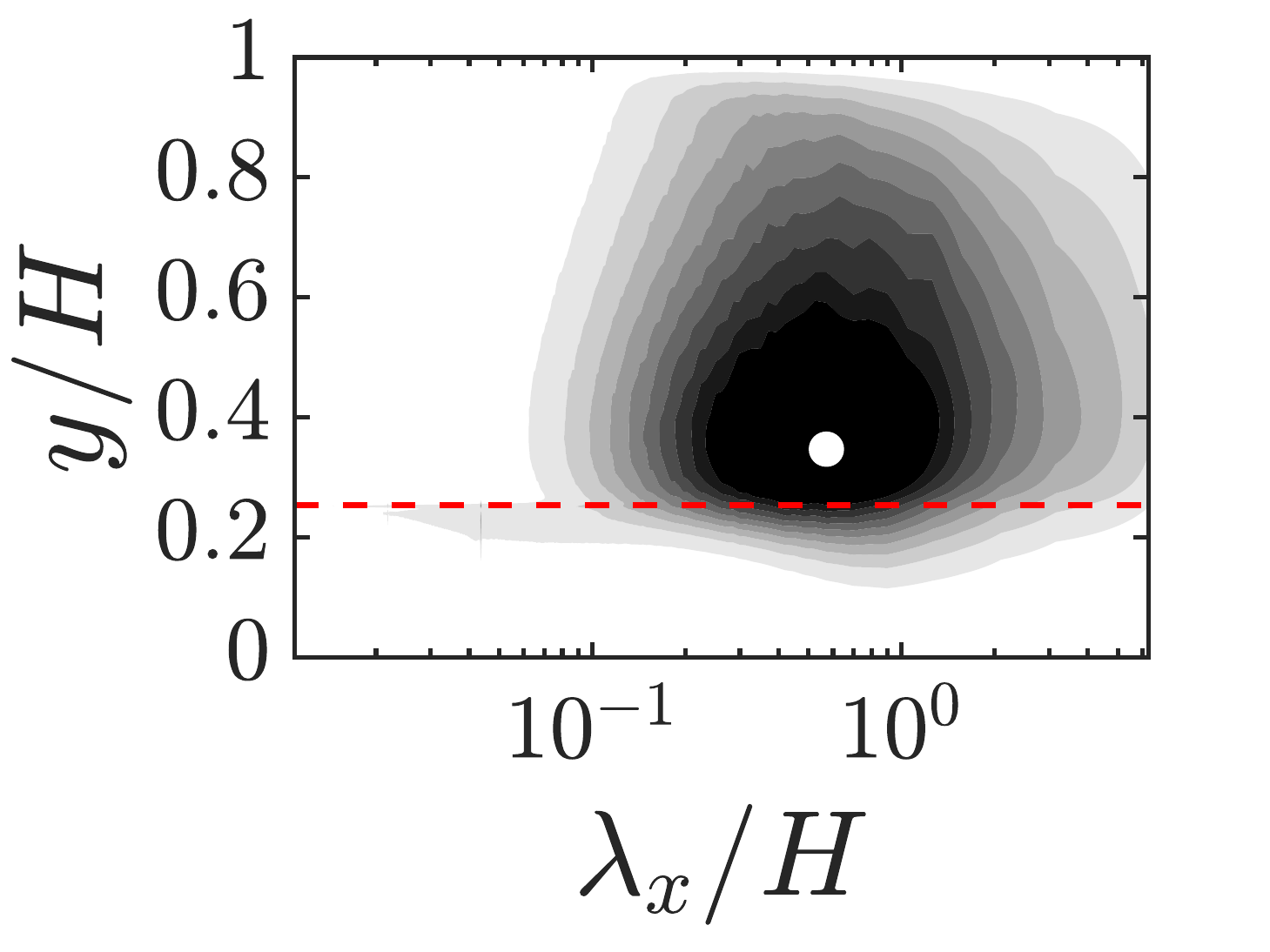}};
    \node[anchor=south west,inner sep=0] (image) at ( 2.7,-2.50) {
    \includegraphics[width=.16\textwidth]{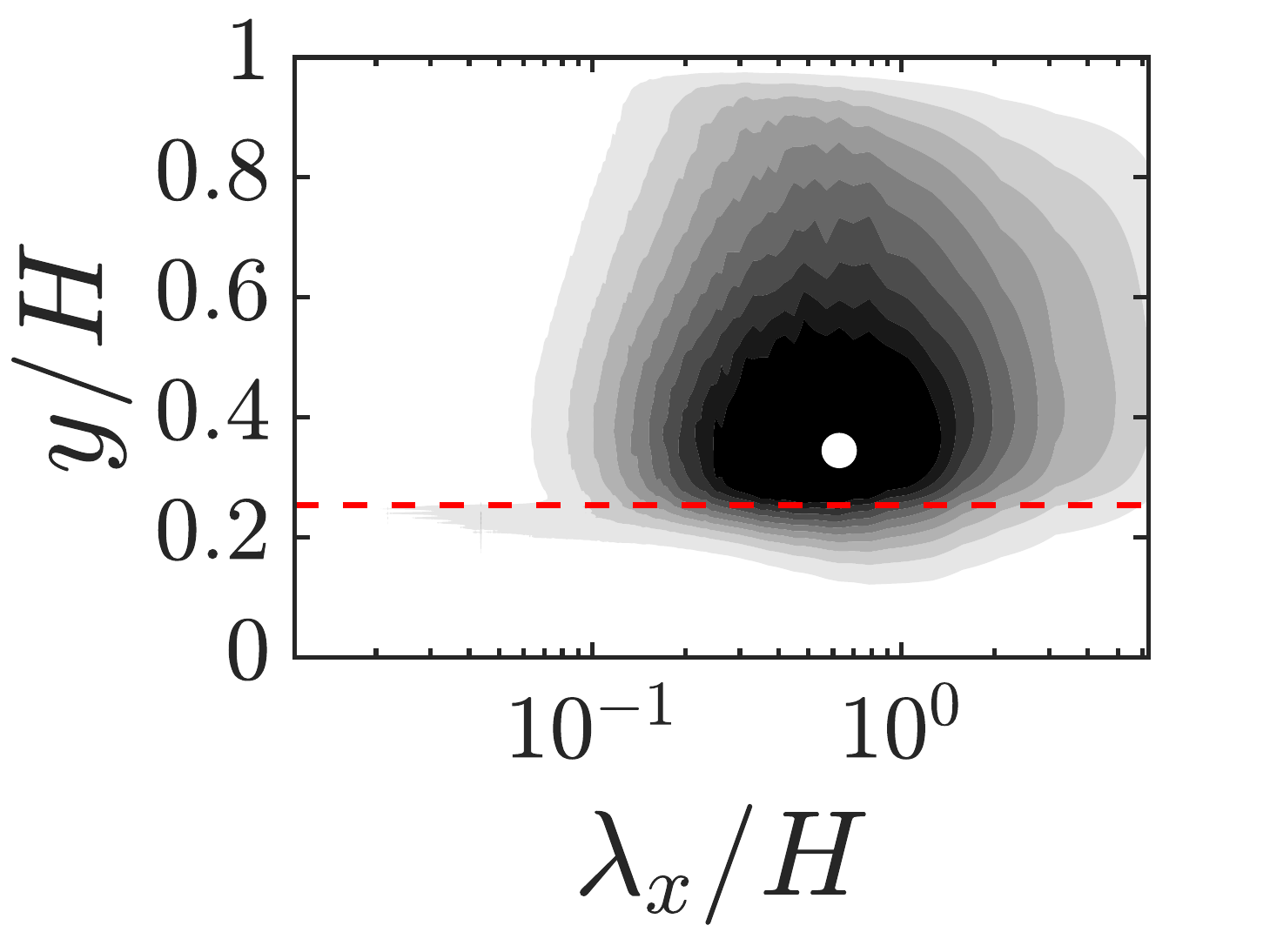}};
    \node[anchor=south west,inner sep=0] (image) at ( 5.4,-2.50) {
    \includegraphics[width=.16\textwidth]{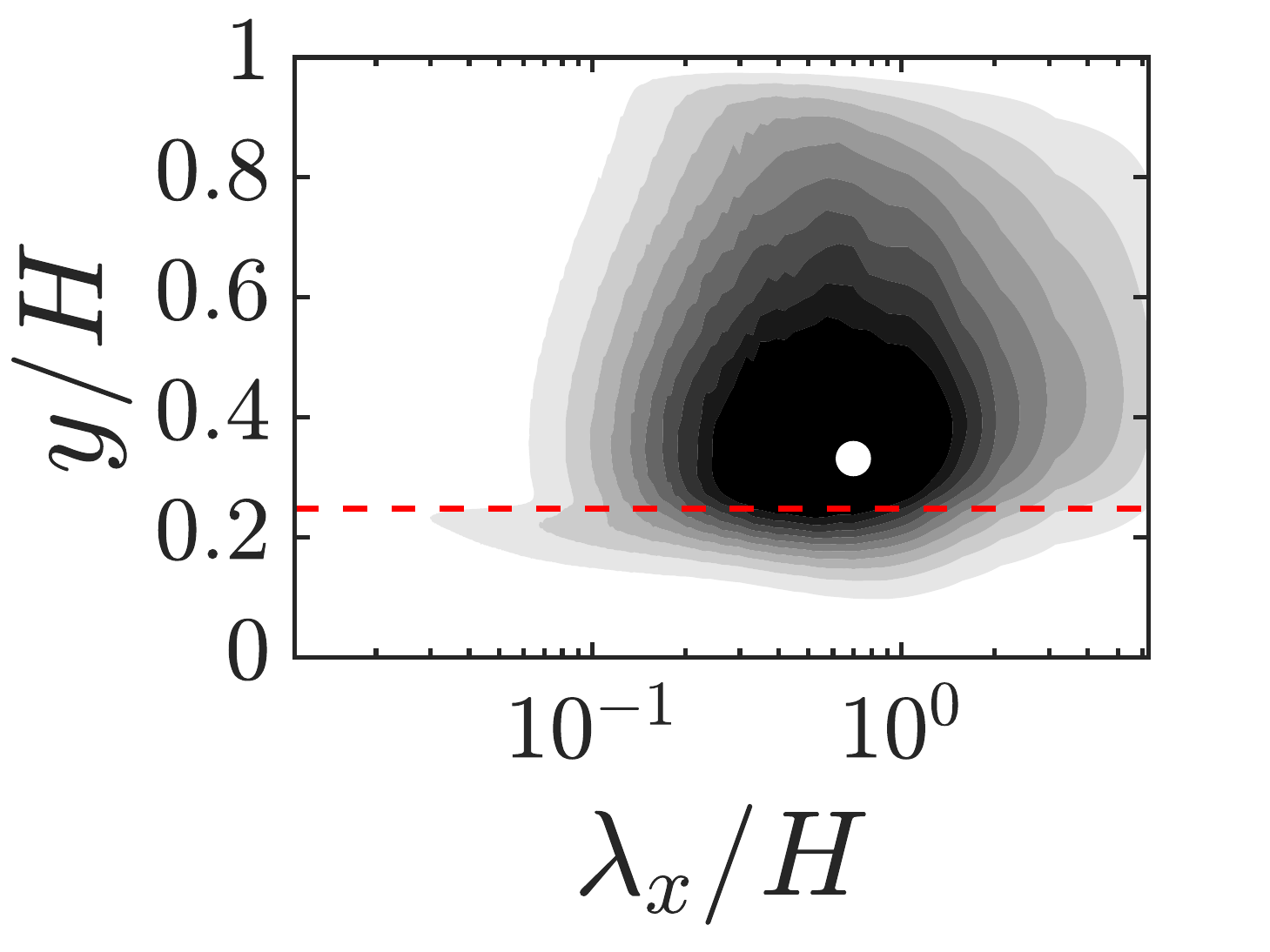}};
    \node[anchor=south west,inner sep=0] (image) at ( 8.1,-2.50) {
    \includegraphics[width=.16\textwidth]{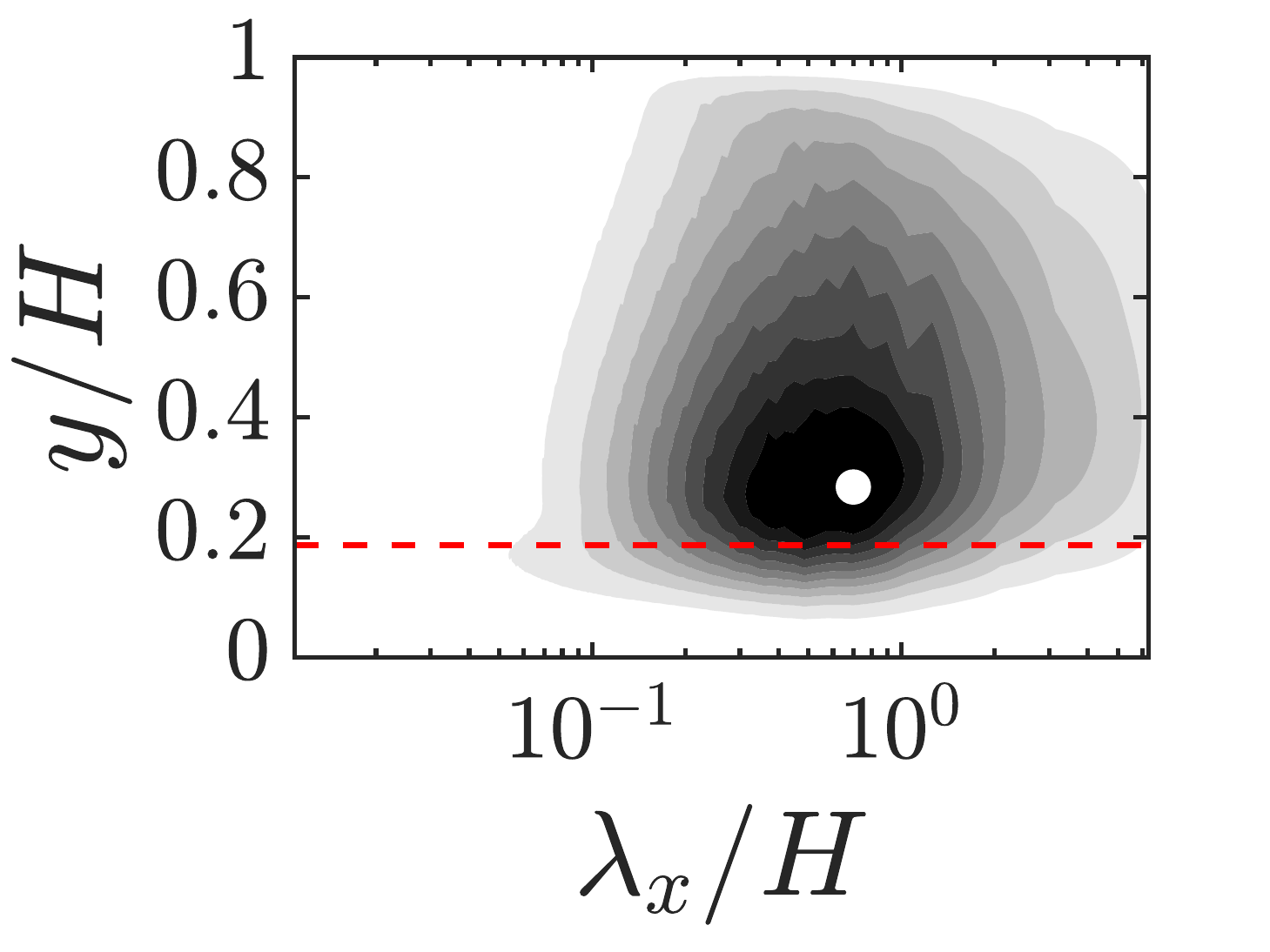}};
    \node[anchor=south west,inner sep=0] (image) at (10.8,-2.50) {
    \includegraphics[width=.16\textwidth]{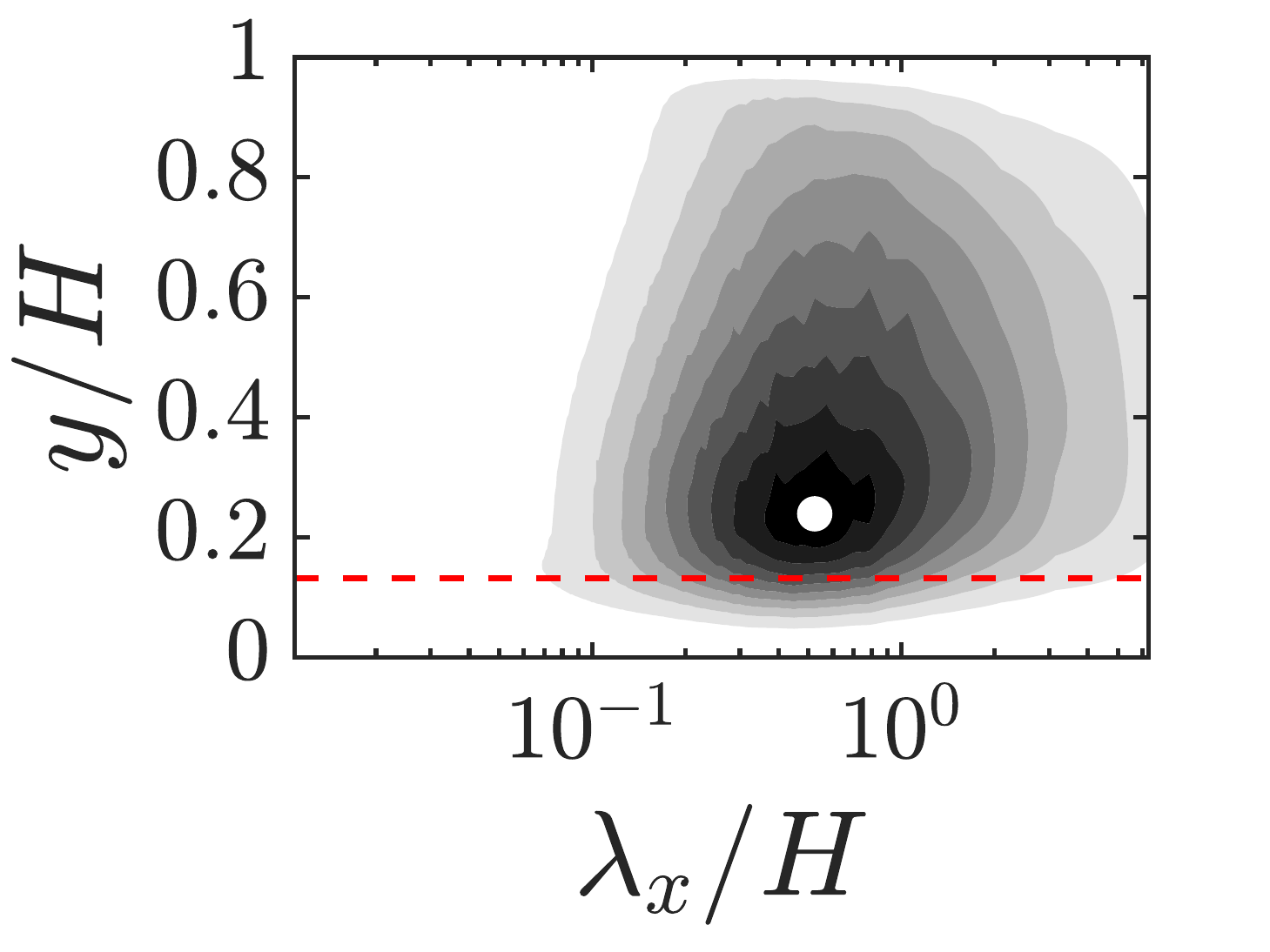}};
    \node[anchor=south west,inner sep=0] (image) at (13.5,-2.50) {
    \includegraphics[width=.16\textwidth]{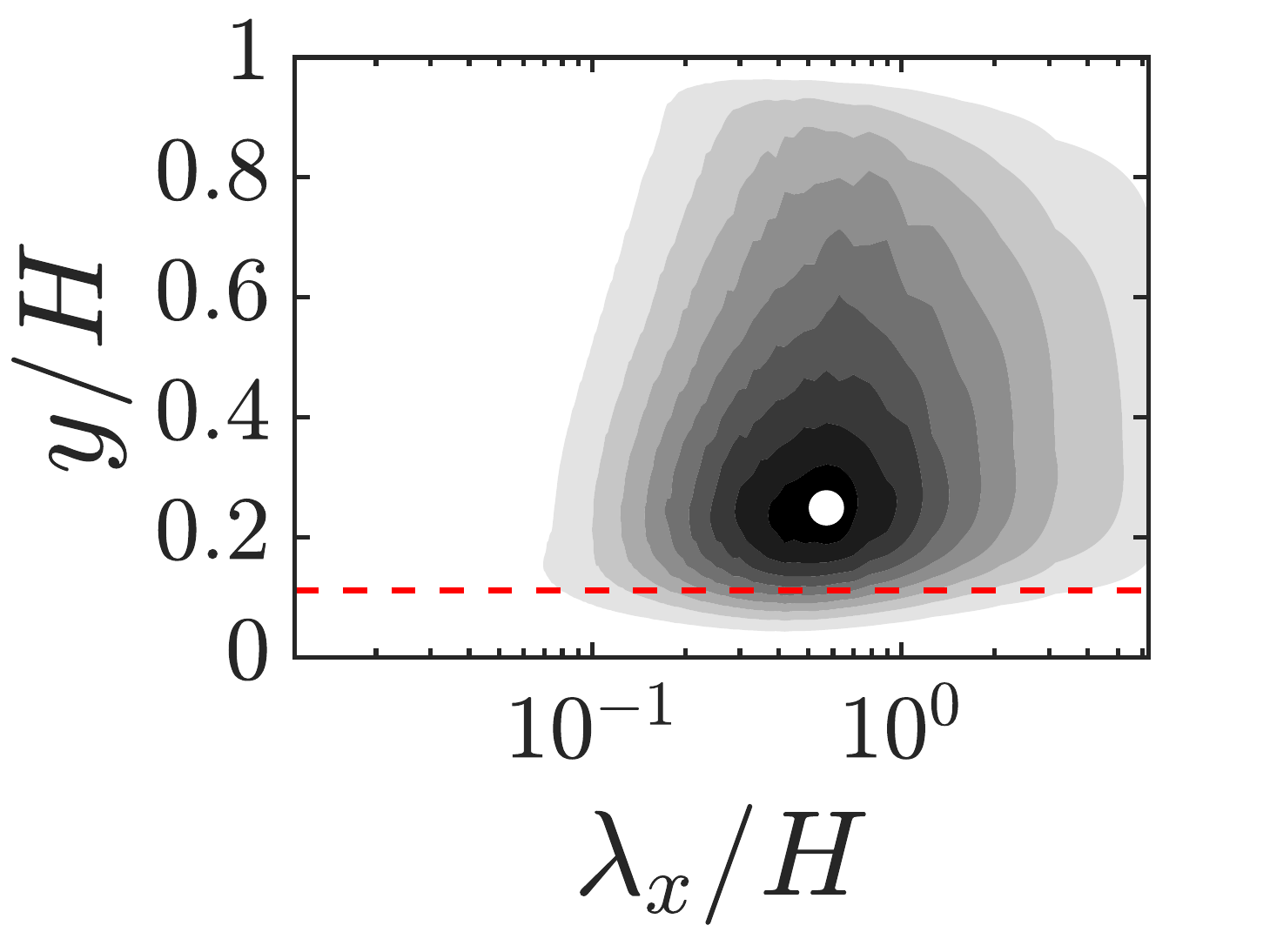}};

    \node[] at (-0.7,-3.7) {\small$\dfrac{2\pi\Phi_{w'w'}}{\lambda_x U^2}$};
    \node[anchor=south west,inner sep=0] (image) at ( 0.0,-5.0) {
    \includegraphics[width=.16\textwidth]{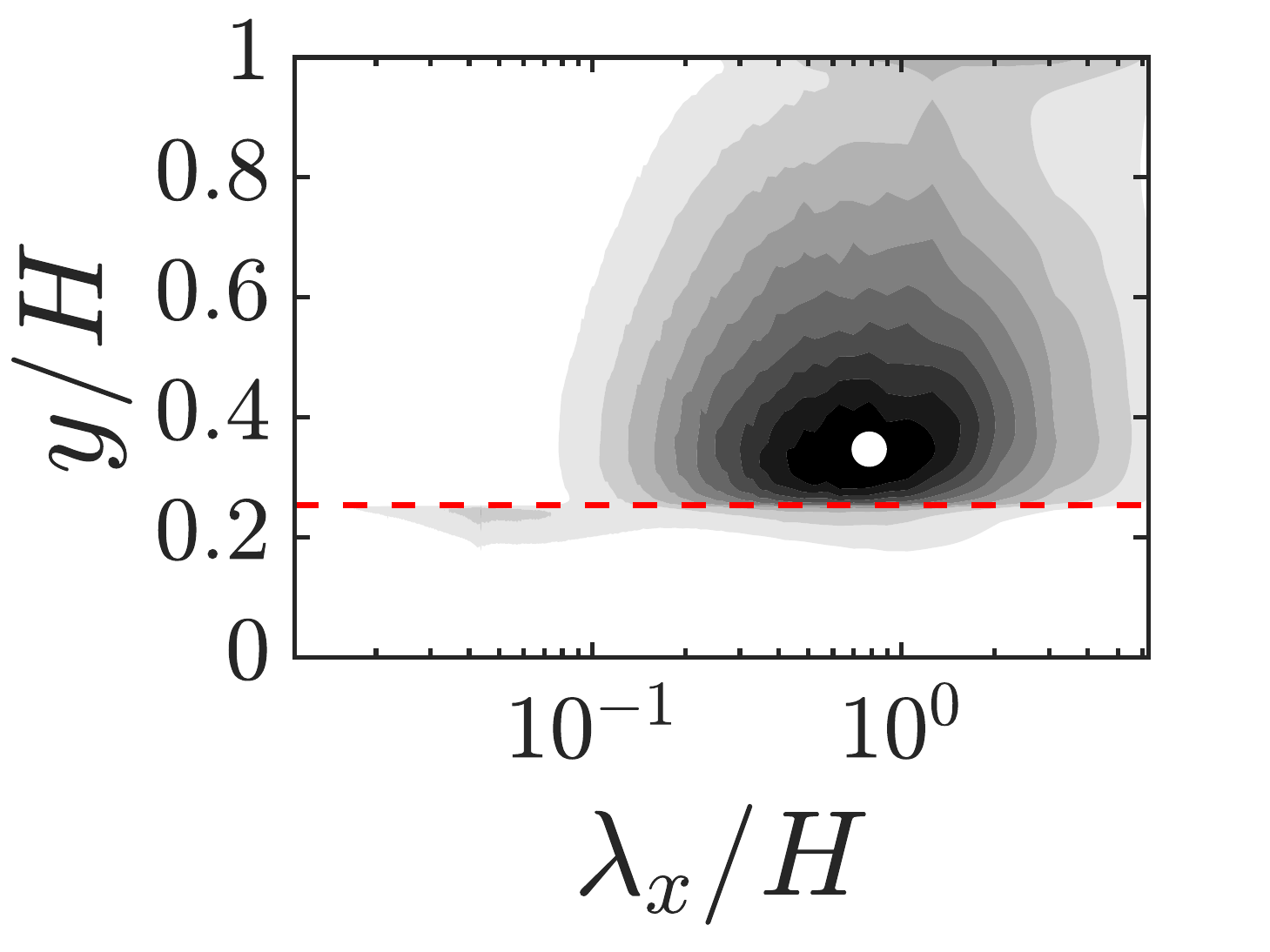}};
    \node[anchor=south west,inner sep=0] (image) at ( 2.7,-5.0) {
    \includegraphics[width=.16\textwidth]{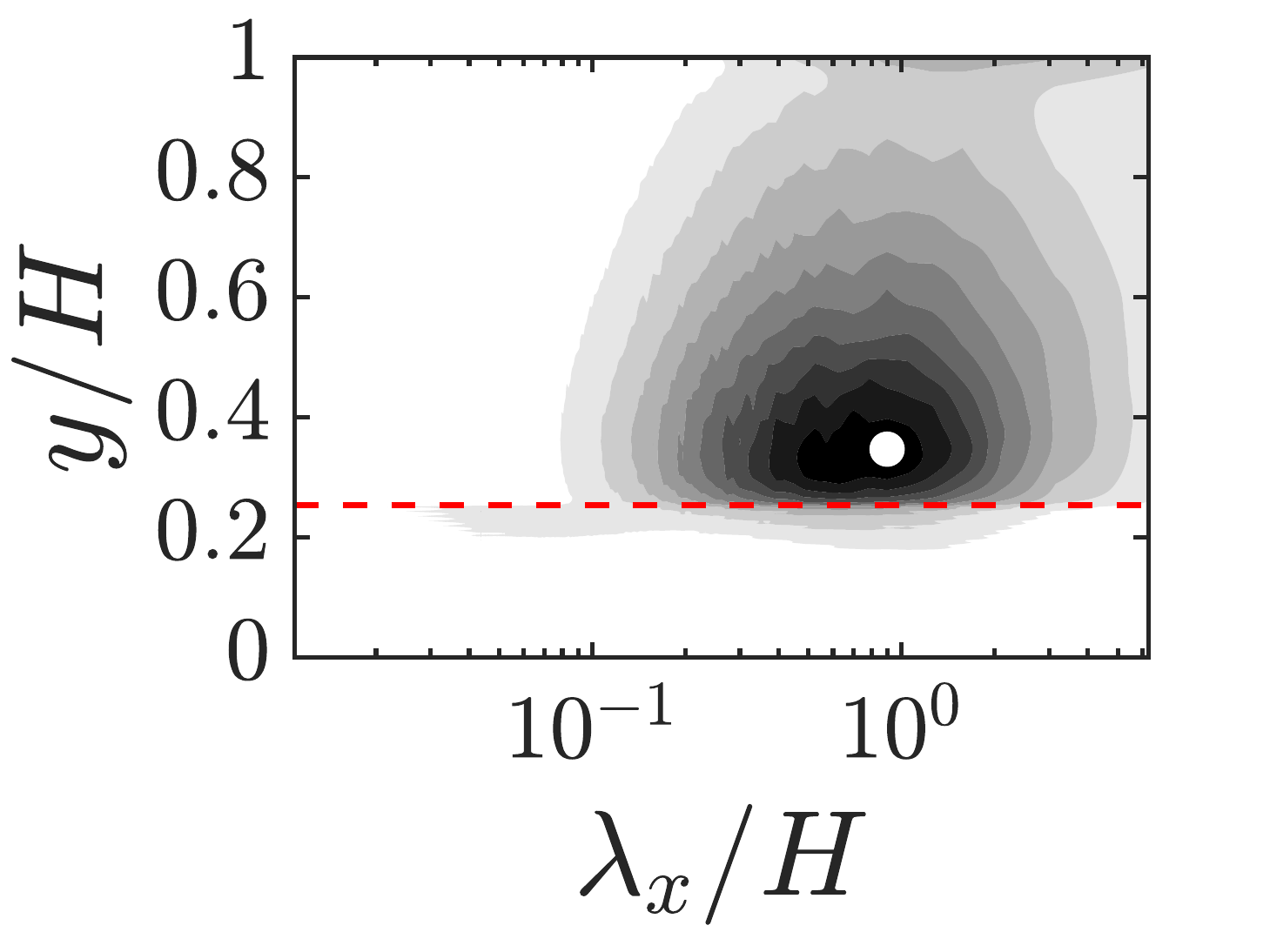}};
    \node[anchor=south west,inner sep=0] (image) at ( 5.4,-5.0) {
    \includegraphics[width=.16\textwidth]{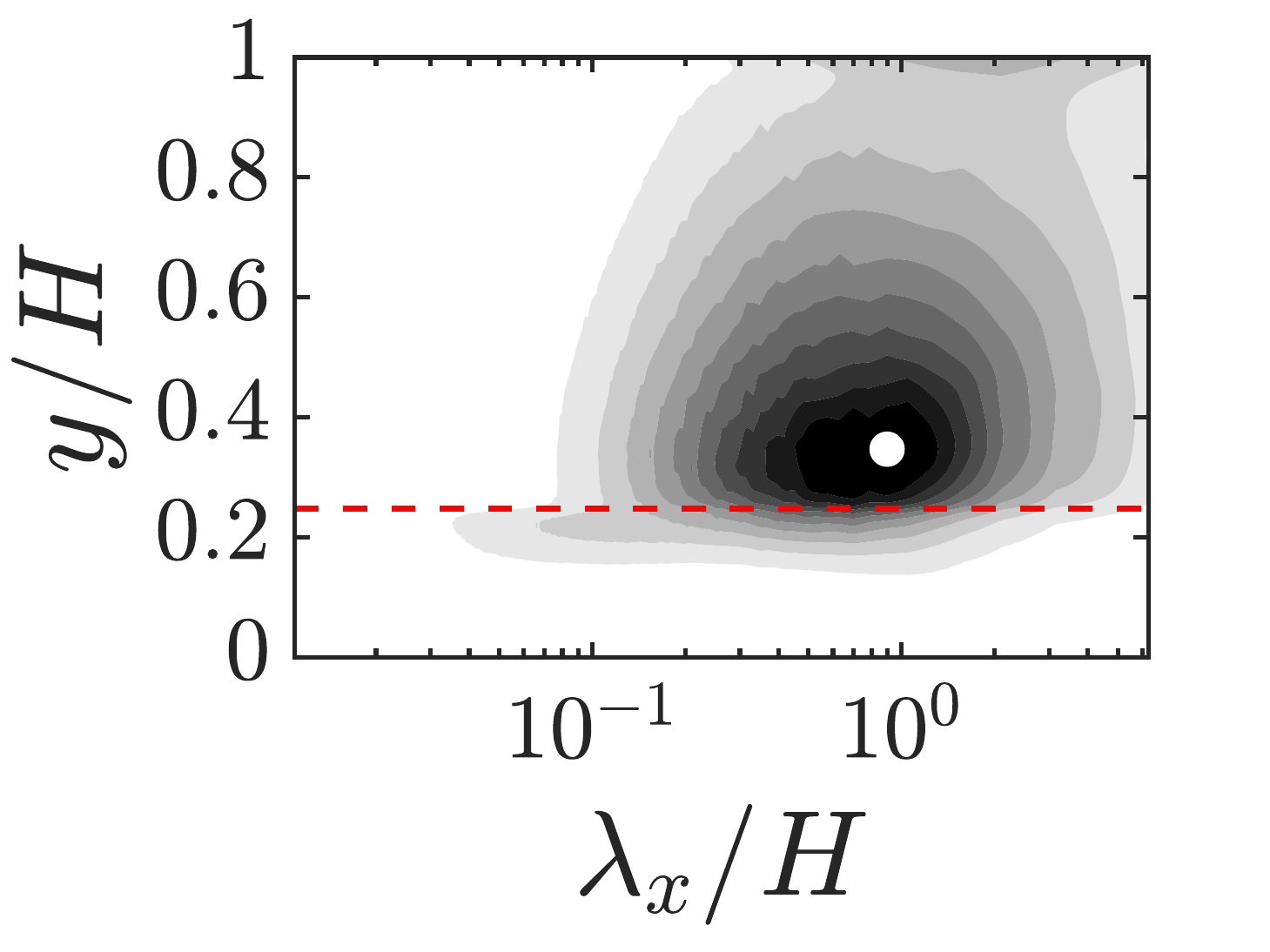}};
    \node[anchor=south west,inner sep=0] (image) at ( 8.1,-5.0) {
    \includegraphics[width=.16\textwidth]{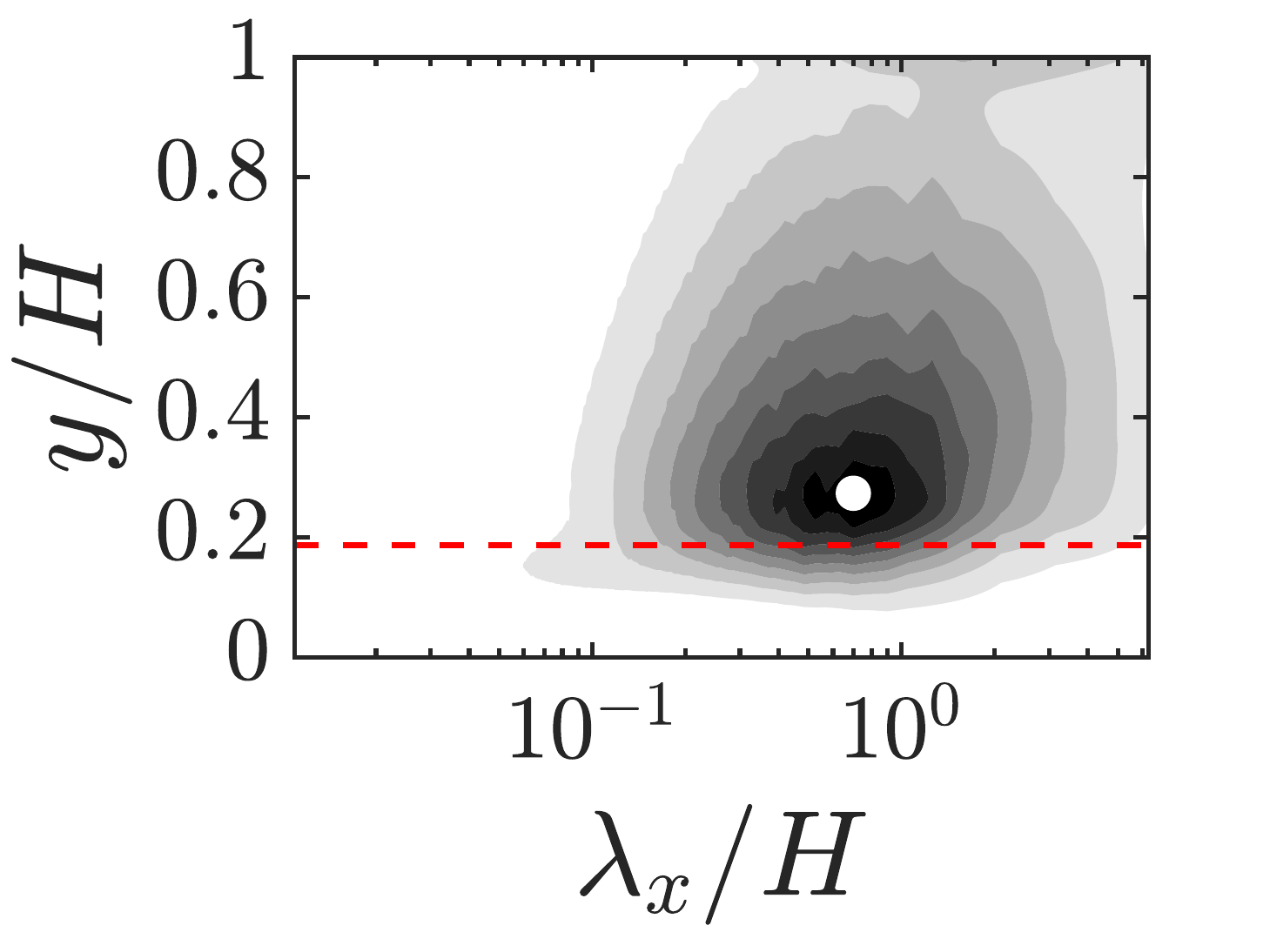}};
    \node[anchor=south west,inner sep=0] (image) at (10.8,-5.0) {
    \includegraphics[width=.16\textwidth]{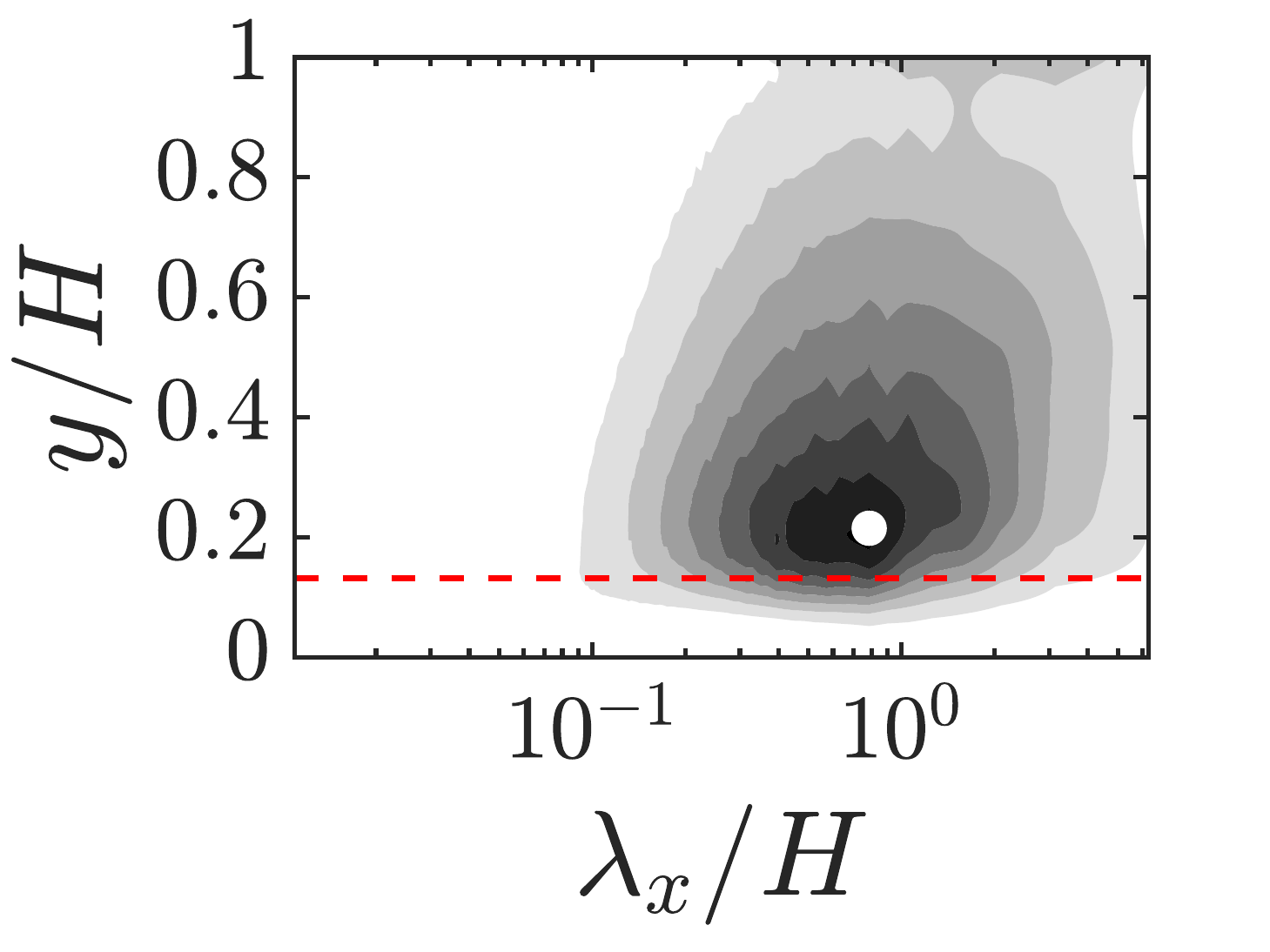}};
    \node[anchor=south west,inner sep=0] (image) at (13.5,-5.0) {
    \includegraphics[width=.16\textwidth]{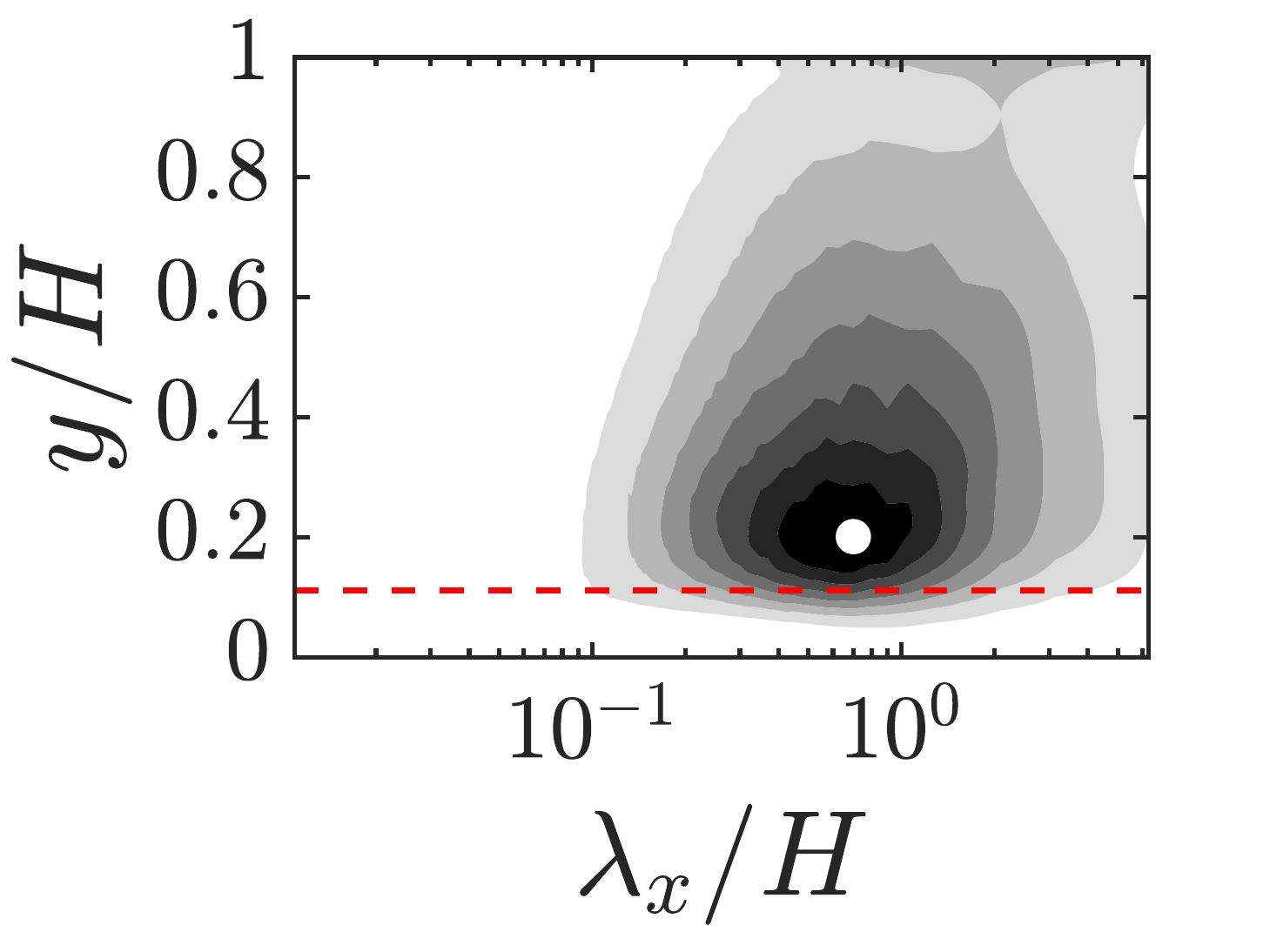}};
    \end{tikzpicture}
    \caption{
    Magnitude of the premultiplied spectra of each fluid velocity component 
$2\pi\Phi_{u'u'}/(U^2\lambda_x)$, where $U$ is the bulk velocity and $u'$ the generic component of the velocity fluctuations,
	(top: streamwise; middle: wall-normal; bottom: spanwise) as a function of the streamwise wavelength $\lambda_x/H$ and wall-normal coordinate $y/H$. Results are shown in different columns as a function of the investigated Cauchy number (from left to right, $\Ca=0, 1, 10, 25, 50, 100$). \added{The white marker denotes the peak of the spectrum, whereas the red horizontal dashed line indicates the averaged height of the filament tips.}
The grey levels range in: $[0, 0.005]$ with a $0.0005$ increment for the streamwise and wall-normal velocity components; $[0, 0.01]$ with a $0.001$ increment for the spanwise velocity component.
\added{Overall, the characteristic peak in the wavelength of the energy-containing scale appears always $\mathcal{O}(H)$ without an appreciable and systematic change associated with $\Ca$.}
    }
    \label{fig3}
\end{figure}

 \begin{figure}
    \centering
    \begin{tikzpicture}
    \node[] at ( 1.6,2.4) {\small$Ca=0$};
    \node[] at ( 4.3,2.4) {\small$Ca=1$};
    \node[] at ( 7.0,2.4) {\small$Ca=10$};
    \node[] at ( 9.7,2.4) {\small$Ca=25$};
    \node[] at (12.4,2.4) {\small$Ca=50$};
    \node[] at (15.1,2.4) {\small$Ca=100$};

    \node[] at (-0.7,1.3) {\small$\dfrac{2\pi\Phi_{u'u'}}{\lambda_z U^2}$};
    \node[anchor=south west,inner sep=0] (image) at ( 0.0,0) {
    \includegraphics[width=.16\textwidth]{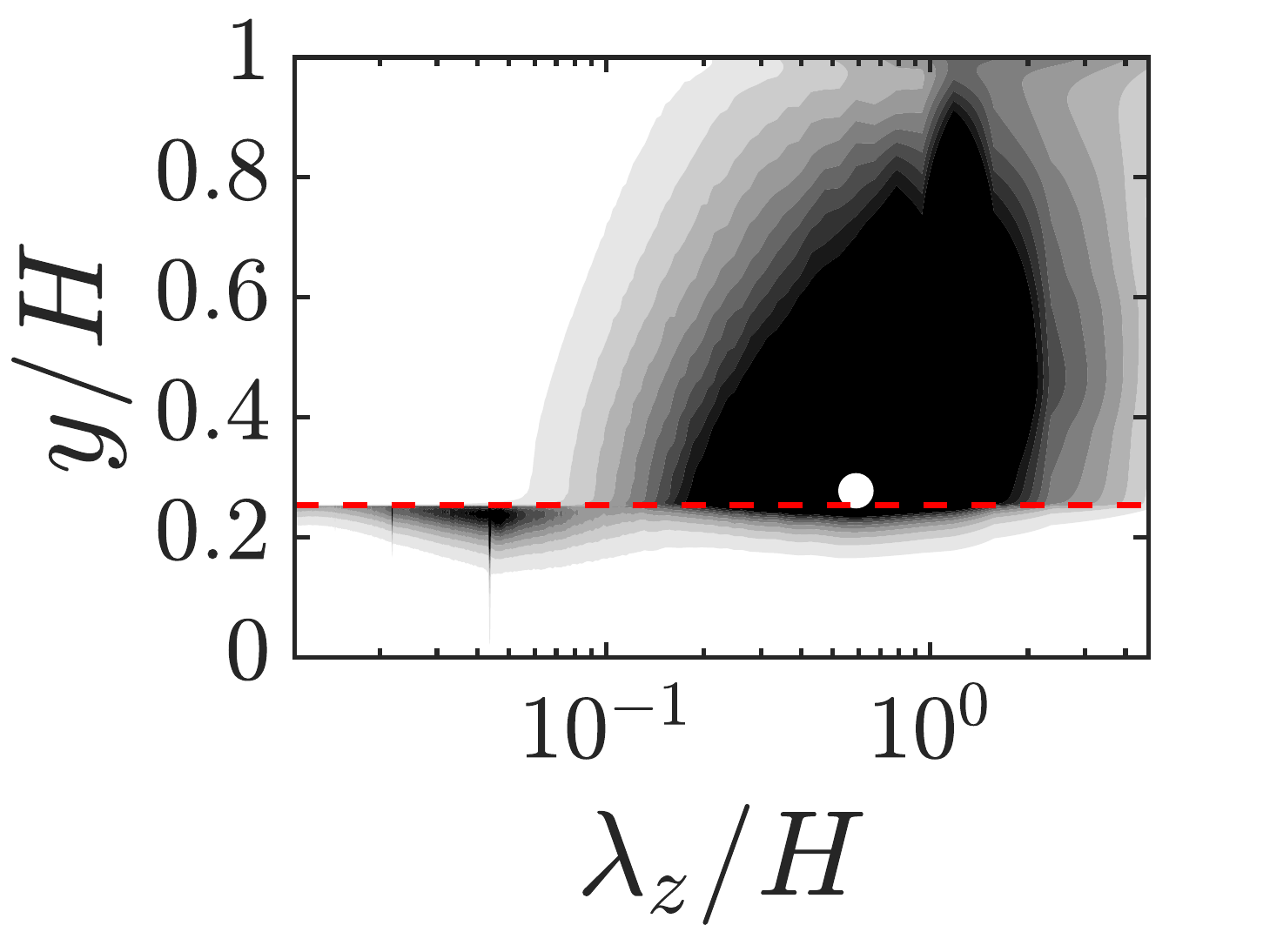}};
    \node[anchor=south west,inner sep=0] (image) at ( 2.7,0) {
    \includegraphics[width=.16\textwidth]{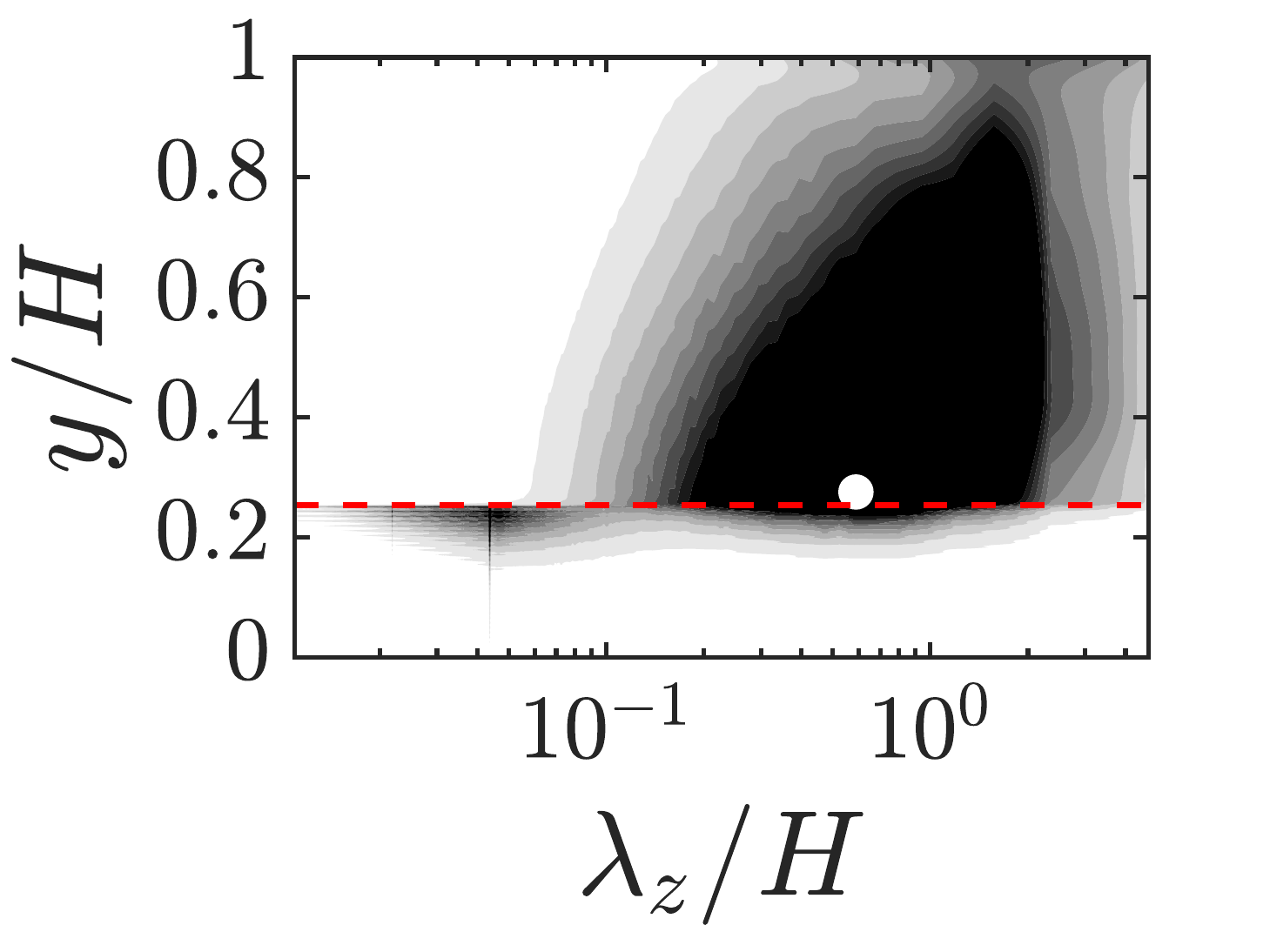}};
    \node[anchor=south west,inner sep=0] (image) at ( 5.4,0) {
    \includegraphics[width=.16\textwidth]{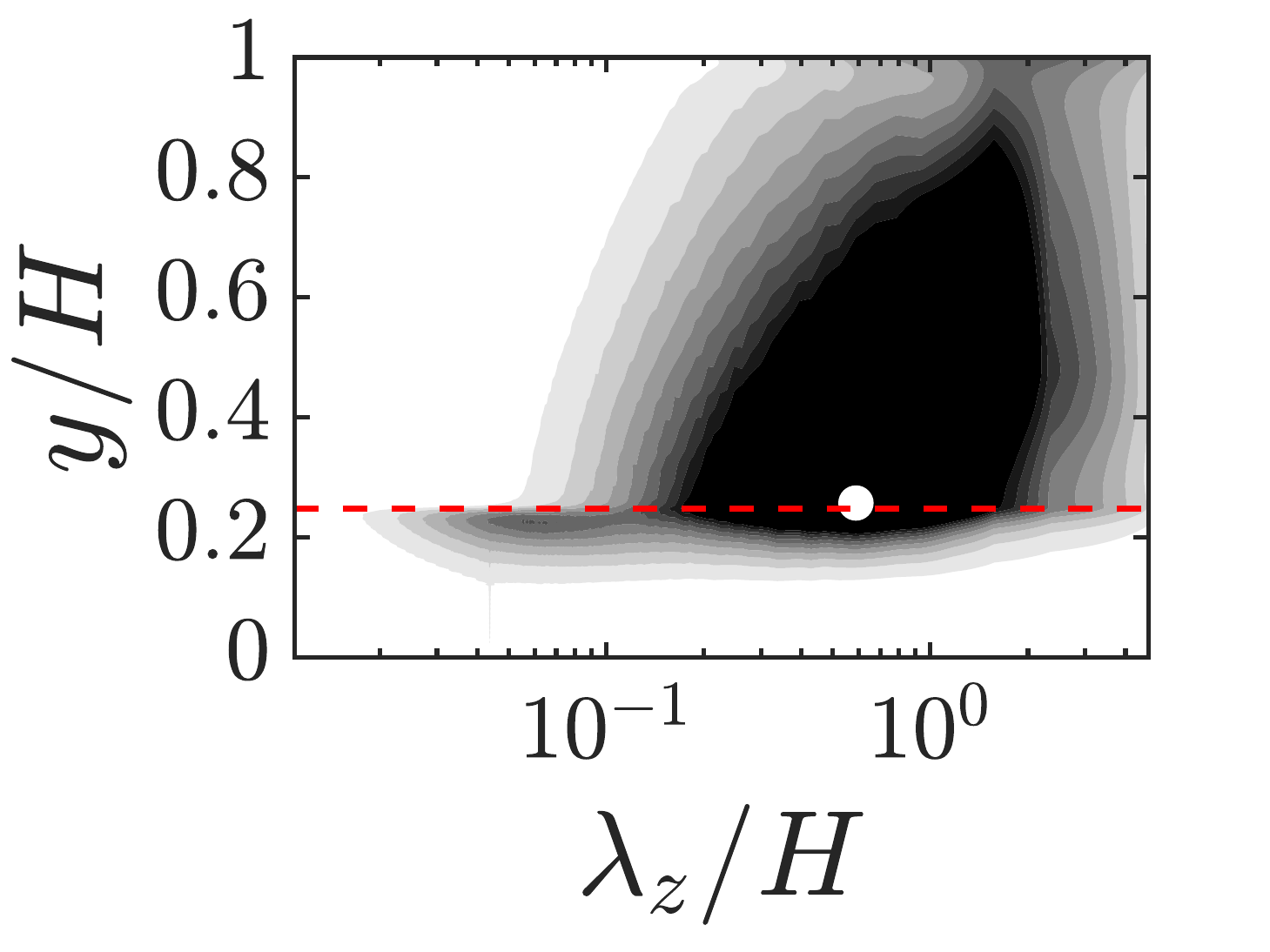}};
    \node[anchor=south west,inner sep=0] (image) at ( 8.1,0) {
    \includegraphics[width=.16\textwidth]{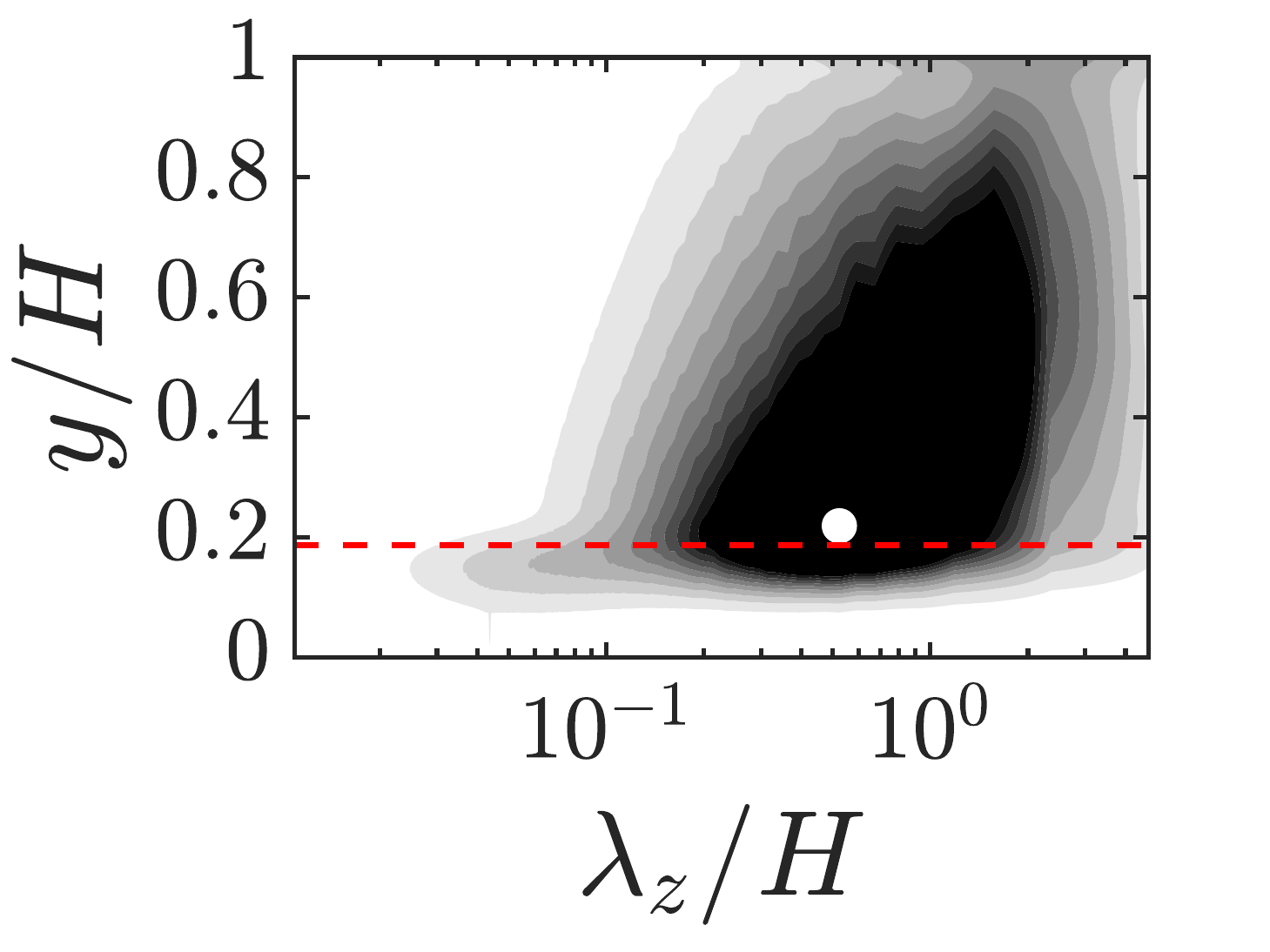}};
    \node[anchor=south west,inner sep=0] (image) at (10.8,0) {
    \includegraphics[width=.16\textwidth]{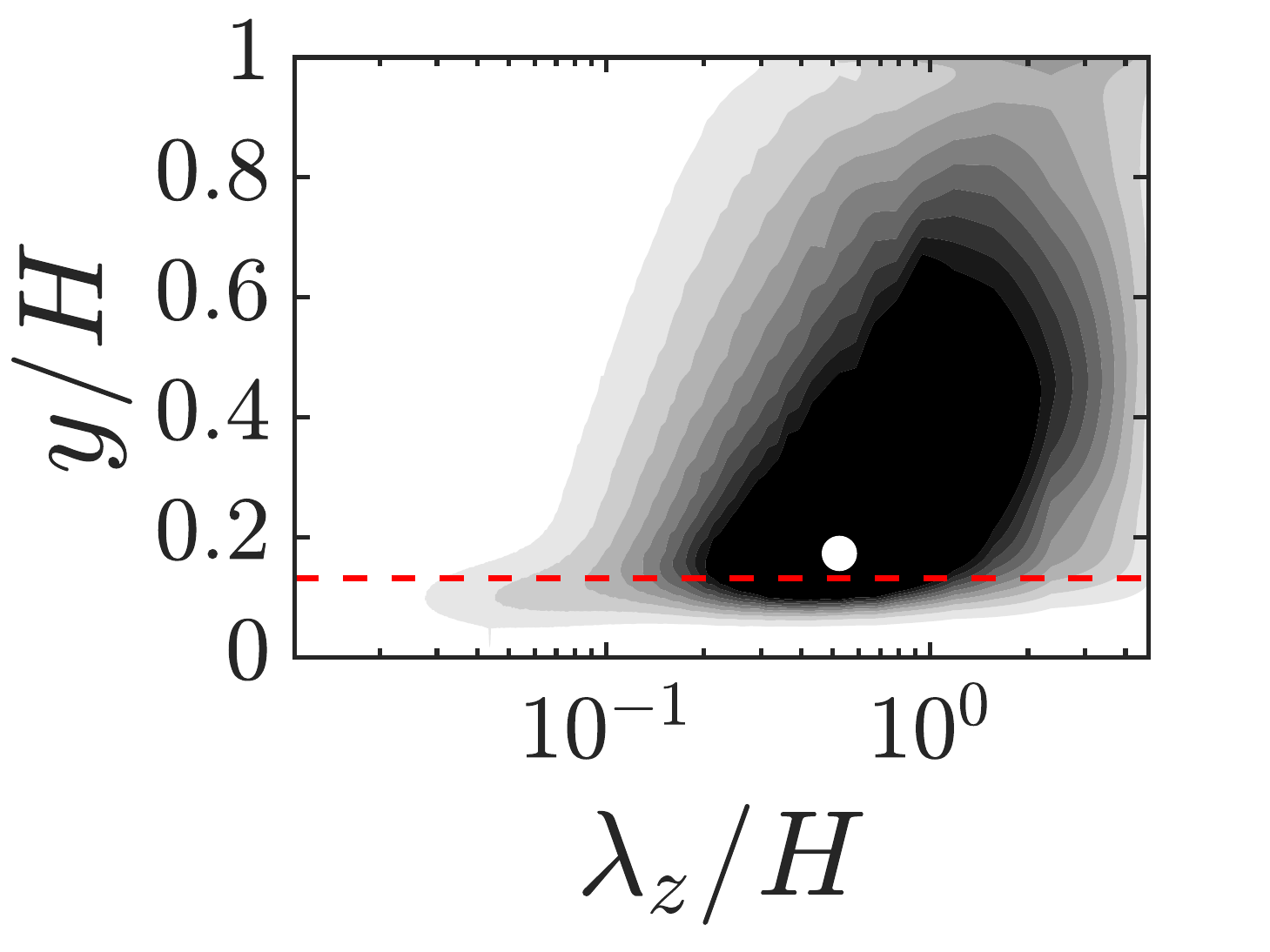}};
    \node[anchor=south west,inner sep=0] (image) at (13.5,0) {
    \includegraphics[width=.16\textwidth]{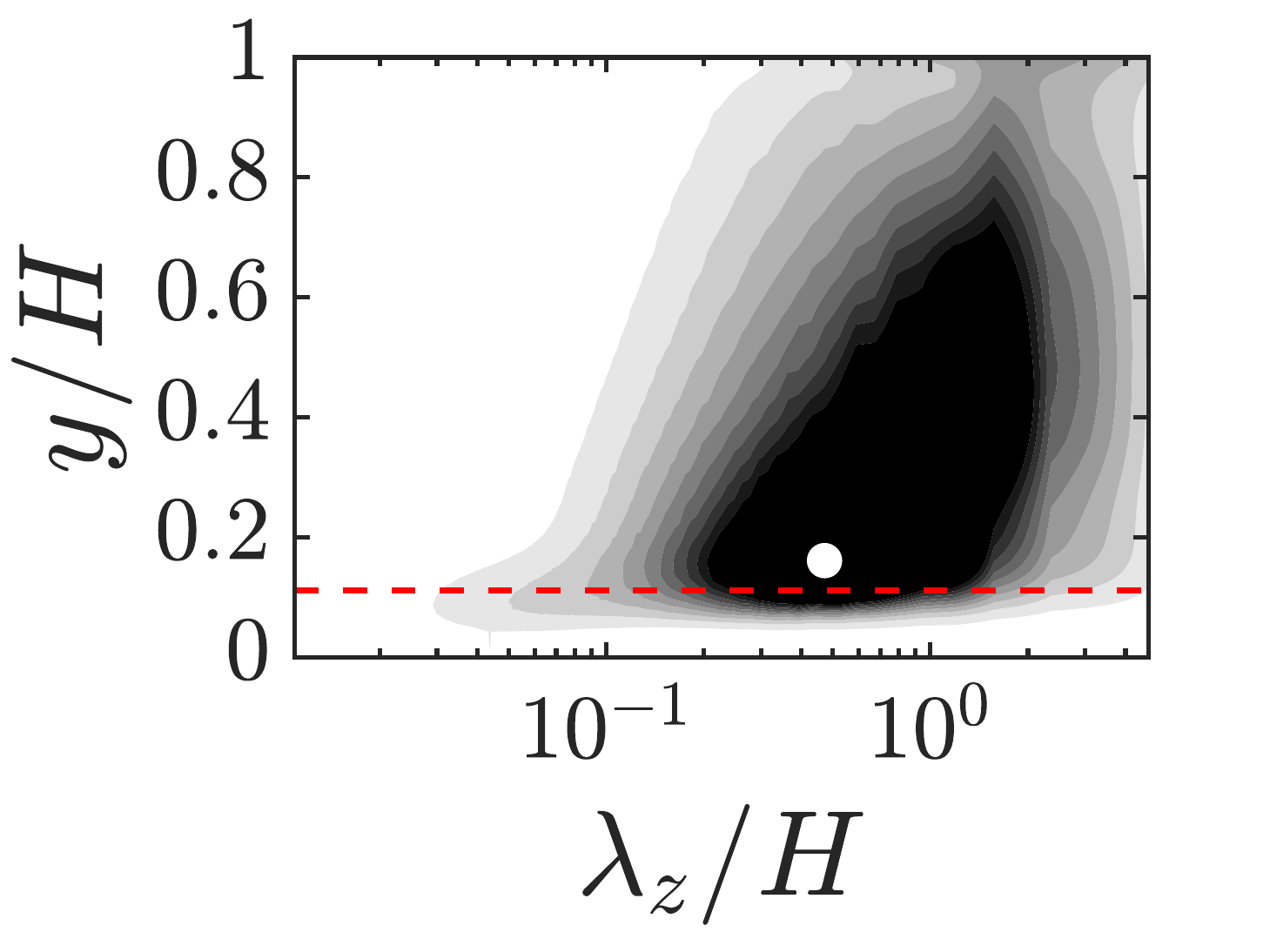}};

    \node[] at (-0.7,-1.2) {\small$\dfrac{2\pi\Phi_{v'v'}}{\lambda_z U^2}$};
    \node[anchor=south west,inner sep=0] (image) at ( 0.0,-2.50) {
    \includegraphics[width=.16\textwidth]{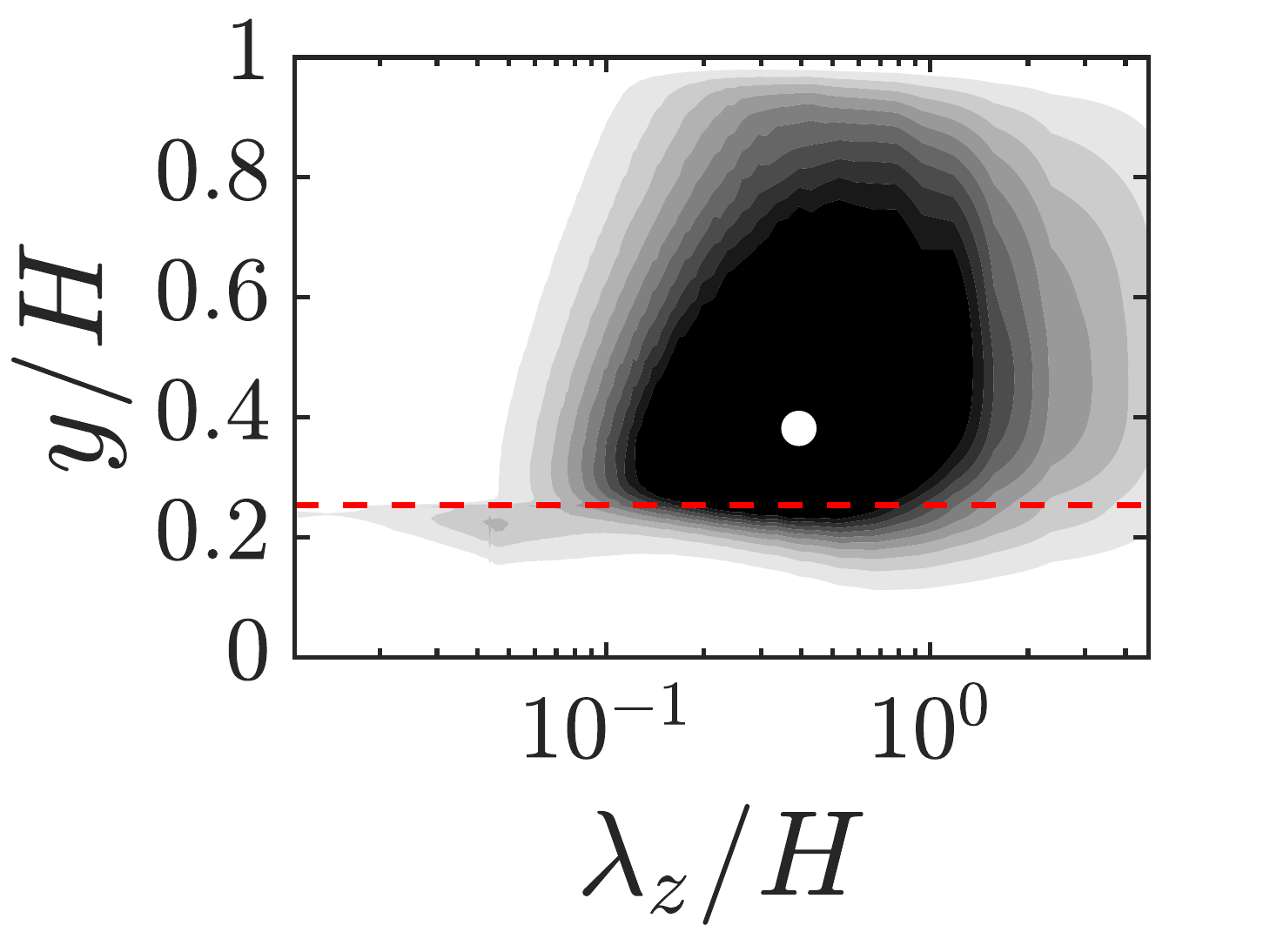}};
    \node[anchor=south west,inner sep=0] (image) at ( 2.7,-2.50) {
    \includegraphics[width=.16\textwidth]{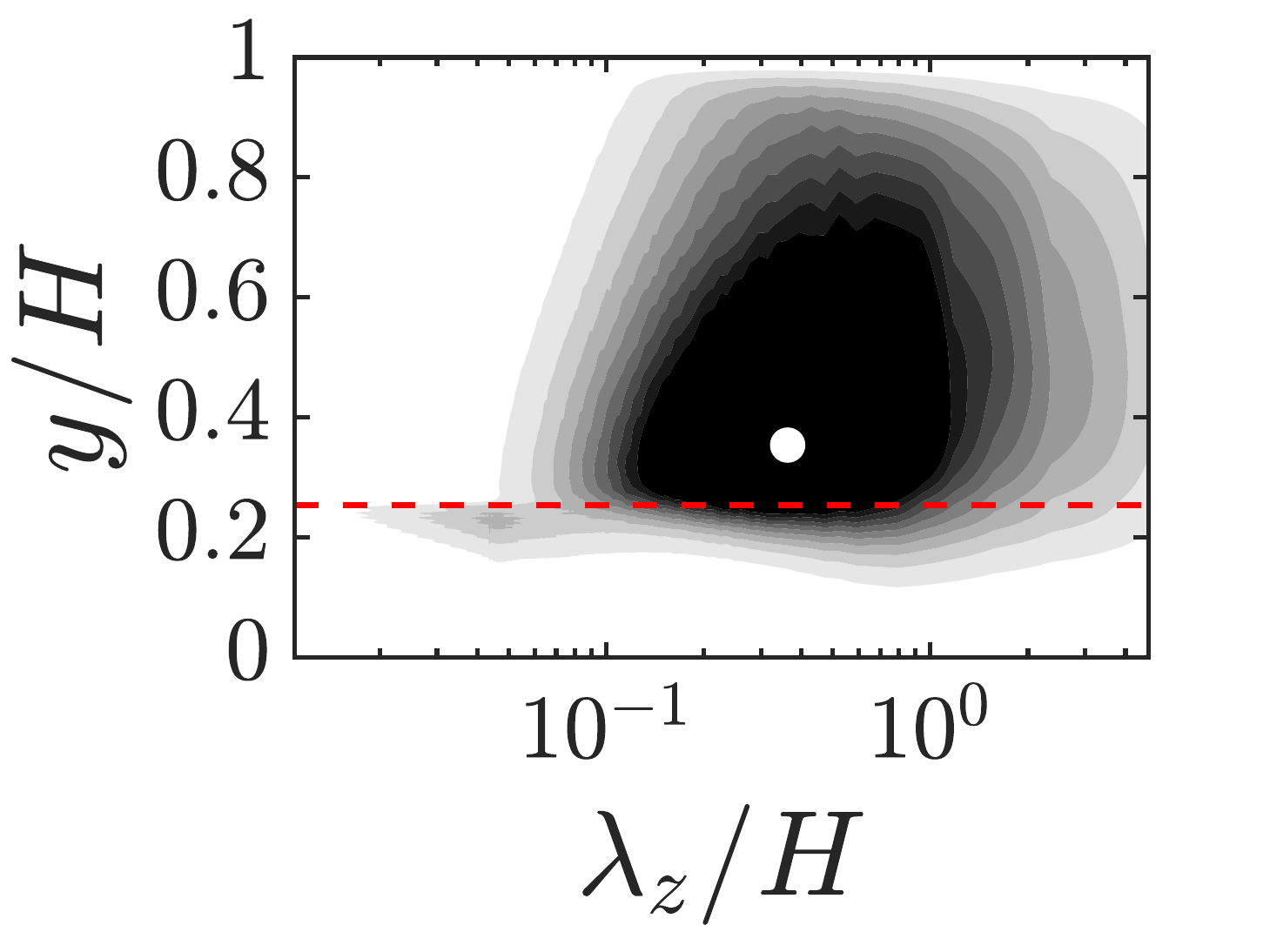}};
    \node[anchor=south west,inner sep=0] (image) at ( 5.4,-2.50) {
    \includegraphics[width=.16\textwidth]{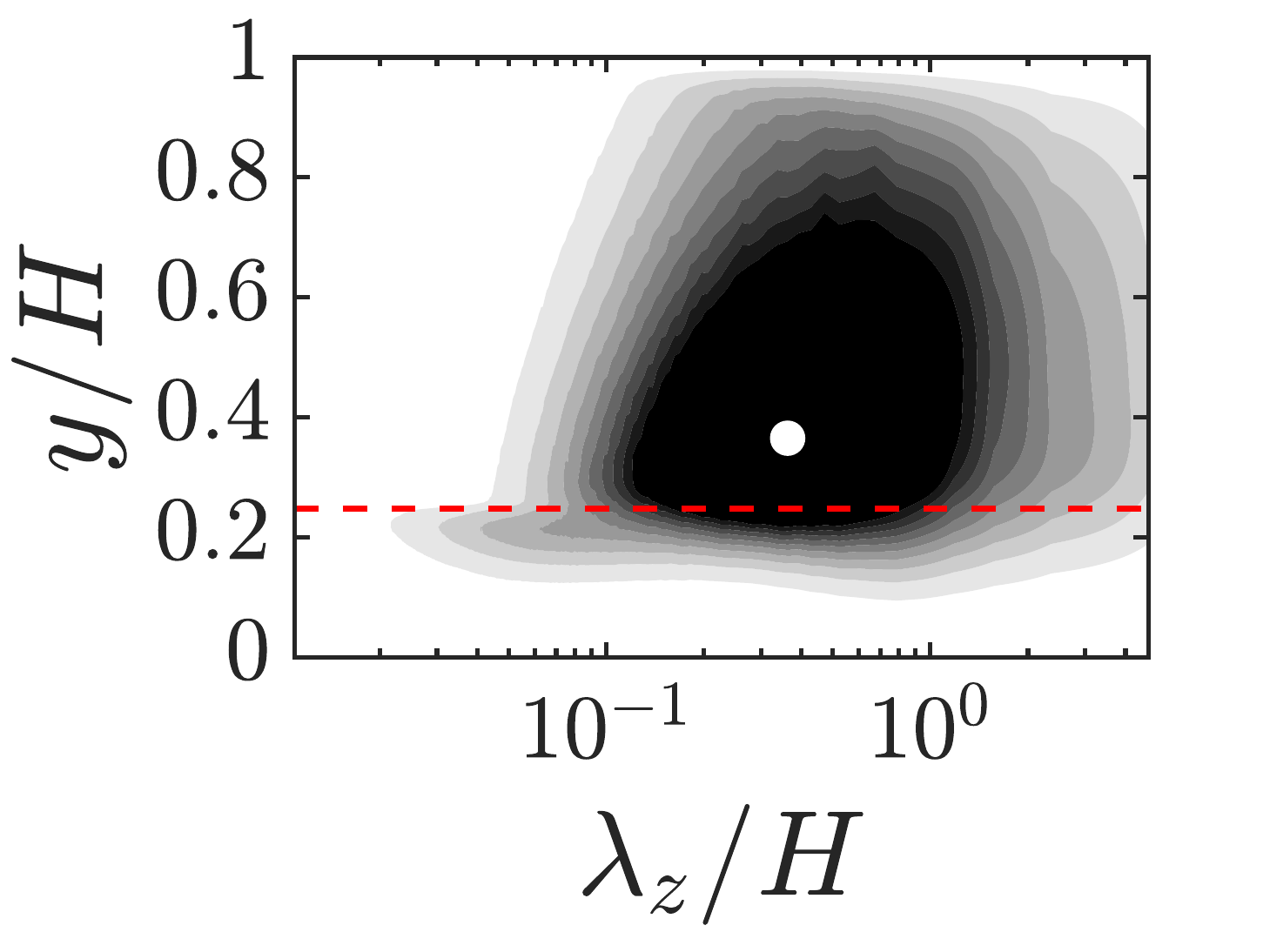}};
    \node[anchor=south west,inner sep=0] (image) at ( 8.1,-2.50) {
    \includegraphics[width=.16\textwidth]{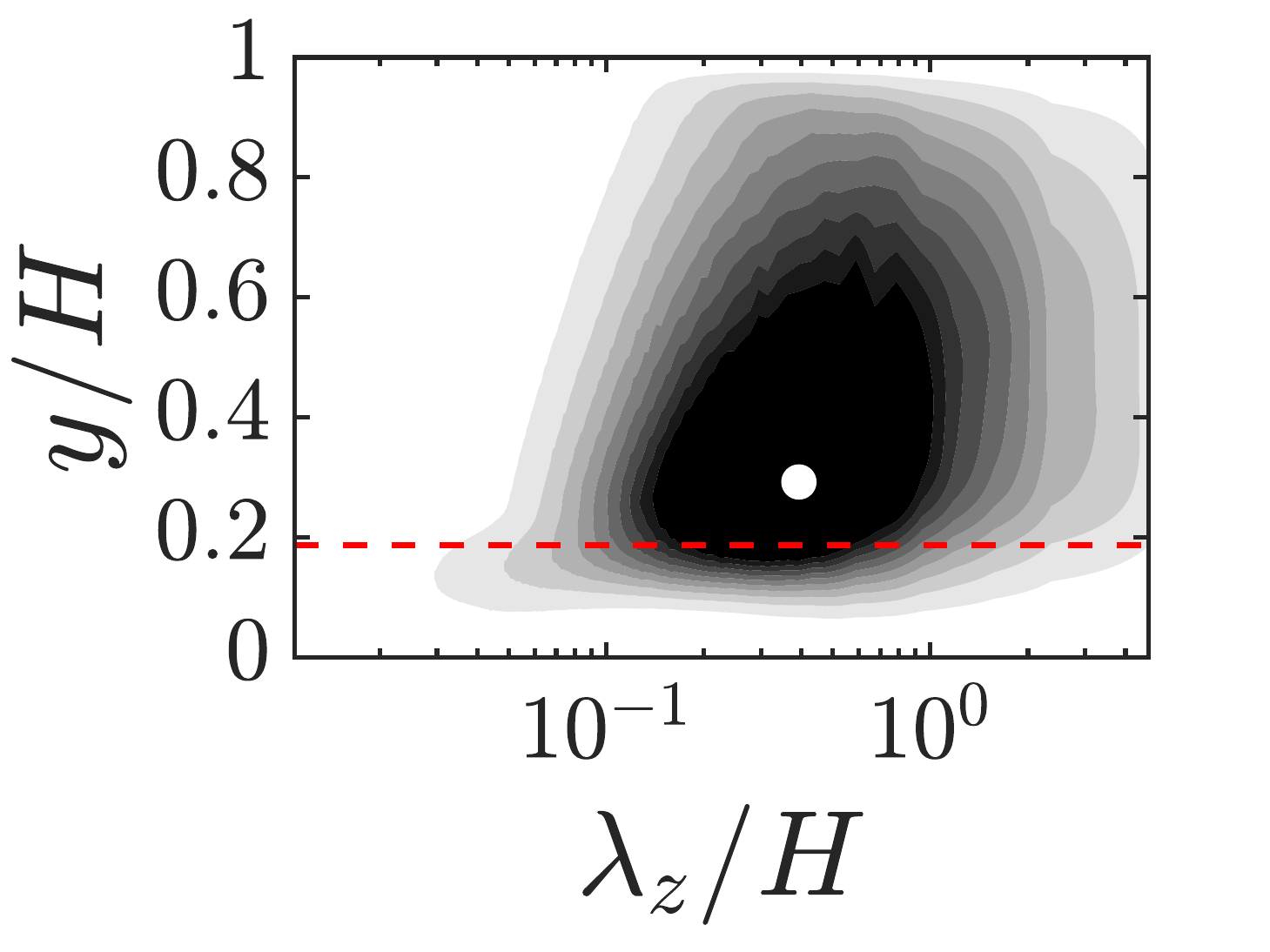}};
    \node[anchor=south west,inner sep=0] (image) at (10.8,-2.50) {
    \includegraphics[width=.16\textwidth]{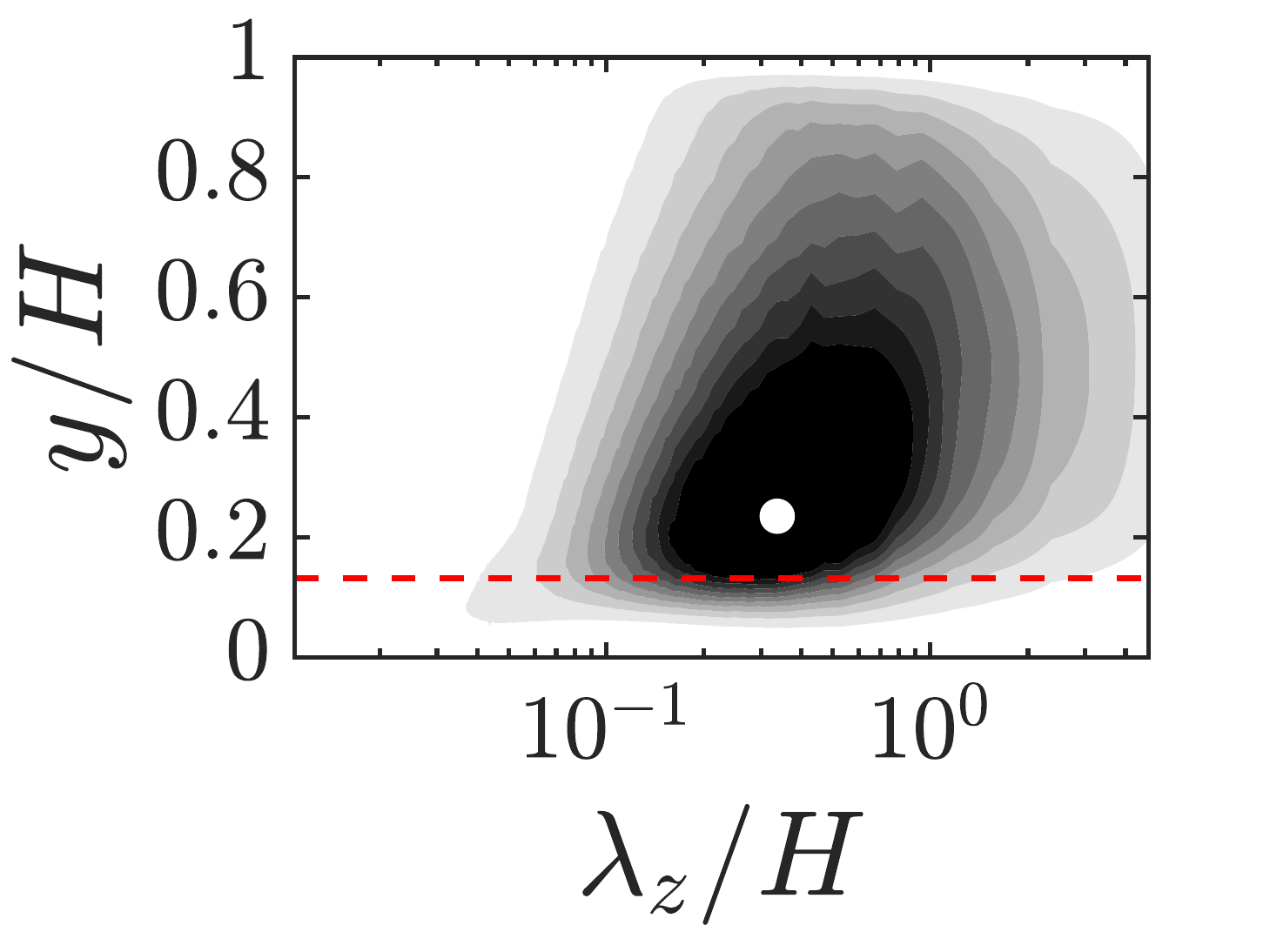}};
    \node[anchor=south west,inner sep=0] (image) at (13.5,-2.50) {
    \includegraphics[width=.16\textwidth]{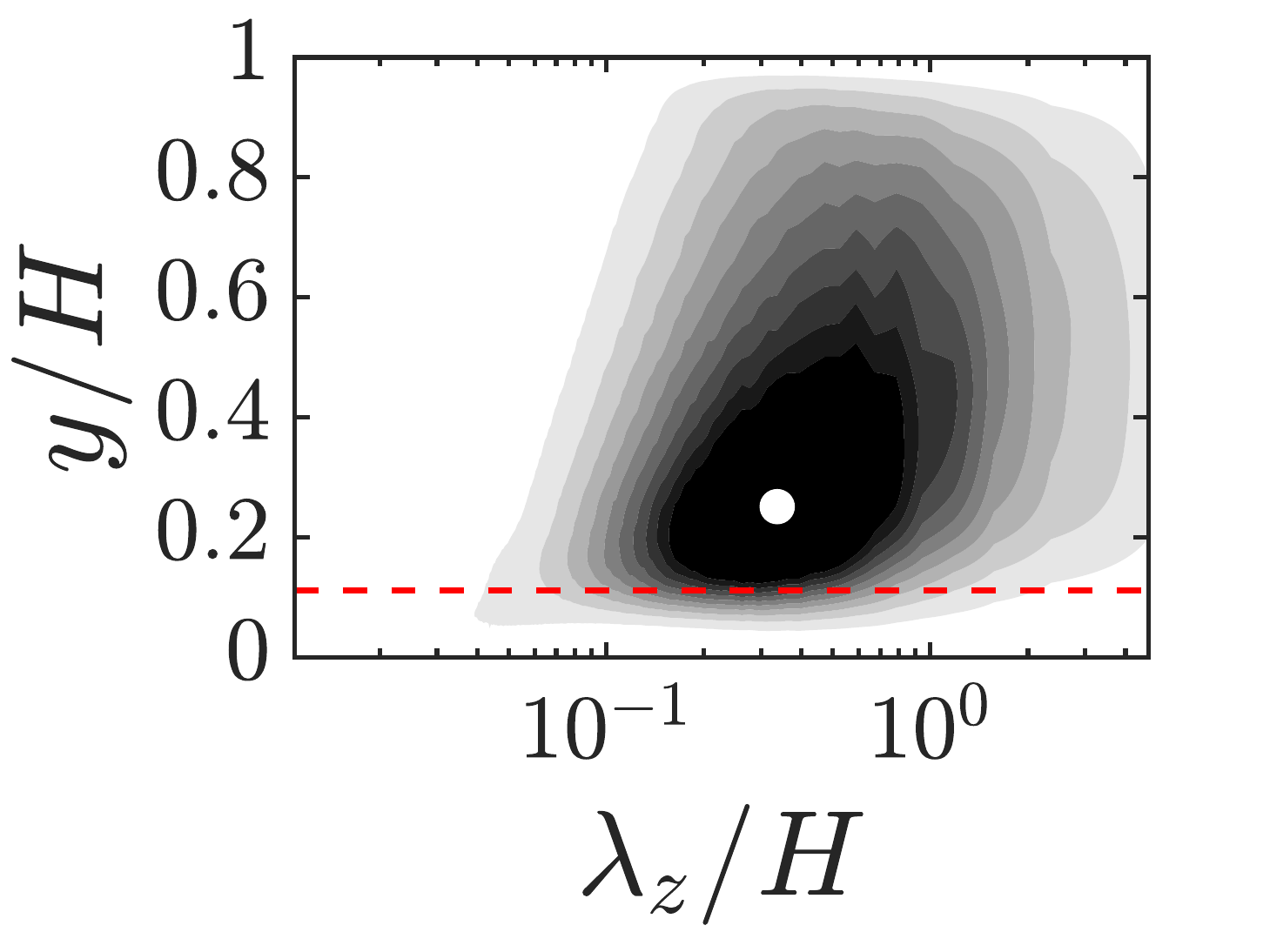}};

    \node[] at (-0.7,-3.7) {\small$\dfrac{2\pi\Phi_{w'w'}}{\lambda_z U^2}$};
    \node[anchor=south west,inner sep=0] (image) at ( 0.0,-5.0) {
    \includegraphics[width=.16\textwidth]{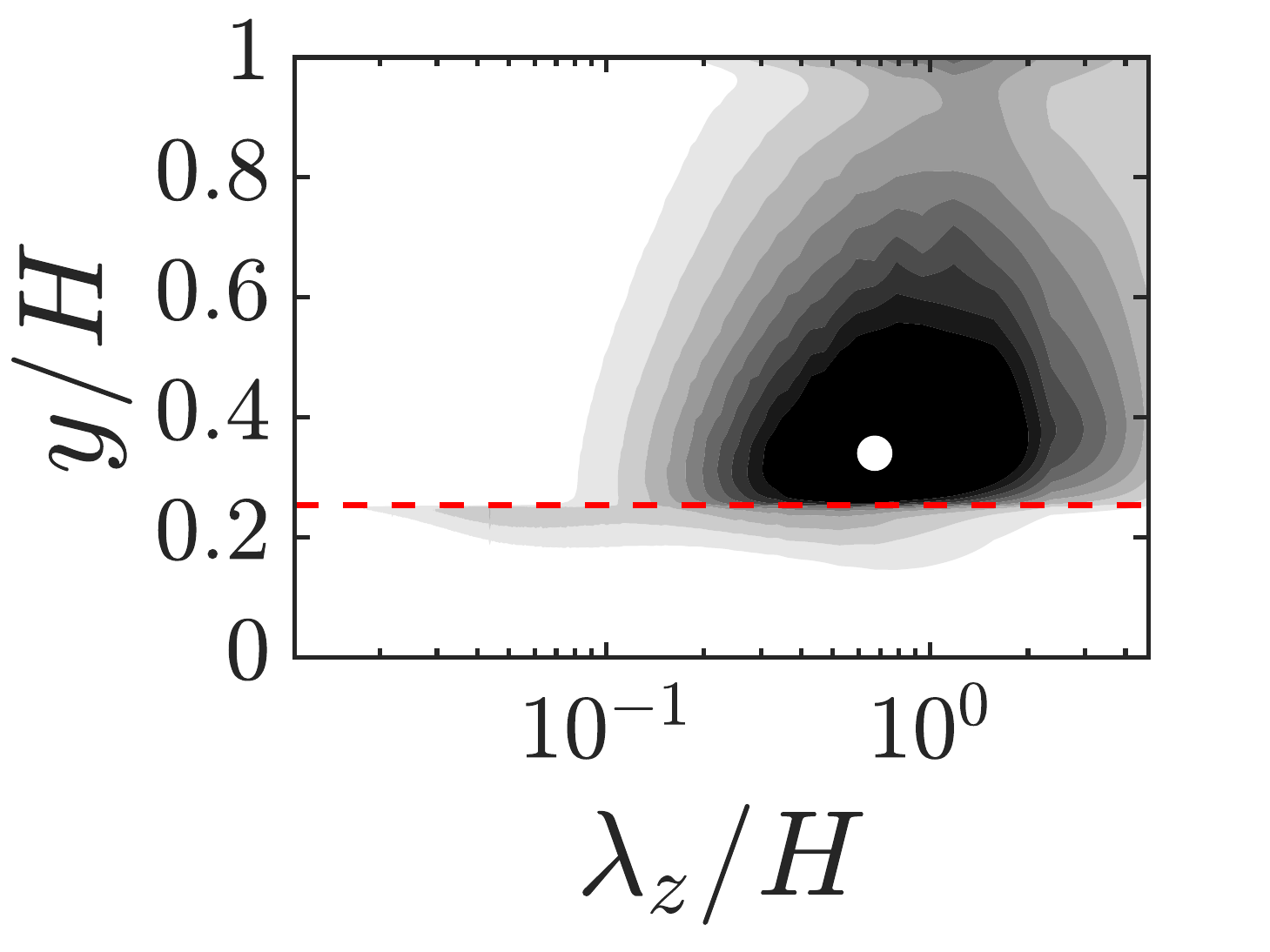}};
    \node[anchor=south west,inner sep=0] (image) at ( 2.7,-5.0) {
    \includegraphics[width=.16\textwidth]{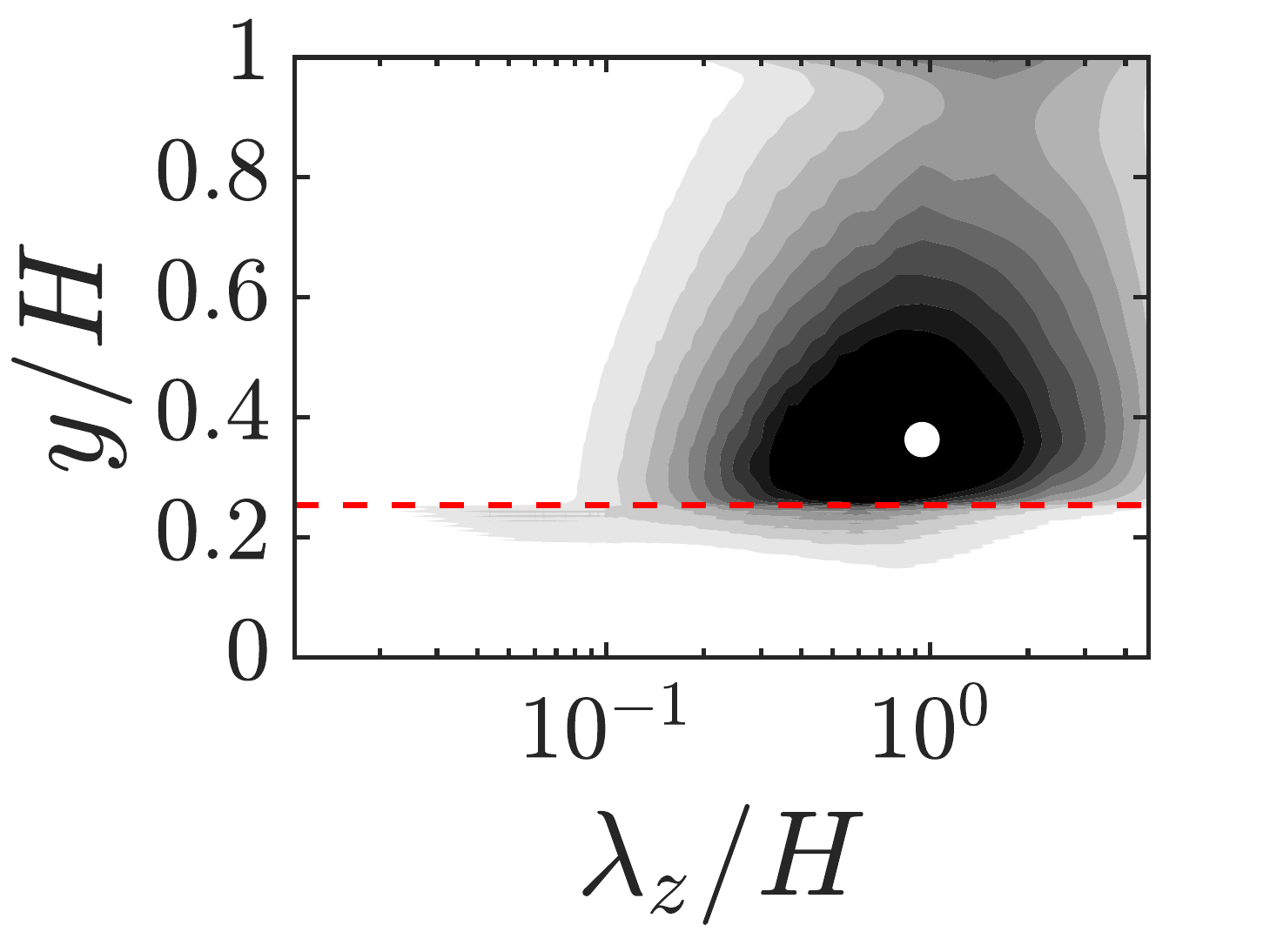}};
    \node[anchor=south west,inner sep=0] (image) at ( 5.4,-5.0) {
    \includegraphics[width=.16\textwidth]{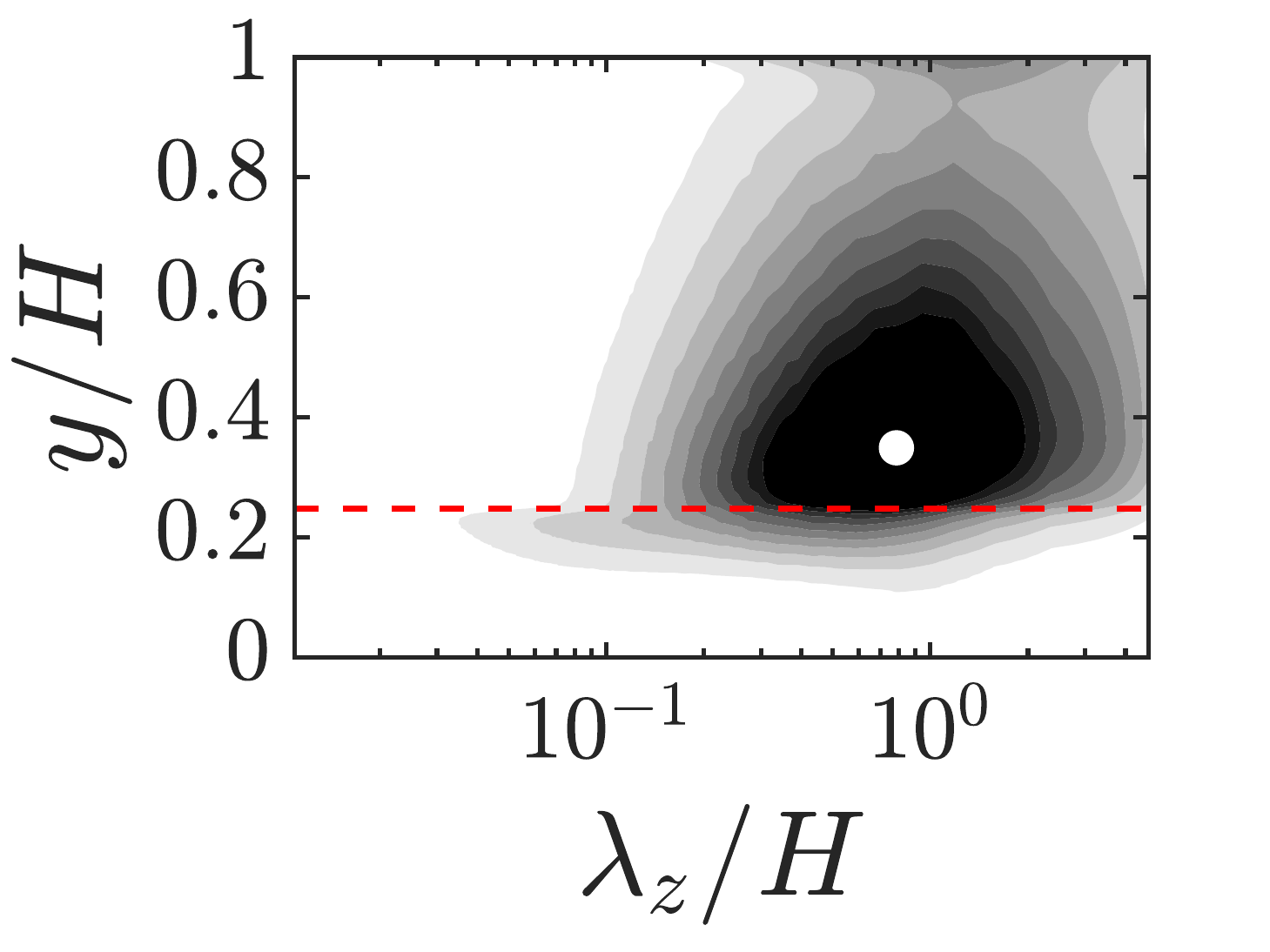}};
    \node[anchor=south west,inner sep=0] (image) at ( 8.1,-5.0) {
    \includegraphics[width=.16\textwidth]{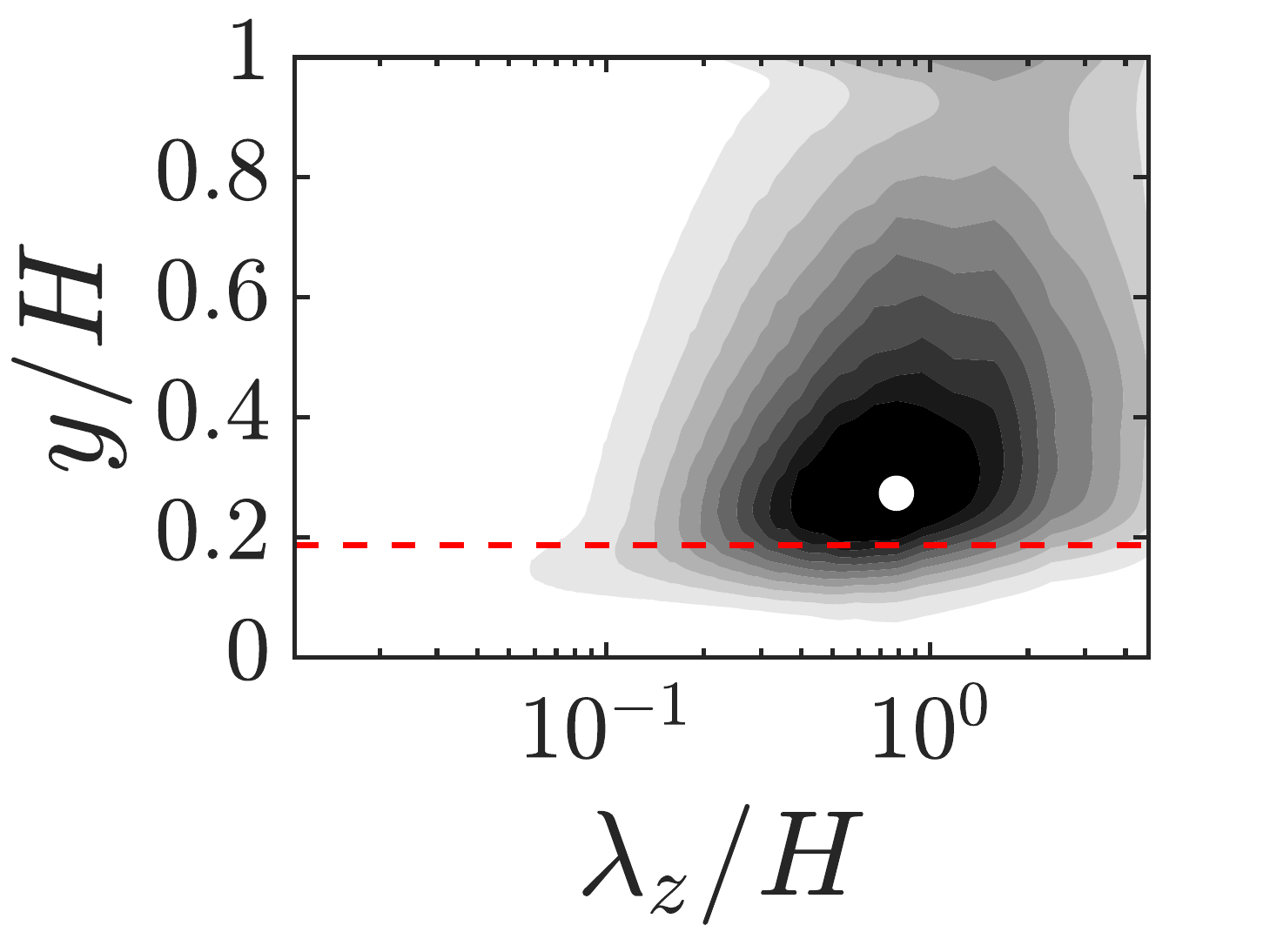}};
    \node[anchor=south west,inner sep=0] (image) at (10.8,-5.0) {
    \includegraphics[width=.16\textwidth]{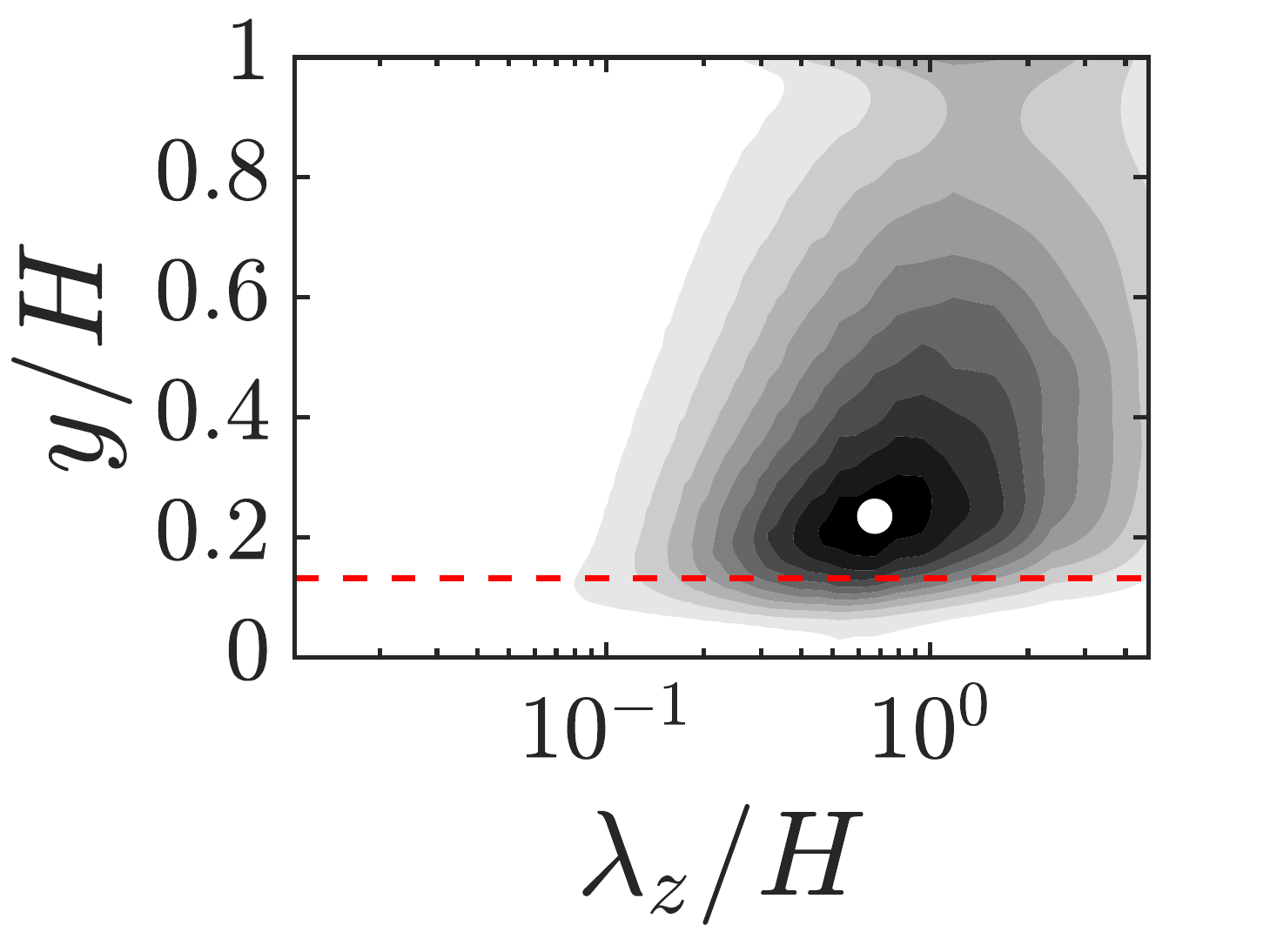}};
    \node[anchor=south west,inner sep=0] (image) at (13.5,-5.0) {
    \includegraphics[width=.16\textwidth]{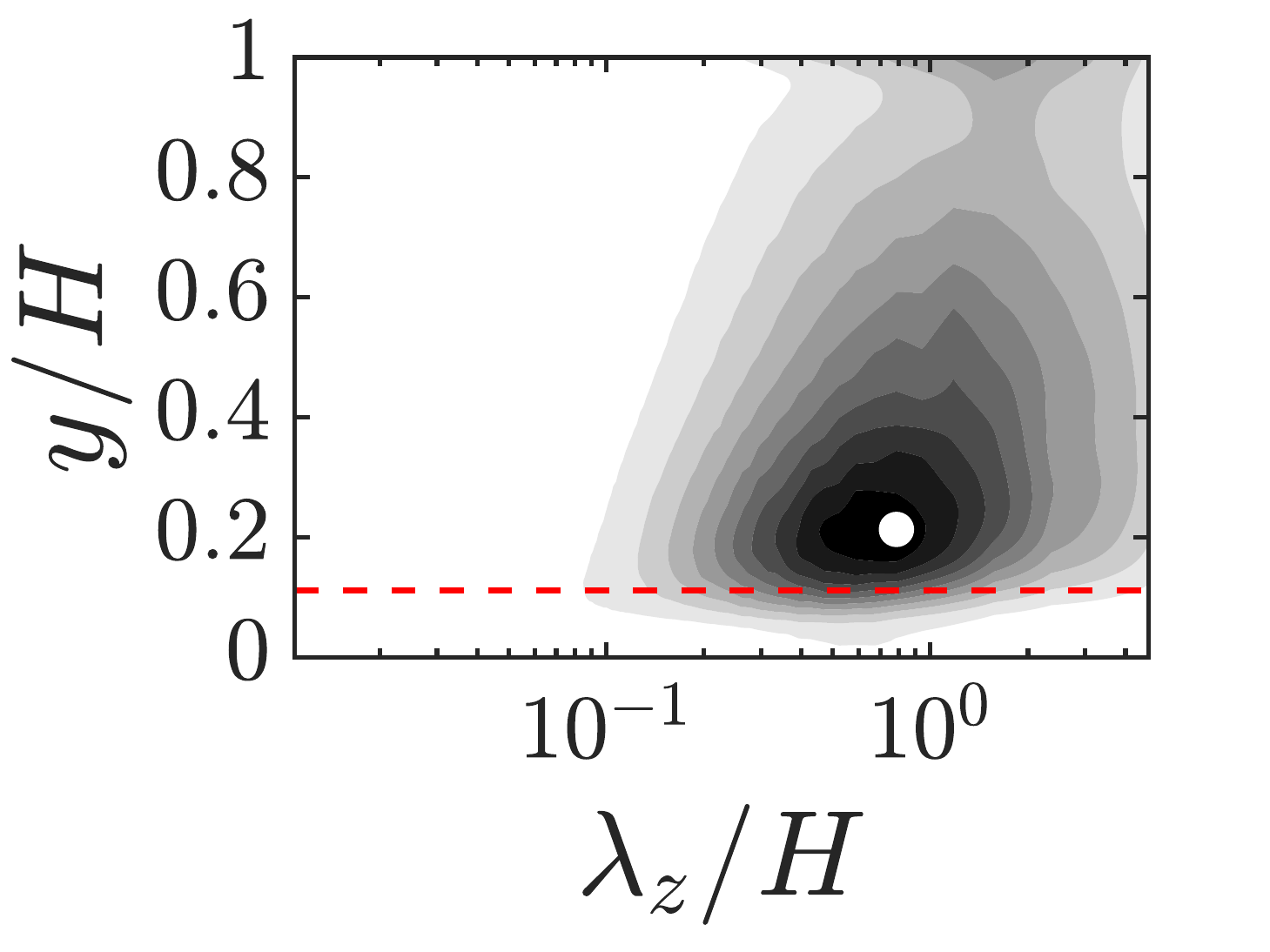}};
    \end{tikzpicture}
    \caption{
    Same as~\cref{fig3}, but as a function of the spanwise wavelength $\lambda_z/H$ and wall-normal coordinate $y/H$.
The grey levels range in: $[0, 0.005]$ with a $0.0005$ increment for the wall-normal velocity component; $[0, 0.01]$ with a $0.001$ increment for the streamwise and spanwise velocity components. \added{Again, the peak of the spectrum is not substantially affected by the variation of $\Ca$.}
    }
    \label{fig4}
\end{figure}
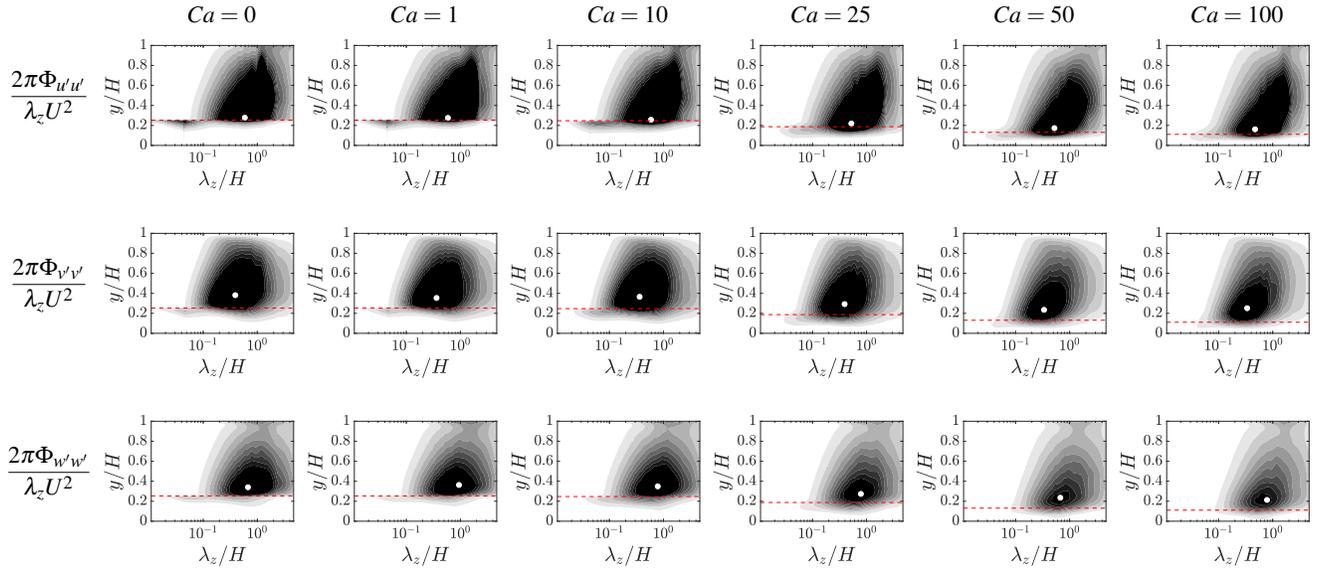

To better characterize the relevant lengthscales of the turbulent flow that are \deleted{consequently} involved in the honami/monami phenomenon, \cref{fig3} shows the magnitude of the fluid velocity spectra for all components (from top to bottom row) as a function of the streamwise wavelength $\lambda_x$ and wall-normal distance $y$.
Similarly, \cref{fig4} reports the same quantities but as a function of the spanwise wavelength $\lambda_z$.
From \cref{fig3}, a clear peak at $\lambda_x \approx \mathcal{O}(H)$ can be observed for all cases and velocity components. Specifically, for the streamwise component (top panels) the peak is found at $\lambda_x/H$ approximately between 1.2 and 1.8, while for the wall-normal component (middle panels) it is found at $\lambda_x/H \approx 0.6 \pm 0.1$, and for the spanwise component (bottom panels) at $\lambda_x/H \approx 0.8 \pm 0.1$. For all these observables, no systematic variation is reported that can be associated with the Cauchy number.
Instead, the peaks in the spectra can be associated with the large-scale coherent structures that dominate the collective motion of the canopy. 
The wall-normal location of the peak appears to gradually decrease while increasing the Cauchy number (from left to right panels), indicating that the dominant turbulent structures remain relatively close to the canopy top when the filaments get bent by the action of the flow (with a maximum offset of about $0.1H$ observed for the wall-normal and spanwise velocity components at $\Ca=100$).
Note that the observed trend consistently complements with experimental evidence recently reported for specific configurations~\cite{houseago2022turbulence}. 
Considering the content in the spanwise wavelength, shown in \cref{fig4}, similar observations can be made. Here, for the streamwise velocity component $\lambda_z/H \approx 0.5 \pm 0.1$, for the wall-normal component $\lambda_z/H \approx 0.35 \pm 0.05$ and for the spanwise component $\lambda_z/H \approx 0.8 \pm 0.1$.
Again, the variation in both the streamwise and spanwise wavelength of the peak is rather minimal and does not appear to follow a clear trend.
This evidence further confirms that the canopy flexibility (and more generally the structural dynamics) does not have a crucial effect in what concerns the spatial features of the flow-structure interaction.
\added{Note that the premultiplied spectra shown in \cref{fig3,fig4} are scaled by the bulk velocity $U$, in order to better show the large-scale peak. An alternative normalization, using the local friction velocity to better visualize the flow structures in the intra-canopy region, is provided in \cref{fig3_frict,fig4_frict}.}

\added{A further confirmation of the marginal role of the Cauchy number on the characterization of the honami/monami comes from the joint probability density function (JPDF) of the streamwise and wall-normal velocity fluctuations components, computed on a plane parallel to the wall at the average canopy tip. As \cref{fig:uvjpdf} shows, the shape of the JPDF remains substantially the same varying the Cauchy number, with more frequent weak ejections (peak in the second quadrant) and fewer strong sweep events (tails in the forth quadrant). The effect of the Cauchy number can only be noticed with an attenuation of the strong sweep events increasing the flexibility of the canopy stems, in agreement with the literature~\cite{ghisalberti2006structure}.}

 \begin{figure}[t!]
    \centering
    \begin{tikzpicture}
    \node[] at ( 2.3,3.4) {\small$2\pi f\Phi_{u'u'}/(U^2\lambda_{x,z})$};
    \node[] at ( 6.3,3.4) {\small$2\pi f\Phi_{v'v'}/(U^2\lambda_{x,z})$};
    \node[] at (10.3,3.4) {\small$2\pi f\Phi_{w'w'}/(U^2\lambda_{x,z})$};
    \node[] at (14.3,3.4) {\small$2\pi f\Phi_{Y'Y'}/(H^2\lambda_{x,z})$};

    \node[anchor=south west,inner sep=0] (image) at ( 0.0,0.0) {
    \includegraphics[width=.23\textwidth]{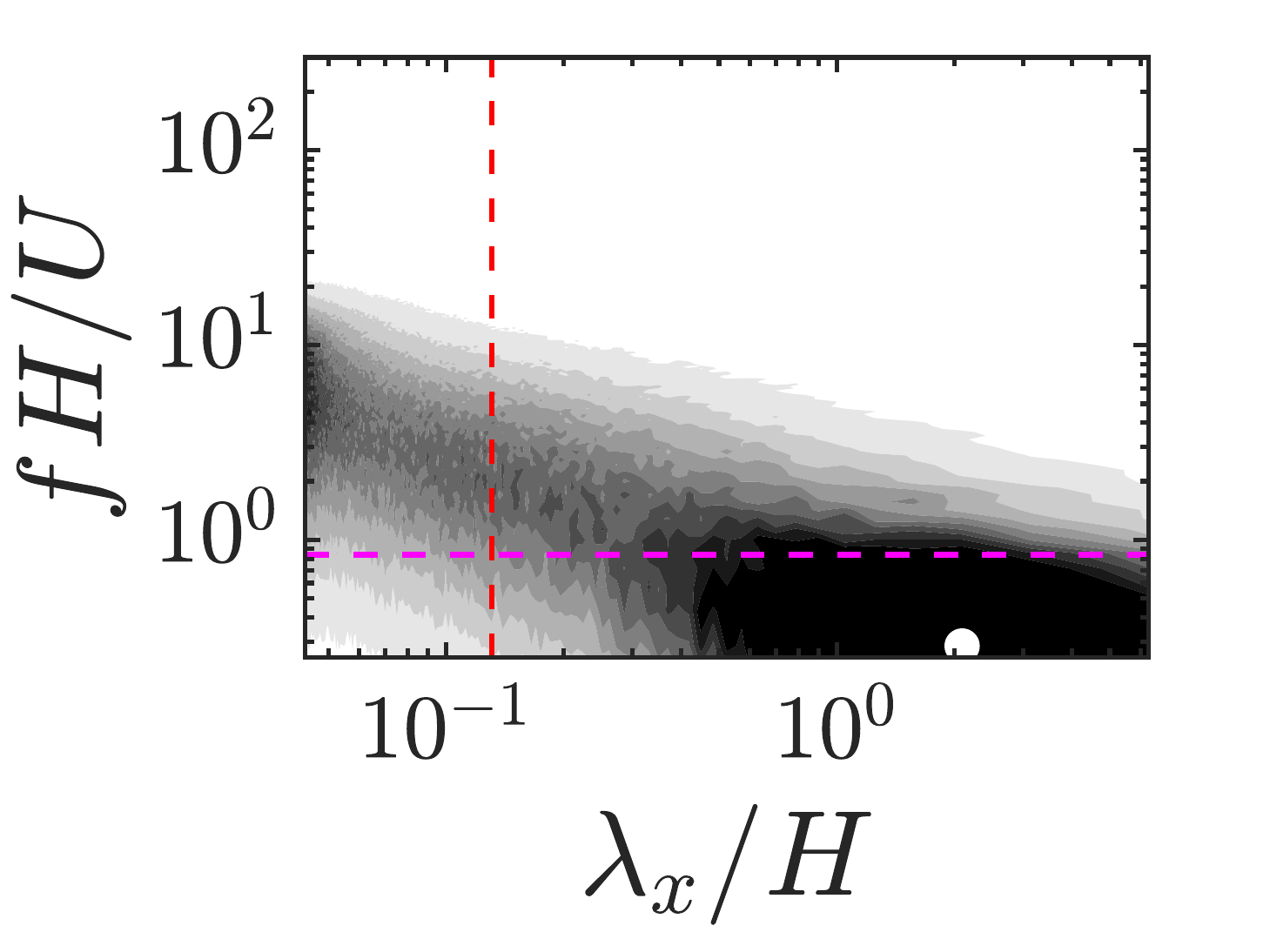}};
    \node[anchor=south west,inner sep=0] (image) at ( 4.0,0.0) {
    \includegraphics[width=.23\textwidth]{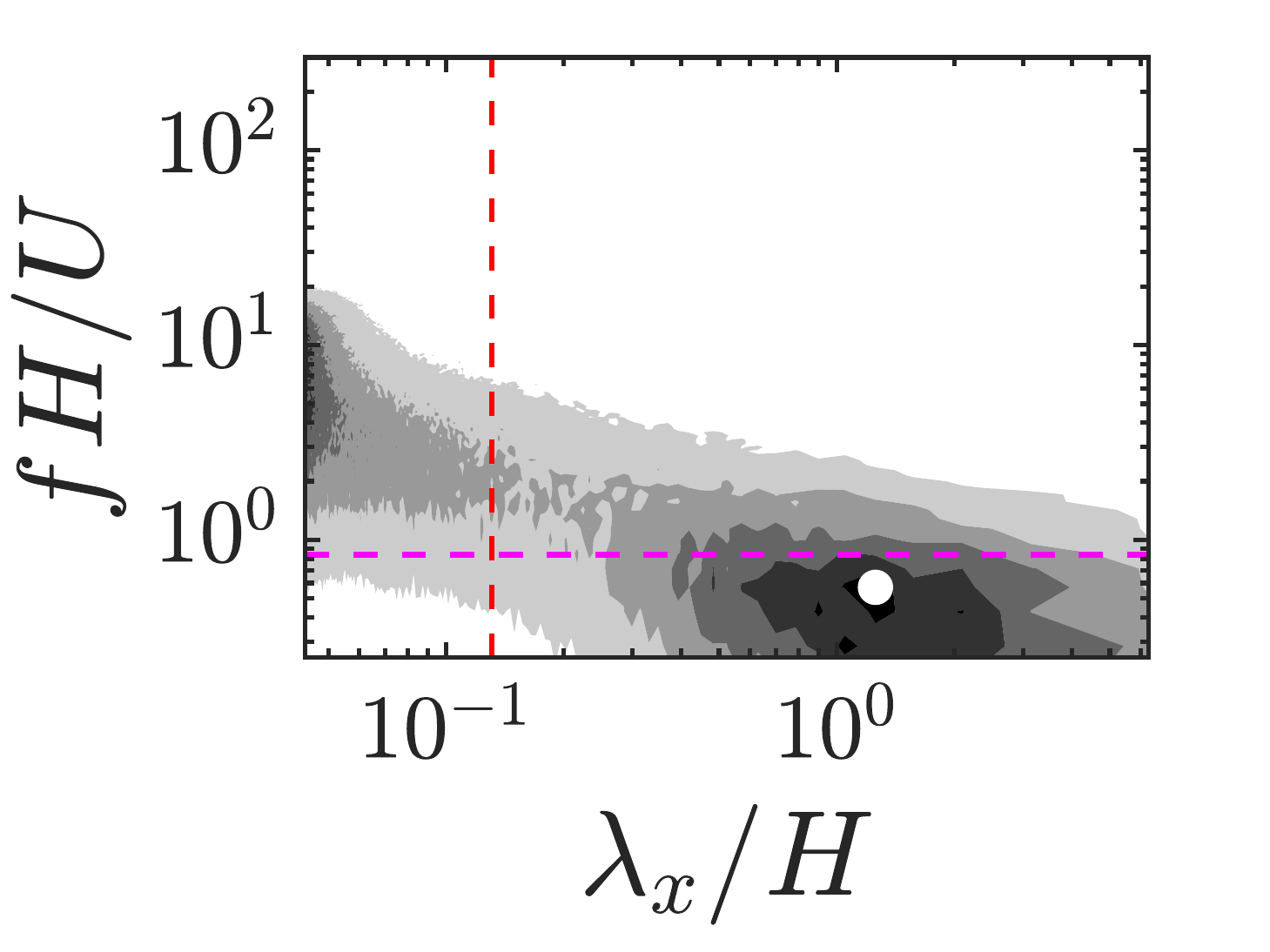}};
    \node[anchor=south west,inner sep=0] (image) at ( 8.0,0.0) {
    \includegraphics[width=.23\textwidth]{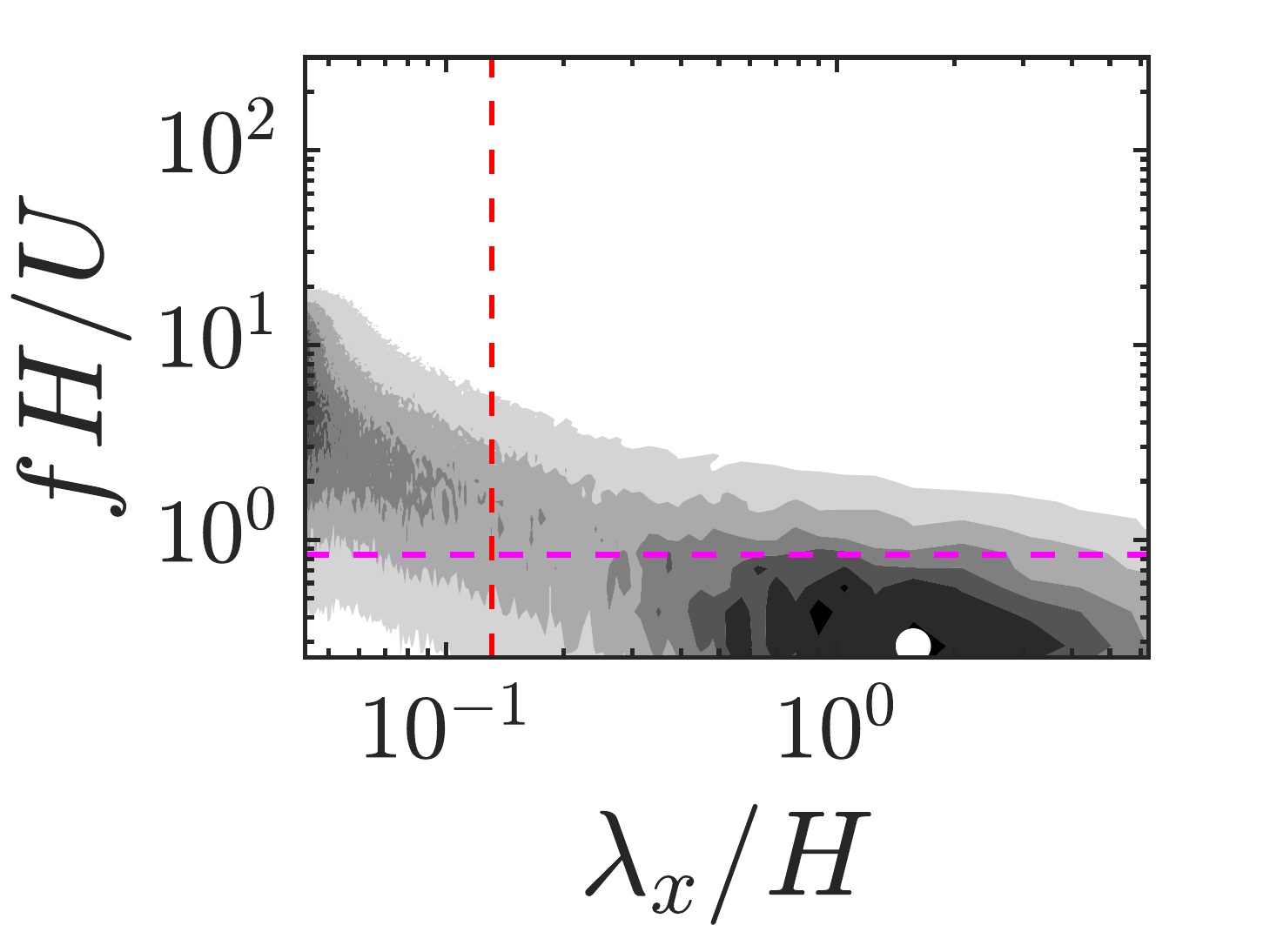}};
    \node[anchor=south west,inner sep=0] (image) at (12.0,0.0) {
    \includegraphics[width=.23\textwidth]{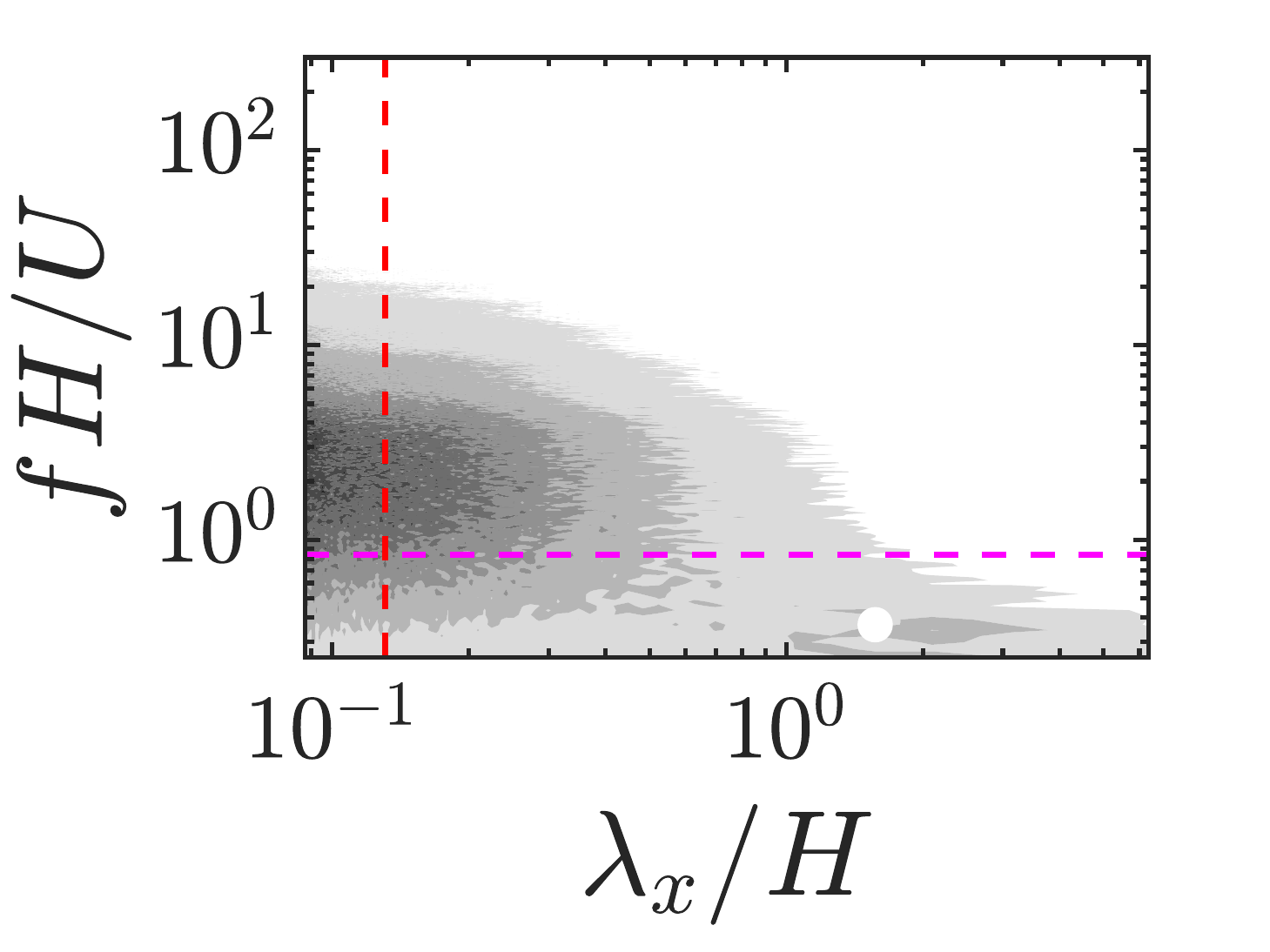}};
    \node[anchor=south west,inner sep=0] (image) at ( 0.0,-3.0) {
    \includegraphics[width=.23\textwidth]{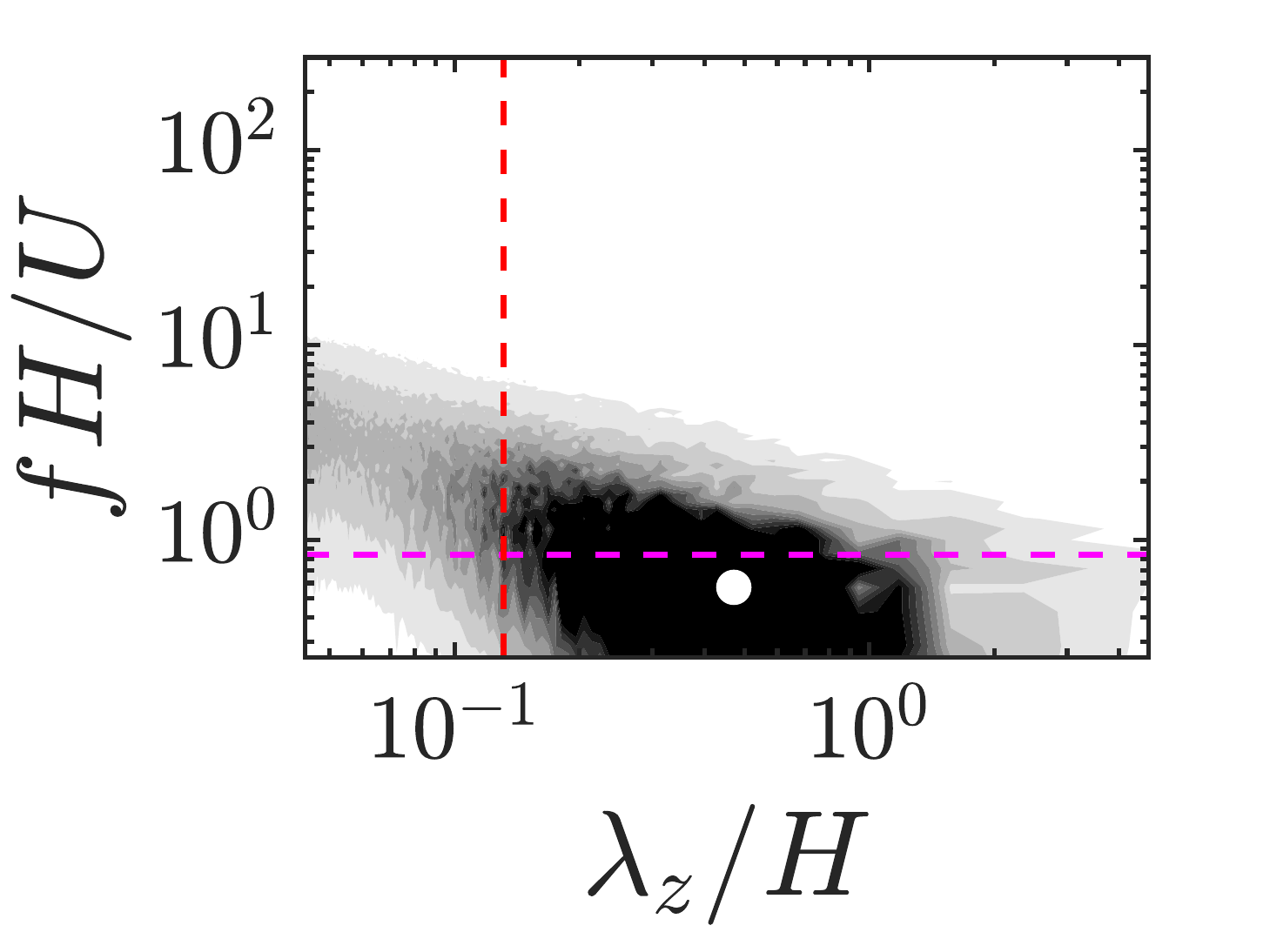}};
    \node[anchor=south west,inner sep=0] (image) at ( 4.0,-3.0) {
    \includegraphics[width=.233\textwidth]{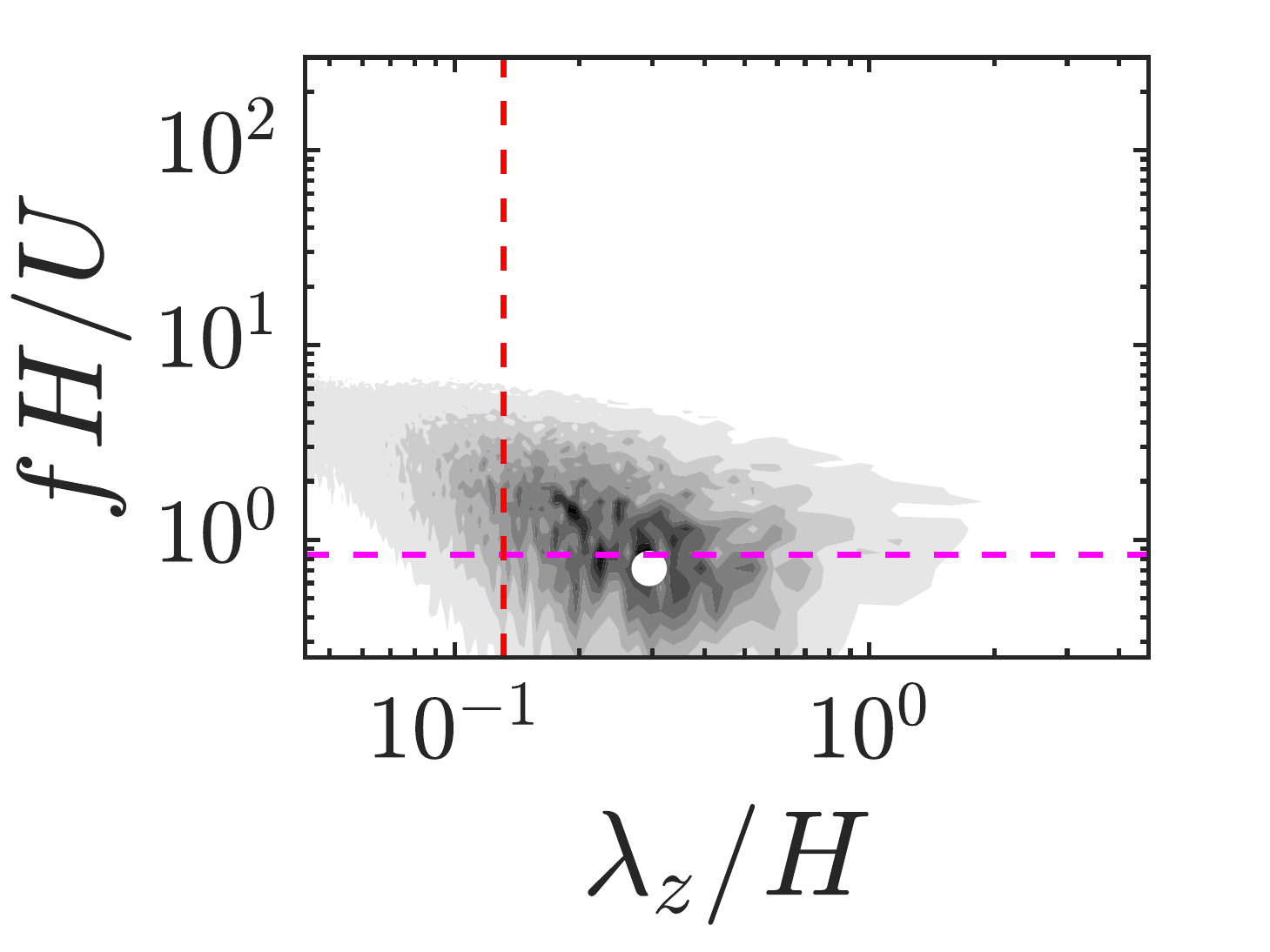}};
    \node[anchor=south west,inner sep=0] (image) at ( 8.0,-3.0) {
    \includegraphics[width=.233\textwidth]{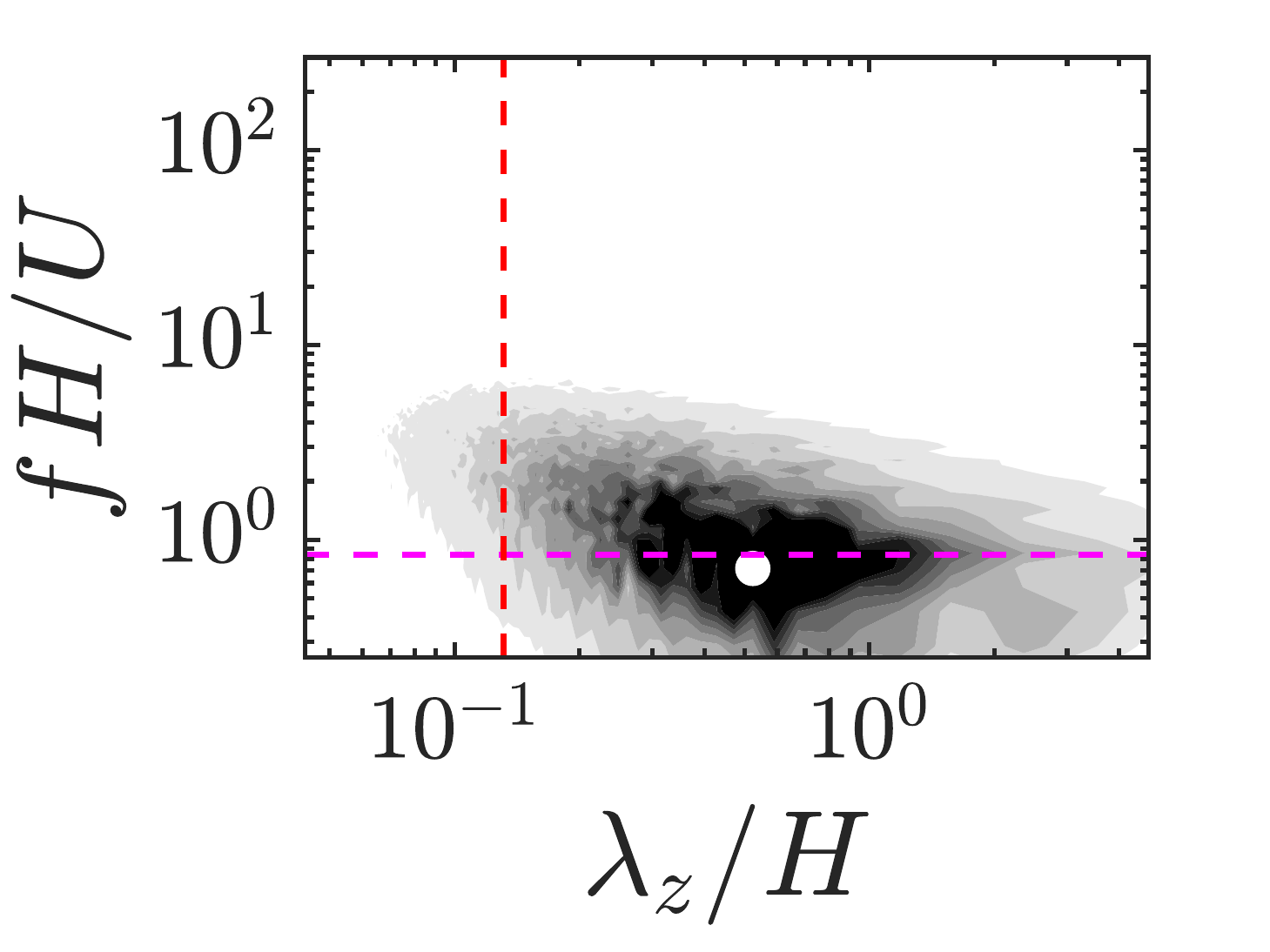}};
    \node[anchor=south west,inner sep=0] (image) at (12.0,-3.0) {
    \includegraphics[width=.23\textwidth]{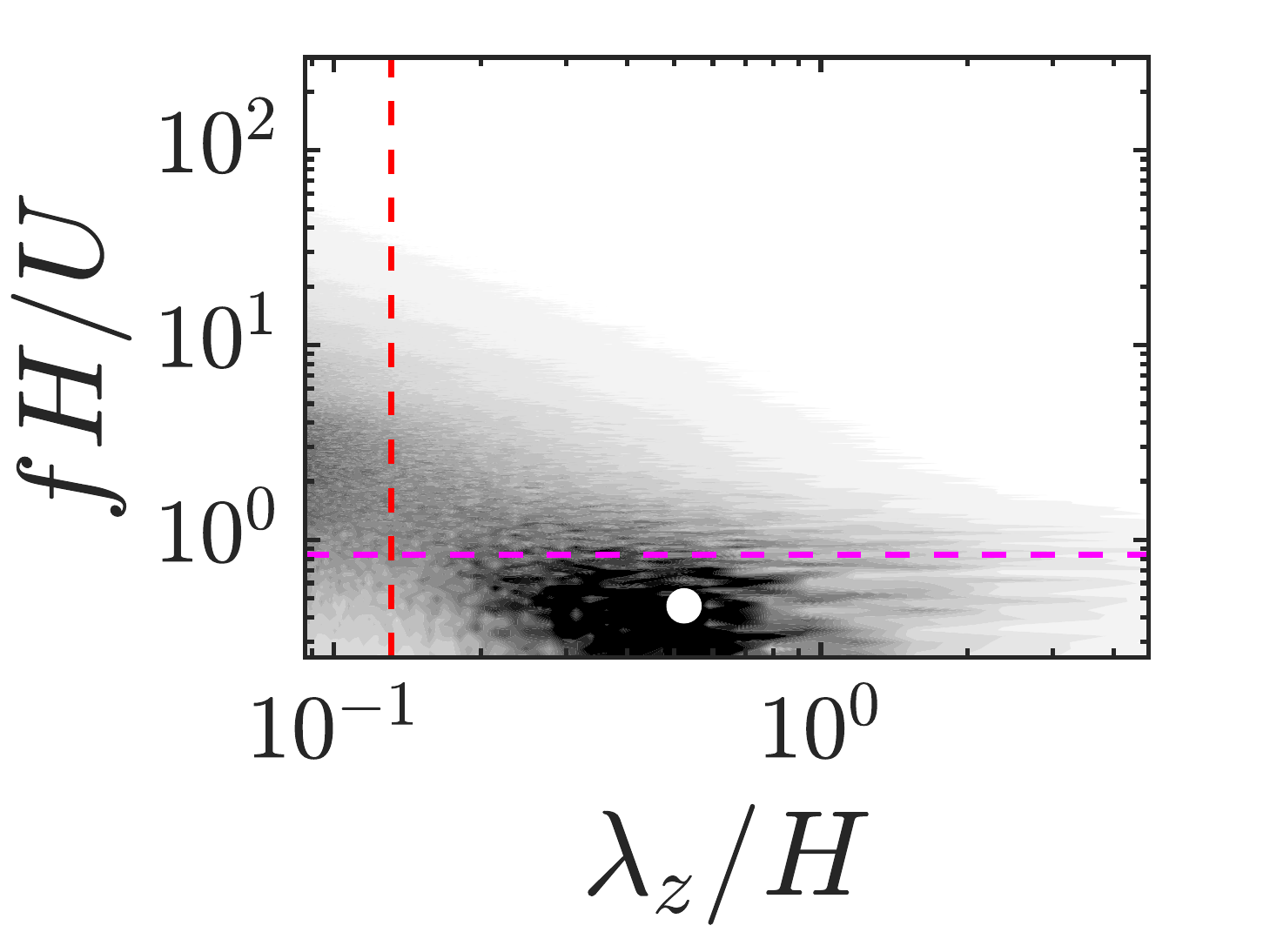}};
    \end{tikzpicture}
    \caption{
    Premultiplied spatio-temporal spectra of the fluid velocity,
	 $2\pi f\Phi_{u'u'}/(U^2\lambda_{x,z})$ (where $U$ is the bulk velocity and $u'$ the generic component of the velocity fluctuations),
	 probed at the canopy average height (from left to right: streamwise, wall-normal and spanwise component) and canopy top wall-normal displacement (rightmost panel), for a representative case with $\Ca=50$. Top and bottom panels show the spatial content in the streamwise and spanwise wavelengths, respectively.
  The vertical red line indicates the wavelength associated to the averaged vertical height of the filament tip, whereas the horizontal violet line corresponds to the first natural frequency of the filament.
  \added{The white marker denotes the peaks of the spectra.}
The grey levels range in: $[0, 2\times 10^{-5}]$ with a $2\times 10^{-6}$ increment for the fluid probed at the canopy average height in the streamwise direction; $[0, 2\times 10^{-4}]$ with a $2\times 10^{-5}$ increment for the fluid probed at the canopy average height in the spanwise direction; $[0, 10^{-6}]$ with a $5\times 10^{-8}$ increment for the canopy top wall-normal displacement of the tips (rightmost panels).
 }
    \label{fig5}
\end{figure}
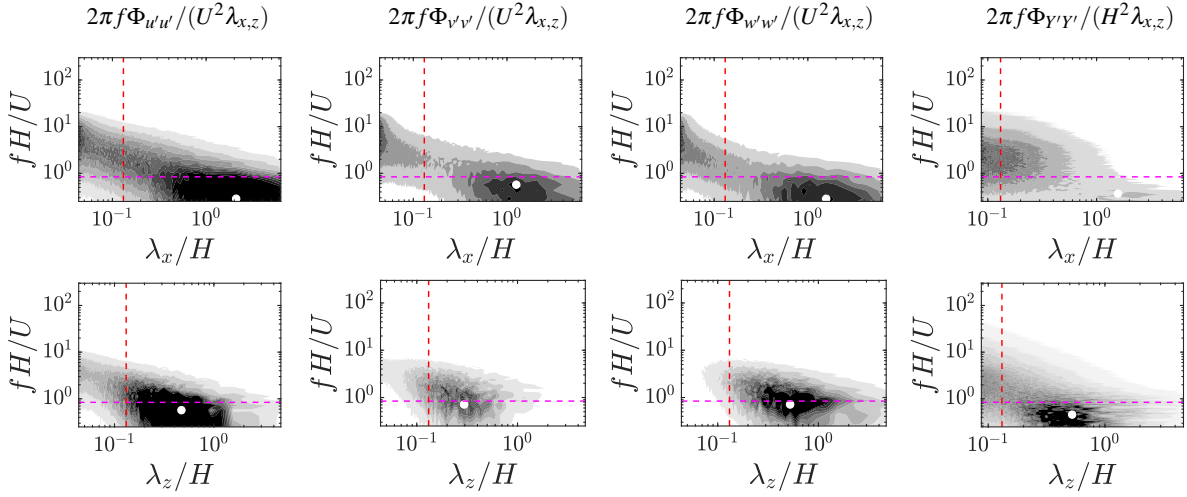

\subsection*{Spatio-\added{temporal} properties of honami/monami}

\begin{figure}
    \centering
    \includegraphics[width=.40\textwidth]{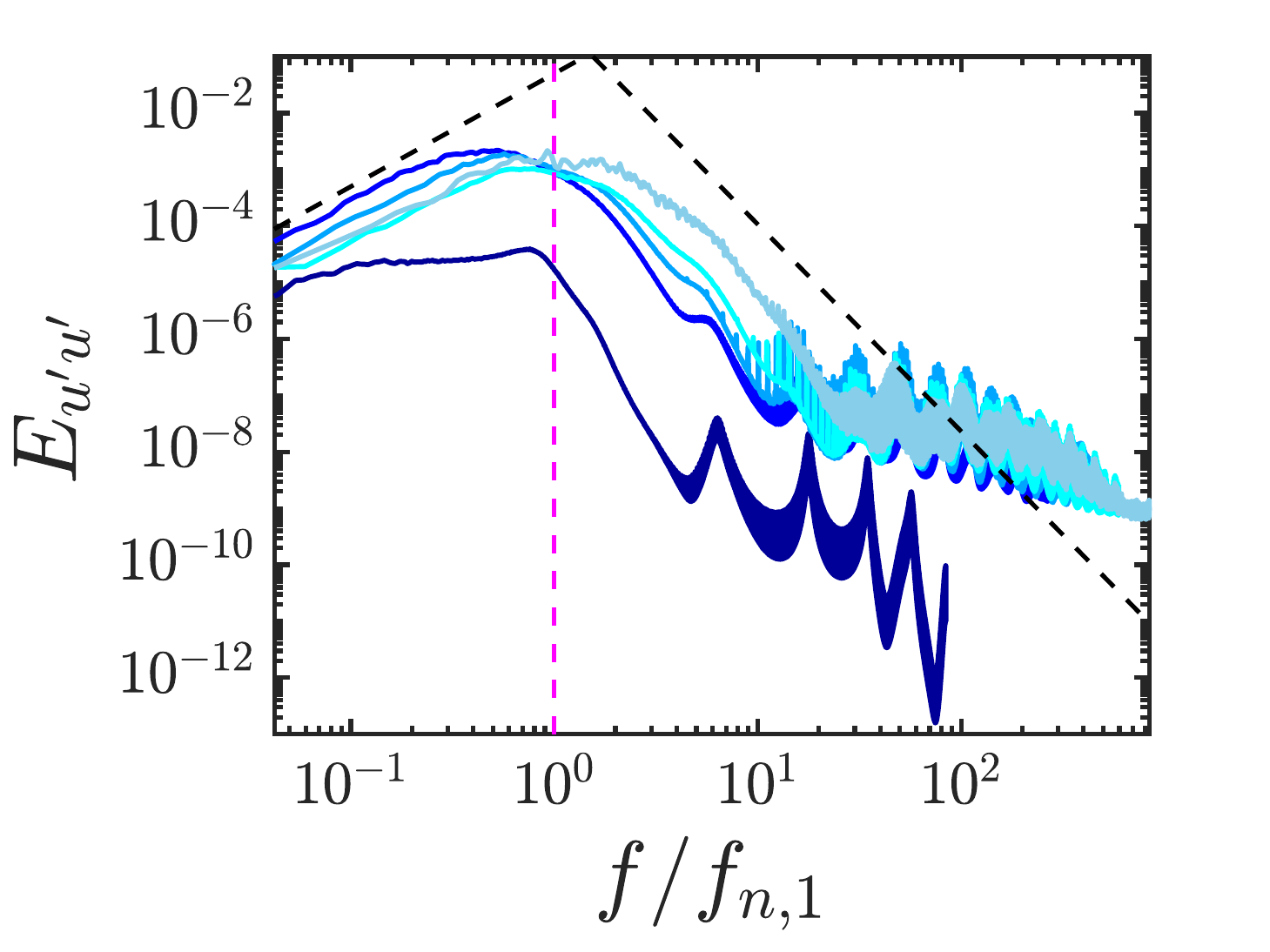}
    \includegraphics[width=.40\textwidth]{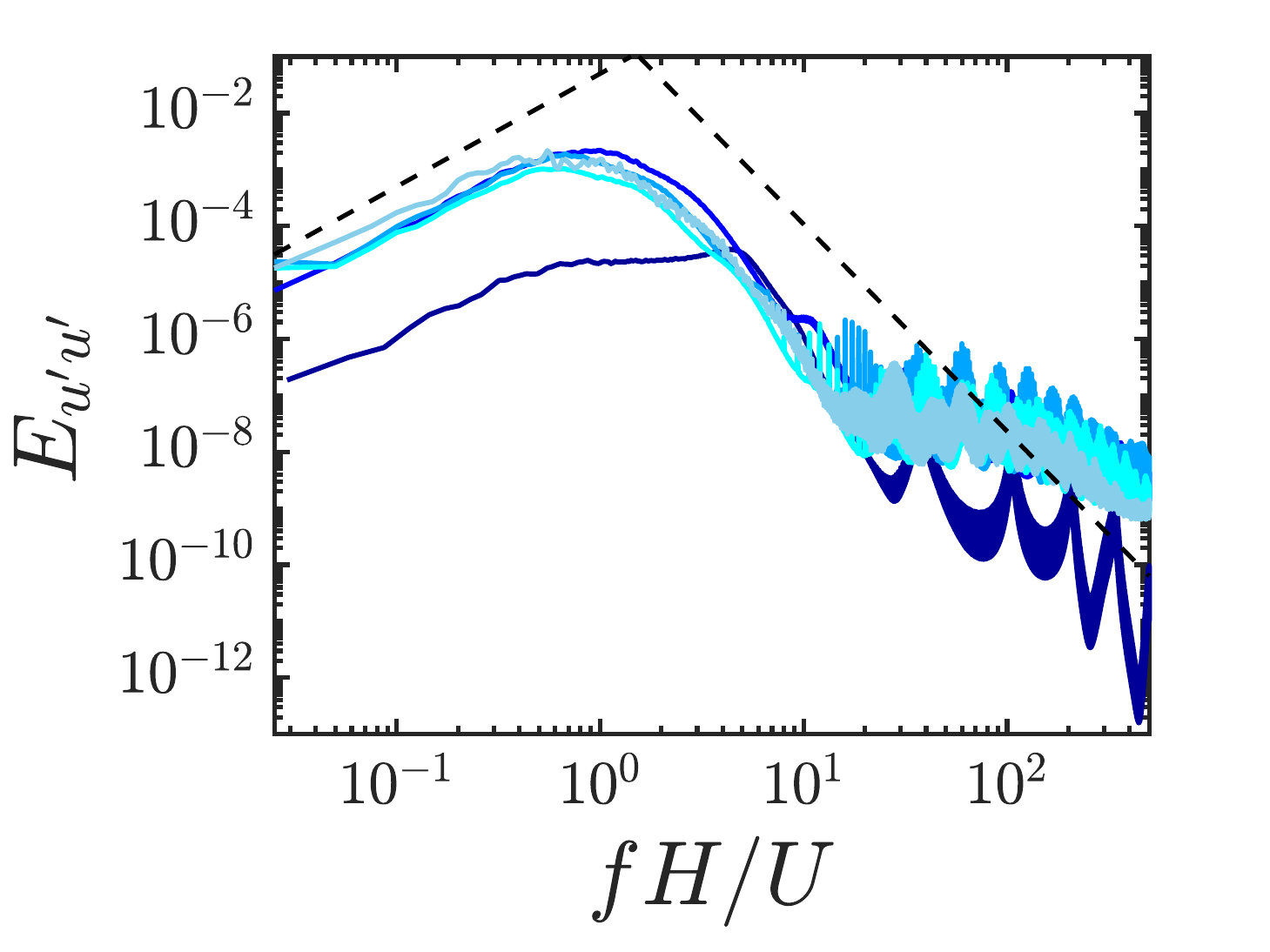}\\
    \includegraphics[width=.40\textwidth]{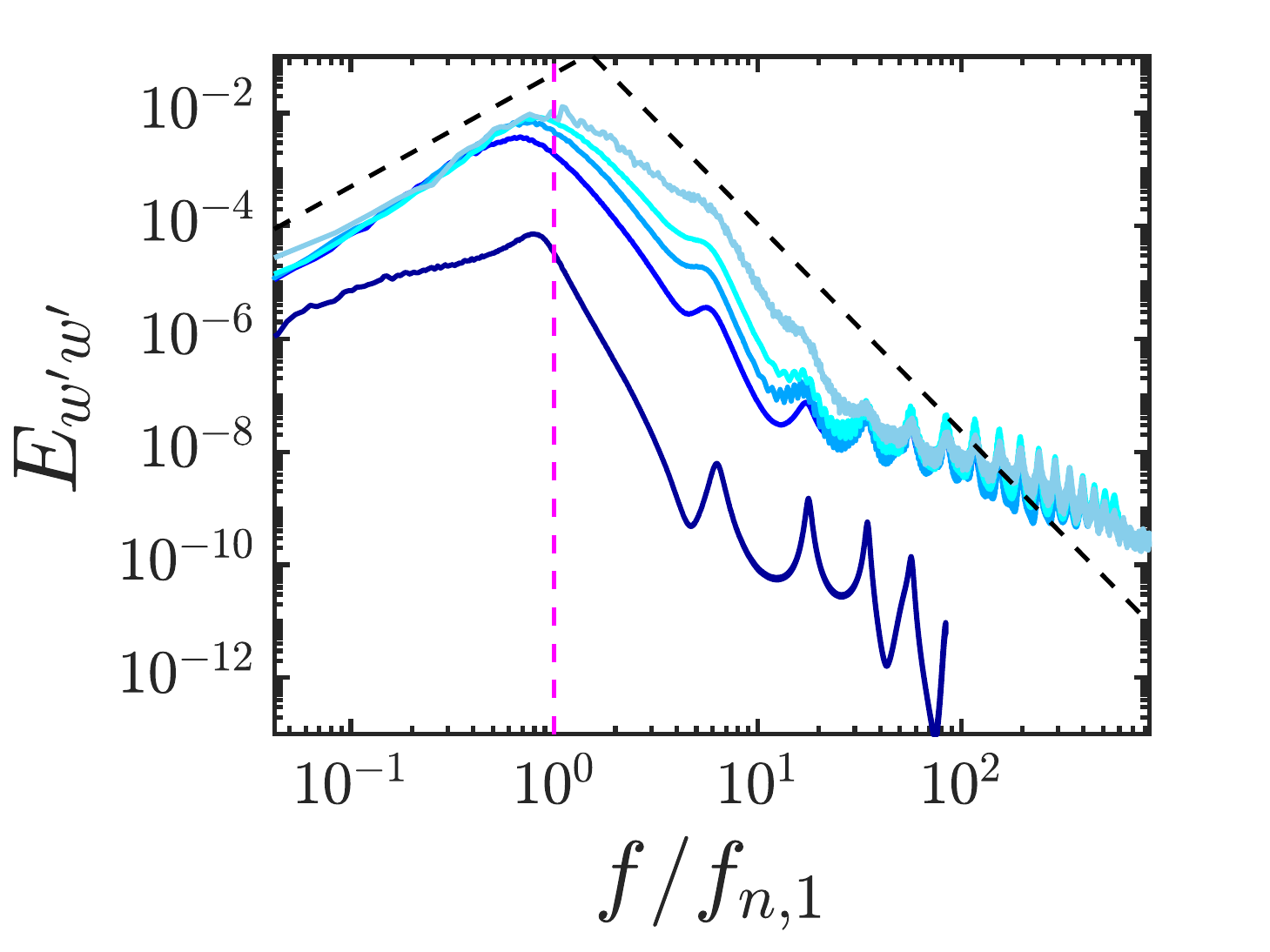}
    \includegraphics[width=.40\textwidth]{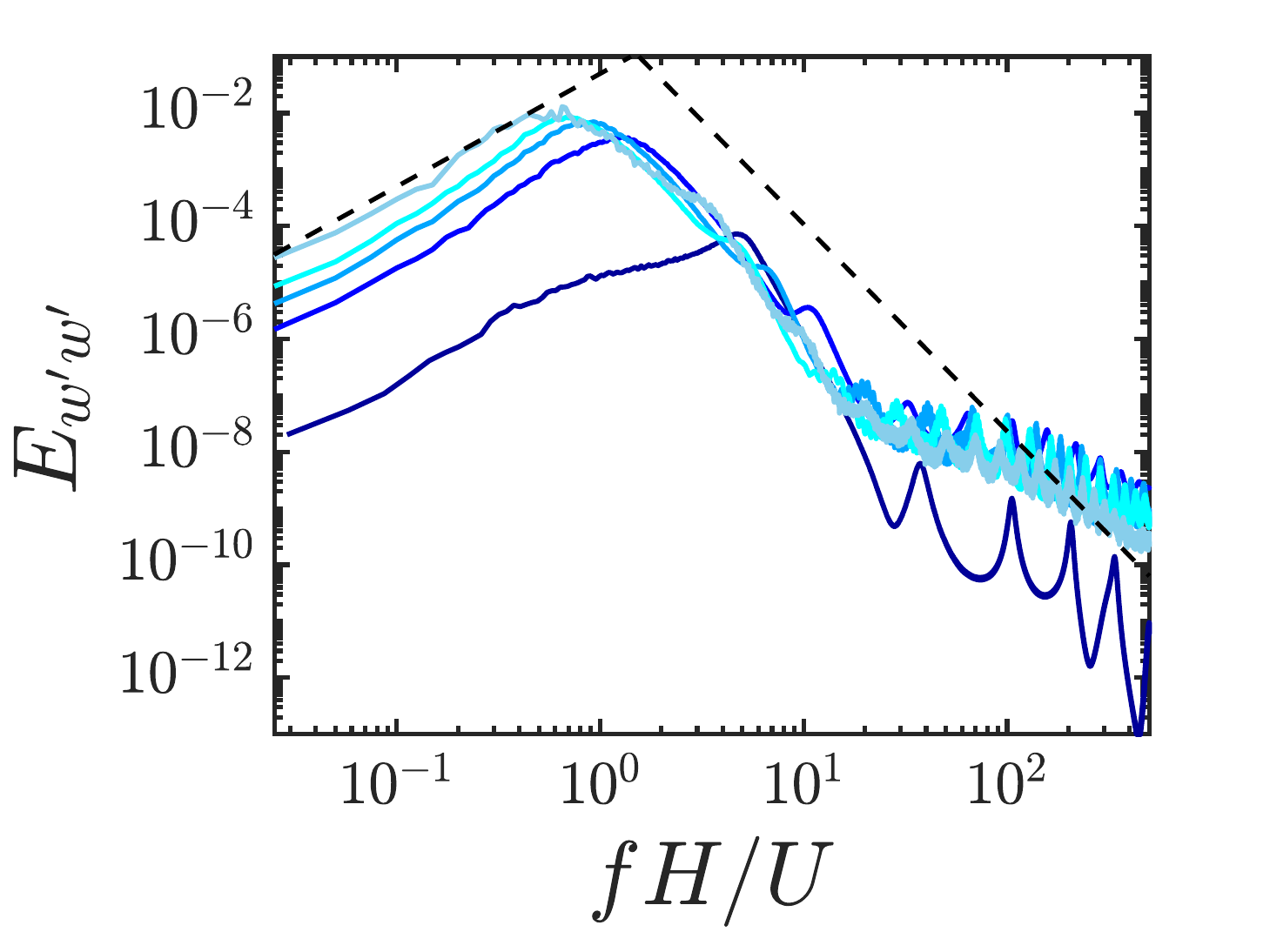}
    \caption{
    Temporal spectra of the filament tip velocity (top: streamwise component; bottom: spanwise component) for different Cauchy numbers (increasing with the curve brightness, $\Ca= 1, 10, 25, 50, 100$). Left panels: normalization with the filament natural frequency; right panels: normalization with bulk flow quantities.
    The former normalization turns out to work better for the spanwise component (i.e., the filaments exhibit their free response behavior), while the streamwise oscillation does not show a similar behavior (except for $\Ca=1$) and is fully controlled by the turbulent flow. The black dashed lines indicate the slopes $f^2$ and $f^{-5/3-2}$ which are obtained by modeling the filament as a harmonic oscillator forced by a fully-developed turbulent flow~\cite{jin2016spectral}, while the vertical magenta lines in the left panels indicate the first natural frequency.
    }
    \label{figInd}
\end{figure}

Expanding our analysis we now consider not only the spatial, but also the \added{temporal content} of the characteristic flow and structural features.
Having previously shown that the large-scale spatial content of the flow is not appreciably altered by the canopy flexibility (i.e., Cauchy number), we
choose as a representative configuration the one at $\Ca=50$  where the collective motion is clearly visible (and can be classified as honami/monami regime in the most classical sense). We can now look at the overall picture by comprehensively comparing, in \cref{fig5}, the spatio-temporal spectra of the flow field (probed in the proximity of the filament tips) and of the canopy tip envelope $Y$.
A low-frequency (approximately ranging between 0.1 and 0.5 $U/H$, depending on the different velocity components), large-wavelength peak can be generally observed at a very similar location for both the flow and canopy motion.
Remarkably, the frequency of the filament tip motion does not appear to agree with the natural frequency (indicated by the violet dashed lines in \cref{fig5}), thus suggesting the absence, at least for the chosen set of governing parameters, of any fingerprint of structural response in the collective dynamics and, in particular, of peculiar phenomena such as lock-in conditions~\cite{py2006frequency,gosselin2009destabilising,gardiner2016wind}.

\subsection*{Individual stem dynamics}

Lastly, we contrast the observation on the canopy collective dynamics by focusing on the motion of the individual filaments composing the canopy. \cref{figInd} shows the power spectra computed from the time histories of the filament tip velocity (averaged over the different filaments), for both the streamwise (top panels) and spanwise (bottom panels) components. To verify the presence of possible scalings for the dominant peak frequency, the horizontal axis is normalized either with the (first mode) natural frequency $f_{n,1}$ (left) or the bulk flow timescale $H/U$ (right panels). In fact, it can be noticed that the latter works better for the streamwise component, while for the spanwise oscillation the dominant frequency scales well with the natural one.
This qualitative difference (to the authors' knowledge so far not reported) can be explained by the presence of a net convective transport of the turbulent fluctuations in the streamwise direction.
However, it can be observed that when decreasing the Cauchy number the dominant frequency eventually becomes comparable with the natural frequency, and it can be consequently argued that a qualitative change occurs at $\Ca=1$, with the flapping frequency being now characteristic of the structural response also in the streamwise component.
Finally, looking at the whole oscillation spectra, it is worth mentioning the possibility of detecting the fingerprint of turbulence over a certain range of timescales. 
Indeed, we can model the filament as a harmonic oscillator subject to a fluctuating aerodynamic force that contains the major features of the incoming turbulent flow, and in particular its characteristic spectral content~\cite{jin2016spectral,olivieri2021direct}. The resulting power spectra associated with the filament oscillation can thus be linked to that of the turbulent forcing by means of a response function dependent on the structural parameters, yielding for the investigated cases the emergence of the two power-law slopes reported in  \cref{figInd}, $f^2$ and $f^{-5/3-2}$, in the lower and higher frequency subranges, respectively. Overall, our numerical results appear in good agreement with such reduced-order model.

\section*{Discussion}

With the goal of unravelling the essential mechanisms in the mutual interaction between a turbulent flow and a dense hairy surface made of flexible filaments, we have performed high-fidelity direct numerical simulations over a range of the Cauchy number (or, equivalently, of the reduced velocity) in order to isolate the effects related to the canopy flexibility and fluid vs structural characteristic timescales.

Our results show that the so-called honami/monami instability and consequent coherent motion is always observed, also at the small Cauchy numbers to which it is commonly attributed the absence of such physical mechanism. When decreasing the Cauchy number, the deformation of the canopy clearly gets smaller as approaching the rigid limit. Yet, the signature of the large-scale flow structures over the canopy collective motion is still present, therefore suggesting that there is not a particular threshold between different dynamical regimes (e.g., gently swaying vs honami/monami). 

\added{The absence of a clear variation in the main features of the flow while varying the Cauchy number similarly supports the idea that the collective dynamics phenomena (i.e., honami/monami) are essentially driven by the resulting turbulent flow, with the flexible canopy having always a passive dynamical behavior in this respect~\cite{ghisalberti2002mixing,tschisgale2021large}.
Here, passive means that the coherent waving of the canopy does not manifest the characteristic structural response of the individual canopy elements and is fully driven by the flow. Conversely, certain features of the individual filament motion may (or may not) show the fingerprint of the structural response (i.e., natural frequency). This is typically the case for the individual spanwise oscillation, while for the streamwise oscillation it can be argued only for sufficiently rigid stems. Nevertheless, our conclusions on the fully-passive nature of the honami/monami collective dynamics are not affected by the particular choice of Cauchy number.}

In conclusion, our findings provide novel physical insight on the nature of the fascinating collective dynamics that can be observed in a multitude of environmental settings as well as engineered textured surfaces with relevant applications to, e.g., drag reduction, mixing and flow control. 
Further work aiming at an extension of our results may incorporate the effect of gravity or consider the role of other parameters such as inertia or canopy geometry, although the main driving mechanism of the collective motion can be expected to remain those highlighted in the present investigation.

\section*{Methods}


\subsection*{Direct numerical simulation of turbulent flows}
The numerical simulations have been performed using the in-house code \textit{Fujin} (\url{https://groups.oist.jp/cffu/code}).
The fluid flow is assumed to be governed by the incompressible Navier-Stokes equations,
\begin{equation}
\partial_t \ub + \ub \cdot \partialb \ub = - \partialb p / {\rho_\mathrm{f}} + \nu \partial^2 \ub +  \fb_\mathrm{fs}, \;\;\;\;\;\;\;\;\;\;\;\; \partialb \cdot \ub = 0,
\label{eq:NS}
\end{equation}
where $\ub \left( \xb,t \right)$ and $p\left( \xb,t \right)$ are the velocity and pressure fields, respectively; $\rho_\mathrm{f}$ is the volumetric fluid density, $\nu$ is the kinematic viscosity,
and $\fb_\mathrm{fs} \left( \xb,t \right)$ is an additional forcing mimicking the presence of the canopy structural elements.
We consider a computational domain with size $L_x/H = 2\pi$, $L_y/H=1$ and $L_z/H=3/2\pi$ \added{with periodic boundary conditions along the streamwise and spanwise directions (i.e., $x$ and $z$), while the no-slip and free-slip conditions are imposed at the bottom ($y=0$) and top ($y=L_y=H$) boundaries, respectively~\cite{monti2020genesis}}. 
\added{The size of the domain is sufficient to contain the largest, energy-containing scales that are developed in the outer flow~\cite{tschisgale2021large}.}
A uniform pressure gradient is imposed in the streamwise direction so that the flow rate is kept constant.

\cref{eq:NS} are solved numerically using the (second-order) central finite-difference method for the spatial discretization and the (second-order) Adams-Bashforth scheme for advancing in time. We employ a fractional step procedure where the Poisson equation for the pressure is solved using an efficient Fast-Fourier-Transform (FFT) based approach. The solver is parallelized using the MPI protocol and the \texttt{2decomp} library (\url{http://www.2decomp.org}).

\subsection*{Flexible filament and canopy model}

The flexible stems composing the canopy are modelled as elastic inextensible filaments by the set of equations
\begin{equation}
\Delta \widetilde{\rho} \,  \ddot{\Xb}= \partial_s \left( T \partial_s \Xb \right) - \gamma \partial^4_s \Xb - \textbf{F}_\mathrm{fs} + \textbf{F}_\mathrm{col}, \;\;\;\;\;\;\;\;\;\;\;\; \partial_s \Xb \cdot \partial_s \Xb = 1,
\label{eq:EB}
\end{equation} 
where $\Delta \widetilde{\rho}$ is the linear density difference between the solid and the fluid,  $\textbf{X} \left( s, t \right)$ is the position of a generic material point of the filament at the curvilinear coordinate $s$ and time $t$; $T \left( s,t \right)$ is the tension enforcing the inextensibility constraint,  $\textbf{F}_\mathrm{fs} \left( s, t \right)$ is the fluid-solid interaction forcing and $\Fb_\mathrm{col} \left( s, t \right)$ is a filament-to-filament or filament-to-wall collision modeling term. The filament's lower end ($s=0$) is clamped to the bottom wall while the other end ($s=h$) is free to oscillate, reflecting into the following set of boundary conditions: 
$ \Xb |_{s=0} = \Xb_0$, $\partial_{s} \Xb |_{s=0} = \varphi_0$,
$ \partial_{ss} \Xb |_{s=h} = 0$, $\partial_{sss} \Xb |_{s=h} = 0$ and $T |_{s=h} = 0$.
From the normal mode analysis for such clamped configuration, the natural frequency of the filament can be evaluated as $f_\mathrm{nat} = (2\pi)^{-1} \alpha \, \sqrt{\gamma / (\Delta \widetilde{\rho} h^4)}$, where $\alpha \approx 3.5160$.
Both a filament-to-filament and filament-to-wall collision model are implemented which prevent that the stems cross each other or the bottom wall while deforming~\cite{snook2012vorticity}. 
However, after extensive testing on different kinds of collision models and their calibration parameters, the influence of the filament-to-filament collision term was found to be very weak, whereas the filament-to-wall interaction model turned out to be necessary only for sufficiently large $\Ca$.
The numerical solution of~\cref{eq:EB} follows the scheme detailed in Ref.~\cite{huang_shin_sung_2007a} with the difference that the bending term is treated implicitly to allow for a larger timestep~\cite{banaei2019numerical,olivieri2021universal,olivieri2022fully}.

\subsection*{Simulation setup}
In the present study, for all cases the fluid flow is described using $n_x = 1152$, $n_y = 384$ and $n_z = 864$ (Eulerian) grid nodes along the streamwise, wall-normal and spanwise direction, respectively. A nonuniform distribution (with a finer and locally uniform resolution in the lower region containing the canopy) is used in the $y$-direction in order to better describe the sharper gradients in the proximity of the canopy top.
The canopy is made of  $N_x \times N_z=15552$ flexible stems, with $N_x = 144$ and $N_z = 108$ stems placed along the streamwise and spanwise direction, respectively. Subdividing the horizontal plane in regular tiles of size $\Delta S = L_x/N_x = L_z/N_z$, each tile is occupied by a stem that is randomly positioned in order to prevent preferential flow channeling effects~\cite{monti2020genesis}.
In the rigid canopy case (i.e., $\Ca=0$), each filament is discretized using $N_\mathrm{L} = 81$ (Lagrangian) points such that the spatial resolution $\Delta s = h / (N_\mathrm{L} -1)$ is approximately equal to the Eulerian grid spacing $\Delta y$ in the wall-normal direction. For the flexible canopy case, using such Lagrangian resolution would impose an excessively small timestep when solving~\cref{eq:EB} and therefore we have set $N_\mathrm{L} = 21$, noting that relaxing the requirement on $\Delta s \approx \Delta y$ is acceptable when the stems are more compliant because of the decrease in the relative velocity between the stem and the flow.

\subsection*{\added{Validation and convergence study}}
\added{The numerical method has extensively been tested in a variety of problems~\cite{rosti2020increase,rosti2020fluid,brizzolara2021fibre,mazzino2021unraveling,hori2022eulerian,olivieri2022fully} and, in particular, the employed solver for the filament dynamics  has been validated in the past with several test cases~\cite{rosti2020flowing,olivieri2022fully}. Results for the representative FSI problem of a flapping filament in laminar flow are reported in \cref{fig:val_huang}.
As another validation, we compare our numerical results with available experimental measurements for the case of a rigid-canopy flow at $\Rey = 7070$, $h/H=0.65$ and $\lambda=0.83$~\cite{shimizu1991experimental}. Results showing good agreement in both the mean velocity profile and Reynolds shear stress distribution are reported in \cref{fig:val}.
Finally, considering for the present parametric study the rigid-canopy case (i.e., $\Ca=0$) where the velocity gradients at the canopy top are more severe, we have assessed the convergence of our results with respect to the spatial resolution along the wall-normal direction $\Delta y$: as shown in \cref{fig:GS},  negligible differences are observed when further refining the grid resolution.}

\section*{Data availability}
All data supporting the study are available from the authors upon reasonable request.



\begin{thebibliography}{10}
\urlstyle{rm}
\expandafter\ifx\csname url\endcsname\relax
  \def\url#1{\texttt{#1}}\fi
\expandafter\ifx\csname urlprefix\endcsname\relax\def\urlprefix{URL }\fi
\expandafter\ifx\csname doiprefix\endcsname\relax\def\doiprefix{DOI: }\fi
\providecommand{\bibinfo}[2]{#2}
\providecommand{\eprint}[2][]{\url{#2}}

\bibitem{ackerman1993reduced}
\bibinfo{author}{Ackerman, J.~D.} \& \bibinfo{author}{Okubo, A.}
\newblock \bibinfo{journal}{\bibinfo{title}{Reduced mixing in a marine
  macrophyte canopy}}.
\newblock {\emph{\JournalTitle{Functional Ecology}}}
  \textbf{\bibinfo{volume}{7}}, \bibinfo{pages}{305--309}
  (\bibinfo{year}{1993}).

\bibitem{abdolahpour2018impact}
\bibinfo{author}{Abdolahpour, M.}, \bibinfo{author}{Ghisalberti, M.},
  \bibinfo{author}{McMahon, K.} \& \bibinfo{author}{Lavery, P.~S.}
\newblock \bibinfo{journal}{\bibinfo{title}{The impact of flexibility on flow,
  turbulence, and vertical mixing in coastal canopies}}.
\newblock {\emph{\JournalTitle{Limnology and Oceanography}}}
  \textbf{\bibinfo{volume}{63}}, \bibinfo{pages}{2777--2792}
  (\bibinfo{year}{2018}).

\bibitem{wong2020shear}
\bibinfo{author}{Wong, C.~Y.}, \bibinfo{author}{Trinh, P.~H.} \&
  \bibinfo{author}{Chapman, S.~J.}
\newblock \bibinfo{journal}{\bibinfo{title}{Shear-induced instabilities of
  flows through submerged vegetation}}.
\newblock {\emph{\JournalTitle{Journal of Fluid Mechanics}}}
  \textbf{\bibinfo{volume}{891}} (\bibinfo{year}{2020}).

\bibitem{tschisgale2021large}
\bibinfo{author}{Tschisgale, S.}, \bibinfo{author}{L{\"o}hrer, B.},
  \bibinfo{author}{Meller, R.} \& \bibinfo{author}{Fr{\"o}hlich, J.}
\newblock \bibinfo{journal}{\bibinfo{title}{Large eddy simulation of the
  fluid--structure interaction in an abstracted aquatic canopy consisting of
  flexible blades}}.
\newblock {\emph{\JournalTitle{Journal of Fluid Mechanics}}}
  \textbf{\bibinfo{volume}{916}} (\bibinfo{year}{2021}).

\bibitem{houseago2022turbulence}
\bibinfo{author}{Houseago, R.~C.} \emph{et~al.}
\newblock \bibinfo{journal}{\bibinfo{title}{On the turbulence dynamics induced
  by a surrogate seagrass canopy}}.
\newblock {\emph{\JournalTitle{Journal of Fluid Mechanics}}}
  \textbf{\bibinfo{volume}{934}} (\bibinfo{year}{2022}).

\bibitem{wang_he_dey_fang_2022}
\bibinfo{author}{Wang, J.}, \bibinfo{author}{He, G.}, \bibinfo{author}{Dey, S.}
  \& \bibinfo{author}{Fang, H.}
\newblock \bibinfo{journal}{\bibinfo{title}{Fluid–structure interaction in a
  flexible vegetation canopy in an open channel}}.
\newblock {\emph{\JournalTitle{Journal of Fluid Mechanics}}}
  \textbf{\bibinfo{volume}{951}}, \bibinfo{pages}{A41} (\bibinfo{year}{2022}).

\bibitem{finnigan2000turbulence}
\bibinfo{author}{Finnigan, J.}
\newblock \bibinfo{journal}{\bibinfo{title}{Turbulence in plant canopies}}.
\newblock {\emph{\JournalTitle{Annu. Rev. Fluid Mech.}}}
  \textbf{\bibinfo{volume}{32}}, \bibinfo{pages}{519--571}
  (\bibinfo{year}{2000}).

\bibitem{nepf2012flow}
\bibinfo{author}{Nepf, H.~M.}
\newblock \bibinfo{journal}{\bibinfo{title}{Flow and transport in regions with
  aquatic vegetation}}.
\newblock {\emph{\JournalTitle{Annu. Rev. Fluid Mech.}}}
  \textbf{\bibinfo{volume}{44}}, \bibinfo{pages}{123--142}
  (\bibinfo{year}{2012}).

\bibitem{brunet2020turbulent}
\bibinfo{author}{Brunet, Y.}
\newblock \bibinfo{journal}{\bibinfo{title}{Turbulent flow in plant canopies:
  historical perspective and overview}}.
\newblock {\emph{\JournalTitle{Boundary-Layer Meteorology}}}
  \textbf{\bibinfo{volume}{177}}, \bibinfo{pages}{315--364}
  (\bibinfo{year}{2020}).

\bibitem{hobeck2012artificial}
\bibinfo{author}{Hobeck, J.} \& \bibinfo{author}{Inman, D.}
\newblock \bibinfo{journal}{\bibinfo{title}{Artificial piezoelectric grass for
  energy harvesting from turbulence-induced vibration}}.
\newblock {\emph{\JournalTitle{Smart Materials and Structures}}}
  \textbf{\bibinfo{volume}{21}}, \bibinfo{pages}{105024}
  (\bibinfo{year}{2012}).

\bibitem{sundin2019interaction}
\bibinfo{author}{Sundin, J.} \& \bibinfo{author}{Bagheri, S.}
\newblock \bibinfo{journal}{\bibinfo{title}{Interaction between hairy surfaces
  and turbulence for different surface time scales}}.
\newblock {\emph{\JournalTitle{Journal of Fluid Mechanics}}}
  \textbf{\bibinfo{volume}{861}}, \bibinfo{pages}{556--584}
  (\bibinfo{year}{2019}).

\bibitem{alvarado2017nonlinear}
\bibinfo{author}{Alvarado, J.}, \bibinfo{author}{Comtet, J.},
  \bibinfo{author}{De~Langre, E.} \& \bibinfo{author}{Hosoi, A.}
\newblock \bibinfo{journal}{\bibinfo{title}{Nonlinear flow response of soft
  hair beds}}.
\newblock {\emph{\JournalTitle{Nature Physics}}} \textbf{\bibinfo{volume}{13}},
  \bibinfo{pages}{1014--1019} (\bibinfo{year}{2017}).

\bibitem{loiseau2020active}
\bibinfo{author}{Loiseau, E.} \emph{et~al.}
\newblock \bibinfo{journal}{\bibinfo{title}{Active mucus--cilia hydrodynamic
  coupling drives self-organization of human bronchial epithelium}}.
\newblock {\emph{\JournalTitle{Nature Physics}}} \textbf{\bibinfo{volume}{16}},
  \bibinfo{pages}{1158--1164} (\bibinfo{year}{2020}).

\bibitem{deblois2023canopy}
\bibinfo{author}{de~Blois, C.}, \bibinfo{author}{Haward, S.~J.} \&
  \bibinfo{author}{Shen, A.~Q.}
\newblock \bibinfo{journal}{\bibinfo{title}{Canopy elastic turbulence:
  Spontaneous formation of waves in beds of slender microposts}}.
\newblock {\emph{\JournalTitle{Physical Review Fluids}}}
  \textbf{\bibinfo{volume}{8}}, \bibinfo{pages}{023301} (\bibinfo{year}{2023}).

\bibitem{py2006frequency}
\bibinfo{author}{Py, C.}, \bibinfo{author}{De~Langre, E.} \&
  \bibinfo{author}{Moulia, B.}
\newblock \bibinfo{journal}{\bibinfo{title}{A frequency lock-in mechanism in
  the interaction between wind and crop canopies}}.
\newblock {\emph{\JournalTitle{Journal of Fluid Mechanics}}}
  \textbf{\bibinfo{volume}{568}}, \bibinfo{pages}{425--449}
  (\bibinfo{year}{2006}).

\bibitem{ghisalberti2006structure}
\bibinfo{author}{Ghisalberti, M.} \& \bibinfo{author}{Nepf, H.}
\newblock \bibinfo{journal}{\bibinfo{title}{The structure of the shear layer in
  flows over rigid and flexible canopies}}.
\newblock {\emph{\JournalTitle{Environmental Fluid Mechanics}}}
  \textbf{\bibinfo{volume}{6}}, \bibinfo{pages}{277--301}
  (\bibinfo{year}{2006}).

\bibitem{raupach1996coherent}
\bibinfo{author}{Raupach, M.~R.}, \bibinfo{author}{Finnigan, J.~J.} \&
  \bibinfo{author}{Brunet, Y.}
\newblock \bibinfo{title}{Coherent eddies and turbulence in vegetation
  canopies: the mixing-layer analogy}.
\newblock In \emph{\bibinfo{booktitle}{Boundary-layer meteorology 25th
  anniversary volume, 1970--1995}}, \bibinfo{pages}{351--382}
  (\bibinfo{publisher}{Springer}, \bibinfo{year}{1996}).

\bibitem{gosselin2009destabilising}
\bibinfo{author}{Gosselin, F.} \& \bibinfo{author}{de~Langre, E.}
\newblock \bibinfo{journal}{\bibinfo{title}{Destabilising effects of plant
  flexibility in air and aquatic vegetation canopy flows}}.
\newblock {\emph{\JournalTitle{European Journal of Mechanics-B/Fluids}}}
  \textbf{\bibinfo{volume}{28}}, \bibinfo{pages}{271--282}
  (\bibinfo{year}{2009}).

\bibitem{singh2016linear}
\bibinfo{author}{Singh, R.}, \bibinfo{author}{Bandi, M.},
  \bibinfo{author}{Mahadevan, A.} \& \bibinfo{author}{Mandre, S.}
\newblock \bibinfo{journal}{\bibinfo{title}{Linear stability analysis for
  monami in a submerged seagrass bed}}.
\newblock {\emph{\JournalTitle{Journal of Fluid Mechanics}}}
  \textbf{\bibinfo{volume}{786}} (\bibinfo{year}{2016}).

\bibitem{zampogna2016fluid}
\bibinfo{author}{Zampogna, G.~A.} \& \bibinfo{author}{Bottaro, A.}
\newblock \bibinfo{journal}{\bibinfo{title}{Fluid flow over and through a
  regular bundle of rigid fibres}}.
\newblock {\emph{\JournalTitle{Journal of Fluid Mechanics}}}
  \textbf{\bibinfo{volume}{792}}, \bibinfo{pages}{5--35}
  (\bibinfo{year}{2016}).

\bibitem{zampogna2016instability}
\bibinfo{author}{Zampogna, G.~A.}, \bibinfo{author}{Pluvinage, F.},
  \bibinfo{author}{Kourta, A.} \& \bibinfo{author}{Bottaro, A.}
\newblock \bibinfo{journal}{\bibinfo{title}{Instability of canopy flows}}.
\newblock {\emph{\JournalTitle{Water Resources Research}}}
  \textbf{\bibinfo{volume}{52}}, \bibinfo{pages}{5421--5432}
  (\bibinfo{year}{2016}).

\bibitem{luminari2016drag}
\bibinfo{author}{Luminari, N.}, \bibinfo{author}{Airiau, C.} \&
  \bibinfo{author}{Bottaro, A.}
\newblock \bibinfo{journal}{\bibinfo{title}{Drag-model sensitivity of
  {{Kelvin-Helmholtz}} waves in canopy flows}}.
\newblock {\emph{\JournalTitle{Physics of Fluids}}}
  \textbf{\bibinfo{volume}{28}}, \bibinfo{pages}{124103}
  (\bibinfo{year}{2016}).

\bibitem{dupont2010modelling}
\bibinfo{author}{Dupont, S.} \emph{et~al.}
\newblock \bibinfo{journal}{\bibinfo{title}{Modelling waving crops using
  large-eddy simulation: comparison with experiments and a linear stability
  analysis}}.
\newblock {\emph{\JournalTitle{Journal of Fluid Mechanics}}}
  \textbf{\bibinfo{volume}{652}}, \bibinfo{pages}{5--44}
  (\bibinfo{year}{2010}).

\bibitem{dijkstra2010modeling}
\bibinfo{author}{Dijkstra, J.} \& \bibinfo{author}{Uittenbogaard, R.}
\newblock \bibinfo{journal}{\bibinfo{title}{Modeling the interaction between
  flow and highly flexible aquatic vegetation}}.
\newblock {\emph{\JournalTitle{Water Resources Research}}}
  \textbf{\bibinfo{volume}{46}} (\bibinfo{year}{2010}).

\bibitem{pan2014strong}
\bibinfo{author}{Pan, Y.}, \bibinfo{author}{Follett, E.},
  \bibinfo{author}{Chamecki, M.} \& \bibinfo{author}{Nepf, H.}
\newblock \bibinfo{journal}{\bibinfo{title}{Strong and weak, unsteady
  reconfiguration and its impact on turbulence structure within plant
  canopies}}.
\newblock {\emph{\JournalTitle{Physics of Fluids}}}
  \textbf{\bibinfo{volume}{26}}, \bibinfo{pages}{2003--2017}
  (\bibinfo{year}{2014}).

\bibitem{marjoribanks2017does}
\bibinfo{author}{Marjoribanks, T.~I.}, \bibinfo{author}{Hardy, R.~J.},
  \bibinfo{author}{Lane, S.~N.} \& \bibinfo{author}{Parsons, D.~R.}
\newblock \bibinfo{journal}{\bibinfo{title}{Does the canopy mixing layer model
  apply to highly flexible aquatic vegetation? {{Insights}} from numerical
  modelling}}.
\newblock {\emph{\JournalTitle{Environmental Fluid Mechanics}}}
  \textbf{\bibinfo{volume}{17}}, \bibinfo{pages}{277--301}
  (\bibinfo{year}{2017}).

\bibitem{monti2019large}
\bibinfo{author}{Monti, A.}, \bibinfo{author}{Omidyeganeh, M.} \&
  \bibinfo{author}{Pinelli, A.}
\newblock \bibinfo{journal}{\bibinfo{title}{Large-eddy simulation of an
  open-channel flow bounded by a semi-dense rigid filamentous canopy: Scaling
  and flow structure}}.
\newblock {\emph{\JournalTitle{Physics of Fluids}}}
  \textbf{\bibinfo{volume}{31}}, \bibinfo{pages}{065108}
  (\bibinfo{year}{2019}).

\bibitem{monti2020genesis}
\bibinfo{author}{Monti, A.}, \bibinfo{author}{Omidyeganeh, M.},
  \bibinfo{author}{Eckhardt, B.} \& \bibinfo{author}{Pinelli, A.}
\newblock \bibinfo{journal}{\bibinfo{title}{On the genesis of different regimes
  in canopy flows: a numerical investigation}}.
\newblock {\emph{\JournalTitle{Journal of Fluid Mechanics}}}
  \textbf{\bibinfo{volume}{891}} (\bibinfo{year}{2020}).

\bibitem{monti2022solidity}
\bibinfo{author}{Monti, A.}, \bibinfo{author}{Nicholas, S.},
  \bibinfo{author}{Omidyeganeh, M.}, \bibinfo{author}{Pinelli, A.} \&
  \bibinfo{author}{Rosti, M.~E.}
\newblock \bibinfo{journal}{\bibinfo{title}{On the solidity parameter in canopy
  flows}}.
\newblock {\emph{\JournalTitle{J. Fluid Mech.}}} \textbf{\bibinfo{volume}{945}}
  (\bibinfo{year}{2022}).

\bibitem{nicholas2022numerical}
\bibinfo{author}{Nicholas, S.}, \bibinfo{author}{Omidyeganeh, M.} \&
  \bibinfo{author}{Pinelli, A.}
\newblock \bibinfo{journal}{\bibinfo{title}{Numerical investigation of regime
  transition in canopy flows}}.
\newblock {\emph{\JournalTitle{Flow, Turbulence and Combustion}}}
  \bibinfo{pages}{1--21} (\bibinfo{year}{2022}).

\bibitem{okamoto2009turbulence}
\bibinfo{author}{Okamoto, T.-A.} \& \bibinfo{author}{Nezu, I.}
\newblock \bibinfo{journal}{\bibinfo{title}{Turbulence structure and
  “{{Monami}}” phenomena in flexible vegetated open-channel flows}}.
\newblock {\emph{\JournalTitle{Journal of Hydraulic Research}}}
  \textbf{\bibinfo{volume}{47}}, \bibinfo{pages}{798--810}
  (\bibinfo{year}{2009}).

\bibitem{jin2016spectral}
\bibinfo{author}{Jin, Y.}, \bibinfo{author}{Ji, S.} \&
  \bibinfo{author}{Chamorro, L.~P.}
\newblock \bibinfo{journal}{\bibinfo{title}{Spectral energy cascade of body
  rotations and oscillations under turbulence}}.
\newblock {\emph{\JournalTitle{Physical Review E}}}
  \textbf{\bibinfo{volume}{94}}, \bibinfo{pages}{063105}
  (\bibinfo{year}{2016}).

\bibitem{gardiner2016wind}
\bibinfo{author}{Gardiner, B.}, \bibinfo{author}{Berry, P.} \&
  \bibinfo{author}{Moulia, B.}
\newblock \bibinfo{journal}{\bibinfo{title}{Wind impacts on plant growth,
  mechanics and damage}}.
\newblock {\emph{\JournalTitle{Plant Sci.}}} \textbf{\bibinfo{volume}{245}},
  \bibinfo{pages}{94--118} (\bibinfo{year}{2016}).

\bibitem{olivieri2021direct}
\bibinfo{author}{Olivieri, S.}, \bibinfo{author}{Viola, F.},
  \bibinfo{author}{Mazzino, A.} \& \bibinfo{author}{Rosti, M.~E.}
\newblock \bibinfo{journal}{\bibinfo{title}{Direct numerical simulation of
  flapping flags in grid-induced turbulence}}.
\newblock {\emph{\JournalTitle{Physics of Fluids}}}
  \textbf{\bibinfo{volume}{33}}, \bibinfo{pages}{085116}
  (\bibinfo{year}{2021}).

\bibitem{ghisalberti2002mixing}
\bibinfo{author}{Ghisalberti, M.} \& \bibinfo{author}{Nepf, H.~M.}
\newblock \bibinfo{journal}{\bibinfo{title}{Mixing layers and coherent
  structures in vegetated aquatic flows}}.
\newblock {\emph{\JournalTitle{Journal of Geophysical Research: Oceans}}}
  \textbf{\bibinfo{volume}{107}}, \bibinfo{pages}{3--1} (\bibinfo{year}{2002}).

\bibitem{snook2012vorticity}
\bibinfo{author}{Snook, B.}, \bibinfo{author}{Guazzelli, E.} \&
  \bibinfo{author}{Butler, J.~E.}
\newblock \bibinfo{journal}{\bibinfo{title}{Vorticity alignment of rigid fibers
  in an oscillatory shear flow: {{Role}} of confinement}}.
\newblock {\emph{\JournalTitle{Phys. Fluids}}} \textbf{\bibinfo{volume}{24}},
  \bibinfo{pages}{121702} (\bibinfo{year}{2012}).

\bibitem{huang_shin_sung_2007a}
\bibinfo{author}{Huang, W.-X.}, \bibinfo{author}{Shin, S.~J.} \&
  \bibinfo{author}{Sung, H.~J.}
\newblock \bibinfo{journal}{\bibinfo{title}{Simulation of flexible filaments in
  a uniform flow by the immersed boundary method}}.
\newblock {\emph{\JournalTitle{J. Comput. Phys.}}}
  \textbf{\bibinfo{volume}{226}}, \bibinfo{pages}{2206 -- 2228}
  (\bibinfo{year}{2007}).

\bibitem{banaei2019numerical}
\bibinfo{author}{Banaei, A.~A.}, \bibinfo{author}{Rosti, M.~E.} \&
  \bibinfo{author}{Brandt, L.}
\newblock \bibinfo{journal}{\bibinfo{title}{Numerical study of filament
  suspensions at finite inertia}}.
\newblock {\emph{\JournalTitle{J. Fluid Mech.}}}
  \textbf{\bibinfo{volume}{882}}, \bibinfo{pages}{A5} (\bibinfo{year}{2020}).

\bibitem{olivieri2021universal}
\bibinfo{author}{Olivieri, S.}, \bibinfo{author}{Mazzino, A.} \&
  \bibinfo{author}{Rosti, M.~E.}
\newblock \bibinfo{journal}{\bibinfo{title}{Universal flapping states of
  elastic fibers in modulated turbulence}}.
\newblock {\emph{\JournalTitle{Phys. Fluids}}} \textbf{\bibinfo{volume}{33}},
  \bibinfo{pages}{071704} (\bibinfo{year}{2021}).

\bibitem{olivieri2022fully}
\bibinfo{author}{Olivieri, S.}, \bibinfo{author}{Mazzino, A.} \&
  \bibinfo{author}{Rosti, M.~E.}
\newblock \bibinfo{journal}{\bibinfo{title}{On the fully coupled dynamics of
  flexible fibres dispersed in modulated turbulence}}.
\newblock {\emph{\JournalTitle{J. Fluid Mech.}}}
  \textbf{\bibinfo{volume}{946}}, \bibinfo{pages}{A34} (\bibinfo{year}{2022}).

\bibitem{rosti2020increase}
\bibinfo{author}{Rosti, M.~E.} \& \bibinfo{author}{Brandt, L.}
\newblock \bibinfo{journal}{\bibinfo{title}{Increase of turbulent drag by
  polymers in particle suspensions}}.
\newblock {\emph{\JournalTitle{Phys. Rev. Fluids}}}
  \textbf{\bibinfo{volume}{5}}, \bibinfo{pages}{041301} (\bibinfo{year}{2020}).

\bibitem{rosti2020fluid}
\bibinfo{author}{Rosti, M.~E.}, \bibinfo{author}{Olivieri, S.},
  \bibinfo{author}{Cavaiola, M.}, \bibinfo{author}{Seminara, A.} \&
  \bibinfo{author}{Mazzino, A.}
\newblock \bibinfo{journal}{\bibinfo{title}{Fluid dynamics of {{COVID-19}}
  airborne infection suggests urgent data for a scientific design of social
  distancing}}.
\newblock {\emph{\JournalTitle{Sci. Rep.}}} \textbf{\bibinfo{volume}{10}},
  \bibinfo{pages}{1--9} (\bibinfo{year}{2020}).

\bibitem{brizzolara2021fibre}
\bibinfo{author}{Brizzolara, S.} \emph{et~al.}
\newblock \bibinfo{journal}{\bibinfo{title}{Fiber tracking velocimetry for
  two-point statistics of turbulence}}.
\newblock {\emph{\JournalTitle{Phys. Rev. X}}} \textbf{\bibinfo{volume}{11}},
  \bibinfo{pages}{031060} (\bibinfo{year}{2021}).

\bibitem{mazzino2021unraveling}
\bibinfo{author}{Mazzino, A.} \& \bibinfo{author}{Rosti, M.~E.}
\newblock \bibinfo{journal}{\bibinfo{title}{Unraveling the secrets of
  turbulence in a fluid puff}}.
\newblock {\emph{\JournalTitle{Physical Review Letters}}}
  \textbf{\bibinfo{volume}{127}}, \bibinfo{pages}{094501}
  (\bibinfo{year}{2021}).

\bibitem{hori2022eulerian}
\bibinfo{author}{Hori, N.}, \bibinfo{author}{Rosti, M.~E.} \&
  \bibinfo{author}{Takagi, S.}
\newblock \bibinfo{journal}{\bibinfo{title}{An {Eulerian-based} immersed
  boundary method for particle suspensions with implicit lubrication model}}.
\newblock {\emph{\JournalTitle{Comput. Fluids}}} \bibinfo{pages}{105278}
  (\bibinfo{year}{2022}).

\bibitem{rosti2020flowing}
\bibinfo{author}{Rosti, M.~E.}, \bibinfo{author}{Olivieri, S.},
  \bibinfo{author}{Banaei, A.~A.}, \bibinfo{author}{Brandt, L.} \&
  \bibinfo{author}{Mazzino, A.}
\newblock \bibinfo{journal}{\bibinfo{title}{Flowing fibers as a proxy of
  turbulence statistics}}.
\newblock {\emph{\JournalTitle{Meccanica}}} \textbf{\bibinfo{volume}{55}},
  \bibinfo{pages}{357--370} (\bibinfo{year}{2020}).

\bibitem{shimizu1991experimental}
\bibinfo{author}{Shimizu, Y.}, \bibinfo{author}{Tsujimoto, T.},
  \bibinfo{author}{Nakagawa, H.} \& \bibinfo{author}{Kitamura, T.}
\newblock \bibinfo{journal}{\bibinfo{title}{Experimental study on flow over
  rigid vegetation simulated by cylinders with equi-spacing}}.
\newblock {\emph{\JournalTitle{Doboku Gakkai Ronbunshu}}}
  \textbf{\bibinfo{volume}{1991}}, \bibinfo{pages}{31--40}
  (\bibinfo{year}{1991}).

\end{thebibliography}



\section*{Acknowledgements} 

The authors acknowledge the computational resources provided by the Scientific Computing Section of the Research Support Division at OIST and the computational time on the Oakbridge-CX and Oakforest-PACS supercomputers at the Information Technology Center, The University of Tokyo, provided through the High Performance Computing Infrastructure (HPCI) System Research Project hp210025. \added{S.O. is supported by grant FJC2021-047652-I funded by MCIN/AEI/10.13039/501100011033 and European Union NextGenerationEU/PRTR.}

\section*{Author contributions}


M.E.R. conceived the original idea, and all authors planned the research and developed the code. A.M. and S.O. performed the numerical simulations. A.M. processed the data and analysed the results with feedback from all authors. A.M. and S.O. outlined the manuscript content. S.O. wrote the manuscript with feedback from all authors.

\section*{Funding} 

The research was supported by the Okinawa Institute of Science and Technology Graduate University (OIST) with subsidy funding from the Cabinet Office, Government of Japan.

\section*{Competing interests}
The authors declare no competing interests.







\renewcommand{\thefigure}{S\arabic{figure}}
\setcounter{figure}{0}

\clearpage

\begin{figure}
    \centering
    \begin{tikzpicture}
    \node[] at ( 1.6,2.4) {\small$Ca=0$};
    \node[] at ( 4.3,2.4) {\small$Ca=1$};
    \node[] at ( 7.0,2.4) {\small$Ca=10$};
    \node[] at ( 9.7,2.4) {\small$Ca=25$};
    \node[] at (12.4,2.4) {\small$Ca=50$};
    \node[] at (15.1,2.4) {\small$Ca=100$};

    \node[] at (-0.7,1.3) {\small$\dfrac{2\pi\Phi_{u'u'}}{\lambda_x u_\tau^2}$};
    \node[anchor=south west,inner sep=0] (image) at ( 0.0,0) {
    \includegraphics[width=.16\textwidth]{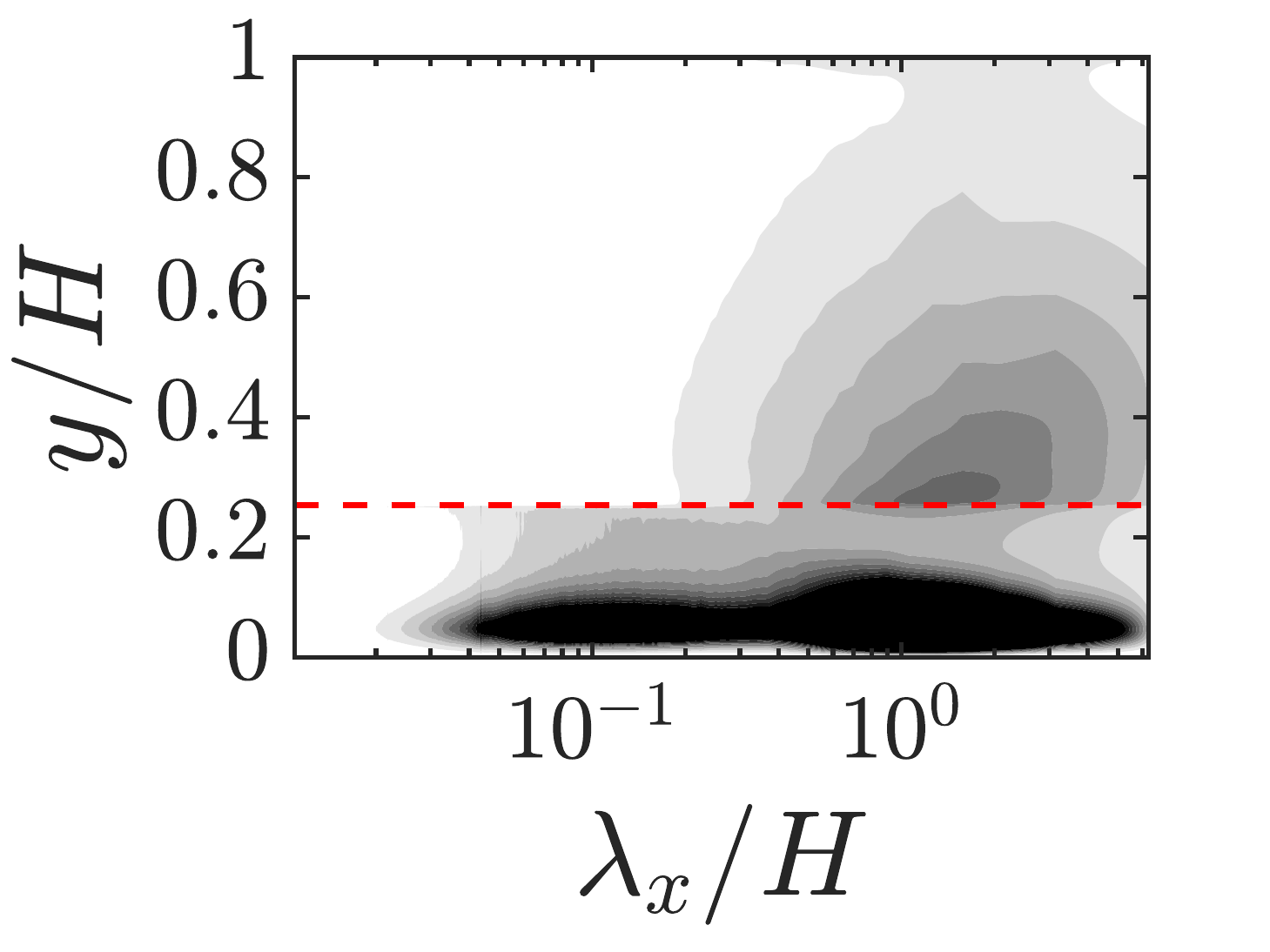}};
    \node[anchor=south west,inner sep=0] (image) at ( 2.7,0) {
    \includegraphics[width=.16\textwidth]{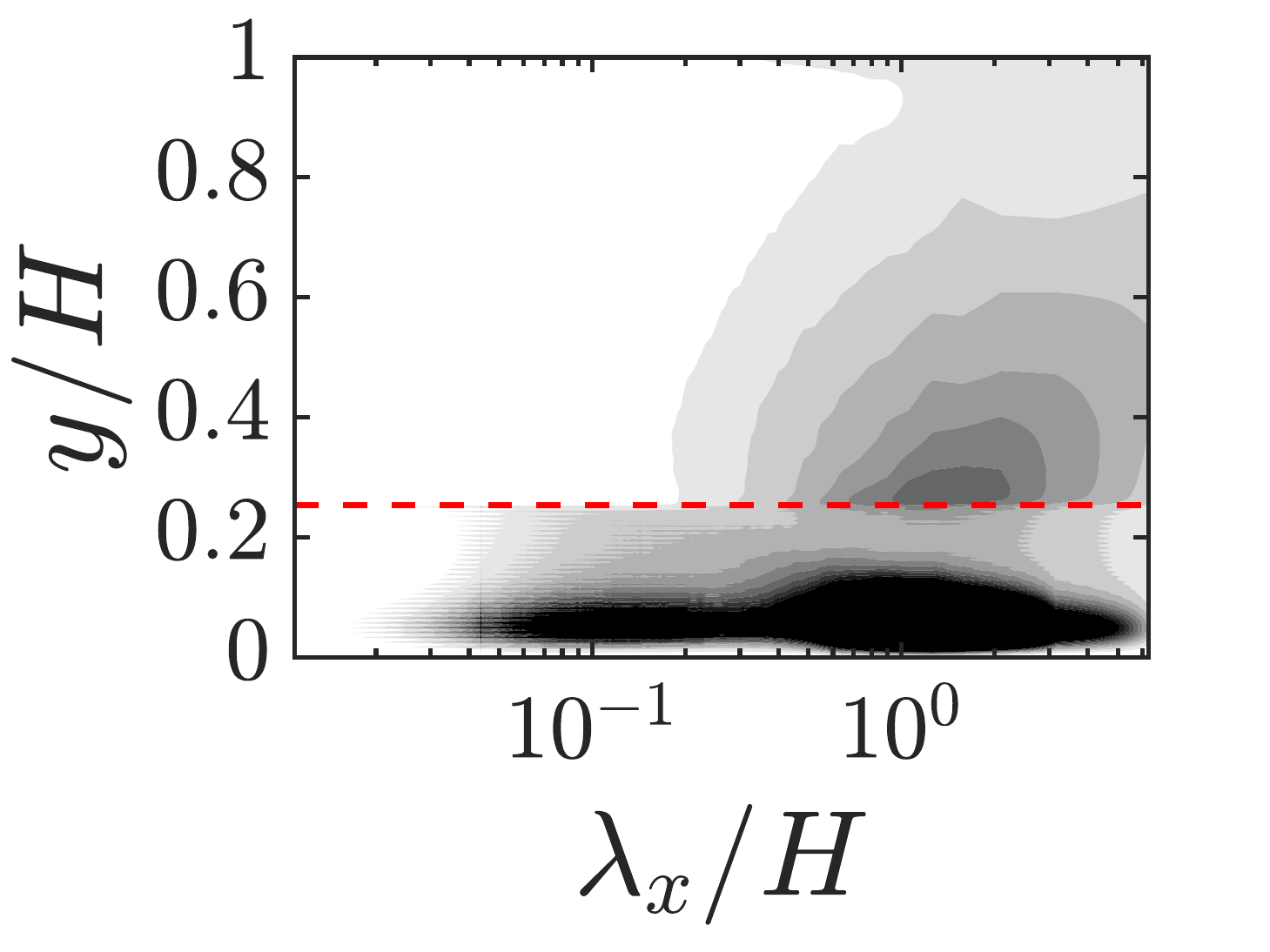}};
    \node[anchor=south west,inner sep=0] (image) at ( 5.4,0) {
    \includegraphics[width=.16\textwidth]{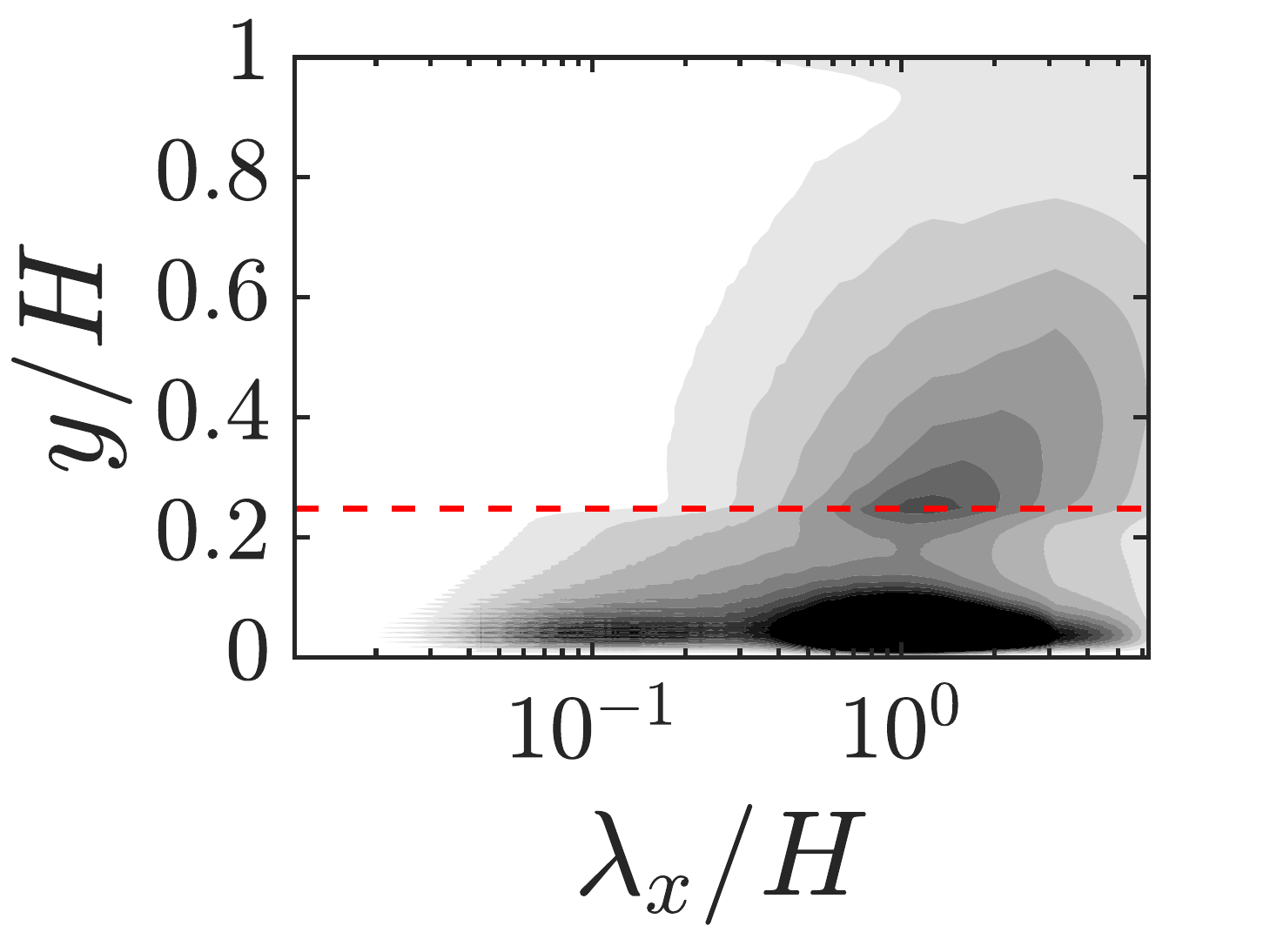}};
    \node[anchor=south west,inner sep=0] (image) at ( 8.1,0) {
    \includegraphics[width=.16\textwidth]{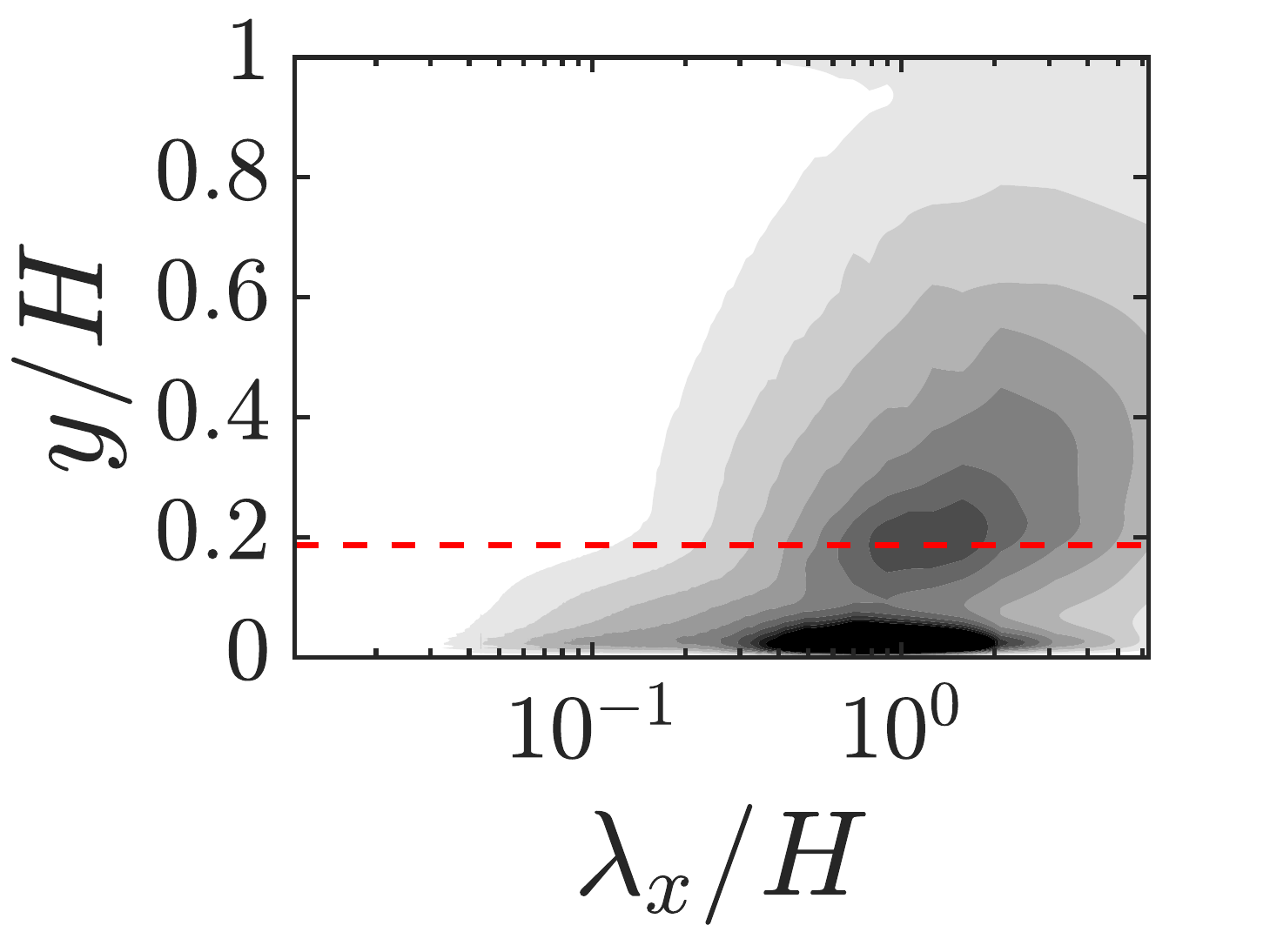}};
    \node[anchor=south west,inner sep=0] (image) at (10.8,0) {
    \includegraphics[width=.16\textwidth]{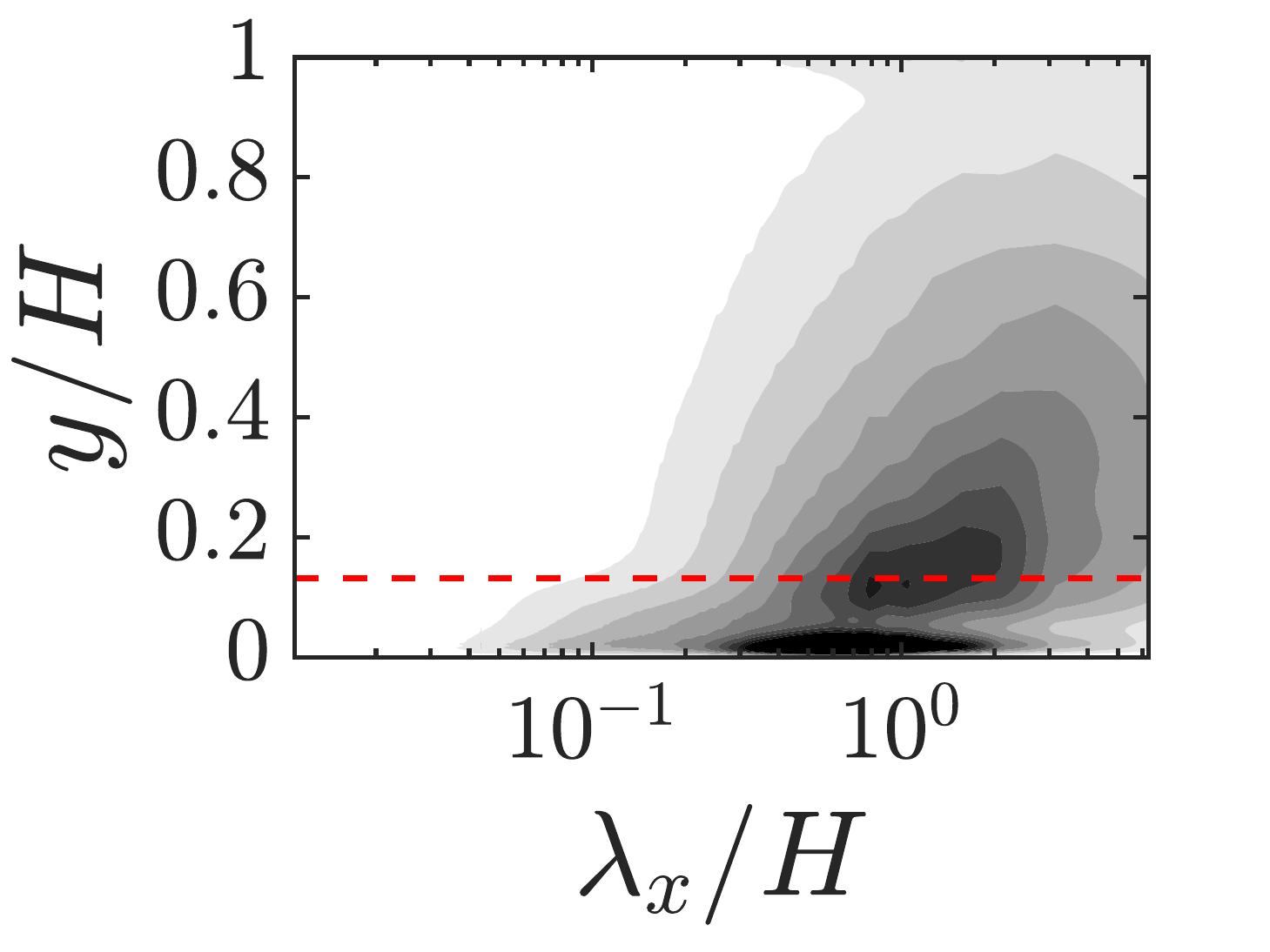}};
    \node[anchor=south west,inner sep=0] (image) at (13.5,0) {
    \includegraphics[width=.16\textwidth]{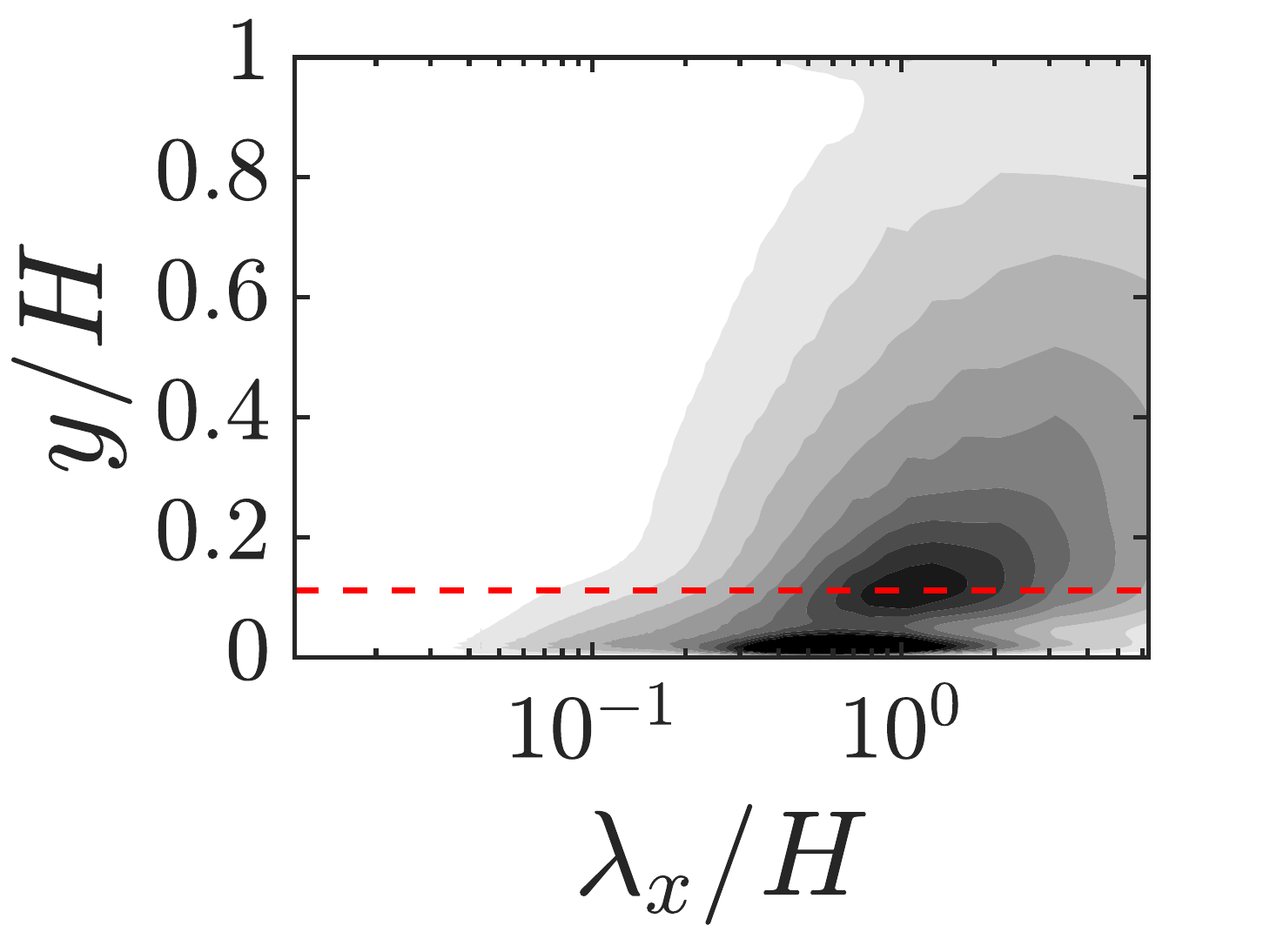}};

    \node[] at (-0.7,-1.2) {\small$\dfrac{2\pi\Phi_{v'v'}}{\lambda_x u_\tau^2}$};
    \node[anchor=south west,inner sep=0] (image) at ( 0.0,-2.50) {
    \includegraphics[width=.16\textwidth]{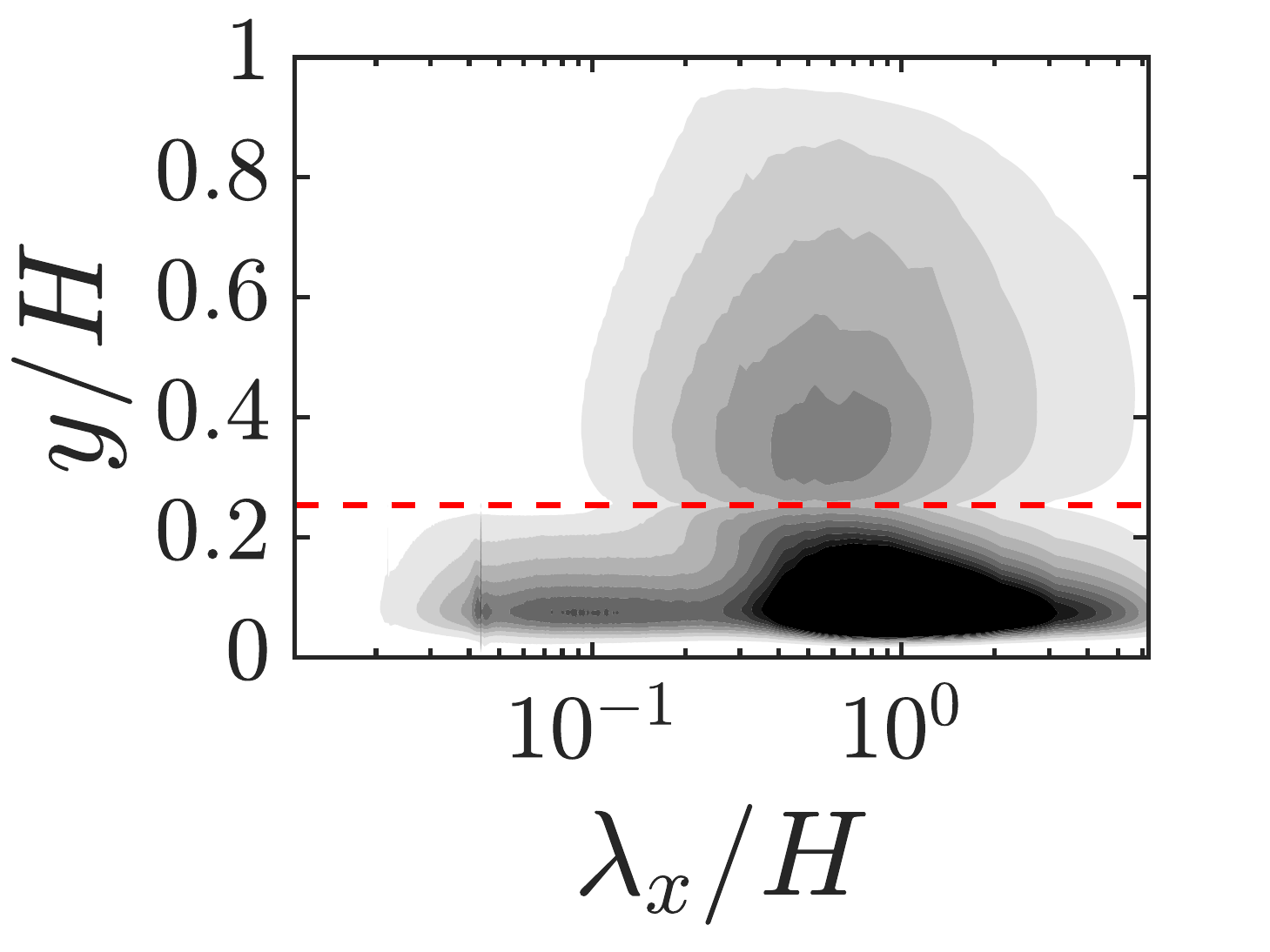}};
    \node[anchor=south west,inner sep=0] (image) at ( 2.7,-2.50) {
    \includegraphics[width=.16\textwidth]{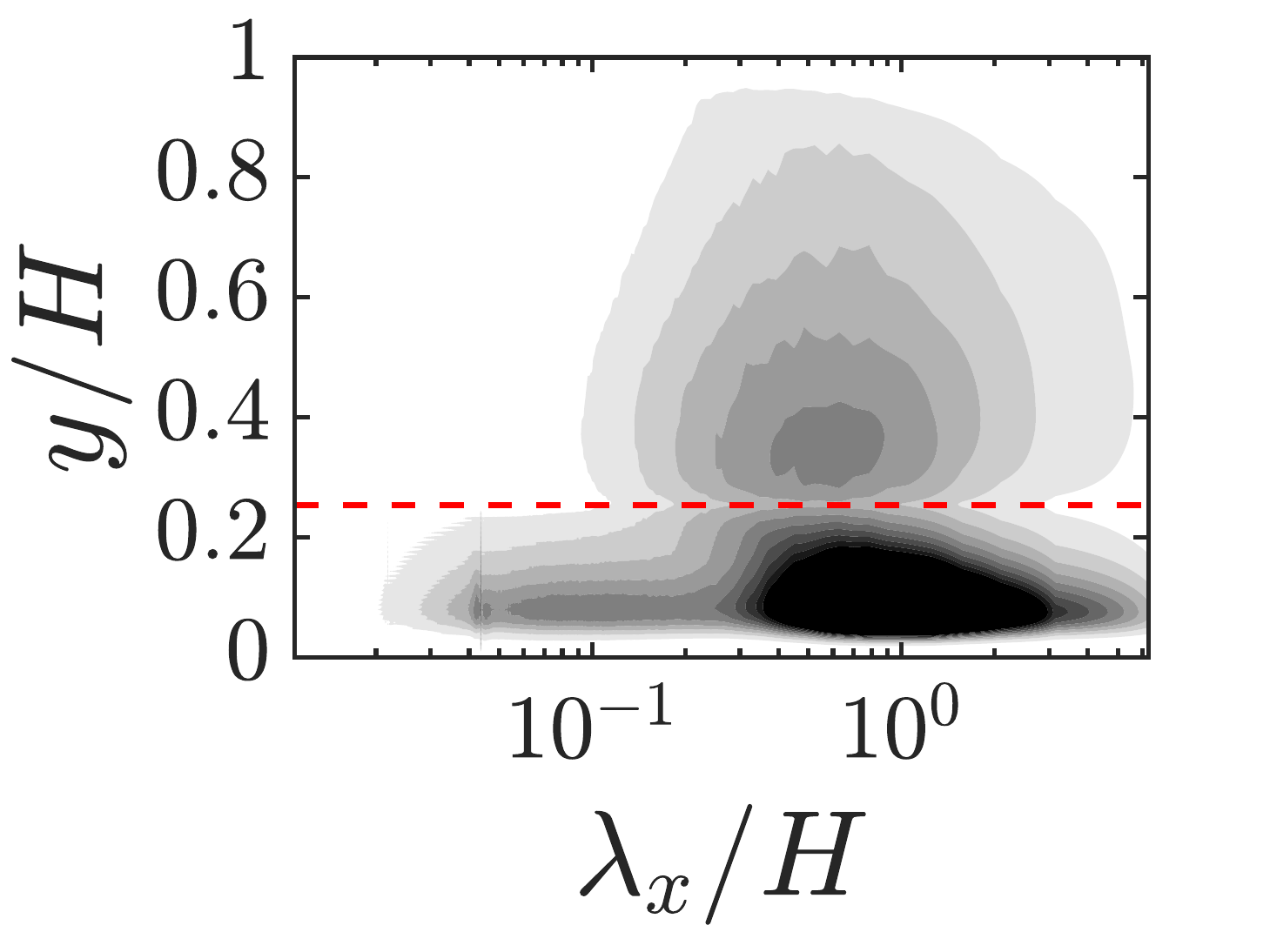}};
    \node[anchor=south west,inner sep=0] (image) at ( 5.4,-2.50) {
    \includegraphics[width=.16\textwidth]{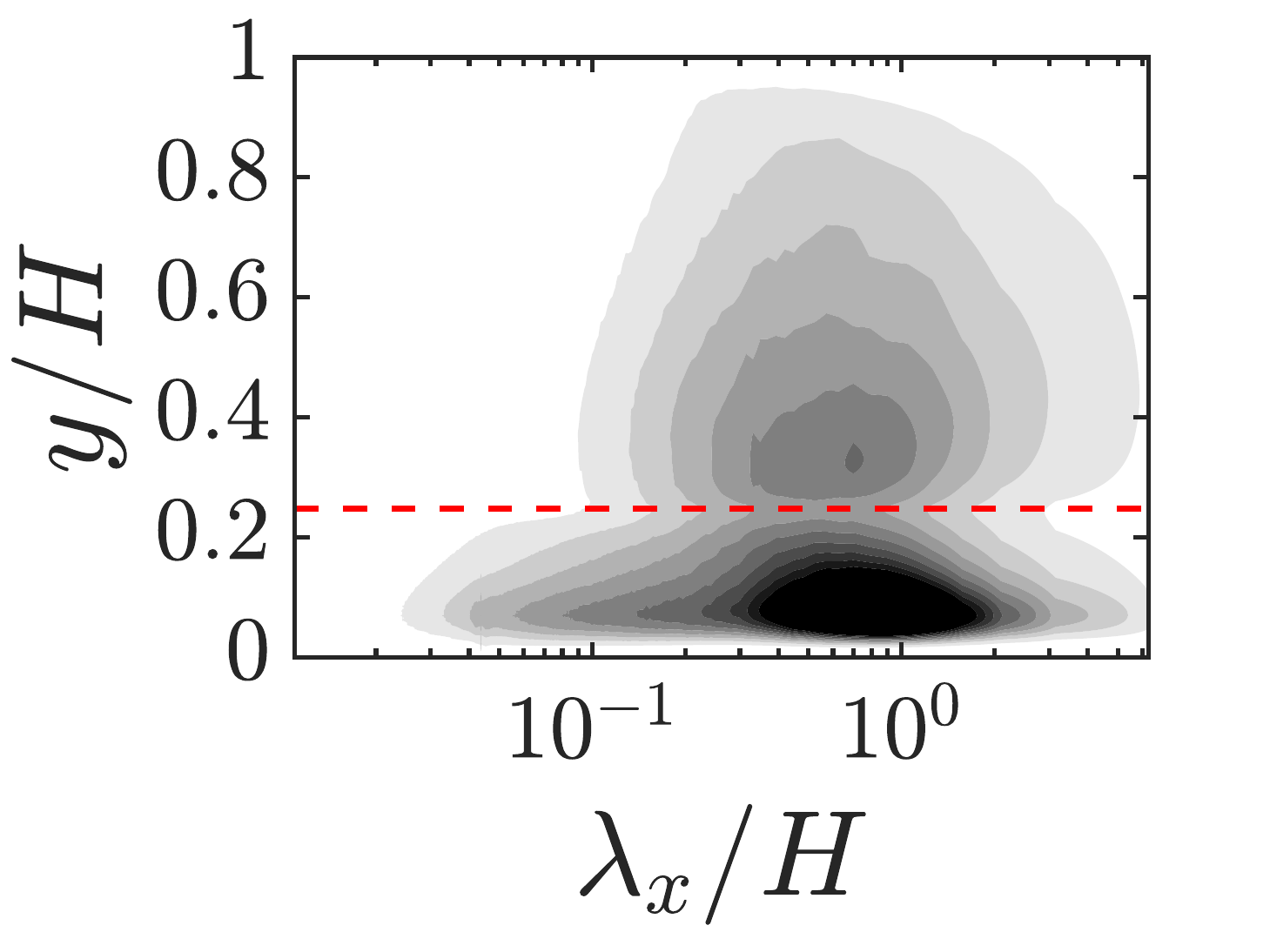}};
    \node[anchor=south west,inner sep=0] (image) at ( 8.1,-2.50) {
    \includegraphics[width=.16\textwidth]{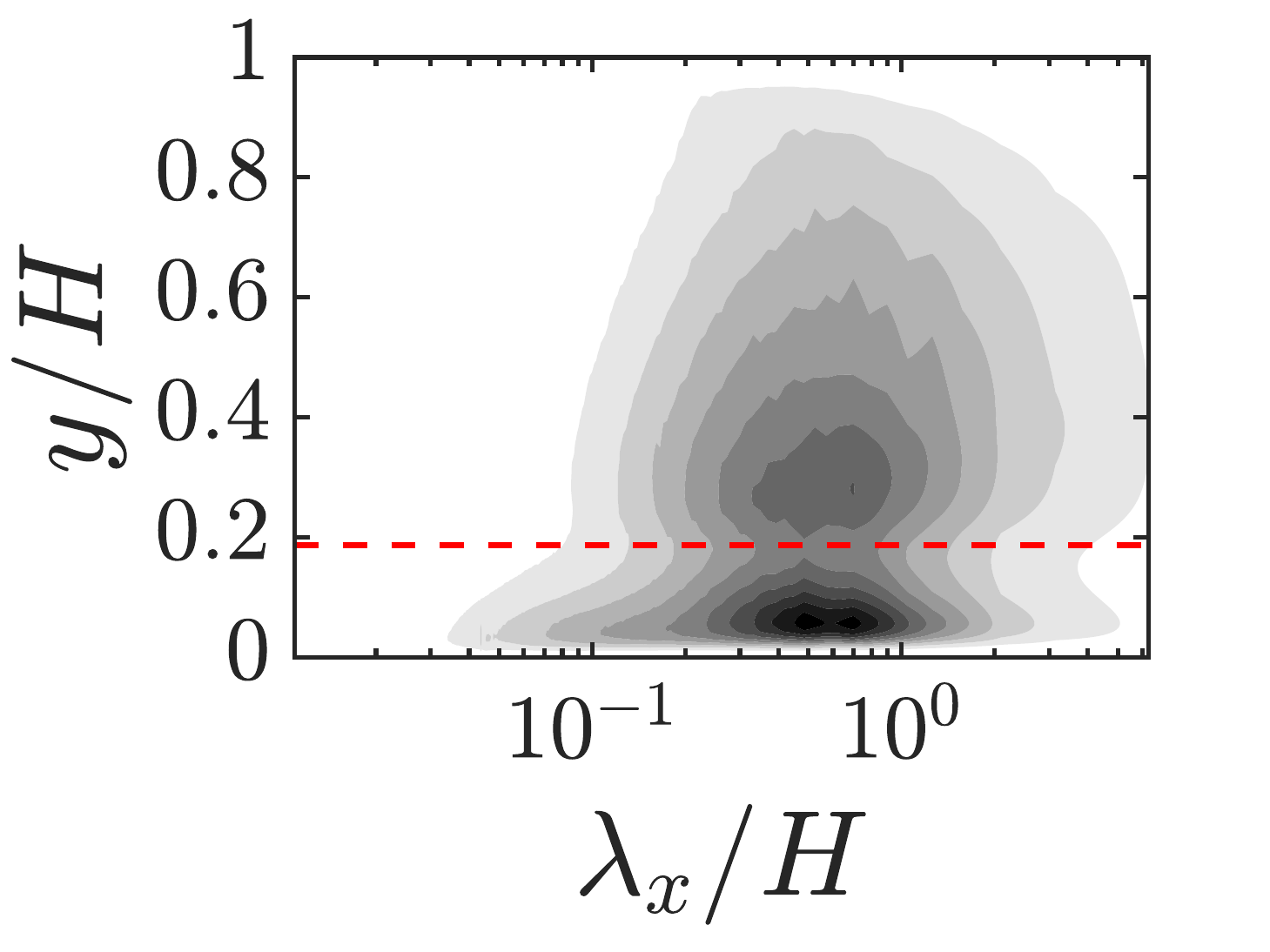}};
    \node[anchor=south west,inner sep=0] (image) at (10.8,-2.50) {
    \includegraphics[width=.16\textwidth]{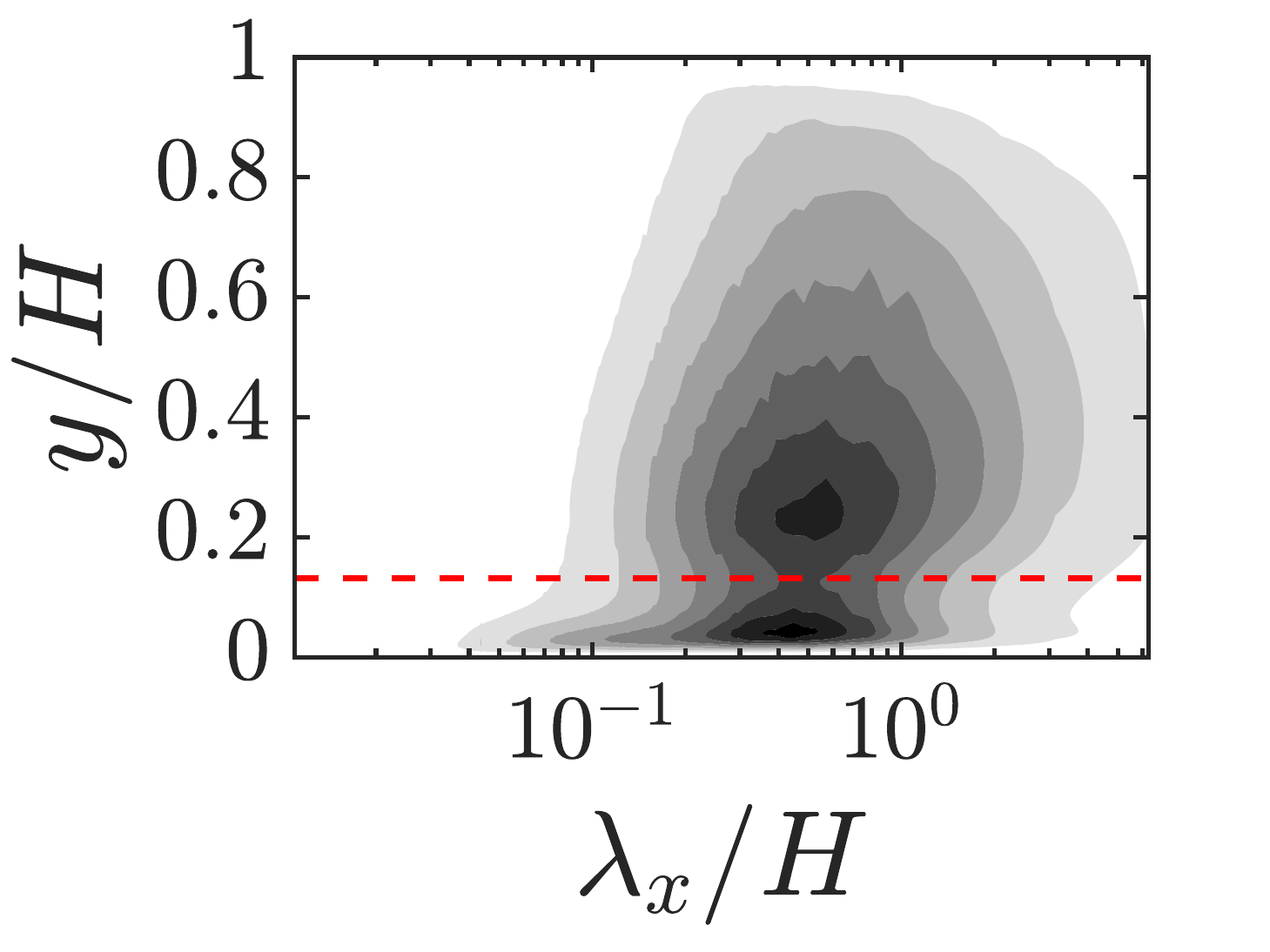}};
    \node[anchor=south west,inner sep=0] (image) at (13.5,-2.50) {
    \includegraphics[width=.16\textwidth]{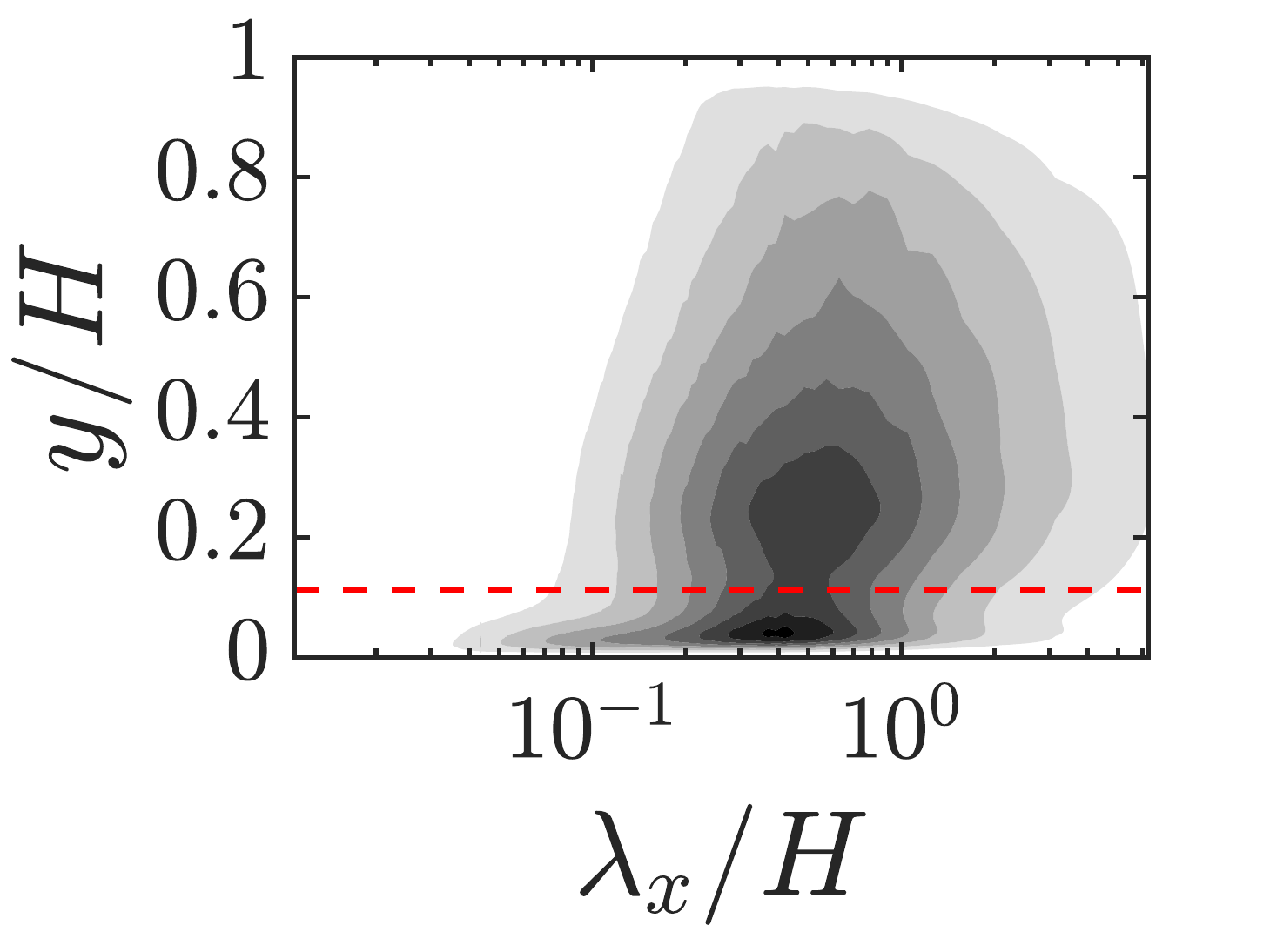}};

    \node[] at (-0.7,-3.7) {\small$\dfrac{2\pi\Phi_{w'w'}}{\lambda_x u_\tau^2}$};
    \node[anchor=south west,inner sep=0] (image) at ( 0.0,-5.0) {
    \includegraphics[width=.16\textwidth]{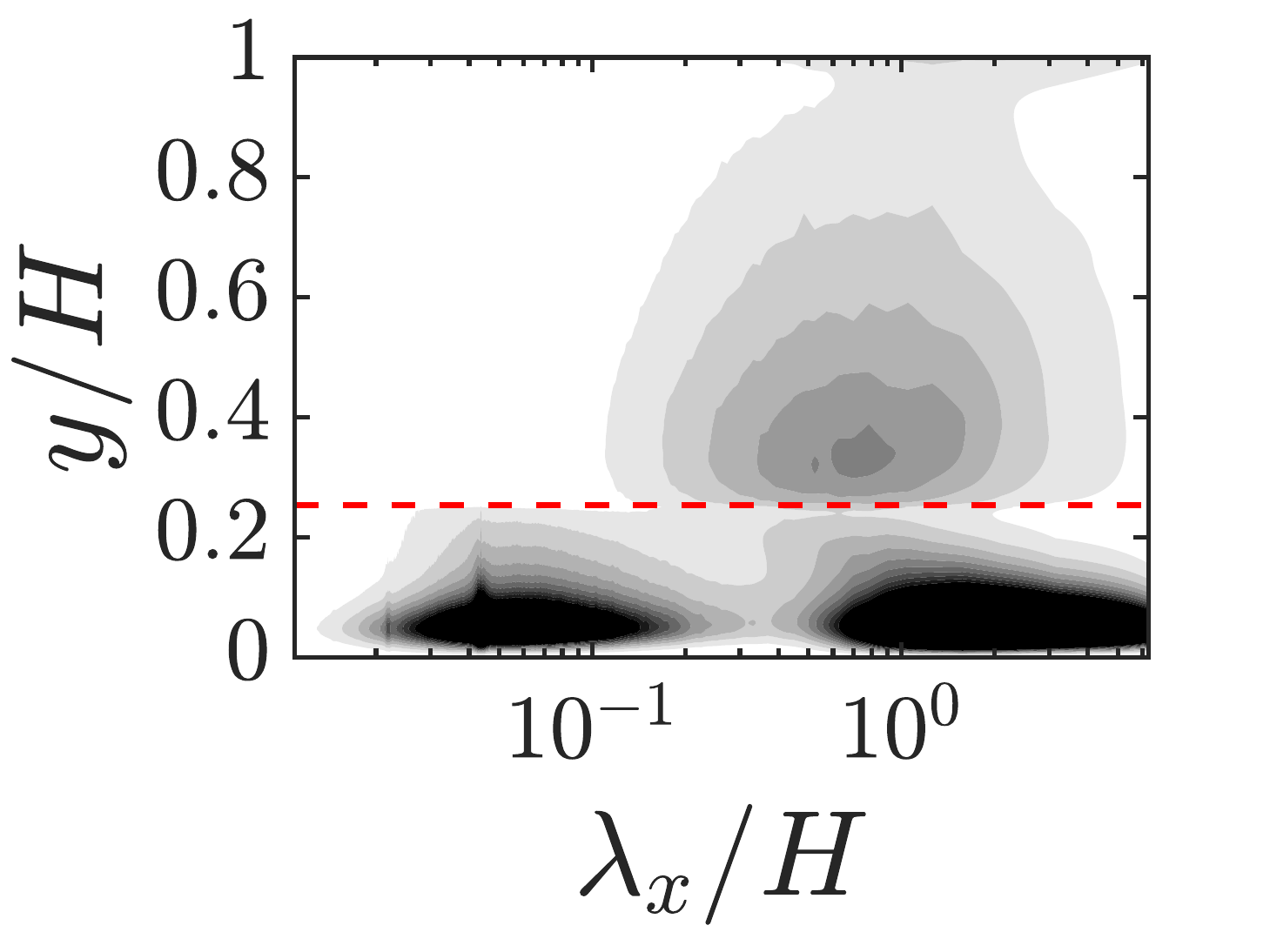}};
    \node[anchor=south west,inner sep=0] (image) at ( 2.7,-5.0) {
    \includegraphics[width=.16\textwidth]{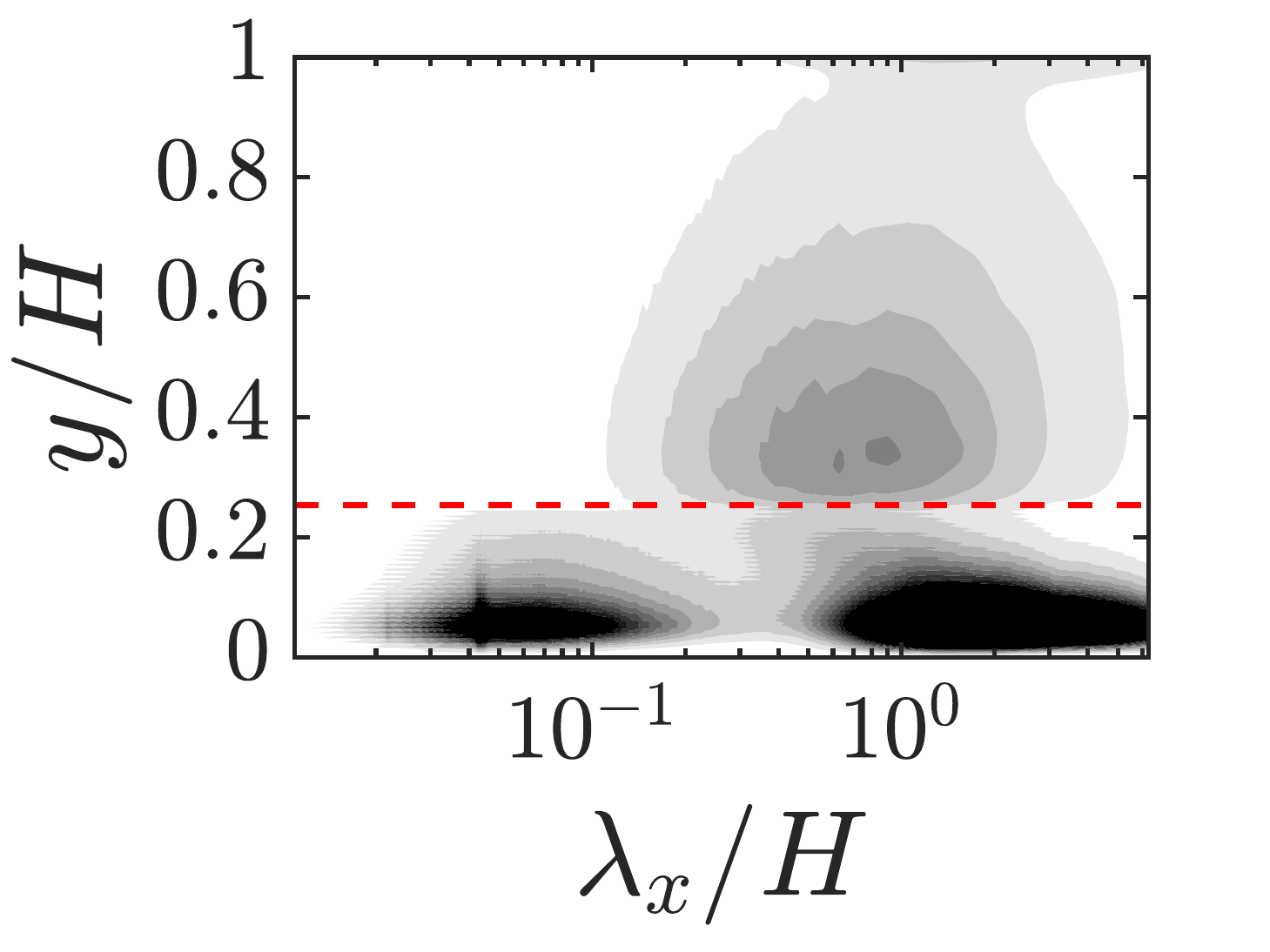}};
    \node[anchor=south west,inner sep=0] (image) at ( 5.4,-5.0) {
    \includegraphics[width=.16\textwidth]{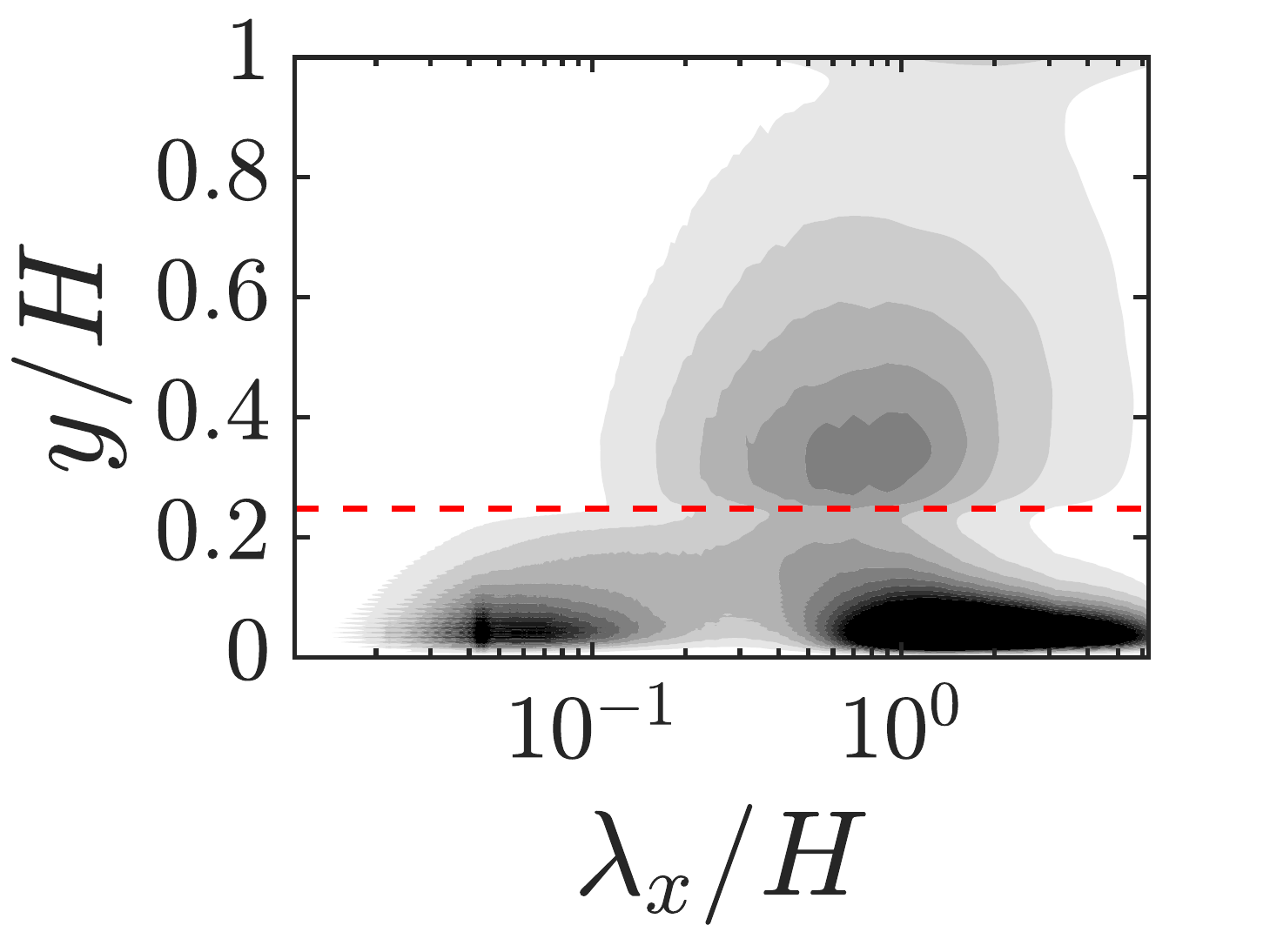}};
    \node[anchor=south west,inner sep=0] (image) at ( 8.1,-5.0) {
    \includegraphics[width=.16\textwidth]{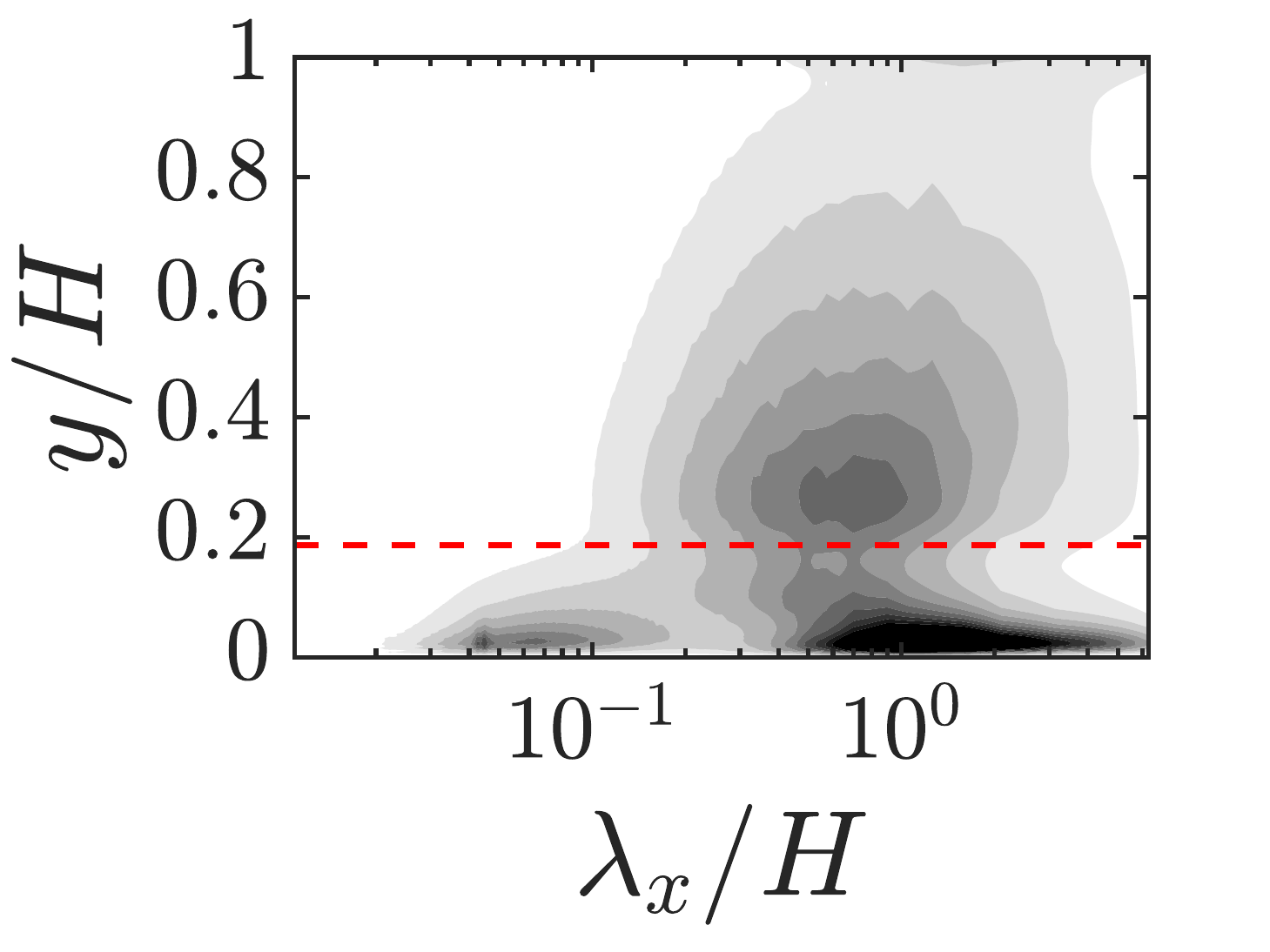}};
    \node[anchor=south west,inner sep=0] (image) at (10.8,-5.0) {
    \includegraphics[width=.16\textwidth]{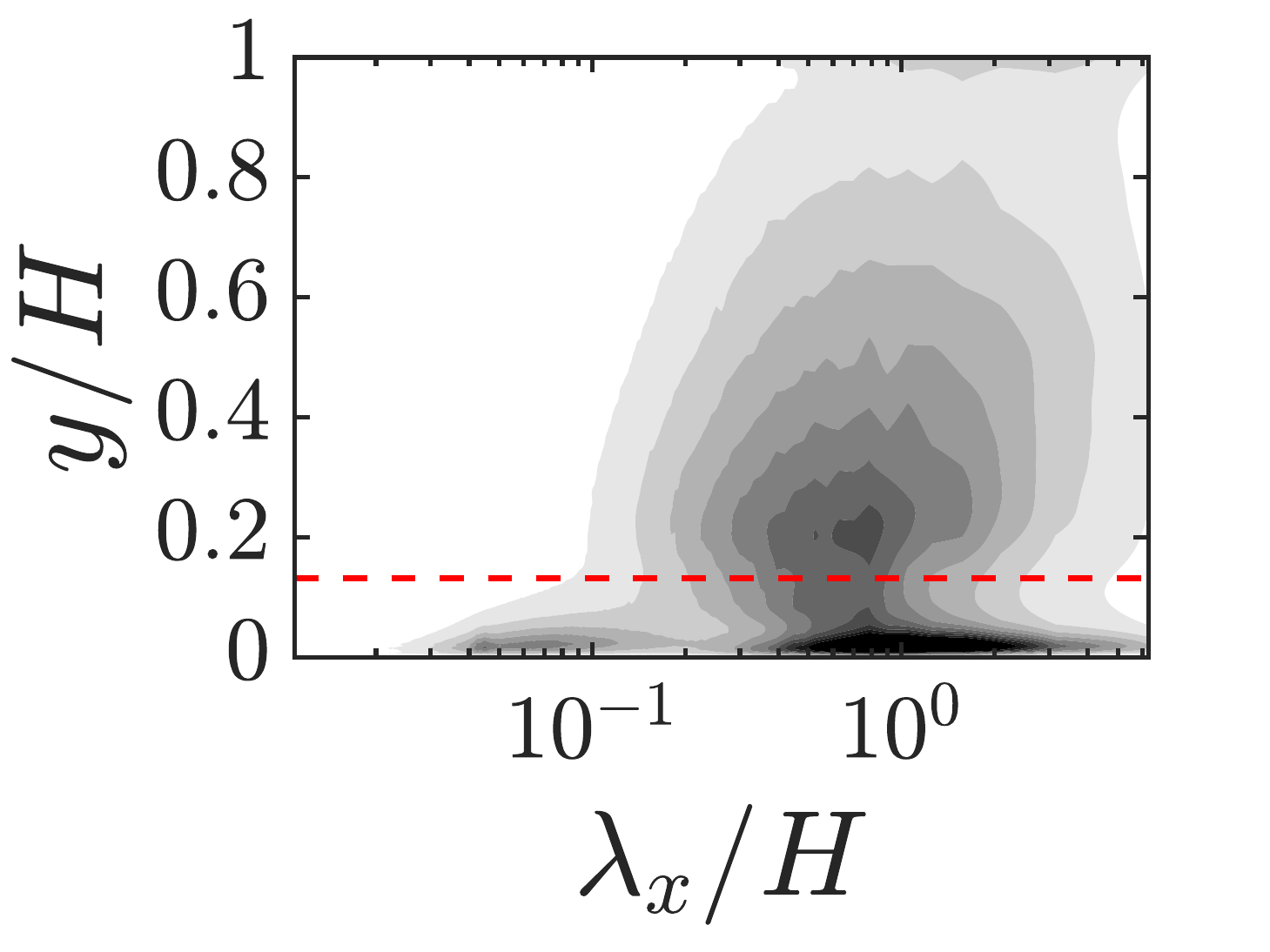}};
    \node[anchor=south west,inner sep=0] (image) at (13.5,-5.0) {
    \includegraphics[width=.16\textwidth]{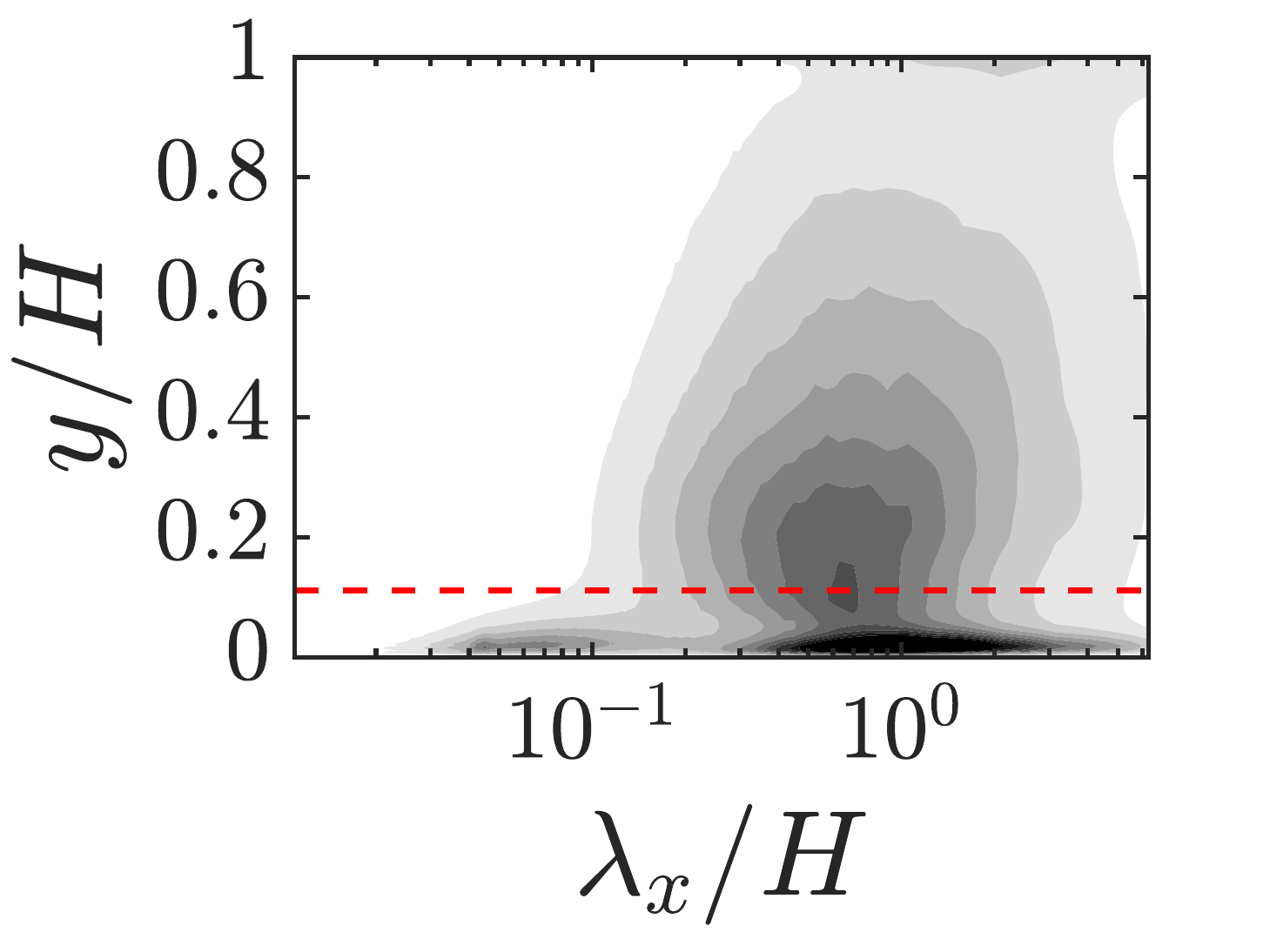}};
    \end{tikzpicture}
    \caption{\added{
    Magnitude of the premultiplied spectra of each fluid velocity component 
	$2\pi\Phi_{u'u'}/(u_\tau^2\lambda_x)$, where 
	$u'$ the generic component of the velocity fluctuations,
	(top: streamwise; middle: wall-normal; bottom: spanwise) as a function of the streamwise wavelength $\lambda_x/H$ and wall-normal coordinate $y/H$. 
	The friction velocity is defined locally as $u_\tau=\sqrt{\tau/[\rho_f(1-y/H)]}$, where $\tau$ is the flow total shear stress, i.e. the sum of the viscous and the turbulent component \cite{monti2020genesis}.
	Results are shown in different columns as a function of the investigated Cauchy number (from left to right, $\Ca=0, 1, 10, 25, 50, 100$). 
	The red horizontal dashed line indicates the averaged height of the filament tips.
	The grey levels range in: $[0, 0.5]$ with a $0.05$ increment for the streamwise and spanwise velocity components; $[0, 0.3]$ with a $0.03$ increment for the wall-normal velocity component.
        }
    }
    \label{fig3_frict}
\end{figure}

\begin{figure}
    \centering
    \begin{tikzpicture}
    \node[] at ( 1.6,2.4) {\small$Ca=0$};
    \node[] at ( 4.3,2.4) {\small$Ca=1$};
    \node[] at ( 7.0,2.4) {\small$Ca=10$};
    \node[] at ( 9.7,2.4) {\small$Ca=25$};
    \node[] at (12.4,2.4) {\small$Ca=50$};
    \node[] at (15.1,2.4) {\small$Ca=100$};

    \node[] at (-0.7,1.3) {\small$\dfrac{2\pi\Phi_{u'u'}}{\lambda_z u_\tau^2}$};
    \node[anchor=south west,inner sep=0] (image) at ( 0.0,0) {
    \includegraphics[width=.16\textwidth]{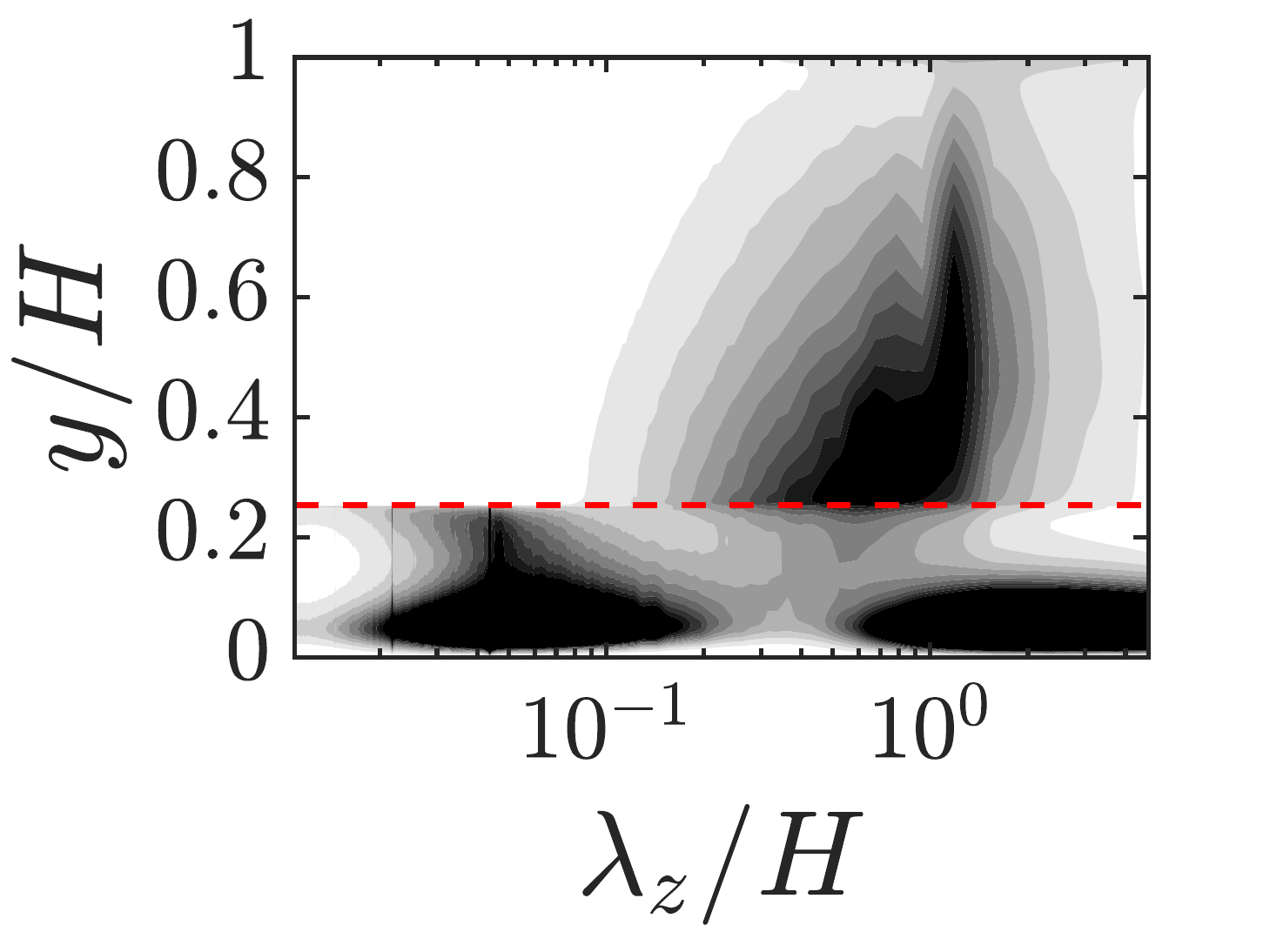}};
    \node[anchor=south west,inner sep=0] (image) at ( 2.7,0) {
    \includegraphics[width=.16\textwidth]{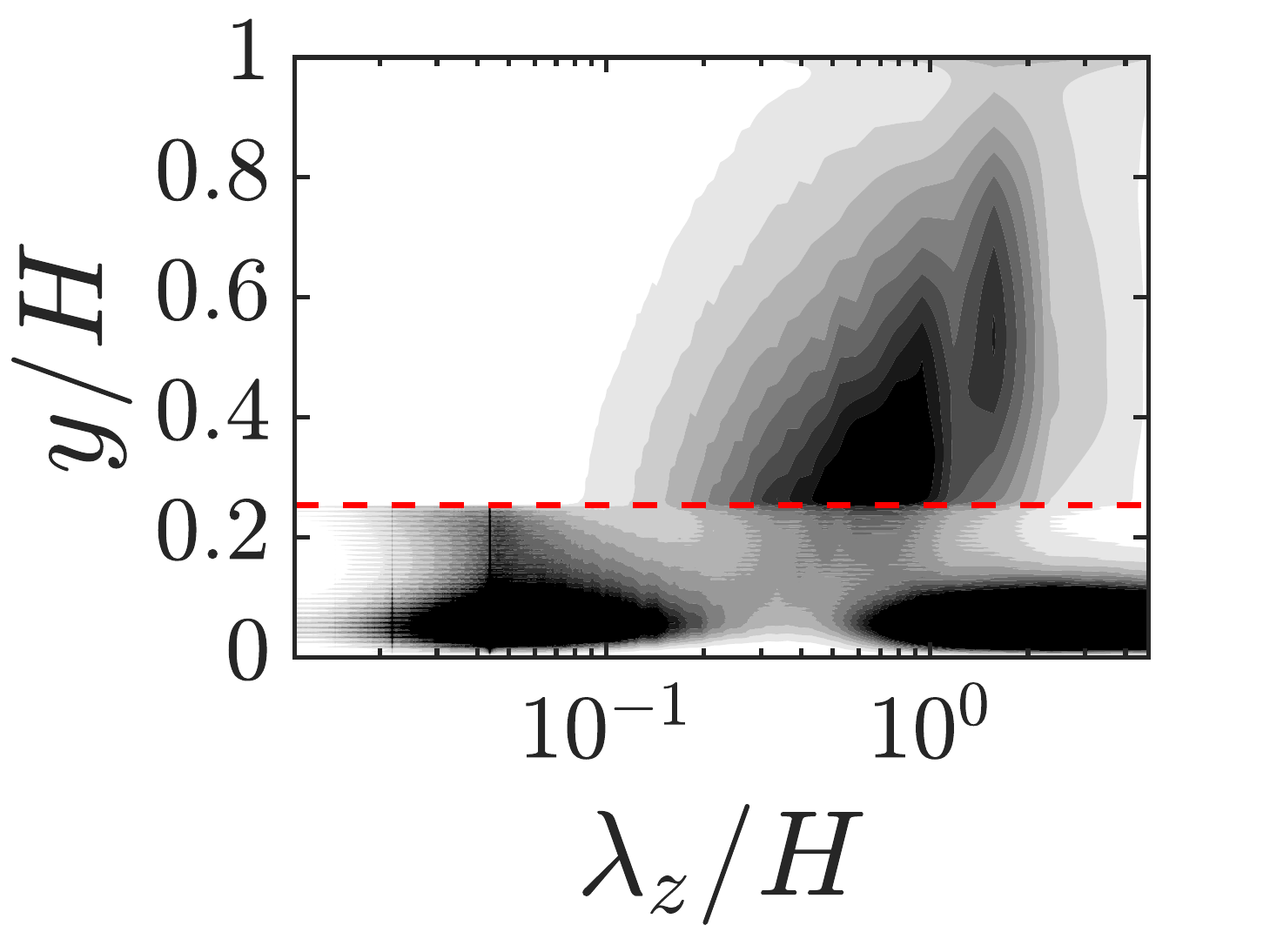}};
    \node[anchor=south west,inner sep=0] (image) at ( 5.4,0) {
    \includegraphics[width=.16\textwidth]{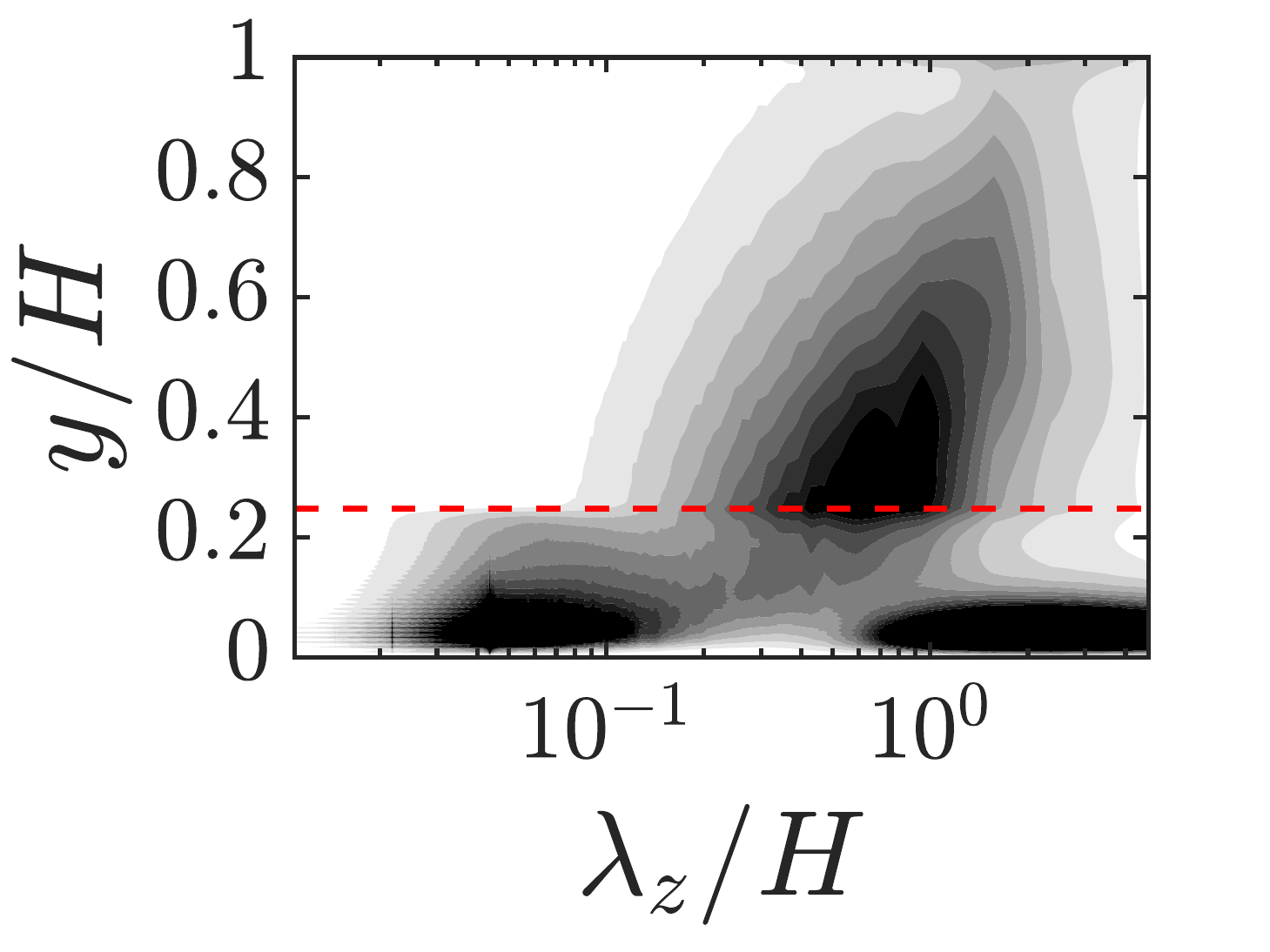}};
    \node[anchor=south west,inner sep=0] (image) at ( 8.1,0) {
    \includegraphics[width=.16\textwidth]{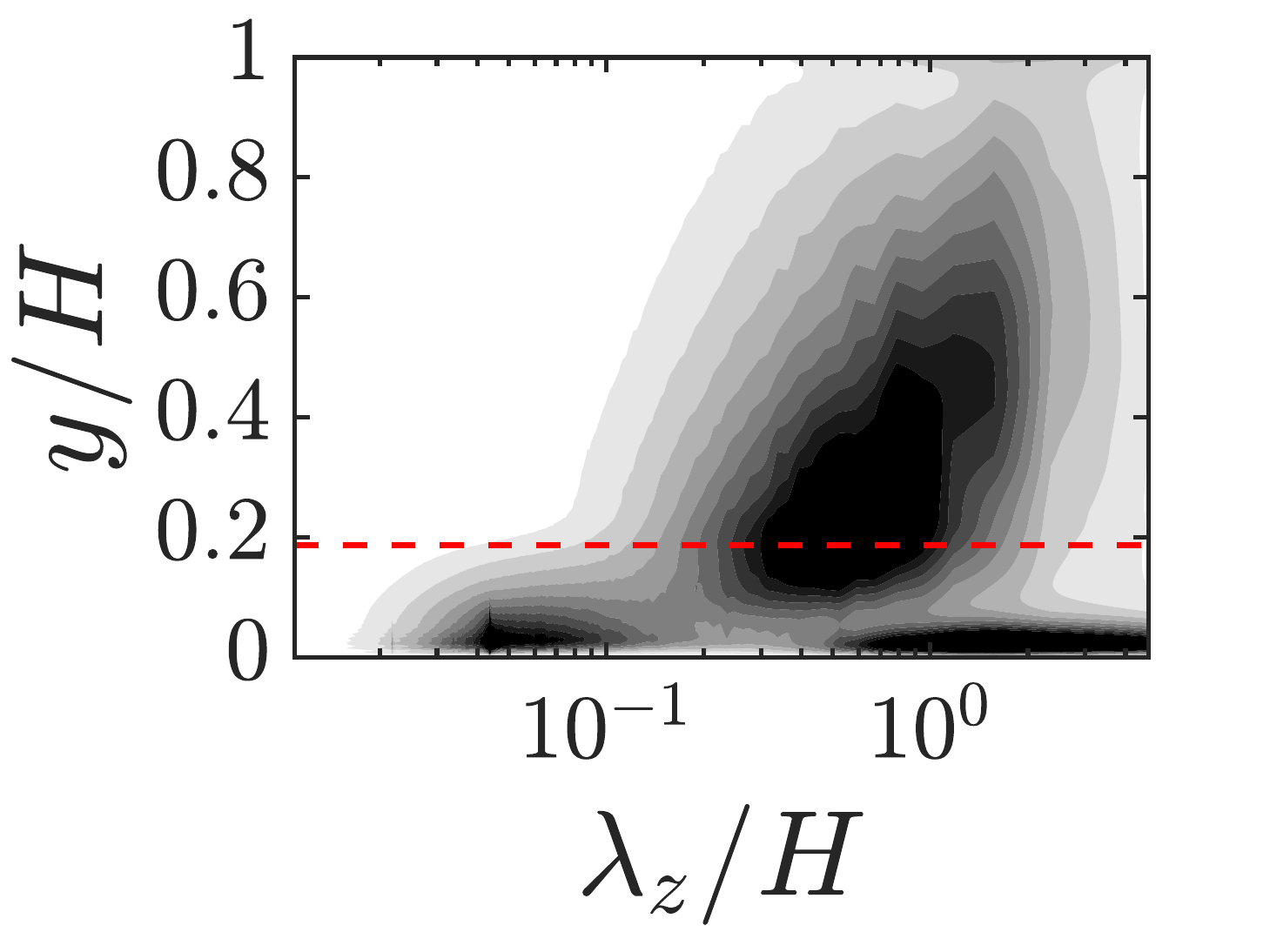}};
    \node[anchor=south west,inner sep=0] (image) at (10.8,0) {
    \includegraphics[width=.16\textwidth]{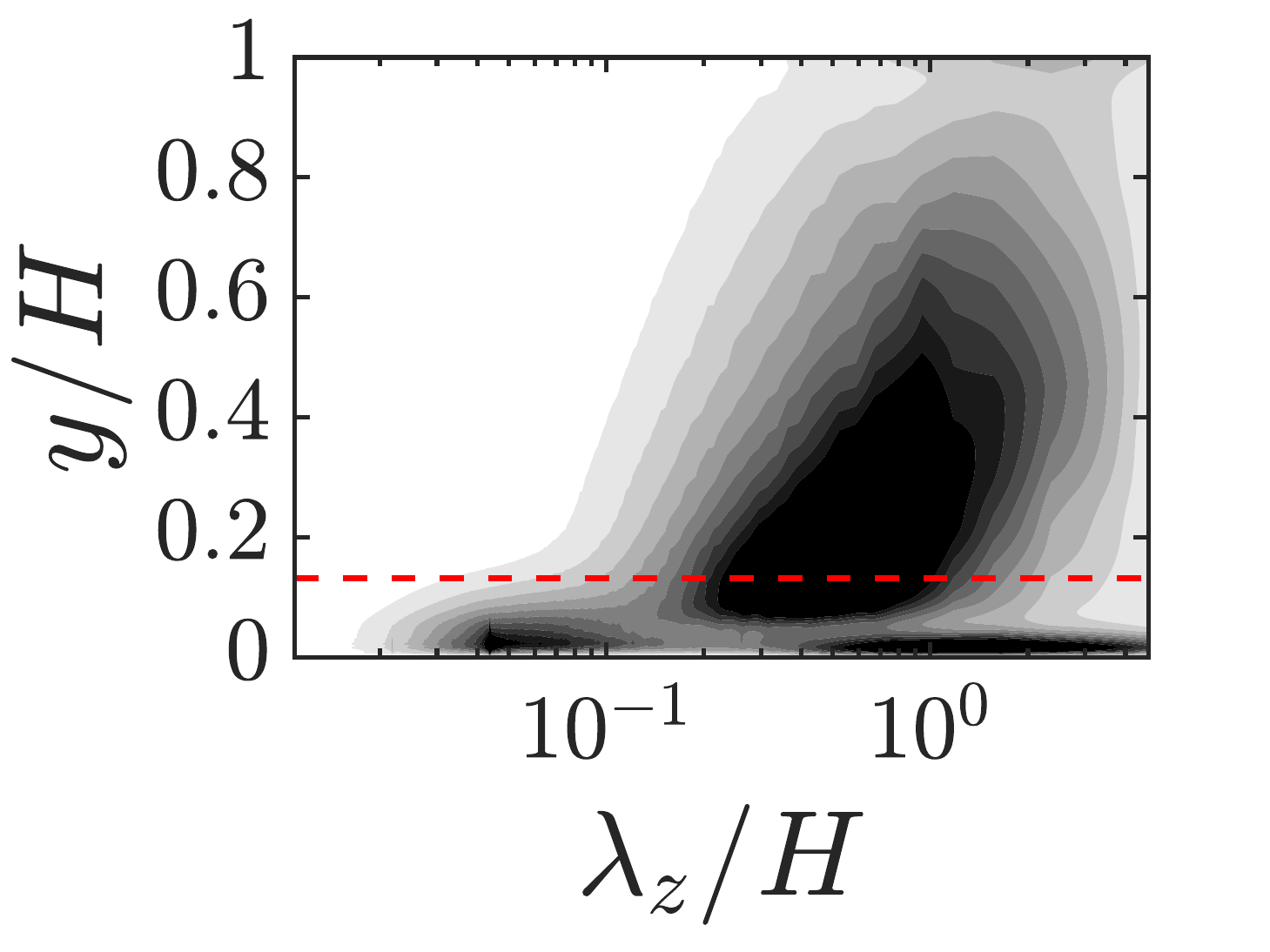}};
    \node[anchor=south west,inner sep=0] (image) at (13.5,0) {
    \includegraphics[width=.16\textwidth]{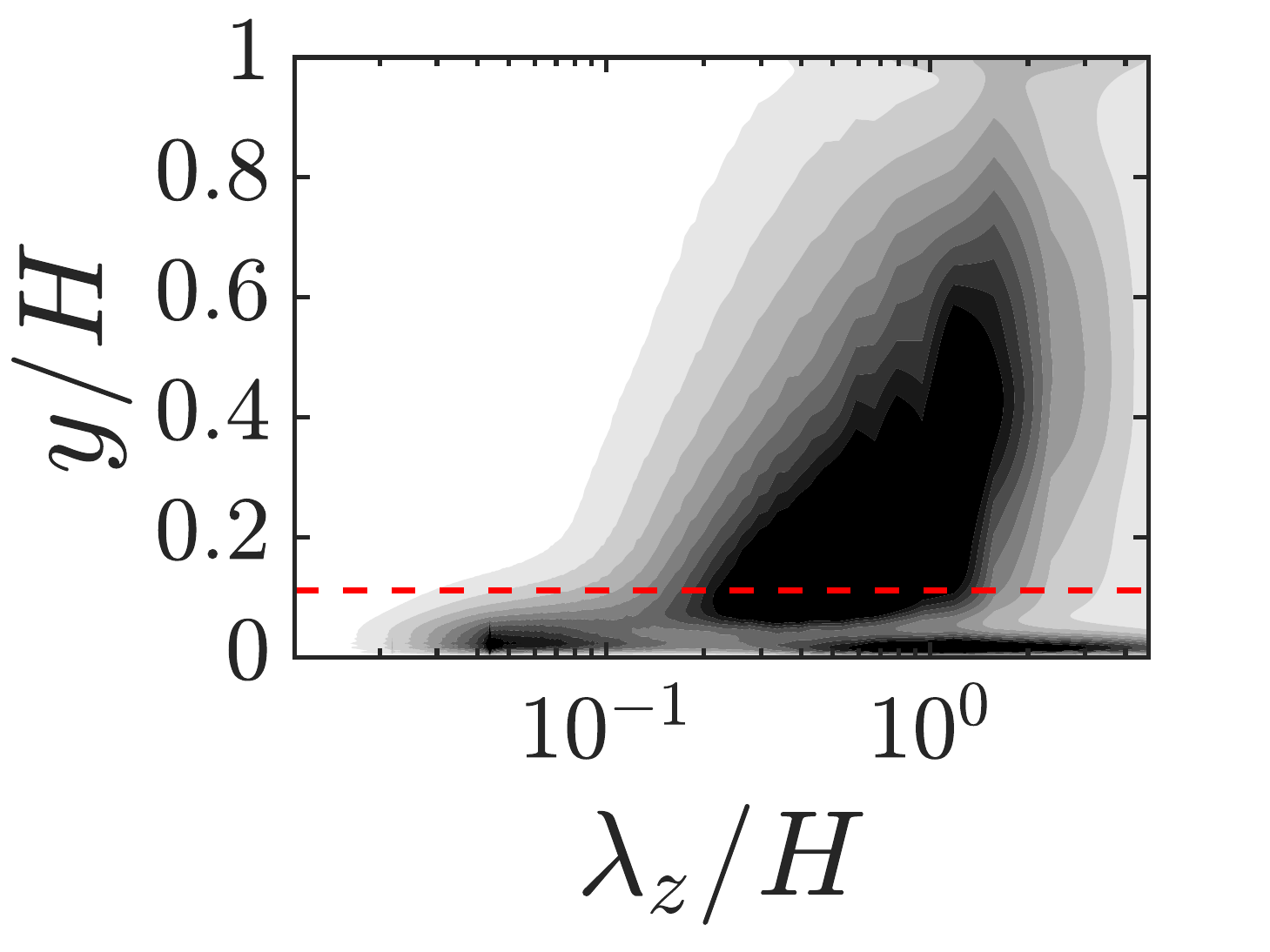}};

    \node[] at (-0.7,-1.2) {\small$\dfrac{2\pi\Phi_{v'v'}}{\lambda_z u_\tau^2}$};
    \node[anchor=south west,inner sep=0] (image) at ( 0.0,-2.50) {
    \includegraphics[width=.16\textwidth]{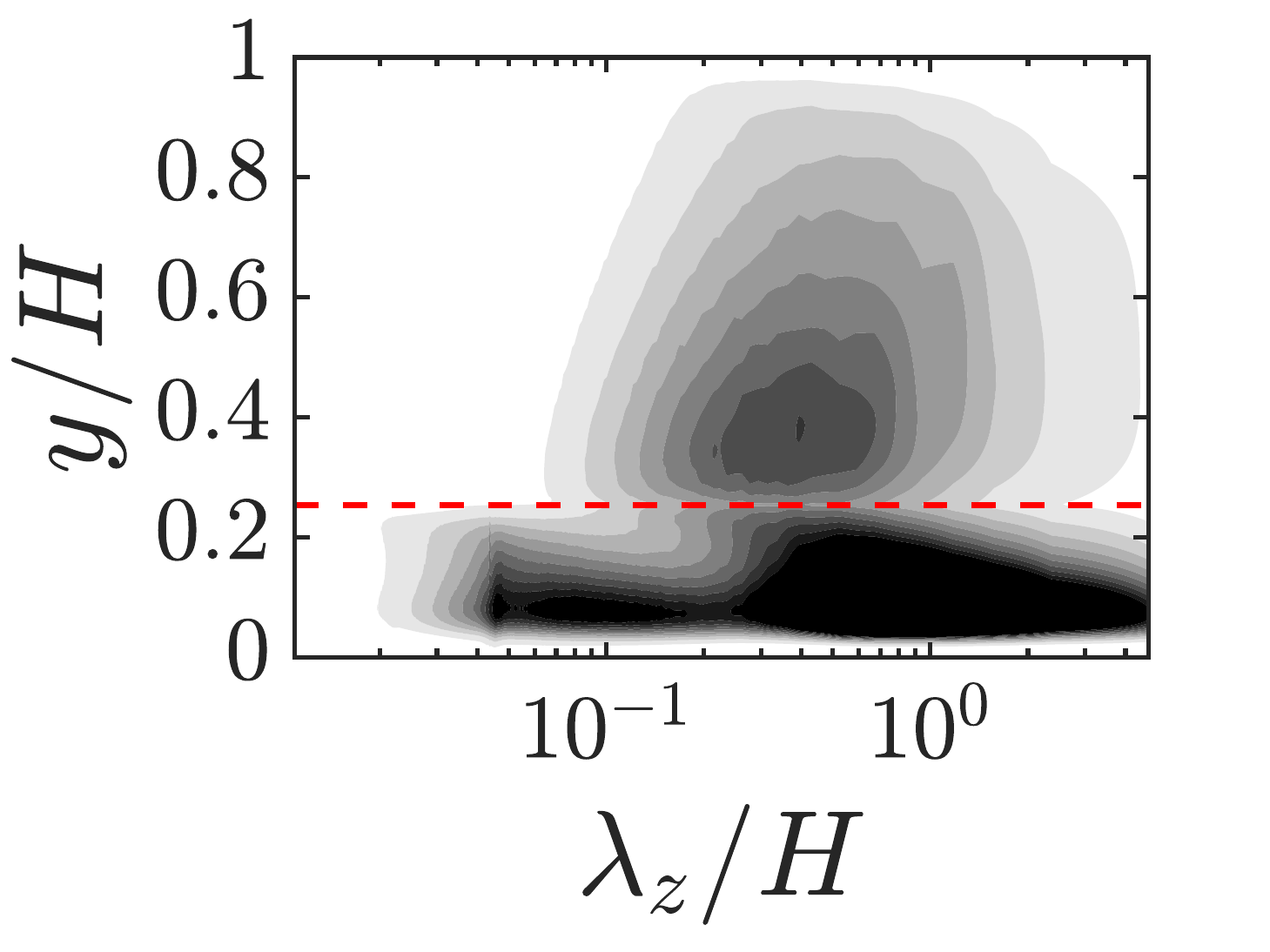}};
    \node[anchor=south west,inner sep=0] (image) at ( 2.7,-2.50) {
    \includegraphics[width=.16\textwidth]{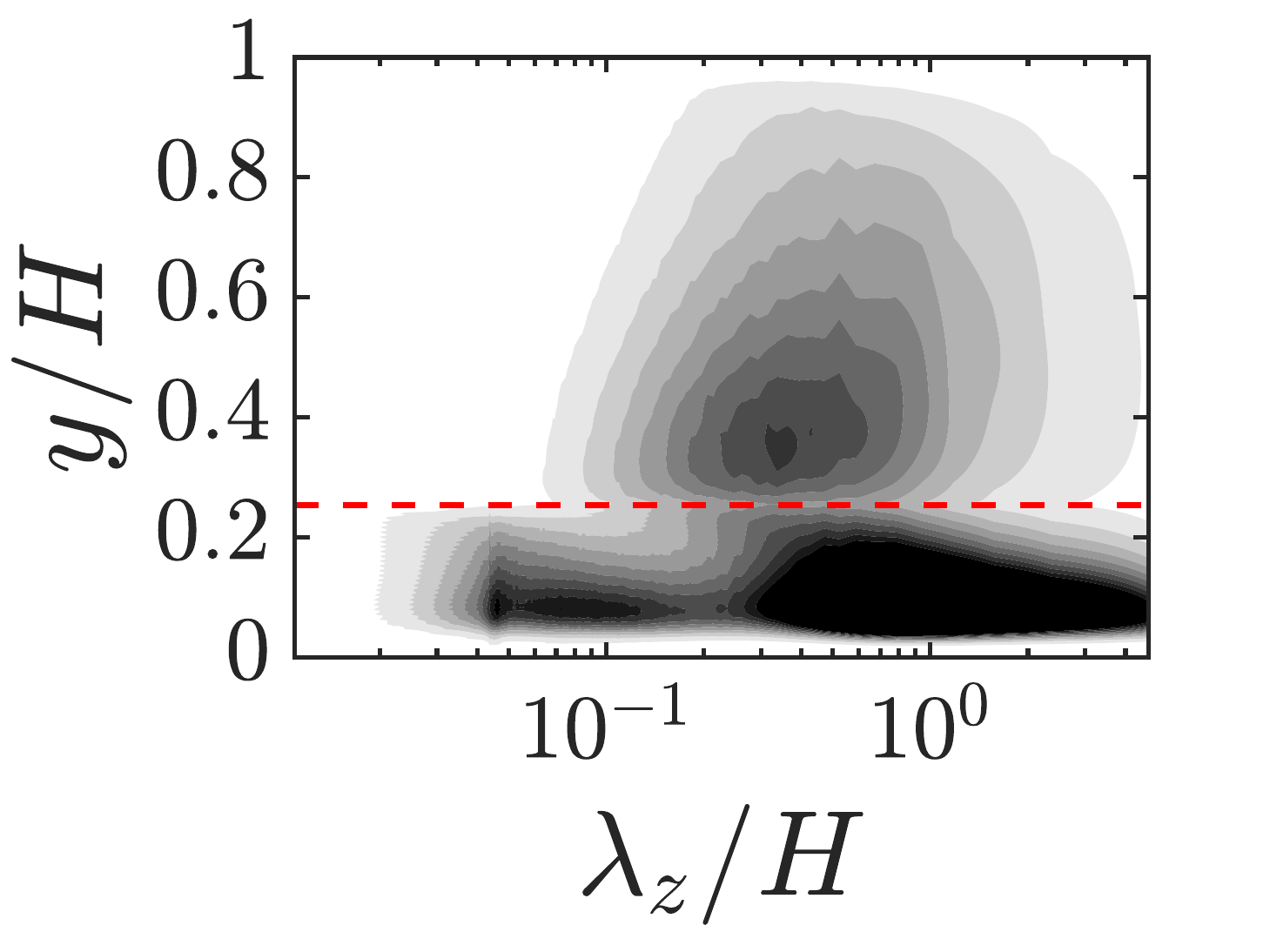}};
    \node[anchor=south west,inner sep=0] (image) at ( 5.4,-2.50) {
    \includegraphics[width=.16\textwidth]{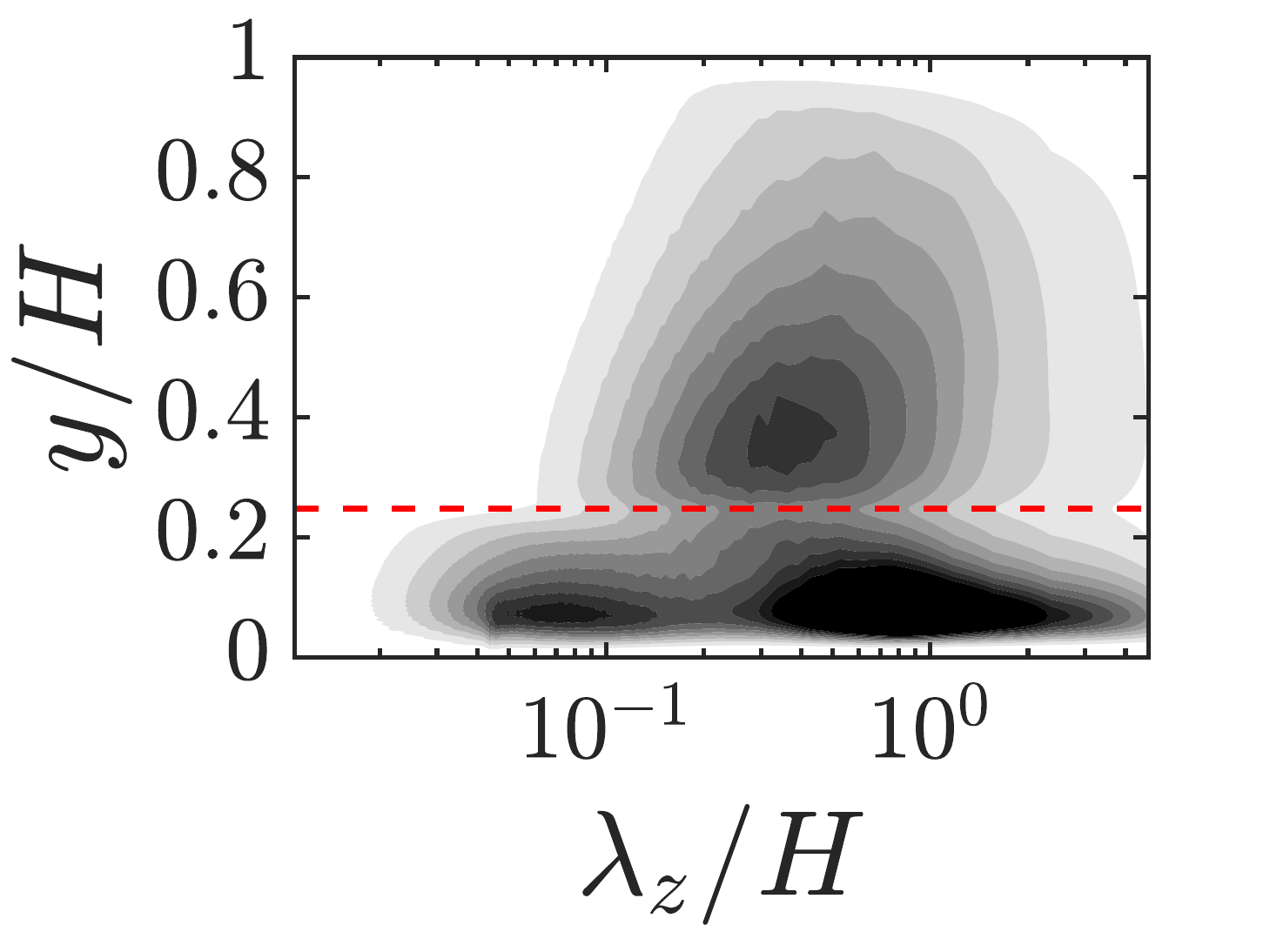}};
    \node[anchor=south west,inner sep=0] (image) at ( 8.1,-2.50) {
    \includegraphics[width=.16\textwidth]{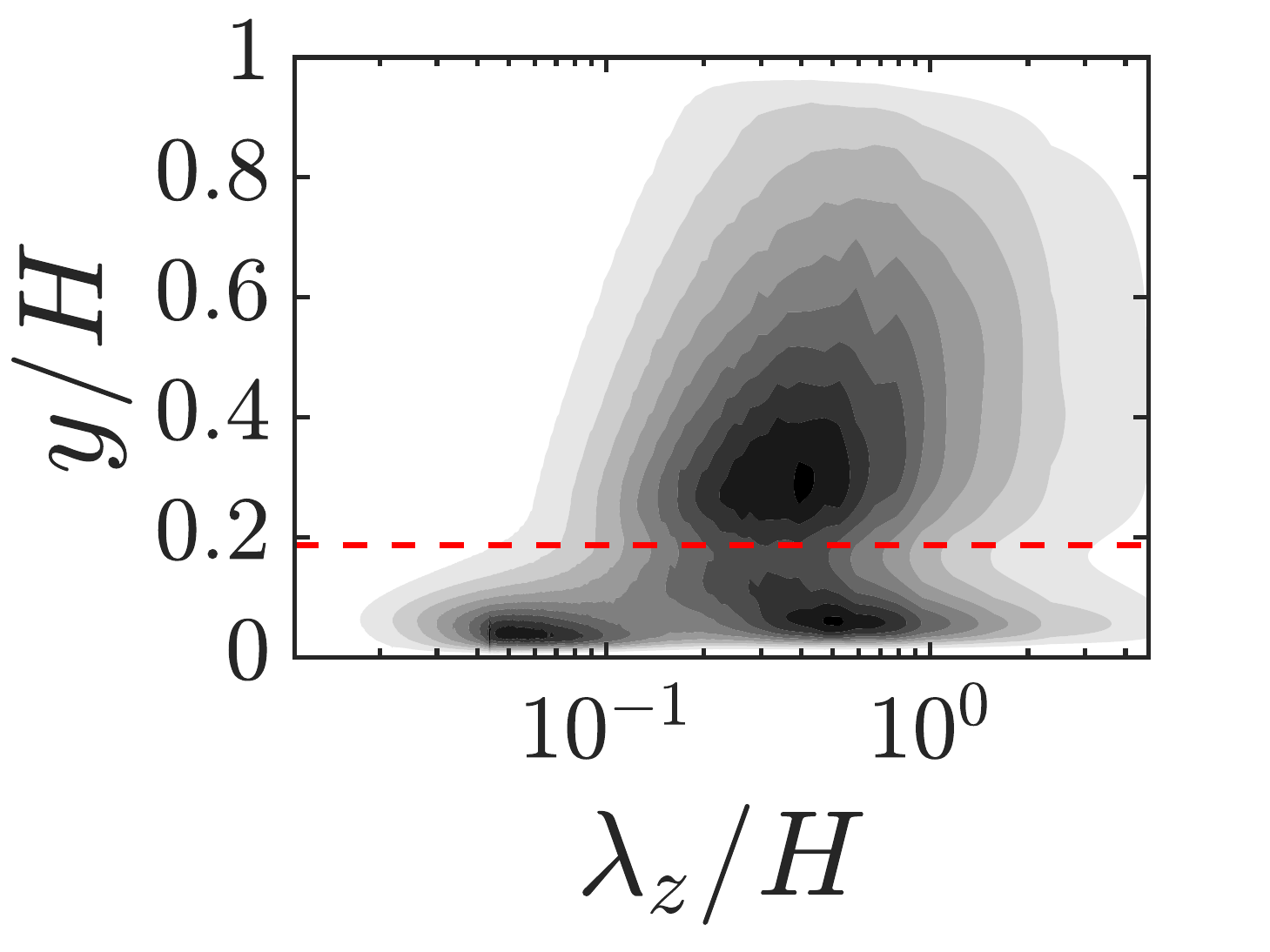}};
    \node[anchor=south west,inner sep=0] (image) at (10.8,-2.50) {
    \includegraphics[width=.16\textwidth]{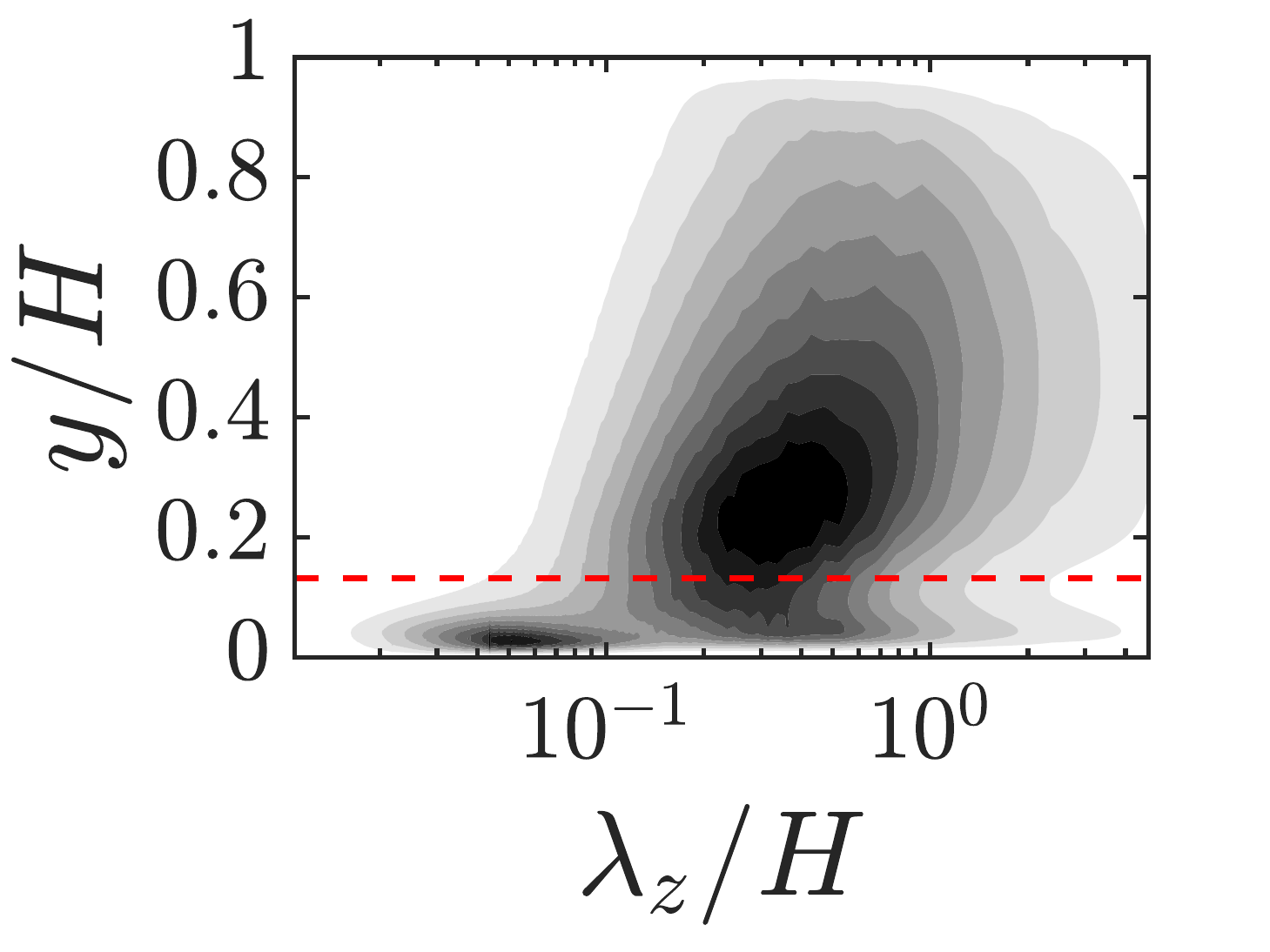}};
    \node[anchor=south west,inner sep=0] (image) at (13.5,-2.50) {
    \includegraphics[width=.16\textwidth]{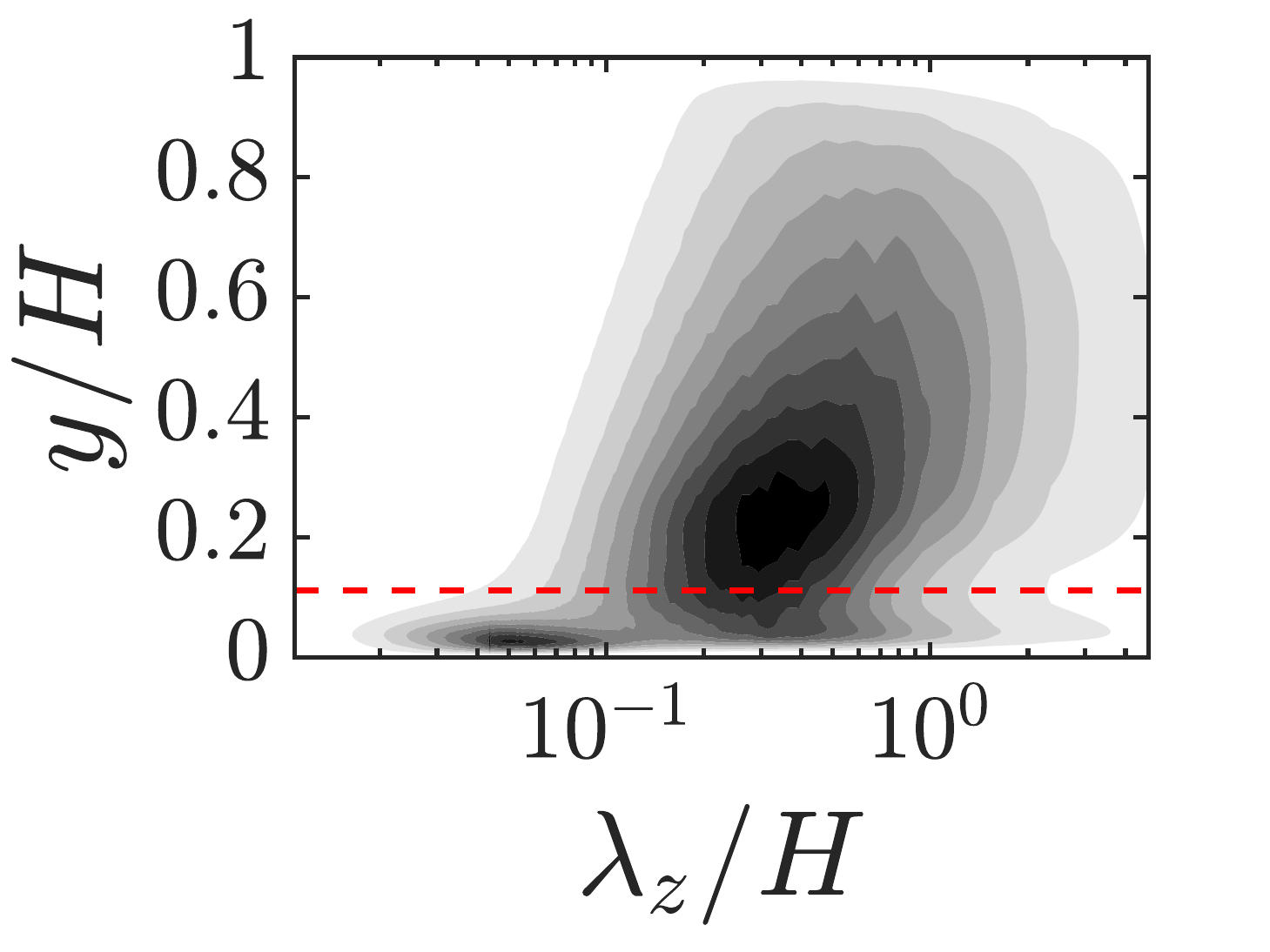}};

    \node[] at (-0.7,-3.7) {\small$\dfrac{2\pi\Phi_{w'w'}}{\lambda_z u_\tau^2}$};
    \node[anchor=south west,inner sep=0] (image) at ( 0.0,-5.0) {
    \includegraphics[width=.16\textwidth]{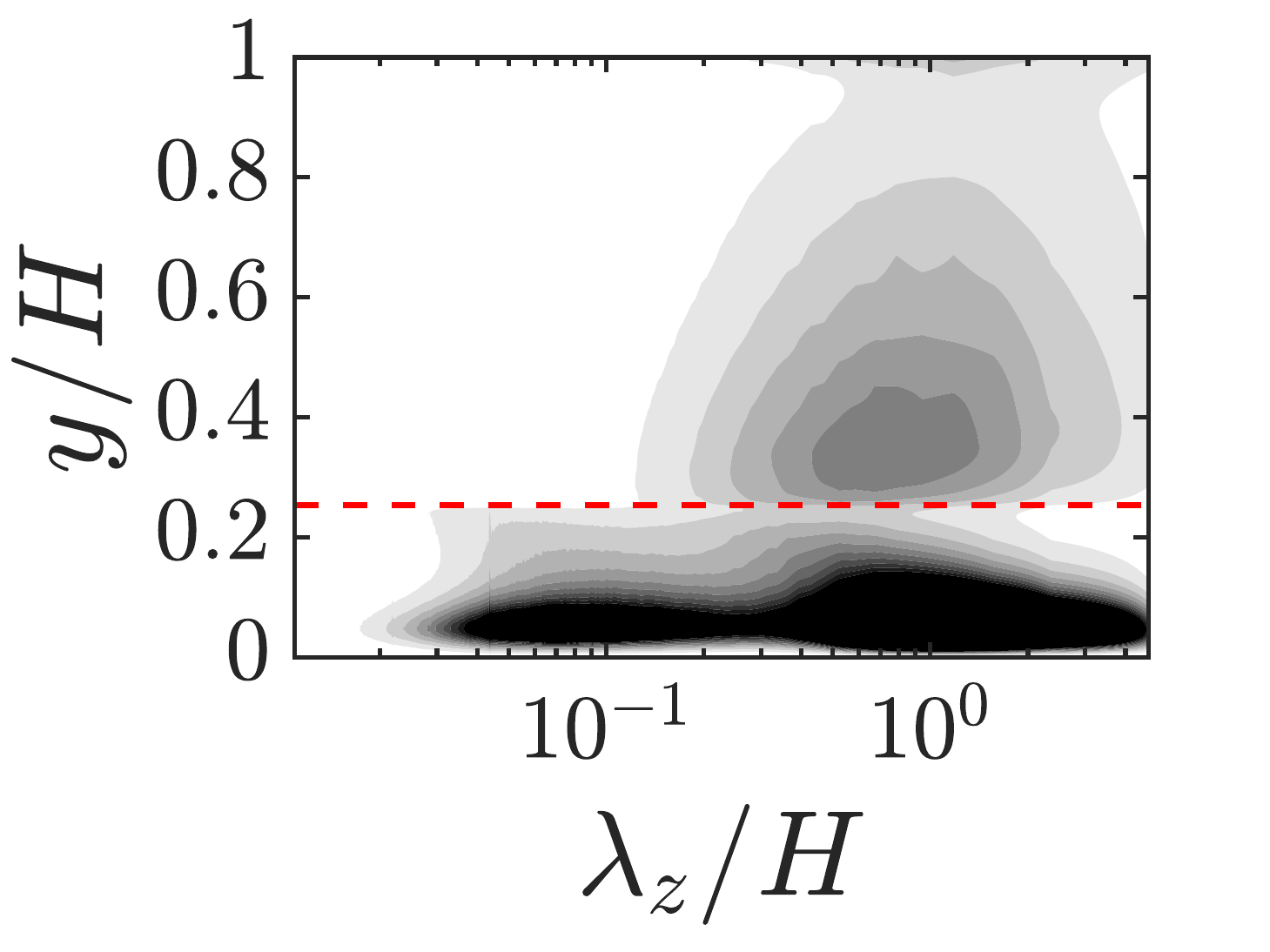}};
    \node[anchor=south west,inner sep=0] (image) at ( 2.7,-5.0) {
    \includegraphics[width=.16\textwidth]{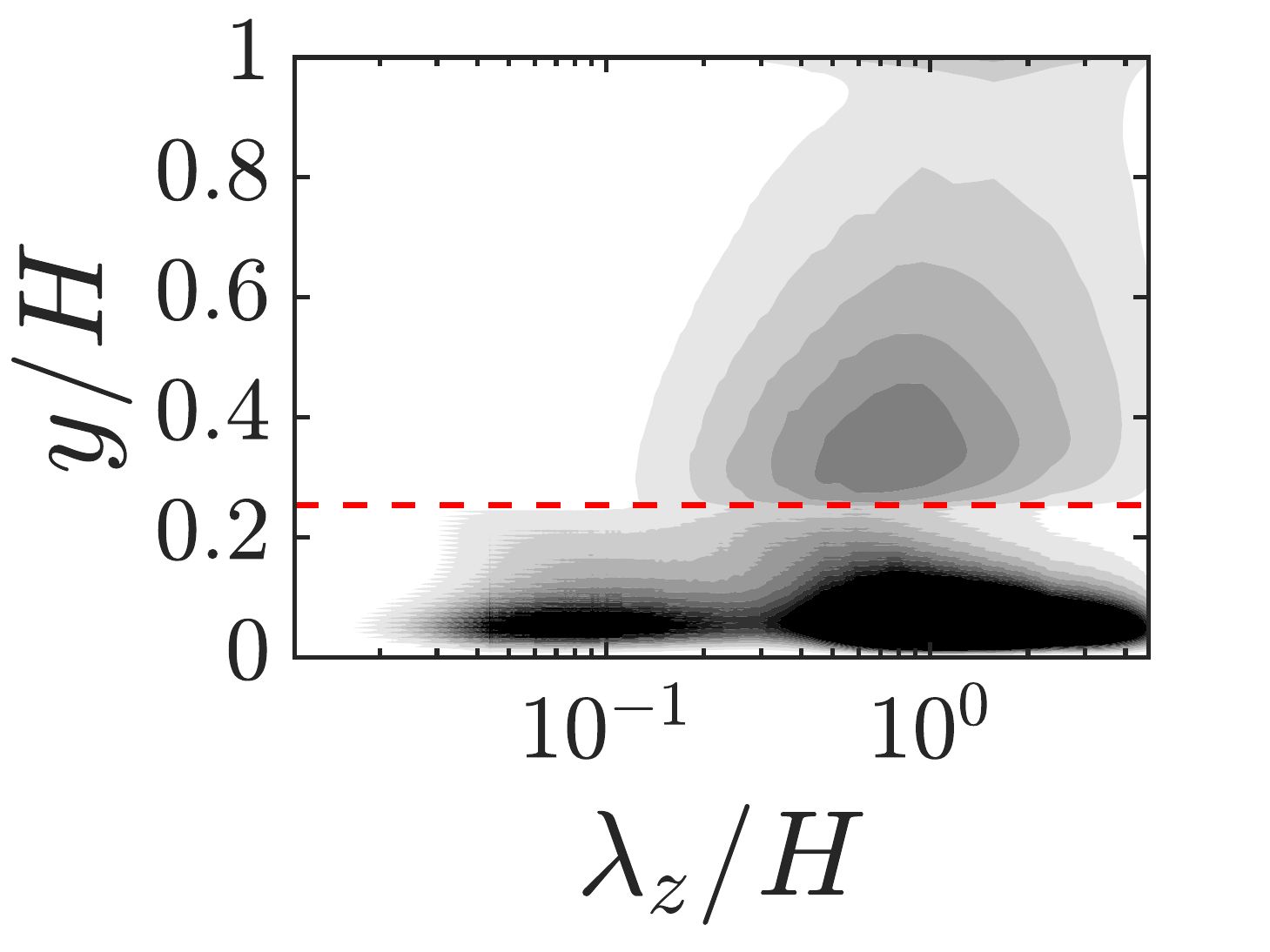}};
    \node[anchor=south west,inner sep=0] (image) at ( 5.4,-5.0) {
    \includegraphics[width=.16\textwidth]{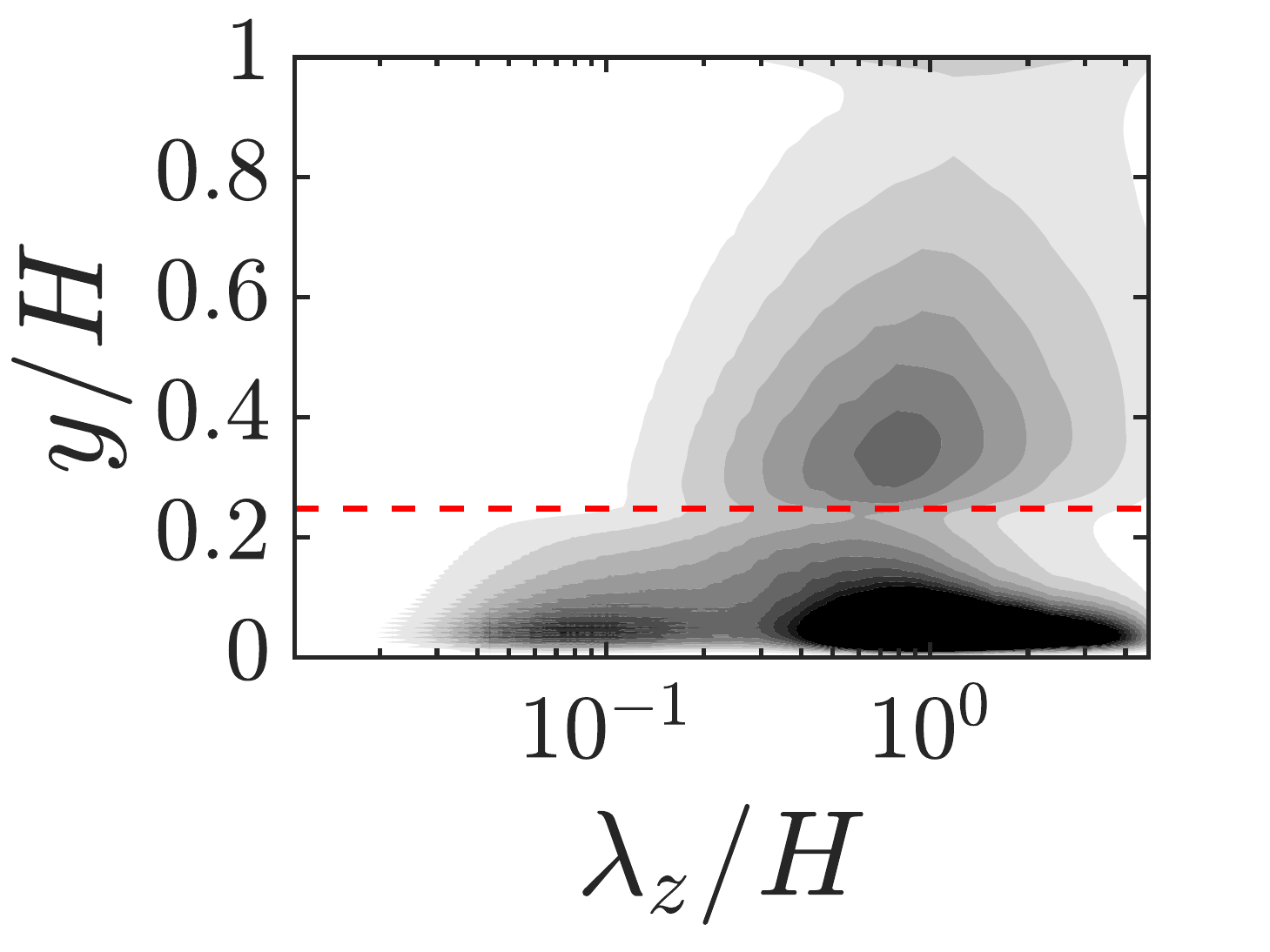}};
    \node[anchor=south west,inner sep=0] (image) at ( 8.1,-5.0) {
    \includegraphics[width=.16\textwidth]{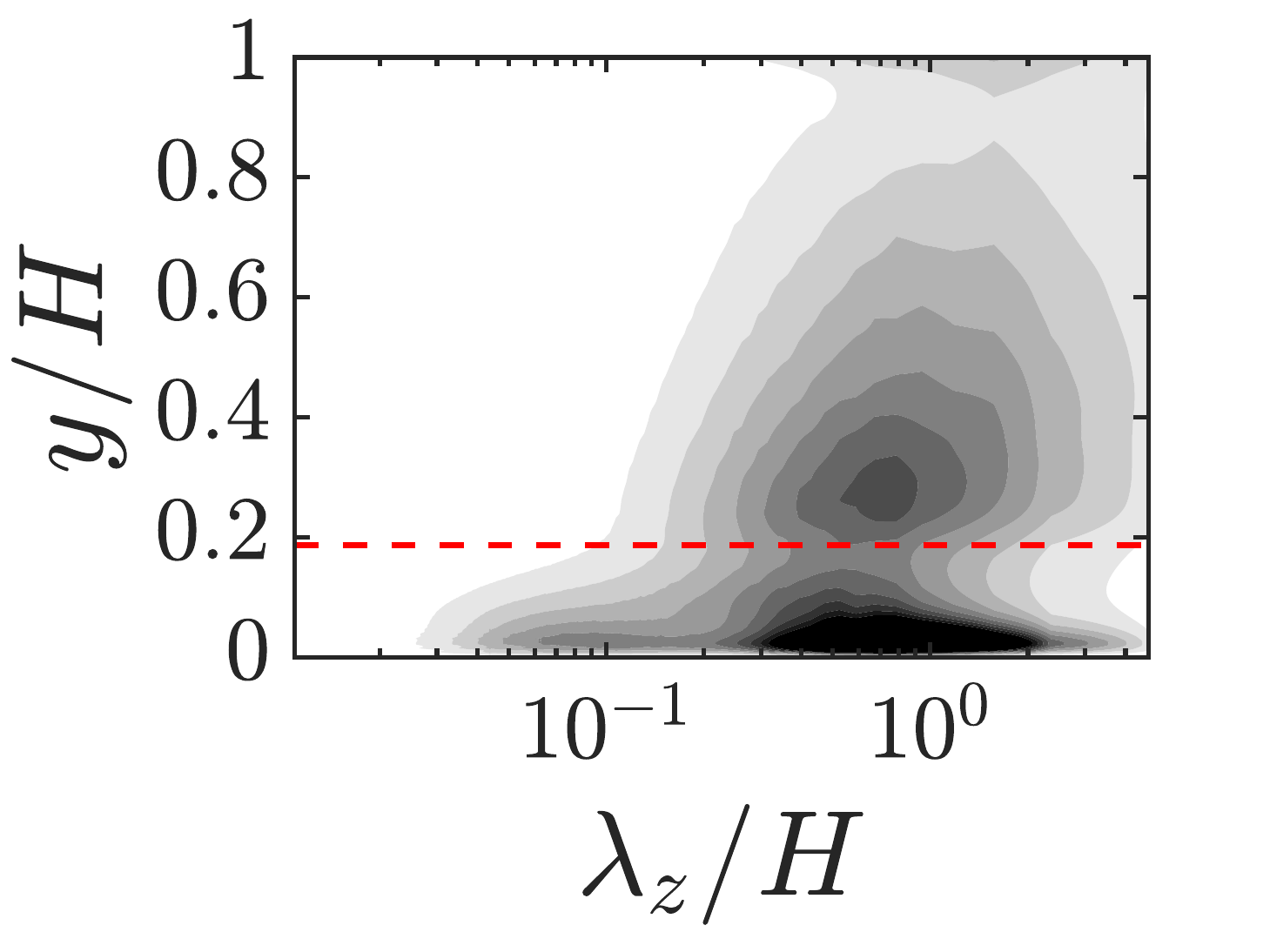}};
    \node[anchor=south west,inner sep=0] (image) at (10.8,-5.0) {
    \includegraphics[width=.16\textwidth]{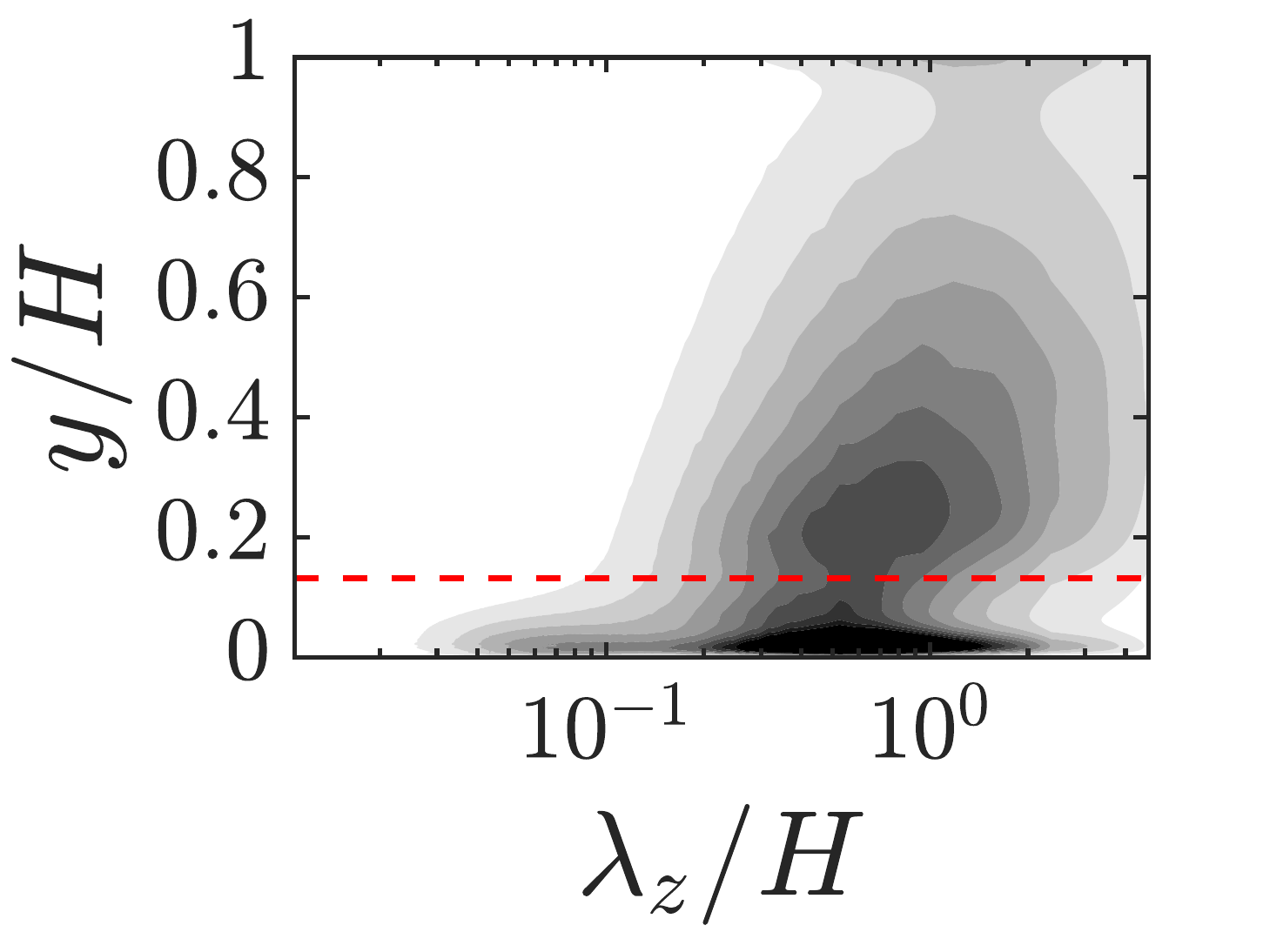}};
    \node[anchor=south west,inner sep=0] (image) at (13.5,-5.0) {
    \includegraphics[width=.16\textwidth]{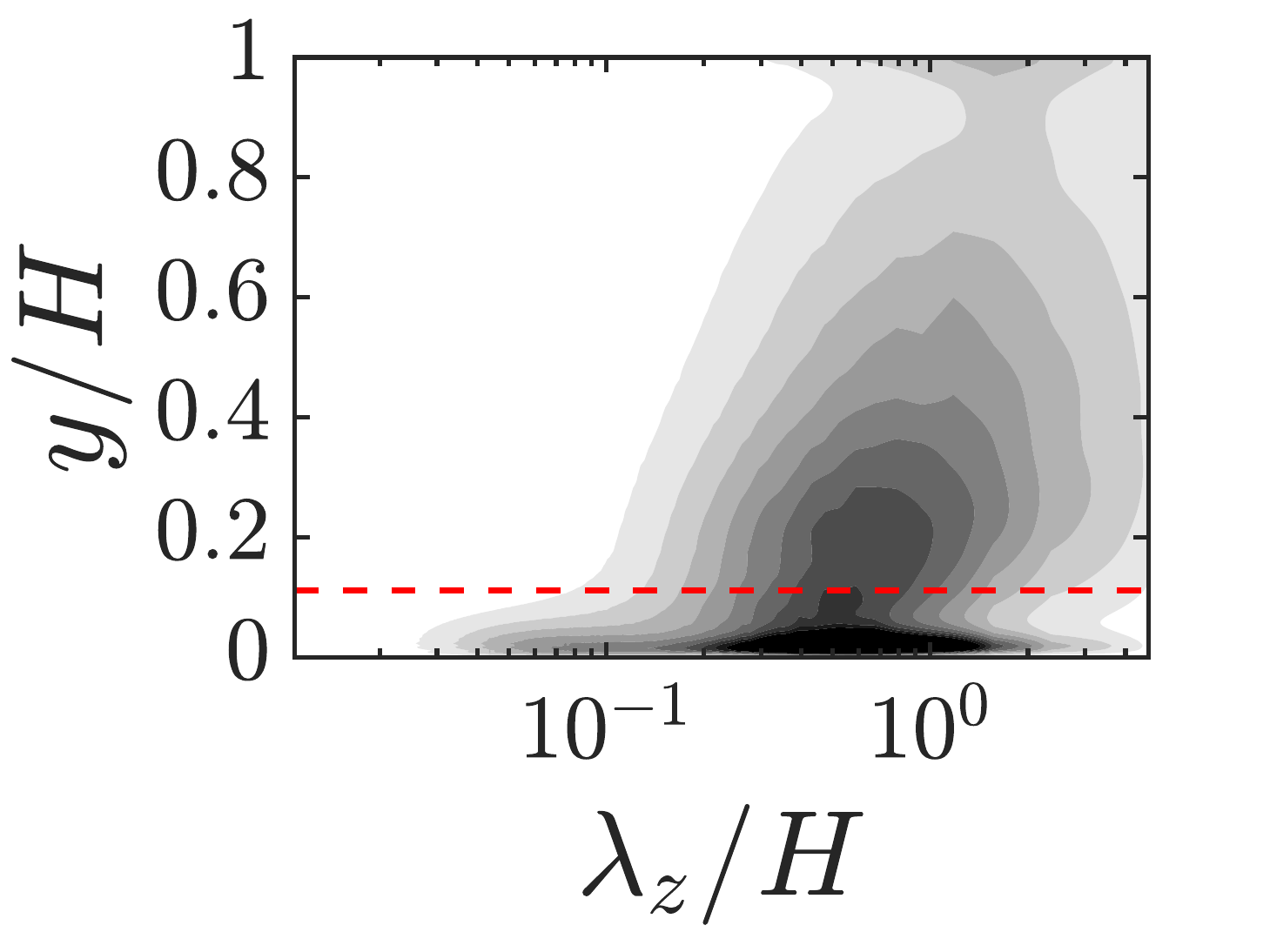}};
    \end{tikzpicture}
    \caption{\added{
    Magnitude of the premultiplied spectra of each fluid velocity component 
$2\pi\Phi_{u'u'}/(u_\tau^2\lambda_z)$, where 
	$u'$ the generic component of the velocity fluctuations,
	(top: streamwise; middle: wall-normal; bottom: spanwise) as a function of the spanwise wavelength $\lambda_z/H$ and wall-normal coordinate $y/H$. 
	The friction velocity is defined locally as $u_\tau=\sqrt{\tau/[\rho_f(1-y/H)]}$, where $\tau$ is the flow total shear stress, i.e. the sum of the viscous and the turbulent component \cite{monti2020genesis}.
	Results are shown in different columns as a function of the investigated Cauchy number (from left to right, $\Ca=0, 1, 10, 25, 50, 100$). 
	The red horizontal dashed line indicates the averaged height of the filament tips.
	The grey levels range in: $[0, 0.5]$ with a $0.05$ increment for the streamwise and spanwise velocity components; $[0, 0.3]$ with a $0.03$ increment for the wall-normal velocity component.
}
    }
    \label{fig4_frict}
\end{figure}

\clearpage

\begin{figure}
    \centering
    \begin{tikzpicture}
    \node[] at ( 3.0, 4.6) {\small$Ca=0$};
    \node[] at ( 8.8, 4.6) {\small$Ca=1$};
    \node[] at (14.6, 4.6) {\small$Ca=10$};
    \node[] at ( 3.0,-5.0) {\small$Ca=25$};
    \node[] at ( 8.8,-5.0) {\small$Ca=50$};
    \node[] at (14.6,-5.0) {\small$Ca=100$};

    \node[anchor=south west,inner sep=0] (image) at ( 0.0,0) {
    \includegraphics[width=.30\textwidth]{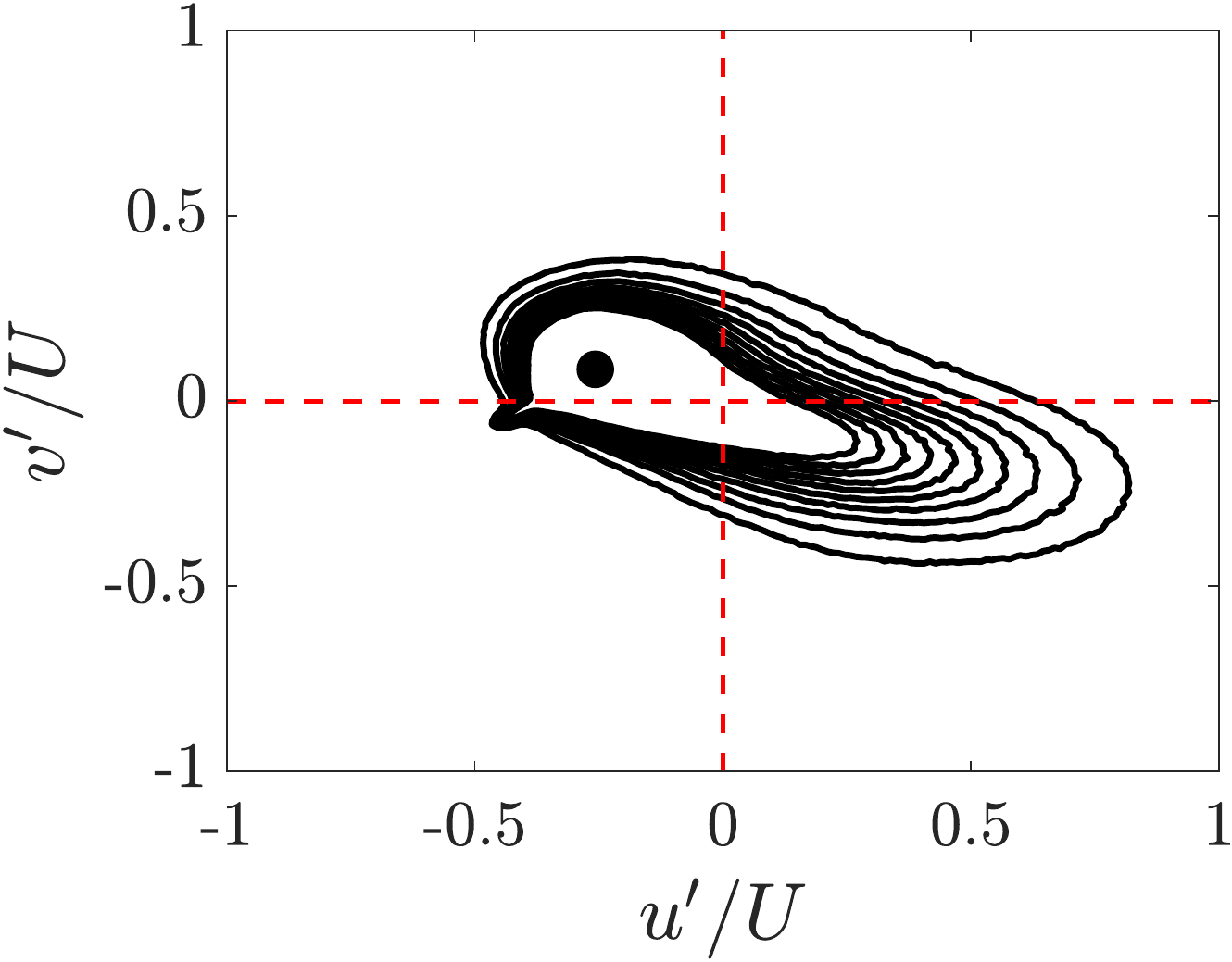}};
    \node[anchor=south west,inner sep=0] (image) at ( 5.8,0) {
    \includegraphics[width=.30\textwidth]{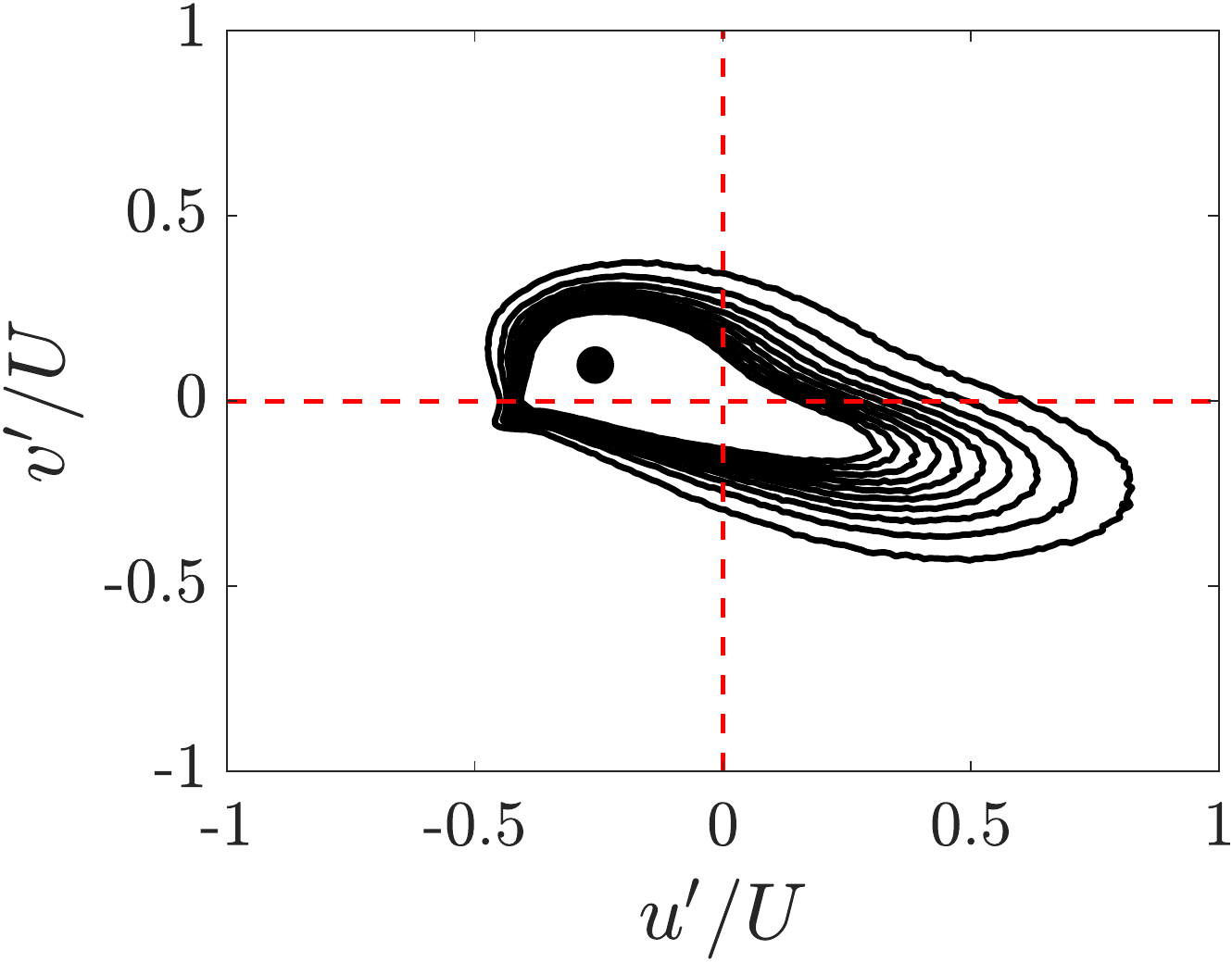}};
    \node[anchor=south west,inner sep=0] (image) at (11.6,0) {
    \includegraphics[width=.30\textwidth]{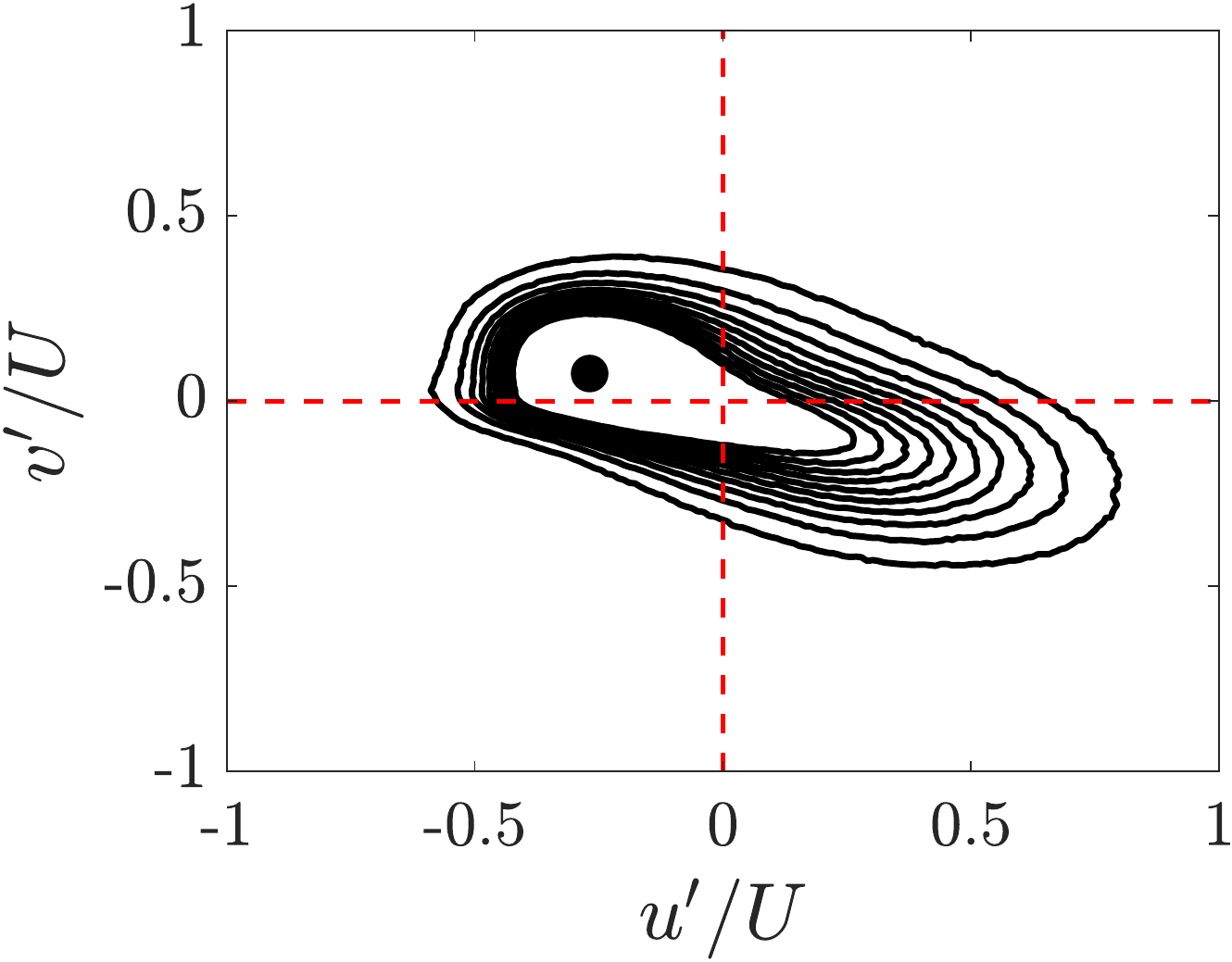}};
    \node[anchor=south west,inner sep=0] (image) at ( 0.0,-4.5) {
    \includegraphics[width=.30\textwidth]{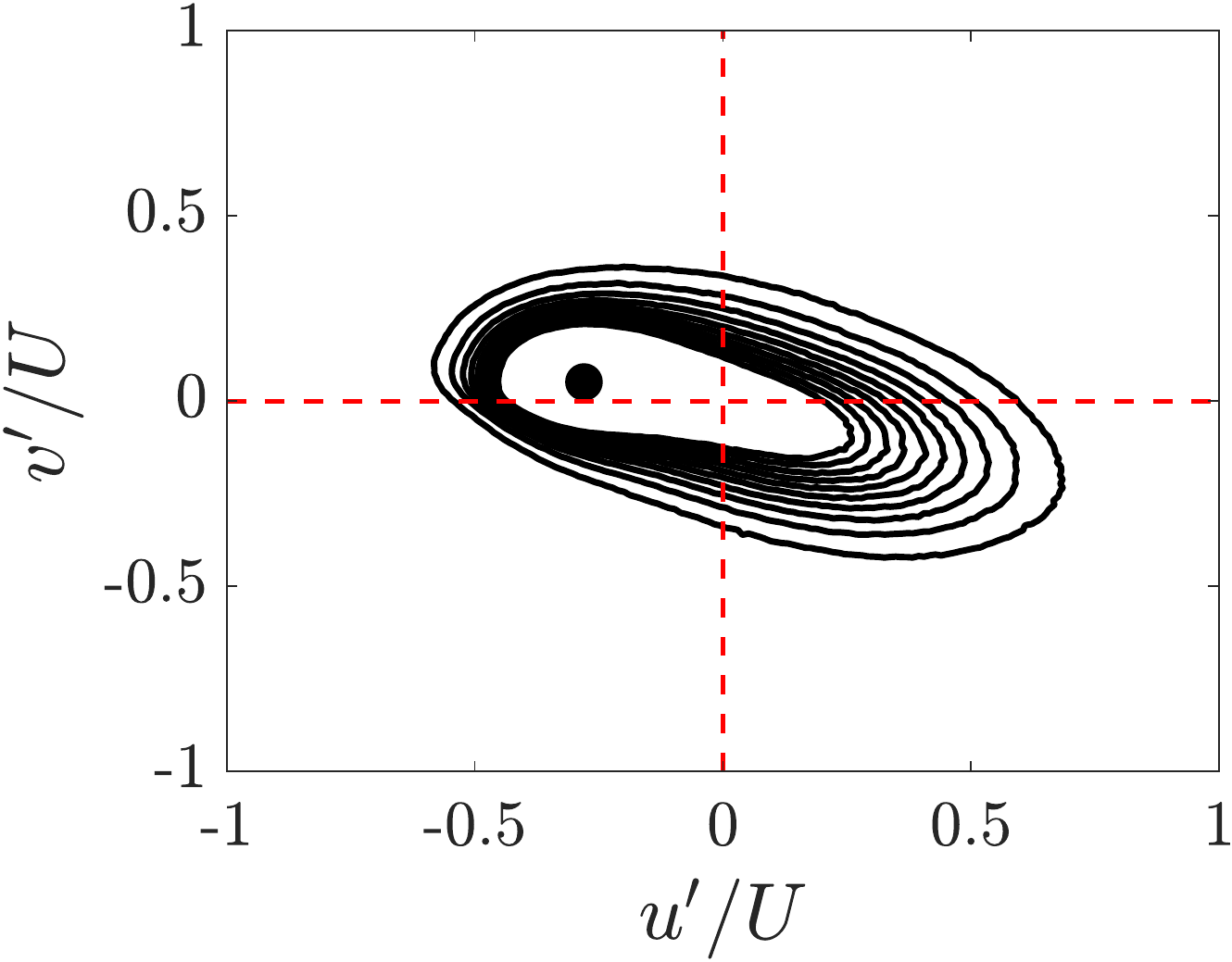}};
    \node[anchor=south west,inner sep=0] (image) at ( 5.8,-4.5) {
    \includegraphics[width=.30\textwidth]{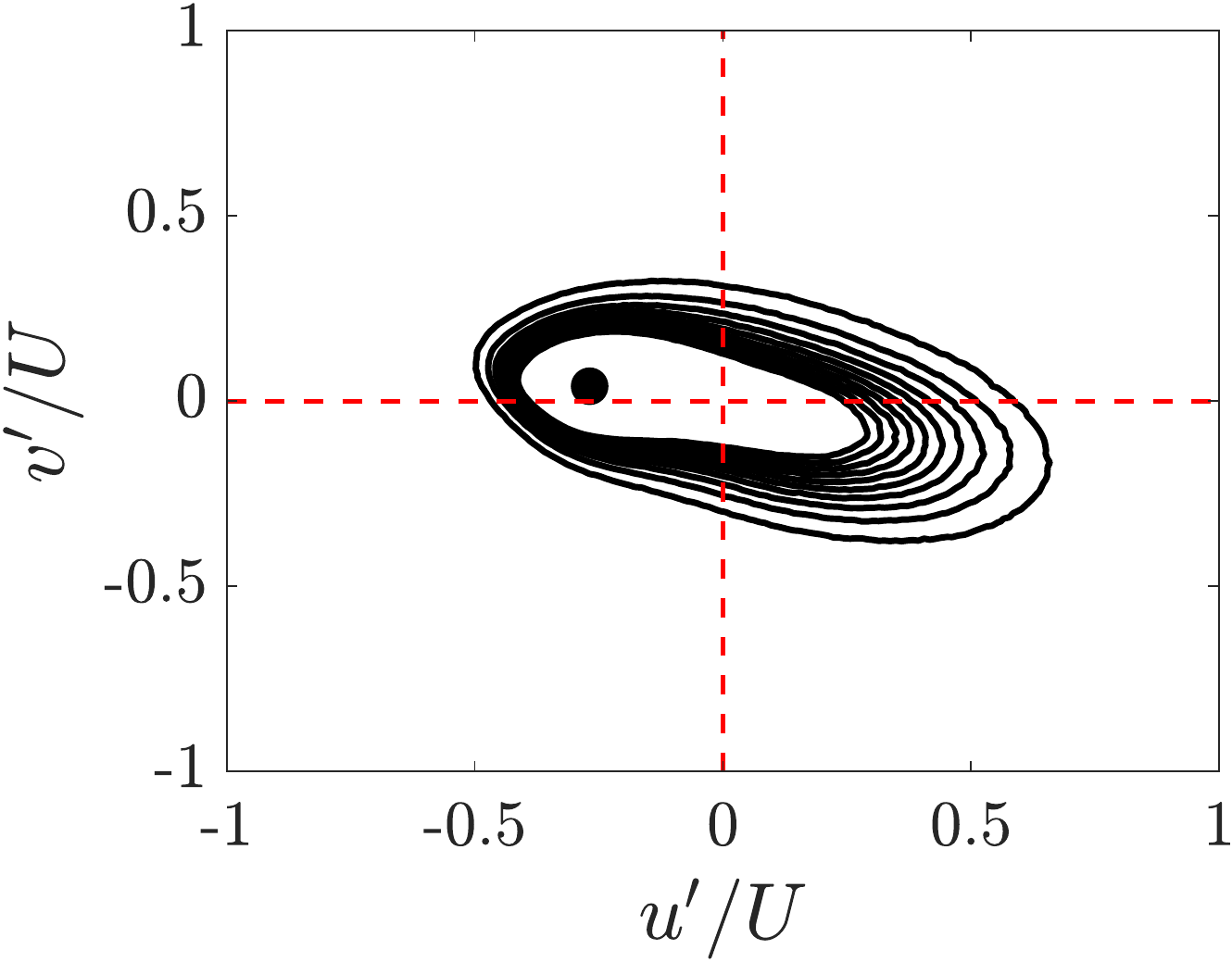}};
    \node[anchor=south west,inner sep=0] (image) at (11.6,-4.5) {
    \includegraphics[width=.30\textwidth]{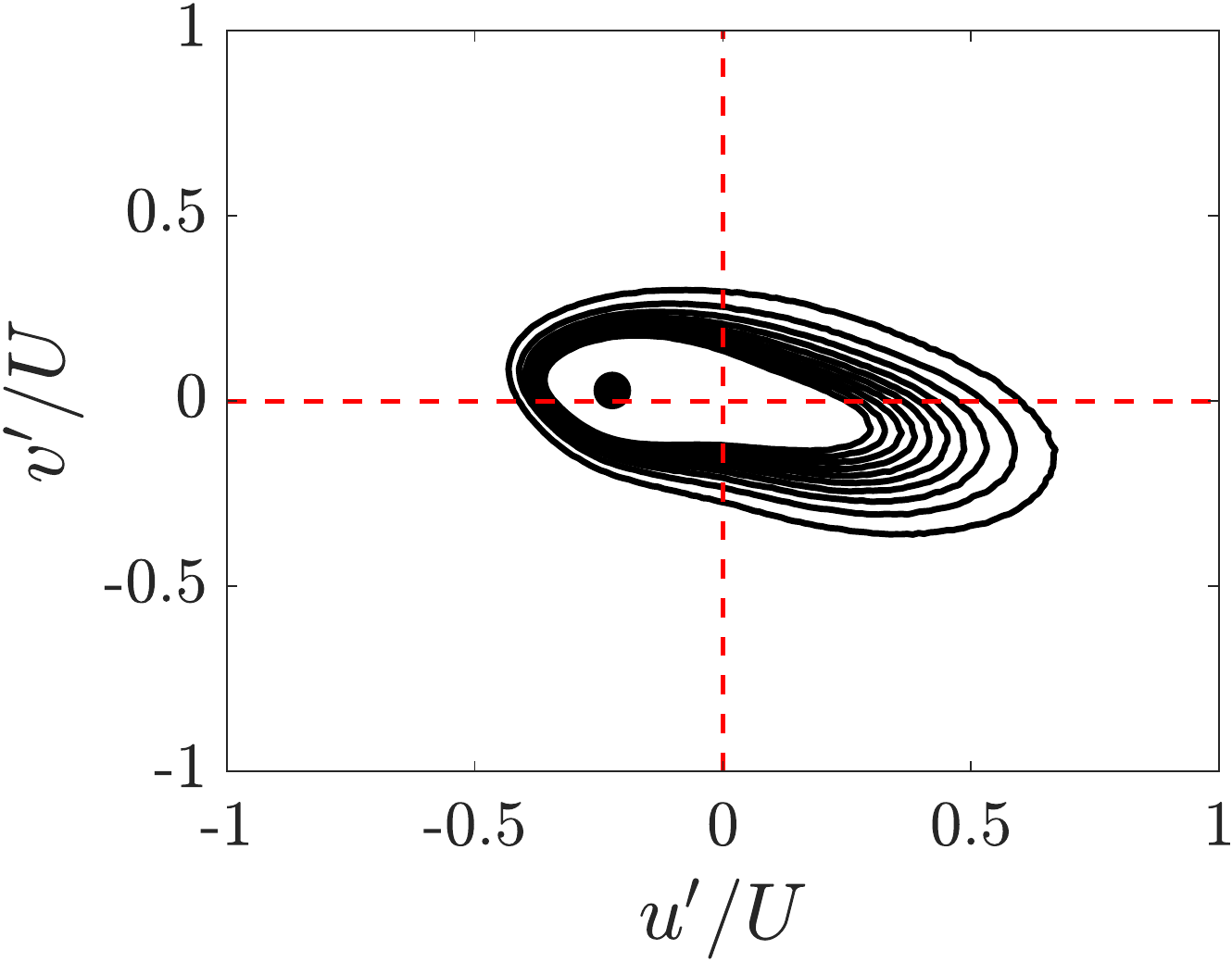}};
    \end{tikzpicture}
    \caption{\added{
           Contours of the joint probability density function of the fluctuations of 
           the streamwise velocity component $u'/U$ and of the wall-normal 
           velocity component $v'/U$ on a plane parallel to the wall, at $y/\bar{Y} = 1$,
	   where $\bar{Y}$ is the average canopy height. The joint probability density function
	   is computed for every Cauchy number. The levels of the contours range $[0:0.2:2]$,
	   with the maximum indicated by the black circular marker (within the second quadrant).
	   Note that the joint probability density functions are normalised such that the integral
	   over the domain is $1$.
	   The red dashed lines represent the axes $u'/U=0$ and $v'/U=0$. Note that the deviation from
	   the classic `oval' shape close to the axis $v'/U=0$ (with $u'/U< 0$) is 
	   caused by the wakes of the filaments.
   }}
    \label{fig:uvjpdf}
\end{figure}

\clearpage

\begin{figure}
  \centering
  \includegraphics[width=0.5\linewidth]{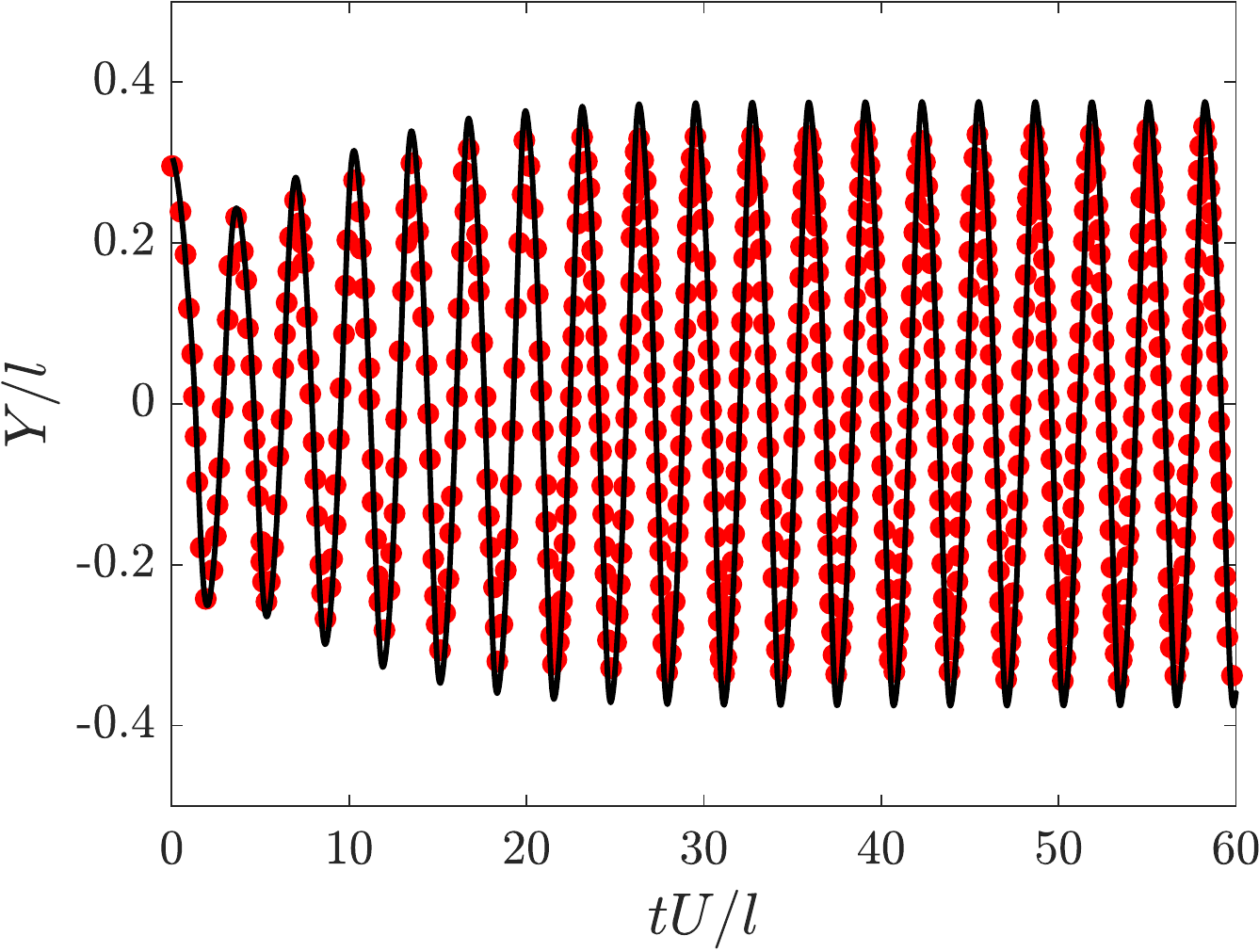}
    \caption{
    \added{
	    Validation of the employed simulation procedure for a flapping filament in uniform flow at $\Rey = 200$ (for more information on the problem setup, see Huang \textit{et al.}\cite{huang_shin_sung_2007a} and, in particular, figure 13 therein). Comparison of the time history of the filament’s trailing point transverse position obtained with \textit{Fujin} (black line) and that of Huang \textit{et al.}\cite{huang_shin_sung_2007a} (red circles). Note that the time is normalised with the inflow velocity $U$ and the length of the filament $l$, while the position of the trailing point is normalised with the length of the filament.
   }
    }
    \label{fig:val_huang}
\end{figure}

\clearpage

\begin{figure}
    \centering
    \includegraphics[width=.40\textwidth]{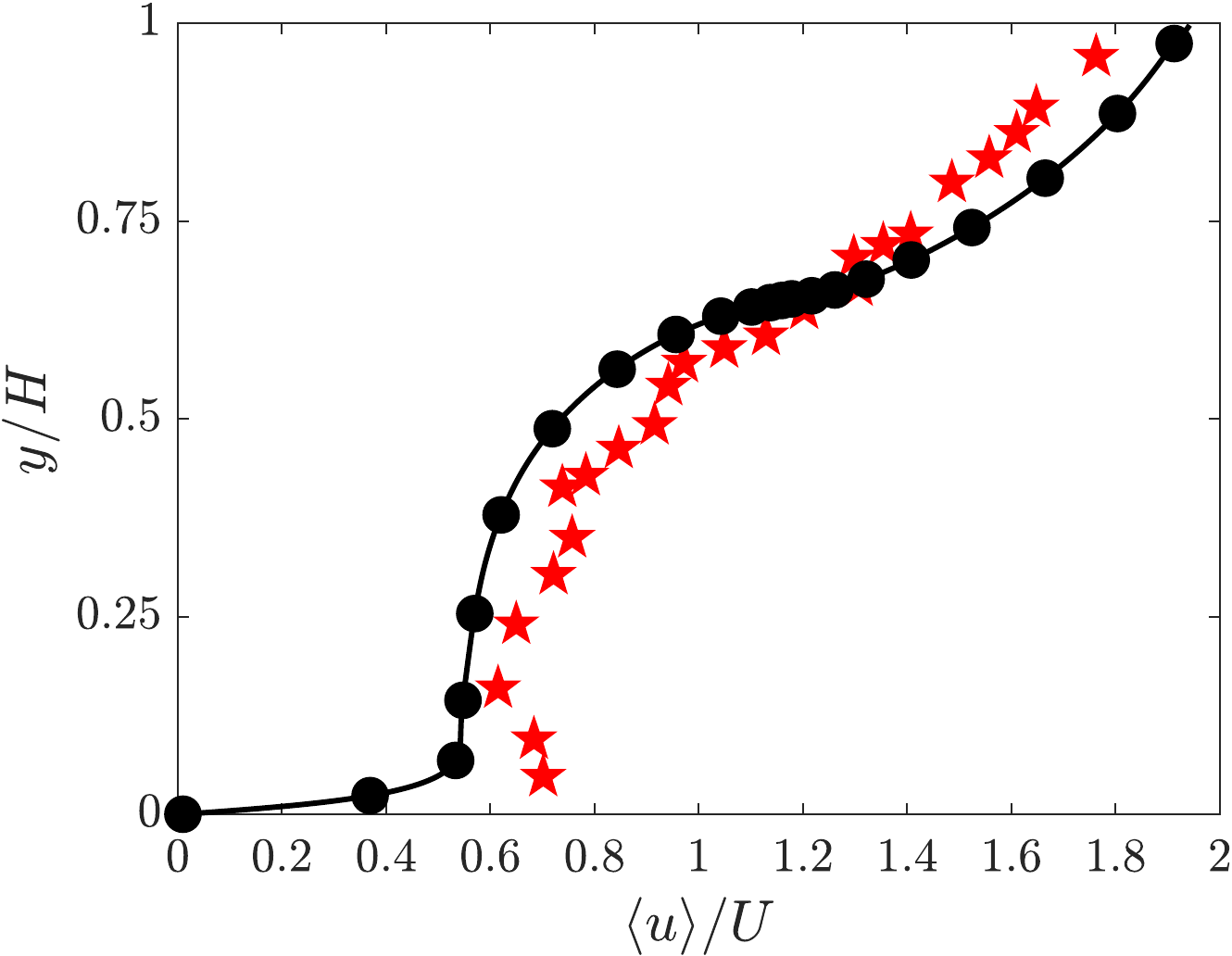}\qquad
    \includegraphics[width=.42\textwidth]{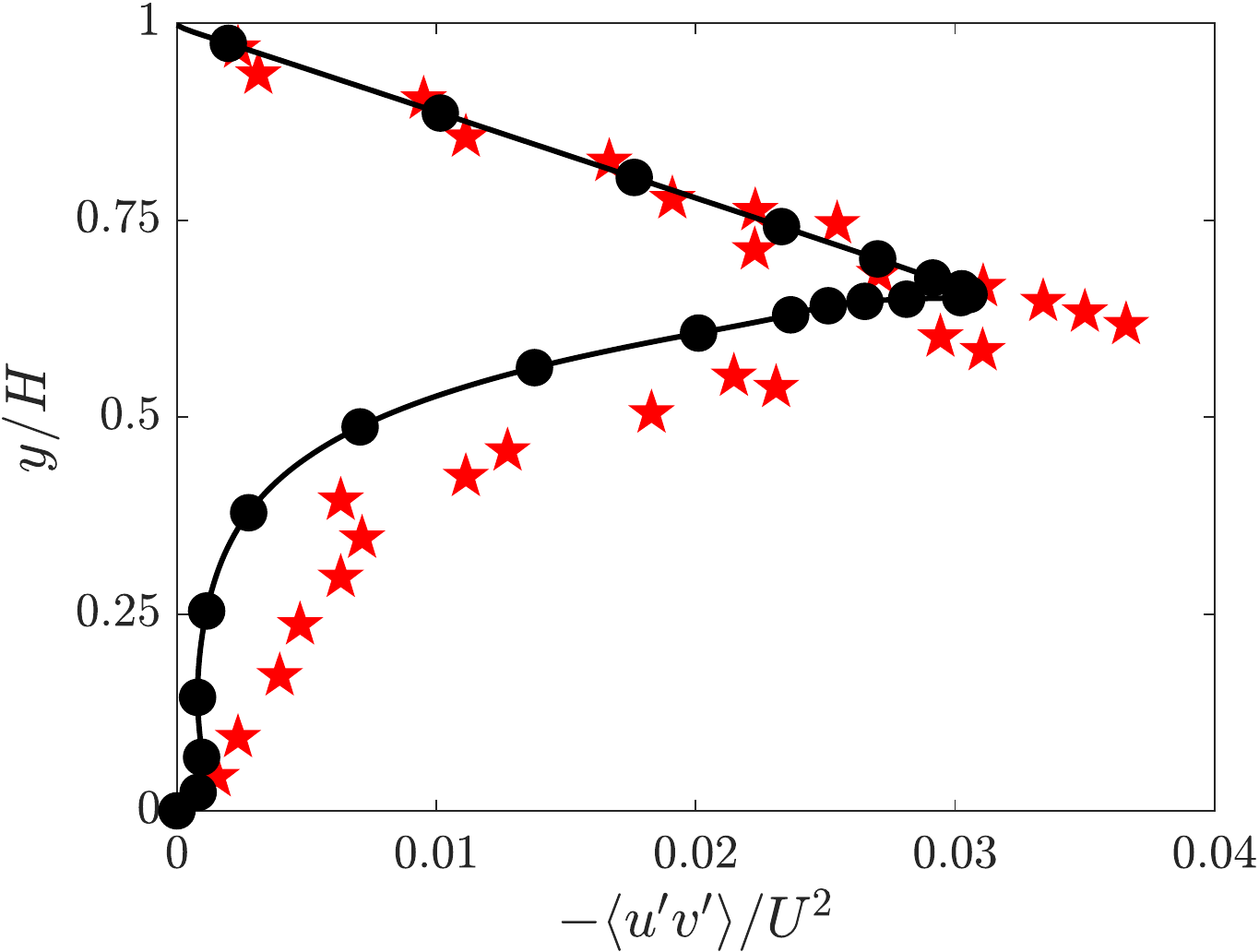}\\
    \caption{
    Validation of the employed simulation procedure for the rigid canopy case (R31) experimentally investigated by Shimizu \emph{et al.}~\cite{shimizu1991experimental}. Comparison between our numerical results (black circles) and the experimental measurements (red stars). Left: mean streamwise velocity profile; right: Reynolds shear stress distribution, as a function of the wall-normal location.
    }
    \label{fig:val}
\end{figure}

\clearpage

\begin{figure}
  \centering
  \setbox1=\hbox{%
  \includegraphics[width=0.42\linewidth,height=4cm]{example-image-b}}
  \subfloat[]{\includegraphics[width=0.4\linewidth]{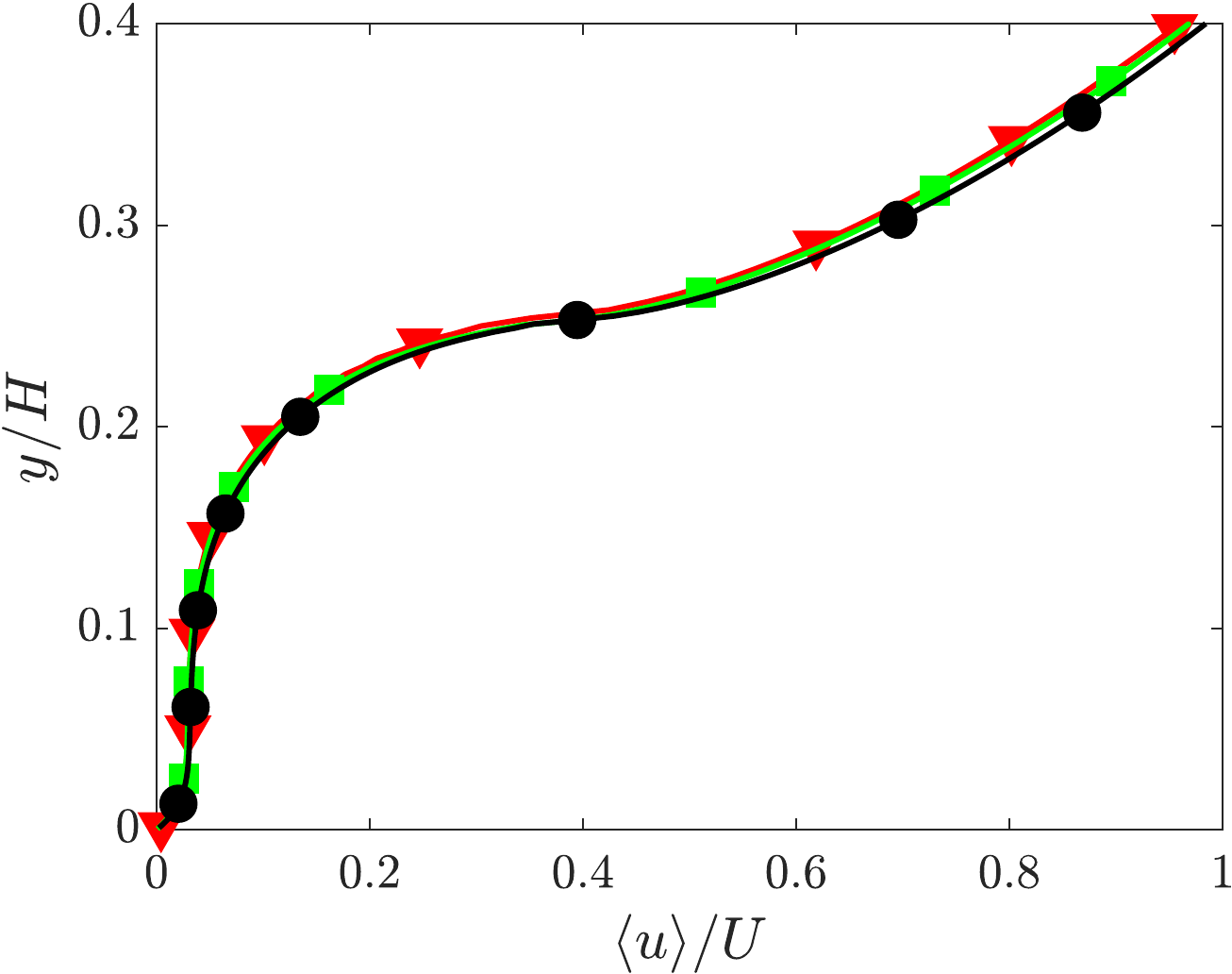}
              \llap{\makebox[\wd1][c]{\raisebox{1.0cm}{%
              \includegraphics[width=0.16\linewidth,height=2.0cm]
              {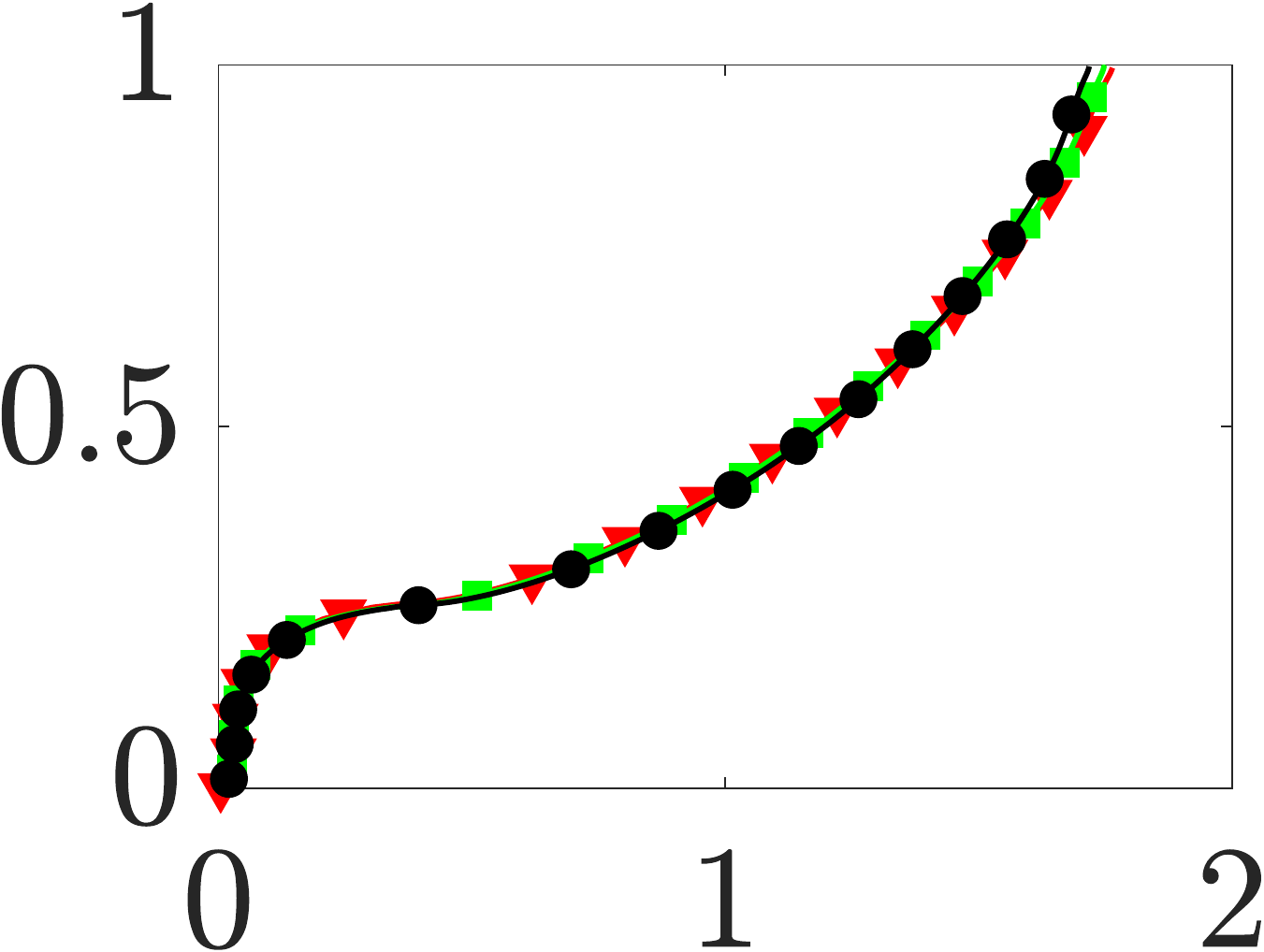}}}}
             }
  \subfloat[]{\includegraphics[width=0.42\linewidth]{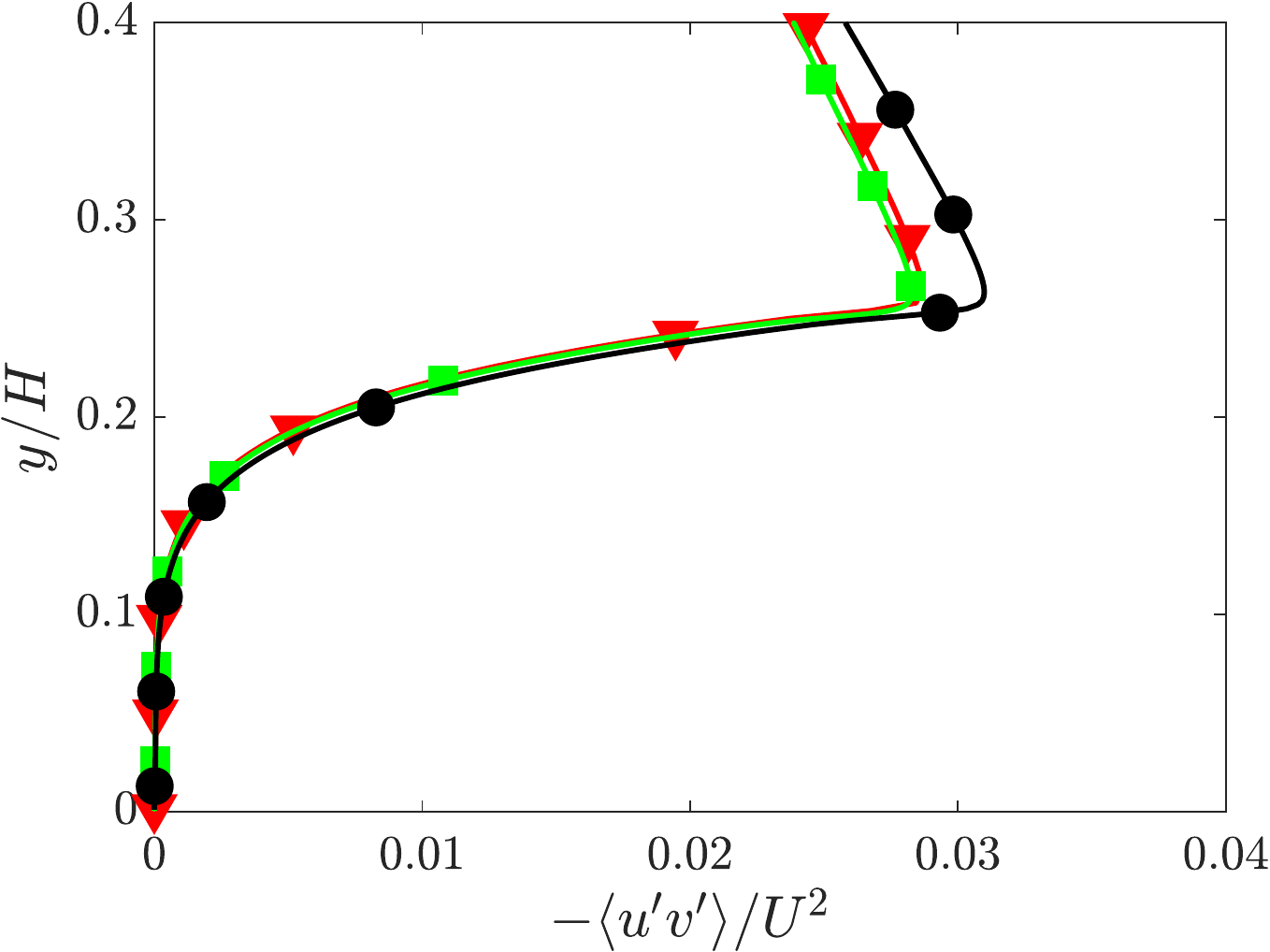}
              \llap{\makebox[\wd1][c]{\raisebox{1.0cm}{%
              \includegraphics[width=0.165\linewidth,height=2.0cm]
              {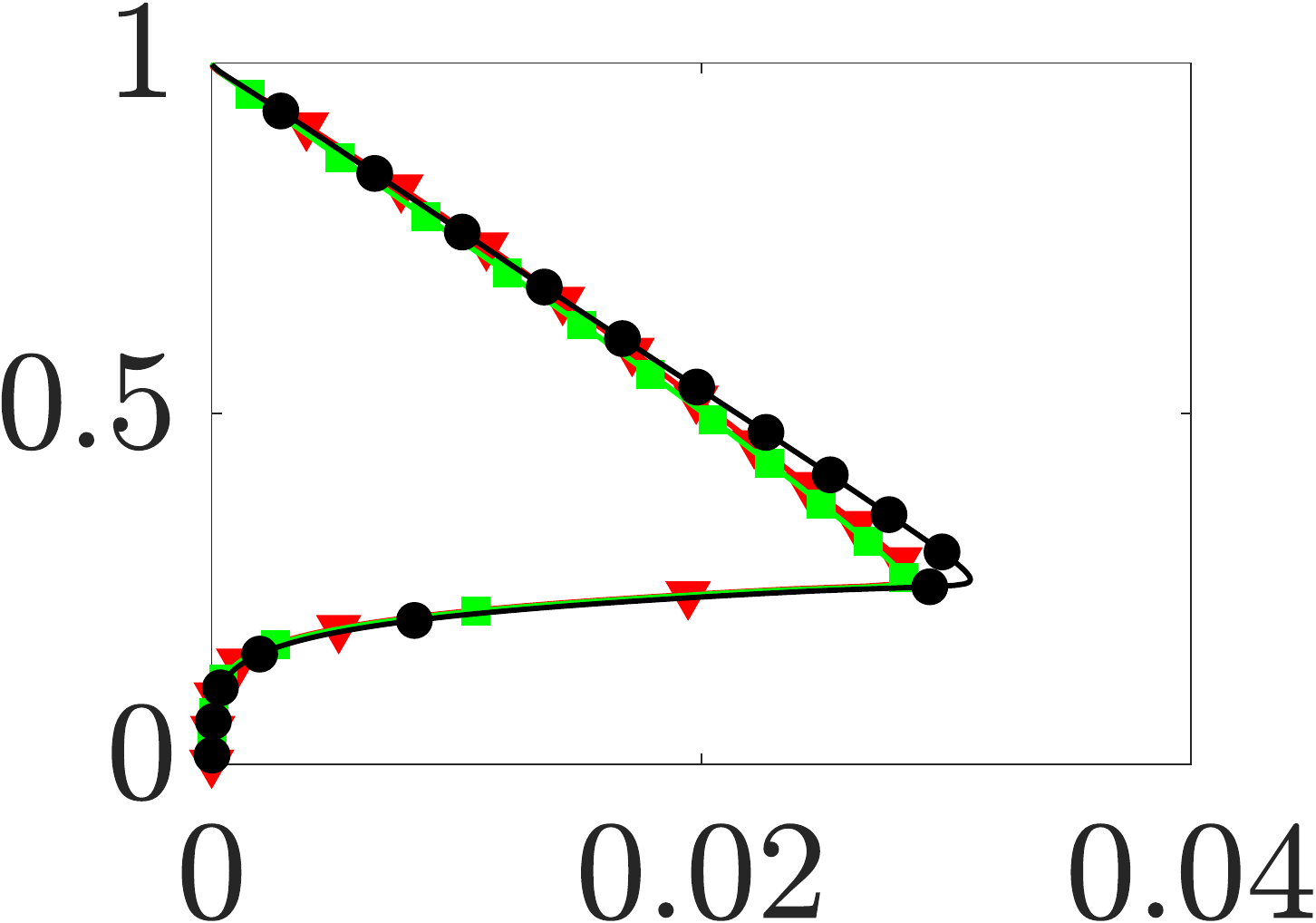}}}}
             }\\
    \caption{
    Convergence study over the wall-normal grid spacing for the rigid canopy case, i.e., $\Ca=0$. \added{Comparison between the solutions obtained with halved (red triangles), baseline (black circles) and doubled (green squares) resolution.} Left: mean streamwise velocity profile; right: Reynolds shear stress distribution, as a function of the wall-normal location. The main panels show a close up of the inner region (up to $y/H=0.4$) whereas the inset panels show the full extension (up to $y/H=1$).  Note that the more evident variation for the Reynolds shear stress is arguably associated to a lack of statistical convergence rather than for the effect of the numerical resolution.
    }
    \label{fig:GS}
\end{figure}

\end{document}